\documentclass[12pt]{article}
\usepackage{epsf,rmpbib}

\newcommand{\intl}{\int\limits}
\newcommand{\eps}{\epsilon}

\renewcommand{\baselinestretch}{1.104}

\newcommand{\beq}{\begin{equation}}
\newcommand{\eeq}{\end{equation}}
\newcommand{\beqa}{\begin{eqnarray}}
\newcommand{\eeqa}{\end{eqnarray}}

\begin{document}

\begin{titlepage}
\thispagestyle{empty}
\renewcommand{\thefootnote}{\fnsymbol{footnote}}
\setcounter{footnote}{1}

\begin{flushright}
CERN-TH/98-233\\
\mbox{July 1998\hspace*{0.6cm}}\\
hep-ph/9807443
\end{flushright}
\vspace*{2cm}

\begin{center}
\Large\bf Renormalons
\end{center}

\vspace{1cm}

\begin{center}
{\sc M. Beneke}\\[0.3cm]
{\em Theory Division, CERN, CH-1211 Geneva 23, Switzerland}\\[0.3cm]
\end{center}

\vspace*{1.5cm}

\centerline{\bf Abstract}
\vspace*{0.2cm}
\noindent 
A certain pattern of divergence of perturbative expansions in 
quantum field 
theories, related to their small and large momentum behaviour, is known 
as renormalons. We review formal and phenomenological aspects of renormalon 
divergence. We first summarize what is known about ultraviolet and 
infrared renormalons from an analysis of Feynman diagrams. Because 
infrared renormalons probe large distances, they are closely connected with  
non-perturbative power corrections in asymptotically free theories such 
as QCD. We discuss this aspect of the renormalon phenomenon in various 
contexts, and in particular the successes and failures of 
renormalon-inspired models of power corrections to hard processes in QCD. 

\end{titlepage}
\renewcommand{\thefootnote}{\arabic{footnote}}
\setcounter{footnote}{0}

\newpage                      
\thispagestyle{empty}
\mbox{}
\newpage
\pagenumbering{roman}
\tableofcontents                    
\newpage
\pagenumbering{arabic}


\section{Introduction}

Quantum field theories seem to be well understood when the interactions 
between elementary degrees of freedom are weak. The rules of field 
theory and renormalization allow us to express observables $R$ as 
series 
\begin{equation}
\label{series}
R = \sum_n r_n \alpha^n
\end{equation}
in the (renormalized) interaction strength $\alpha$. Almost invariably, 
however, these series are divergent for any $\alpha$, 
\begin{equation}
\label{asympgeneral}
r_n \stackrel{n\to\infty}{\sim} K a^n n!\,n^b,
\end{equation}
and it is not at all obvious how the equality sign in (\ref{series}) 
should be interpreted. In this report we will be concerned with a 
particular source of divergence that has become known as renormalon 
divergence. Originally discovered in the 1970s \cite{GN74,L77,tH77}, 
it has continued 
to receive attention in a much more phenomenological context 
since about 1992. Indeed, the divergent behaviour of perturbative 
expansions is more than a mathematical curiosity. It often 
indicates profound physics such as a non-trivial, non-perturbative 
structure of the vacuum and its excitations. 

Many of the early studies 
of large-order behaviour in perturbation theory, starting from the 
work of \cite{D52} and others \cite{H52,T53,P53}, have hence 
focused on the 
question of whether a quantum field theory can be constructed 
non-perturbatively from the 
perturbative expansions and analyticity properties of their Green 
functions. This turns out not to be the case for quantum 
field theories of phenomenological relevance. 
The renaissance period of large-order behaviour, and renormalons 
in particular, dating from \cite{BY92,Z92,M92}, addresses different 
questions. From the 1970s to 1992 quantum chromodynamics (QCD) had been 
growing from a qualitative to a quantitative theory of strong 
interaction phenomena. The first third order perturbative calculations 
had just become available for $e^+ e^-$ annihilation \cite{GKL91,SS91} 
and deep inelastic scattering \cite{LTV91,LV91}, and experiments were 
reaching a precision that had to be matched by theoretical accuracy.
It was therefore natural to ask how much could be learned about the 
parameters that enter the asymptotic formula (\ref{asympgeneral}) 
and whether asymptotic estimates could have anything to do with 
exact multi-loop results, that is, whether they could be 
extrapolated to $n\sim 2$-$3$. If so, one could estimate yet higher 
orders and improve the theoretical precision. Another aspect has 
drawn more attention later. 
As will be discussed 
at length, renormalon divergence is a direct consequence of the short- 
and long-distance behaviour of field theories. The long-distance 
behaviour is especially interesting in theories like QCD, whose 
coupling $\alpha_s$ grows with distance and eventually eludes 
a perturbative treatment. Sensitivity to non-perturbative 
long-distance/large-time 
behaviour is inevitable to some degree in any measurement, that  
refers to asymptotic states, even if the fundamental scattering 
process occurs at small distances such as in high-energy 
electron-positron annihilation or deep inelastic lepton-nucleon 
scattering at large momentum transfer. Perturbative factorization allows 
us to separate the short-distance part, characterized by a large 
momentum scale $Q$, from the long-distance part, characterized by a small 
momentum scale $\Lambda\sim 1\,$GeV, up to power corrections. 
Schematically,\footnote{The long-distance part at leading power, 
$\langle {\cal O}\rangle(\mu,\Lambda)$, may vanish altogether in 
$e^+ e^-$ annihilation observables, for example event shape variables.} 
\begin{equation}
\label{factschematic}
R(Q,\Lambda) = C(Q,\mu) \otimes \langle {\cal O}\rangle(\mu,\Lambda) + 
\,\,\mbox{power corrections}\,\,\left(\frac{\Lambda}{Q}\right)^p,
\end{equation}
with $\mu$ a factorization scale. 
But perturbative factorization tells us little about the form of power 
corrections. Power corrections can be large at intermediate energies, 
sometimes up to $M_Z\sim 90\,$GeV, or they are important to ascertain the 
parametric accuracy that could at best be achieved perturbatively. 
Most of the interest in renormalons derives from the fact 
that the (infrared) renormalon behaviour of $C(Q,\mu)$ 
is related to power corrections. Strictly speaking, only the 
scaling behaviour (in $Q$) of the power 
correction can be inferred through renormalon divergence. 
However, it is also interesting to take one step further and to 
construct models that quantify the absolute magnitude of power 
corrections. Models of this kind, inspired by renormalons, profit from 
being consistent with the short-distance behaviour of QCD, but suffer 
from being somewhat unspecific as far as non-perturbative properties 
of hadrons are concerned.

Not all expectations at some time connected with the subject 
have been fulfilled. It may be fair to say that the 
conceptual progress remained little compared to the pioneering work 
of \cite{tH77,P78,P79,D84,Mue85}. On the other hand, while the early 
discussions of renormalons refer almost exclusively to the two-point 
function of electromagnetic currents and its operator product expansion, 
the generality of the phenomenon, and its usefulness for observables 
that do not admit an operator product expansion,  
has been appreciated only recently. This development has reached the 
point where it has inspired new experimental QCD studies. 

This report reflects this development in that it puts emphasis on results 
with potential phenomenological implications. It is divided roughly in two 
parts. The first part is more theoretical and collects what is known 
about renormalon divergence from a general point of view. The second part 
addresses applications to specific processes. The report is not intended 
to be comprehensive in details regarding this 
second part. Rather, the idea is that it summarizes, for 
each topic, the principal ideas and results, and that it could serve 
as a guide to the original literature.

In Section~2 we begin with basic concepts and terminology related to 
divergent series and renormalons. We embark on an introductory tour 
through the Borel plane and treat an example of ultraviolet (UV) and 
infrared (IR) renormalon divergence. We then explain the connection 
of renormalons with operator production expansions and, more generally, 
perturbative factorization. This connection is crucial.  
In fact, many of the results on power corrections 
summarized in the phenomenology part could have equally been 
obtained from extending perturbative factorization 
without ever using the concept of renormalons. This section could 
be read as a basic introduction to the subject, summarizing the status 
prior to and around 1993.

In Section~3 we deal with renormalons from an entirely diagrammatic point 
of view. Since it is the asymptotic behaviour of perturbative coefficients 
in large orders which is under discussion, one should, after all, be 
able to extract it from Feynman graphs. Treating separately UV and IR 
renormalons, we discuss how the values of $a$ and $b$ in 
(\ref{asympgeneral}) are computed and why $K$ cannot be computed. The 
starting point is an expansion in the number of flavours in QED and QCD, 
which allows us to check our expectations for `real QCD'. The 
perturbative coefficients $r_n$ depend on renormalization conventions 
to define the coupling $\alpha$ (and, possibly, other relevant 
parameters) and are arbitrary to a large extent. Section~3 concludes 
with a discussion of how scheme dependence is reflected in the 
large-order behaviour of the $r_n$ and an overview of methods 
to calculate `bubble graphs', which play a prominent role in applications 
of renormalons. 

In Section~4 we ask what the divergence of perturbative series tells us 
about non\--per\-tur\-bative 
effects and explain the relation of IR renormalons 
and power corrections. This is first studied in first orders of the 
$1/N$ expansion of the two-dimensional $O(N)$ $\sigma$-model, 
which, contrary to flavour expansions in QED and QCD, provides a 
non-perturbative set-up for the problem. We shall learn that 
the existence of 
IR renormalons is specific to performing infrared factorization 
in dimensional regularization: they are indirect manifestations of 
power-like factorization scale dependence, which is otherwise absent in 
dimensional renormalization.\footnote{At least in its conventional 
usage, that is, if one does not subtract poles in dimensions other 
than $n$ for theories in $n$ dimensions.} As a consequence IR 
renormalons are related to the 
UV renormalization properties, power divergences, to be precise, 
of operators that parametrize power corrections, if such can be 
identified. This interpretation of IR renormalons in terms of 
operator mixing between operators of different dimension also clarifies 
that without additional assumptions IR renormalons can tell us 
little about the matrix elements of these 
operators. We exemplify the matching 
between IR renormalons and UV behaviour of power corrections 
for twist-four corrections to deep inelastic scattering, using 
the flavour expansion as a toy model.

Section 5 constitutes the second part in its entirety; it  
reviews applications 
of ideas based on or related to renormalon behaviour to processes of 
phenomenological interest. We identify three main strains of 
applications: related to the size and estimation of perturbative 
coefficients, related to the scaling behaviour of power corrections, and 
related to modelling the absolute magnitude of power corrections. 
Because several of these aspects can be interesting for any given process, 
the section is divided by processes. The first set of processes consists 
of those where the large, perturbative momentum scale is given by 
a large momentum transfer. Inclusive observables in high-energy 
$e^+ e^-$ annihilation and $\tau$ decay, structure functions in 
deep-inelastic scattering, and hadronic reactions such as Drell-Yan 
production belong to this class. Power corrections of order $1/Q$ to 
event shape observables in $e^+ e^-$ annihilation are reviewed in 
some detail because of their considerable experimental interest. For 
the second set of observables the large scale is given by the mass of 
a heavy quark, of a bottom quark in practice. Beginning with the quark mass 
parameter itself, we then consider exclusive and inclusive heavy quark 
decays and, finally, systems of two heavy quarks, described by 
non-relativistic QCD.

The problem of power UV divergences mentioned above 
is even more acute in lattice 
computations of power-suppressed effects. Renormalons enter here mainly 
to remind us that power divergences have to be subtracted 
non-perturbatively. Section~6 gives a brief account of activities 
in this direction.

In Section~7 we summarize and collect open questions.


\section{Basic concepts}
\setcounter{equation}{0}
\label{sectbascon}

In this section we briefly introduce some concepts that appear in 
connection with renormalons. We begin with the notions of 
divergent/asymptotic series and the Borel transform. We then compute 
as an elementary example the leading IR and UV renormalon singularity 
of the vector current-current correlation function in the bubble chain 
approximation. This approximation is already sufficient to work out the 
main aspects of renormalons, with generalizations and refinements being 
delegated to later sections. Because the concepts of factorization and 
the operator product expansion (OPE) are crucial in this context and will 
lead as a red thread through this review, a separate subsection expands 
on the relation between the OPE and renormalons. We then return to 
the current-current correlation function and discuss its singularities 
in the Borel plane.

This section may be read 
as a first overview of basic ideas, which will recur in more general 
treatments or further examples later. The section is fairly self-contained 
on an elementary level, but points to later sections for more details. 
A more detailed and formal discussion of the divergent series 
problem in the context of renormalons can be found in \cite{F97}. The 
reprint volume \cite{LeGZ90} collects many of the early papers on 
divergent series in quantum field theories, with emphasis on 
instanton-induced divergence, and provides an introduction to the 
subject.

\subsection{Divergent series}
\label{divseries}

Divergent series are common in applied mathematics and there is 
nothing `wrong' with them. However, given the divergent series expansion 
$R\sim \sum_n r_n \alpha^n$ of $R$, the following questions arise:
\begin{enumerate}
\item How does one assign a numerical value (`sum') to the series?
\item How is the series or its sum related to the original (`exact') 
function $R(\alpha)$? Is the sum of the series identical to $R$?
\end{enumerate}
There is little to say about the second question for series expansions 
that occur in renormalizable field theories realized in nature, because 
we do not know how to define $R$ non-perturbatively.\footnote{Lattice 
regularization provides the exception. In this case, one has to deal with 
the continuum and infinite volume limit. We adhere to continuum 
definitions at this point.}  

In order that a divergent series be useful as an approximation to $R$, 
it should be {\em asymptotic} to $R$ in a region ${\cal C}$ 
of the complex $\alpha$-plane. Then there exist numbers $K_N$ 
such that
\begin{equation}
\label{asympdef}
\left| R(\alpha)-\sum_{n=0}^N r_n \alpha^n \right| < K_{N+1} \alpha^{N+1}
\end{equation}
for all $\alpha$ in ${\cal C}$ and the truncation error at order $N$ 
is uniformly bounded to be of order $\alpha^{N+1}$. If 
\begin{equation}
\label{rnagain}
r_n \stackrel{n\to\infty}{\sim} K a^n n!\,n^b
\end{equation}
with constants $K$, $a$, $b$, one often 
finds that also $K_N\propto a^N N!\,N^b$. The truncation error 
follows the same pattern as the terms of the series themselves. It first 
decreases until
\begin{equation}
N_\star\sim \frac{1}{|a| \alpha},
\end{equation}
beyond which the approximation of $R$ does not improve through 
the inclusion 
of further terms in the series. If $N_\star \gg 1$, the approximation 
is good up to terms of order
\begin{equation}
\label{bestapprox}
K_{N_\star} \alpha^{N_\star} \sim e^{-1/(|a|\alpha)}.
\end{equation}
Provided $r_n\sim K_n$, the best approximation is achieved when  
the series is truncated at its minimal term and the truncation error is 
roughly given by the minimal term of the series.

Since there is no rigorous non-perturbative 
definition of $R$ in theories such as QED and QCD, 
we cannot even ask whether series expansions are 
asymptotic. It is usually assumed that they are. The justification is 
that if QED (QCD) {\em is} the theory of 
electromagnetic (strong) interactions, 
non-perturbative results are provided by (ideal) measurements. The fact 
that independent determinations of the coupling constant $\alpha$ are 
consistent with each other indicates that the series which enter these 
determinations are not entirely arbitrary. It is also usually assumed 
that $r_n\sim K_n$.

Note that if nothing is known of $R$ but its series expansion, there 
is actually no difference between a divergent and convergent series 
regarding the second question above. The sum of a convergent series 
may still differ from $R$ by exponentially small terms 
$\exp(-1/\alpha)$. In turn, while a divergent series implies that 
$R$ is non-analytic at $\alpha=0$, non-analyticity does not 
imply divergence. The answer to the second question is trivial 
only if $R$ is analytic in $\alpha=0$.

To improve over the best approximation (\ref{bestapprox}), the 
divergent series 
has to be summed. There may be many ways of doing this. For factorially 
divergent series, {\em Borel summation} is most useful. We first 
define the {\em Borel transform}\footnote{It is convenient to denote 
by $r_n$ the coefficient of $\alpha^{n+1}$ rather than $\alpha^n$. 
Without loss of generality we can assume that $R$ has no constant 
term or we can treat the constant term  separately.} as 
\begin{equation}
\label{defborelt}
R\sim\sum_{n=0}^\infty r_n\alpha^{n+1} \,\,\Longrightarrow \,\,
B[R](t) = \sum_{n=0}^\infty r_n\,\frac{t^n}{n!}.
\end{equation}
If $B[R](t)$ has no singularities for real 
positive $t$ and does not increase too 
rapidly at positive infinity, we can define the 
{\em Borel integral} ($\alpha$ positive) as 
\begin{equation}
\label{borelint}
\tilde{R} = \int\limits_0^\infty \,dt\,e^{-t/\alpha}\,B[R](t),
\end{equation}
which has the same series expansion as $R$. The integral 
$\tilde{R}$, if it exists, 
gives the Borel sum of the original divergent series.

To determine whether the Borel sum equals $R$ non-perturbatively 
requires that we know more about $R$ than its formal series expansion. 
The Watson-Nevanlinna-Sokal theorem \cite{S80} guarantees this 
equality, provided $R$ meets certain analyticity requirements in 
addition to satisfying asymptotic estimates of the form 
(\ref{asympdef}). These requirements are too strong for 
renormalizable theories \cite{tH77}.  

Returning to the Borel transform, assume that 
\begin{equation}
\label{div}
r_n=K a^n\Gamma(n+1+b)
\end{equation}
exactly. Unless $b$ is a negative integer, the Borel 
transform of the series is given by
\begin{equation}
\label{bpole}
B[R](t) = \frac{K\Gamma(1+b)}{(1-a t)^{1+b}}.
\end{equation}
For $b=-m$ a negative integer (in which case the first few $r_n$ are 
discarded), it follows from (\ref{defborelt}) that
\begin{equation}
\label{bpolelog}
B[R](t) = \frac{(-1)^m}{\Gamma(m)}\,(1-a t)^{m-1}\,\ln(1-a t) + 
\,\mbox{polynomial in }t.
\end{equation}
Hence non-sign-alternating series ($a>0$), which as we shall see 
are expected in QED and QCD,  yield singularities at positive 
$t$. It follows that already the Borel integral does not exist. 

Nevertheless, the Borel transform and Borel integral are useful 
concepts. The Borel transform can be considered as a generating 
function for the series coefficients $r_n$. As seen from 
(\ref{div}, \ref{bpole}) the divergent behaviour of the original series is 
encoded in the singularities of its Borel transform. Hence, divergent 
behaviour is often referred to through poles/singularities in 
the {\em Borel plane}. This language is particularly convenient for 
subleading divergent behaviour. Note that larger $a$, i.e. faster 
divergence, leads to singularities closer to the origin $t=0$ of the 
Borel plane.

When there are singularities at positive $t$, the Borel integral may  
still be defined by moving the contour above or below the singularities. 
For the series (\ref{div}) with $a>0$, the so-defined Borel integral 
acquires an imaginary part
\begin{equation}
\label{im}
\mbox{Im}\,\tilde{R}(\alpha) = \mp\frac{\pi K}{a}\,e^{-1/(a\alpha)}\,
(a\alpha)^{-b},
\end{equation}
where the sign depends on whether the integration is taken in the upper 
or lower complex plane. The difference between the two definitions is 
often called `ambiguity of the Borel integral'. It is exponentially 
small in the expansion parameter $\alpha$ and in this 
sense non-perturbative. It is also parametrically of 
the same order as the minimal term (\ref{bestapprox}) of the series.  
(We did not keep track of pre-exponential factors in (\ref{bestapprox}).) 

It is customary to take these ambiguities in the Borel integral as 
an indication that exponentially small terms of the same form as (\ref{im}) 
must be added explicitly to the series expansion, after which 
ambiguities in defining the sum of the perturbative series cancel and 
an improved approximation to the exact function is 
obtained.\footnote{Because the coupling 
$\alpha_s(Q)$ depends logarithmically 
on $Q$, exponentially small terms (in $\alpha_s(Q)$) are referred to as 
{\em power corrections} (in $Q$) in QCD applications.} As a 
simplistic example of how this is supposed to work, let us assume that 
the `exact' result is given by  
\begin{equation}
\label{exta1}
R(\alpha)\equiv \sum_{n=0}^\infty (-1)^n\,\frac{\Psi(n)}{n!\alpha^n}, 
\end{equation}
which defines an analytic function in the entire complex plane 
except for $\alpha=0$. 
($\Psi$ is the logarithmic derivative of the $\Gamma$-function.) 
Its complete asymptotic expansion, for $\alpha>0$, is given by 
a divergent series and an exponentially small term:
\begin{equation}
\label{exta2}
R(\alpha) = -\sum_{n=0}^\infty n! \alpha^{n+1} + \,e^{-1/\alpha}\, 
\left(-\ln\alpha\mp i\pi\right).
\end{equation}
If the divergent sum is understood as the Borel integral in the 
upper complex plane (upper sign) or lower plane (lower sign), (\ref{exta2}) 
is exactly equal to (\ref{exta1}) and the ambiguity in the Borel 
integral of the divergent series is indeed cancelled by the 
twofold ambiguity in the exponential term.
Without more knowledge of the exact function than what 
is usually available 
in field theories, this is a heuristic line of thought. It also assigns 
a privileged role to Borel summation, as sign-alternating series ($a<0$) 
are then believed not to require adding exponentially small terms, 
while from the point of view of (\ref{bestapprox}) there is no 
difference between sign-alternating and fixed-sign series. As will 
be seen later, the chain 
\begin{equation}
\begin{array}{l}\mbox{fixed-sign factorial}\\[-0.1cm]
\mbox{divergence}
\end{array}\,\,\,\,
\Longrightarrow\,\,\,\,
\begin{array}{l}\mbox{ambiguity of the}\\[-0.1cm]\mbox{Borel integral}
\end{array}\,\,\,\,
\Longrightarrow\,\,\,\,
\begin{array}{l}\mbox{addition of exponentially}\\[-0.1cm]\mbox{
\hspace*{0.7cm} small terms}
\end{array}
\end{equation}
is supported by physics arguments and calculations in toy models. However, 
it is important to bear in mind that it is not rigorous.

\subsection{Renormalons}
\label{renormalonexample}

This section provides a first, non-technical introduction to 
renormalon divergence. We begin with a short and classic calculation 
and interpret it afterwards. 

\begin{figure}[t]
   \vspace{-3.3cm}
   \epsfysize=30cm
   \epsfxsize=20cm
   \centerline{\epsffile{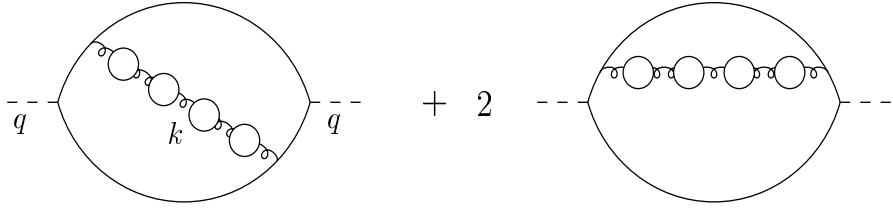}}
   \vspace*{-22.8cm}
\caption[dummy]{\small The simplest set of `bubble' diagrams for the 
Adler function consists of all diagrams with any number of fermion 
loops inserted into a single gluon line. \label{fig1}}
\end{figure}

Consider the correlation functions of two vector currents $j_\mu=
\bar{q}\gamma_\mu q$ of massless quarks
\begin{equation}
\label{currentcorr}
(-i)\int\,d^4x\,e^{-i q x}\,\langle 0|T\,(j_\mu(x) j_\nu(0))|0\rangle 
= \left(q_\mu q_\nu-q^2 g_{\mu\nu}\right)\,\Pi(Q^2)
\end{equation}
with $Q^2=-q^2$. We now compute the contribution of the fermion 
bubble diagrams shown in Fig.~\ref{fig1} to the Adler function 
\begin{equation}
\label{adlerdef}
D(Q^2)=4 \pi^2\,\frac{d\Pi(Q^2)}{dQ^2}.
\end{equation}
The set of selected diagrams is gauge-invariant, but it is not 
the only set of diagrams that contributes to renormalon divergence. 
It is selected here for illustration and a systematic investigation 
is postponed to Section~\ref{sectrenfeyn}.  
Renormalons were originally found in 
bubble diagrams \cite{GN74,L77,tH77}, and these diagrams still 
feature so prominent in discussions of renormalons that sometimes 
they are even identified with them. 

The Adler function requires no additional subtractions beyond those 
contained in the renormalized QCD Lagrangian. Therefore no regularization 
is needed, provided the fermion loop insertions are renormalized. The 
renormalized fermion loop is given by
\begin{equation}
\label{simpleloop}
-\beta_{0f}\alpha_s\left[\ln(-k^2/\mu^2) + C\right]
\end{equation}
with a scheme-dependent constant $C$ and $\beta_{0f}=N_f T/(3\pi)$ 
the fermion contribution to the one-loop 
$\beta$-function.\footnote{Unless otherwise stated, $\alpha_s$ denotes the 
strong coupling renormalized in the modified minimal 
subtraction ($\overline{\rm MS}$) scheme 
\cite{BBDT78} at the subtraction point $\mu$. 
We use the following convention for 
the $\beta$-function:
\begin{equation}
\label{betadef}
\beta(\alpha_s) = \mu^2\frac{\partial\alpha_s}{\partial \mu^2} = 
\beta_0\alpha_s^2+\beta_1\alpha_s^3+\ldots.
\end{equation}
The $\beta$-function is scheme-dependent, but the first two coefficients 
are scheme-independent in the class of massless subtraction schemes. 
We will often need 
\begin{equation}
\label{beta0}
\beta_0 = \beta_{0NA}+\beta_{0f} = -\frac{1}{4\pi}\left(\frac{11 C_A}{3}-
\frac{4N_f T}{3}\right),
\end{equation}
where $C_A=N_c=3$, $T=1/2$ and $N_f$ the number of massless quark 
flavours. For future use we recall that $C_F=(N_c^2-1)/(2 N_c)=4/3$.}
In the $\overline{\rm MS}$ scheme $C=-5/3$. 

\begin{figure}[t]
   \vspace{-1.2cm}
   \epsfysize=9cm
   \epsfxsize=7cm
   \centerline{\epsffile{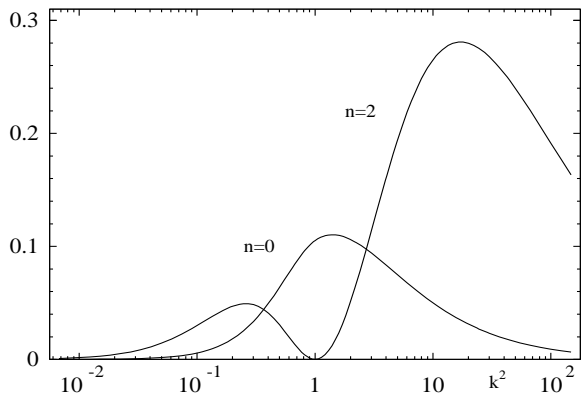}}
   \vspace*{-1.7cm}
\caption[dummy]{\small The integrand of (\ref{basint}) for 
$n=0$ and $n=2$ as function of $\hat{k}^2$. 
The vertical scale is arbitrary.\label{fig2}}
\end{figure}

Proceeding with the 
diagrams of Fig.~\ref{fig1}, we integrate over the loop momentum 
of the `large' fermion loop and the angles of the gluon momentum 
$k$. Defining $\hat{k}^2=-k^2/Q^2$, we obtain
\begin{equation}
\label{basint}
D = \sum_{n=0}^\infty \,\alpha_s\int\limits_0^\infty \,\frac{d\hat{k}^2}
{\hat{k}^2}\,F(\hat{k}^2)\,\left[\beta_{0f}\alpha_s\ln\left(
\hat{k}^2\frac{Q^2e^{-5/3}}{\mu^2}\right)\right]^n.
\end{equation}
The exact expression for $F$ can be found in \cite{N95}, but we do not 
need it for our present 
purpose.\footnote{The function $F(\hat{k}^2)/(4\pi \hat{k}^2)$ 
is called $\hat{w}_D$ in \cite{N95}.} Rather 
than calculating the final integral exactly, we evaluate it 
approximately for $n\gg 1$. Provided the renormalization scale 
$\mu$ is kept fixed with order of perturbation theory and is taken 
of order $Q$, the dominant contributions to the integral come from 
$k\gg Q$ and $k\ll Q$, because of the large logarithmic 
enhancements in these regions. Hence, it is sufficient to know the 
small-$\hat{k}$ and large-$\hat{k}$ behaviour of $F$:
\begin{eqnarray}
\label{smallk}
F(\hat{k}^2) &=& \frac{3 C_F}{2\pi}\,\hat{k}^4 + {\cal O}(\hat{k}^6 
\ln \hat{k}^2),\\
\label{largek}
F(\hat{k}^2) &=& \frac{C_F}{3\pi}\,\frac{1}{\hat{k}^2}
\left(\ln\hat{k}^2+\frac{5}{6}\right) +
{\cal O}\!\left(\frac{\ln\hat{k}^2}{\hat{k}^4}\right).
\end{eqnarray}
Note that UV and IR finiteness of the Adler function 
implies that $F$ must have a power-like approach to zero for both  
large and small $\hat{k}^2$. The integrand of (\ref{basint}) is shown 
in Fig.~\ref{fig2} for $n=0$ and $n=2$. It is clearly seen how the 
integrand is dominated by loop momentum of order $Q$ for $n=0$, but 
peaks at large and small $\hat{k}^2$ for $n$ as small as~2. Splitting 
the integral (\ref{basint}) at $\hat{k}^2=\mu^2/(Q^2 e^{-5/3})$ and 
inserting (\ref{smallk}) for the small-$\hat{k}^2$ interval and 
(\ref{largek}) for the large-$\hat{k}^2$ interval, one obtains
\begin{equation}
\label{dex}
D = \frac{C_F}{\pi} \sum_{n=0}^\infty \alpha_s^{n+1} 
\left[\frac{3}{4}\left(\frac{Q^2}{\mu^2} e^{-5/3}\right)^{-2}
\left(-\frac{\beta_{0f}}{2}\right)^n n! + 
\frac{1}{3}\,\frac{Q^2}{\mu^2} e^{-5/3}\,
\beta_{0f}^n\,n!\,\left(n+\frac{11}{6}\right)\right],
\end{equation}
where the first term comes from small $\hat{k}$ and the second from 
large $\hat{k}$. Accordingly, the factorial divergence exhibited by the 
two series components is called {\em infrared (IR) renormalon} and 
{\em ultraviolet (UV) renormalon}.\footnote{Some etymology: the word 
`renormalon' first appeared in \cite{tH77}. Apparently it was chosen, 
because the only other known source of divergent behaviour, related 
to instantons, had been called `instanton divergence'. The divergent 
behaviour discussed here was then novel and is characteristic of  
renormalizable field theories.} Eq.~(\ref{dex}) is accurate up to 
relative corrections of order $n\,(2/3)^n$ from the infrared and 
$(1/2)^n$ from the ultraviolet region. The corresponding singularities in 
the Borel plane lie at $t=-2/\beta_{0f}$ (IR renormalon) and 
$t=1/\beta_{0f}$ (UV renormalon). Using (\ref{div}, {\ref{bpole}), 
the Borel transform obtained from (\ref{dex}) reads
\begin{eqnarray}
\label{firstpolesD}
B[D](u) &=& \frac{3 C_F}{2 \pi}
\left(\frac{Q^2}{\mu^2} e^{-5/3}\right)^{-2}\frac{1}{2-u}
\qquad\mbox{(first IR renormalon)} 
\nonumber\\
&&\hspace*{-1.2cm}\,+ \,\frac{C_F}{3\pi}
\,\frac{Q^2}{\mu^2} e^{-5/3}\,\left[
\frac{1}{(1+u)^2}+\frac{5}{6}\frac{1}{1+u}\right]
\qquad\mbox{(first UV renormalon)},
\end{eqnarray}
where we defined $u=-\beta_{0f} t$. 
The large-order behaviour of the Adler function is dominated 
by the UV renormalon. The UV renormalon singularity is a 
double pole \cite{Ben93a}, which is equivalent to the additional factor of 
$n$ in (\ref{dex}) and can be traced back to the logarithm of $\hat{k}^2$ 
in (\ref{largek}). Eq.~(\ref{firstpolesD}) 
provides us with the singularities 
closest to the origin of the Borel plane. The exact Borel transform 
of the set of diagrams of Fig.~\ref{fig1} is known 
\cite{Ben93a,Bro93} and we return to it in Section~\ref{inclepem}. 
One finds an infinite sequence of IR (UV) renormalon poles at positive 
(negative) integer $u$ with the exception of $u=1$. 

In the following we define the term `renormalon' as a singularity 
of the Borel transform related to large or small loop momentum 
behaviour.\footnote{Note that the recent literature is not always 
precise on this point. For example, `renormalon' can be found used 
as a synonym for `power correction', especially 
in the context of QCD applications.} The set of bubble graphs provides 
an approximation to renormalon singularities. 

We have seen how renormalon divergence arises technically. Let us now 
collect some observations on the calculation, which are essential to its 
understanding:

1. The Adler function is UV and IR finite and hence depends only on 
one scale, $Q$. Hence we expect that the loop integrals should be 
dominated by $k\sim Q$. Renormalon divergence is related to the 
fact that this is not the case when the number of loops becomes large. 
The leading contributions to (\ref{basint}) arise from 
\begin{eqnarray}
\label{saddles}
k_{\rm IR}^2 &\sim& \mu^2\,e^{5/3}\,e^{-n/2},
\\
k_{\rm UV}^2 &\sim& \mu^2\,e^{5/3}\,e^n.
\end{eqnarray}
Hence, each logarithm of $\hat{k}^2$ 
counts as a factor of $n$. The presence 
of two very different scales and `large logarithms' suggests a 
renormalization group treatment. In contrast with more familiar 
applications of renormalization group methods, the hierarchy of 
scales is not fixed by external parameters, but generated by the 
loop diagrams themselves. All results on renormalon divergence that  
are independent of special classes of diagrams follow, in one way or 
another, from renormalization group methods or simply from the fact that 
there exist two different scales.

2. To compute the leading divergent behaviour, only the expansion 
at small or large $\hat{k}^2$ of the integrand of the skeleton diagrams  
(Fig.~\ref{fig1} without the fermion loop insertions) was needed. 
One can turn this statement around and say that the fermion loop 
insertions (and hence renormalon divergence) probe the large and 
small momentum tails of $F(\hat{k}^2)$, which would otherwise give 
a small contribution to the integral of $F$, see the case $n=0$ in 
Fig.~\ref{fig2}. The possibility to use IR renormalons to keep track of 
IR sensitivity of Feynman integrals will be essential in the 
analysis of power corrections in QCD. In this respect the absence of 
a $\hat{k}^2$-term in (\ref{smallk}) (and, hence, the absence 
of a singularity 
at $u=1$ in (\ref{firstpolesD})) has significance and corresponds to 
the absence of a dimension-2 operator in the operator product expansion 
of the current-current correlation function as we discuss later in 
this section.

For UV renormalons we observe 
a similarity to ordinary UV renormalization, for 
quad\-ra\-tic (logarithmic) UV divergences would be in correspondence with 
a $\hat{k}^2$ ($\ln\hat{k}^2$) term in the large momentum expansion 
(\ref{largek}) of $F$. Hence the suggestion of \cite{P78} that the 
leading UV renormalon at $u=-1$ can be 
compensated by dimension-6 counterterms. 
This will be explained in detail in Section~\ref{uvren}.

3. Renormalons are often associated with the notion of the 
`running coupling'. Interchanging the sum over $n$ and the integration 
in (\ref{basint}), one obtains
\begin{equation}
\label{drun}
D = \int\limits_0^\infty\,\frac{d\hat{k}^2}
{\hat{k}^2}\,F(\hat{k}^2)\,\alpha_s\!\left(k e^{-5/6}\right),
\end{equation}
where
\begin{equation}
\label{onelooprun}
\alpha_s(k) = \frac{\alpha_s(\mu)}{1-\alpha_s(\mu)\beta_{0}\ln(k^2/\mu^2)} 
\equiv \frac{1}{-\beta_{0}\ln k^2/\Lambda^2}
\end{equation}
is the familiar one-loop running coupling which follows from 
(\ref{betadef}). Hence the set of diagrams with a single chain of 
fermion loops can be obtained by integrating the skeleton diagram 
with the one-loop running coupling at the vertices.

In writing the previous two equations, we have in fact taken the 
first step beyond the set of bubble graphs. It is evident that in 
QCD the fermion bubble graphs give (\ref{onelooprun}) with the fermion 
contribution $\beta_{0f}$ to the $\beta$-function only. We may 
add the gluon and ghost bubbles, but the resulting coefficient 
would be gauge-dependent. The integral over the running coupling 
(\ref{drun}) with (\ref{onelooprun}) literally implicitly incorporates 
{\em some} contributions from vertex diagrams.

The substitution of $\beta_{0f}$ by $\beta_0$ has profound consequences, 
because it changes the location of renormalon singularities. Since the 
signs of $\beta_{0f}$ and $\beta_0$ are different, UV renormalons move 
to the negative real axis in the Borel plane (implying sign-alternating 
factorial divergence), while IR renormalons move to the 
positive real axis and obstruct (naive) Borel summation. 
According to the discussion in Section~\ref{divseries}, this 
implies that in QCD IR renormalons indicate that 
non-perturbative corrections should be 
added to define the theory unambiguously, while the same is true for 
UV renormalons in QED. This is of course exactly what one expects, 
because the coupling becomes strong in the infrared (ultraviolet) 
in QCD (QED).

Nevertheless, the extrapolation to the full non-abelian $\beta_0$ 
at this stage seems to be {\em ad hoc} and has often been shrouded in 
mystery. We will argue in Section~\ref{sectrenfeyn} that the 
substitution of $\beta_{0f}$ by $\beta_0$ can be justified 
diagrammatically, so that indeed renormalon singularities 
are located at multiples of $1/\beta_0$ in QCD. We have already seen 
that renormalon divergence is related to the counting of logarithms 
of loop momentum. Since for an observable like the Adler function, 
the $\beta$-function (broken scale invariance) is the only source 
of logarithms, it seems clear that one must end up with 
$\beta_0$ also in the non-abelian theory (QCD). Eventually we will 
see that the location of renormalon singularities is fixed by 
renormalization group arguments alone \cite{P78,Mue85} once 
UV and IR factorization is established for the loop momentum regions 
from which renormalons arise. For the further discussion we 
will therefore assume that the location of renormalon singularities 
is dictated by the first coefficient of the $\beta$-function 
also in QCD.\footnote{Hence, in QCD, $u=-\beta_0 t$ is understood 
in (\ref{firstpolesD}). 
Then, in QCD, $u$ is positive when $t$ is positive and this is the 
reason for the minus sign in the definition of $u$.}

4. In spite of what has been said, the running 
coupling is of minor importance once one is interested in 
IR renormalons as probes of power corrections. This point is 
often not well understood. The physics of 
power corrections resides in the small-momentum behaviour of 
the skeleton diagram, see (\ref{smallk}), and the running coupling 
is unrelated to it. The running coupling turns the 
small momentum behaviour into factorial divergence and makes 
it visible in the perturbative expansion. 
From the formulae of Section~\ref{divseries}, we find, using 
(\ref{onelooprun}), that the first IR renormalon pole in 
(\ref{firstpolesD}) yields an ambiguity in the definition of the 
Adler function that scales as 
\begin{equation}
\label{Q4}
\delta D(Q^2) \propto e^{2/(\beta_0\alpha_s(Q))} \sim 
\left(\frac{\Lambda}{Q}\right)^4,
\end{equation}
where $\Lambda$ is the QCD scale parameter.\footnote{The scale parameter 
is scheme-dependent. Without qualification we have in mind a scale of 
order 0.5-1$\,$GeV.} The power behaviour follows from 
(\ref{smallk}). If $F(k^2)\sim k^a$ at small $k$, an ambiguity of order 
$(\Lambda/Q)^a$ would have followed, together with a leading IR 
renormalon singularity at $u=a/2$. There is a simple way to understand 
this: the minimal term of the series associated with 
the IR renormalon occurs at $n$ such that $k_{\rm IR}\sim \Lambda$ 
in (\ref{saddles}). Hence if $F(k^2)\sim k^a$ the contribution from 
such $k$ scales as $\Lambda^a$.

5. The interchange of summation and integration that led to 
(\ref{drun}) is actually not justified, because in QCD (QED) the one-loop 
running coupling has a {\em Landau pole} in the infrared (ultraviolet) 
region. The problem this causes in defining the integral (\ref{drun}) 
is technically equivalent to the problem of defining the sum of 
the divergent series expansion of the integral. However, it is 
important to note that the renormalon and Landau pole phenomenon 
are logically disconnected in general. Whether a Landau pole 
exists or not is a strong-coupling problem and it depends on higher 
coefficients $\beta_{1}$, etc., of the $\beta$-function and on 
power corrections to the running of the coupling. On the other hand, 
renormalons always exist as seen from the fact that the location of 
renormalon singularities does not depend on higher coefficients 
of the $\beta$-function. (It is a simple exercise to convert 
(\ref{drun}) with {\em two-loop} running coupling into an 
expression for the Borel transform by a change of variables and to 
check what happens whether or not the $\beta$-function has a 
fixed point.) More details on this point are found in 
\cite{Gru96,DU96,PdR96}.

\subsection{Factorization and operator product expansions}
\label{fope}

We have already alluded to the fact that the ideas of factorization, 
the operator product expansion (OPE) and the renormalization group could 
be applied to renormalons, because there exist two very different 
scales in the problem. Mathematically OPEs 
amount to constructing an expansion in powers and logarithms of the small 
ratio of the two scales; so it seems that this could (almost) always 
be done. But there is more to factorization and OPEs, 
because the quantity under consideration should be 
broken into different pieces each of which depends on only 
one of the two scales.

The simplest and earliest example of factorization is renormalization 
itself. To define QCD or any other renormalizable field theory, one has 
to introduce an ultraviolet cut-off $\Lambda_{\rm UV}$. Renormalizability 
guarantees that all cut-off dependence can be absorbed into 
universal renormalization constants. These constants being universal, 
i.e. independent of external momenta of Green functions, they disappear 
from relations of physical quantities, thus rendering them cut-off 
insensitive up to terms that scale with inverse powers of the cut-off. 
The residual cut-off dependence could be further reduced by adding 
higher-dimension operators to the Lagrangian together with their 
respective set of renormalization/coupling constants. Ultraviolet 
renormalons, which originate from loop momentum larger than external 
momenta, can be understood entirely in terms of such renormalization 
theory methods. This will be explained in detail in Section~\ref{uvren}.

In QCD, which is strongly coupled in the infrared, the concept of 
{\em infrared factorization} is crucial. In this case, once factorization 
is achieved, the short-distance contributions can be computed and the 
long-distance contributions parametrized. Since the latter do not 
depend on the short-distance scale, they drop out in relations 
of physical quantities which differ only in their short-distance 
set-up. Infrared factorization was first applied in QCD to 
deep inelastic scattering \cite{CHM72}, based on the OPE of 
\cite{Wil69}.

The OPE is a powerful method, but it applies to a 
restricted class of observables. Most of QCD phenomenology, 
from jet physics to hadron-hadron collisions, relies on  
{\em perturbative factorization}, developed from the late 1970s on and 
reviewed in \cite{CSS89}. The idea of factorization is the same as 
in the OPE, but the approach is 
different in that one inspects the factorization properties of Feynman 
diagrams. It is more difficult in this approach 
to go beyond the leading power in the ratio of the two disparate 
scales and it has rarely been done \cite{EFP83,BB91,QS91}. In every 
case, the procedure is to identify and isolate the IR-sensitive 
regions in Feynman integrals and then to substitute them by non-perturbative 
and process-independent parameters. (As an example one may have 
in mind how parton densities are introduced in the perturbative 
factorization approach to deep inelastic scattering and compare this with 
the OPE treatment of deep inelastic scattering.)
IR renormalons are a useful addition to this strategy. As mentioned above, 
IR renormalons cause ambiguities/prescription dependences in summing 
the associated divergent series and we expect them to be cancelled only 
after exponentially small terms in $\alpha_s$ have been added, or, according 
to (\ref{onelooprun}), power corrections in $Q$. 

Let us  return to the 
Adler function to illustrate how IR renormalons lead us 
to non-perturbative parameters for power corrections. First, the 
sequence of IR renormalons is related to terms in the small-momentum 
expansion in the gluon momentum. The only scale $Q$ can be factored 
out and hence the IR parameter must be the matrix element of 
a local operator. Since there are no external hadrons, 
one needs a vacuum matrix element. It is a single gluon line that is 
soft in Fig.~\ref{fig1} 
which requires the operator to be bilinear in the gluon fields. 
The Adler function is a Lorentz scalar, and gauge-invariance excludes 
$A_\mu^A A^{A,\mu}$, where $A_\mu^A$ denotes the gluon field. This 
leaves covariant derivatives acting on the product of two field 
strength tensors with all Lorentz indices contracted. Thus, starting 
with the operator of lowest dimension (four), one is uniquely 
led to introduce the {\em gluon condensate}
\begin{equation}
\langle 0| G_{\mu\nu}^A G^{A,\mu\nu}|0\rangle
\end{equation}
as a parameter for the leading infrared contributions to the 
Adler function. (The argument that leads to this conclusion is 
worked out more thoroughly in \cite{Mue85}.) The gluon 
condensate adds to the Adler function a non-perturbative contribution 
of order $(\Lambda/Q)^4$, in coincidence with (\ref{Q4}). We also 
see that the potential IR renormalon at $u=1$ can be excluded 
because we would not be able to write down any operator 
matrix element of dimension two for it.
 
The gluon condensate contribution to current-current correlation 
functions could have been discovered in this way. Historically, \cite{SVZ79} 
were led to introduce it when they considered the OPE of the 
correlation function. The 
connection with IR renormalons was noted soon after by \cite{P79}. 
The OPE for the current-current correlation 
function reads
\begin{eqnarray}
\label{wilson}
D(Q) &=& C_0(Q^2/\mu^2) +  \frac{1}{Q^4}\left[C_{GG}(Q^2/\mu^2)\,
\langle 0| G_{\mu\nu}^A G^{A,\mu\nu}|0\rangle(\mu) + 
C_{\bar{q} q}(Q^2/\mu^2)\,m_q\langle 0|\bar{q}{q}|0\rangle(\mu)\right] 
\nonumber\\
&&\,+ \,{\cal O}(1/Q^6),
\end{eqnarray}
where we assumed that the fermion in the large fermion loop in 
Fig.~\ref{fig1} has mass $m_q\ll Q$. Starting from (\ref{wilson}), 
we conclude this section with a few general remarks regarding 
the relation of IR renormalons and parameters for power corrections. 
Most of these remarks are taken up again in Section~\ref{nonp} in 
a more concrete context. There we will compute explicit examples, 
non-perturbatively for the non-linear $\sigma$ model, and perturbatively 
for twist-4 corrections to deep-inelastic scattering.

In constructing the OPE one introduces a factorization scale 
$\mu$. This is often controversially discussed in the context of renormalons, 
although the problem seems to be one of semantics. 
The loop momentum region $k\sim Q\gg \mu$ is part of the 
coefficient functions, while the low momentum region 
$k\sim \Lambda\ll\mu$ is factored into the condensates. From this 
conceptually strict point of view the Wilson coefficients have 
no IR renormalons. Since UV renormalons are Borel summable, we may say 
that the Wilson coefficients can be defined unambiguously. The IR 
renormalons are part of the condensates, because the divergence 
sets in when $k\sim \Lambda$ as we saw above. If one introduces 
a rigid cut-off in the way described, the gluon condensate does 
not just scale as $\Lambda^4$, but also contains a power-like cut-off 
dependence beginning with $\mu^4$. Note that the IR renormalon 
contribution to (\ref{dex}) matches this 
cut-off dependence exactly. The interpretation of the first IR renormalon 
in current-current correlation functions as a perturbative contribution 
to the gluon condensate is developed further in \cite{Z92,BZ93}. 

A rigid cut-off is impractical for calculations beyond leading order 
and one uses dimensional regularization to implement factorization. 
In this scheme, only non-analytic terms (logarithms) are unambiguously 
factorized, while the Feynman integrals that contribute to the 
coefficient functions are integrated over all $k$. The 
operator matrix elements are only logarithmically $\mu$-dependent 
and the factorially divergent IR renormalon series resides in the 
coefficient function $C_0$. Conceptually this may seem more awkward, 
because $C_0$ and the gluon condensate separately are 
prescription-dependent, so that only the sum of both contributions 
to (\ref{wilson}) is unique. If we could compute everything, both, 
rigid-cut-off factorization and dimensional factorization, which in 
the present context are discussed in \cite{NSVZ85} and \cite{D82,D84}, 
respectively, would result in the same asymptotic expansion of the 
Adler function in powers and logarithms of $\Lambda/Q$. 

Although rigid-cut-off factorization results in a physically more 
intuitive picture, the terminology adopted in the literature 
on renormalons largely follows the one suggested 
by dimensional regularization. 
Thus, we will often say that IR renormalons in coefficient functions 
indicate that certain power-suppressed terms should exist. One might 
have equally considered the IR renormalon as part of these 
power-suppressed terms themselves and discarded it from the coefficient 
function. In this sense, an IR renormalon `problem', as it is sometimes 
stated, does not exist. Whichever point of view is preferred, since 
IR renormalons can be assigned to coefficient functions or operator 
matrix elements, they are related to mixing of operators of different 
dimension. Note that IR renormalons are IR contributions to coefficient 
functions, but {\em ultraviolet} contributions to operator matrix 
elements as indicated by their power-like $\mu$-dependence. To be 
precise, IR renormalons are related to properties of higher-dimension 
{\em operators} and not of their {\em matrix elements}. This is why, 
without additional assumptions, renormalons give us little quantitative 
insight into non-perturbative effects, but tell us much about 
their scaling with the large scale $Q$. A useful analogy is provided 
by the leading-twist formalism for deep inelastic scattering. The 
(logarithmic) $Q$-dependence of parton distributions can be computed 
perturbatively, but the parton distributions themselves cannot. 
Except that one refers to power-like $Q$-dependence, the situation with 
IR renormalons is just the same.

We have kept the quark mass in (\ref{wilson}) to make the following 
important point: while IR renormalons lead one to introduce 
non-perturbative parameters for power corrections, the gluon condensate 
(and higher dimension gluonic operators with derivatives) in case 
of (\ref{wilson}), one cannot be sure that one obtains all of them. 
In (\ref{wilson}) one would obviously miss the quark condensate, 
because it is the order parameter for chiral symmetry breaking, 
which does not occur to any (finite) order in perturbation 
theory.\footnote{If $m_q=0$, one can instead find 
dimension-6 four-fermion operators protected by 
chiral symmetry.} In general, those operators will be missed that 
are protected from mixing with lower-dimensional ones, which usually 
means that their matrix elements are unambiguous and 
physical.\footnote{In the case of the quark condensate, 
$m_q\langle 0|\bar{q} q|0\rangle$ is physical, as follows for 
example from the Gell-Mann--Oakes--Renner relation.} 
In particular, there is the possibility that 
power corrections parametrically larger than those found through 
IR renormalons are missed. However, since operators do mix unless 
there is a particular reason that they should not (such as a symmetry), 
such cases can 
often be identified. Still, it requires some understanding of 
the form of operators, which one does not have in all applications 
considered to date. 

IR renormalons (and condensates) evidently 
refer to power corrections that originate from long distances. The 
OPE, which factorizes long and short distances,  
does not exclude power corrections/non-perturbative contributions 
from short distances, which are logically part of the coefficient 
functions (contrary to IR renormalon contributions, there is no 
ambiguity in this assignment). Very little is known about such 
contributions and the only known source of such 
contributions is small-size instantons. While the power-suppressed 
terms discussed in this report are typically of order $1/Q^{1-4}$, 
small-size instantons give rise to terms of order 
$(1/Q^2)^{-2\pi\beta_0}$ or smaller, which are strongly suppressed 
in comparison. For this reason, we will ignore them altogether. 

For the current-current correlation function the IR renormalon
phenomenon reinforces that the notion of perturbative and non-perturbative 
effects is ambiguous and requires a prescription. On the other hand, 
one does not learn from IR renormalons anything new 
about power corrections 
beyond the content of the OPE treatment of \cite{SVZ79}. The situation 
is very different for observables that do not admit an operator product 
expansion, even though they may be treated at leading power with standard 
perturbative methods, for instance fragmentation processes in $e^+ e^-$ 
annihilation and the related event shape variables. Power corrections 
to these processes do not lend themselves easily to an operator 
interpretation, and IR renormalons turned out to be very useful in taking 
the step beyond leading power. In these cases renormalon-based 
methods are conceptually connected to an extension of perturbative 
factorization techniques beyond the leading power.

Many of the questions concerning the large-order behaviour of the 
series expansion in $\alpha_s$ can also be asked about the operator 
product expansion, i.e. the expansion in $\Lambda/Q$. But much less 
is known for the latter. There is good reason to believe that 
the OPE is also divergent \cite{Shi94}, but the precise behaviour is 
not known, not even whether the divergence is sign-alternating 
or not. It is not known whether the OPE is asymptotic and whether 
exponentially small terms in $\Lambda/Q$ have to be added to recover 
the exact result. If the OPE is asymptotic the important question arises 
in what region in the complex $Q^2$ plane it is asymptotic. For 
example, the expansion might be asymptotic in the euclidian region 
($Q^2$ real and positive), but the bound on the remainder may not 
be analytically continued to the cuts at negative $Q^2$ or may degrade 
as the domain of validity in the complex plane increases. In this case 
further calculation of power-suppressed terms would not improve the 
approximation of minkowskian quantities. The fact that the OPE may 
not provide an asymptotic expansion for minkowskian quantities provides 
a mathematical definition of what is usually referred to as 
`violations of parton-hadron duality', although the terminology is 
not homogeneous in the literature. The question has so far been 
addressed only in models \cite{CDSU97,GL98,BSZ98,BSUV98}. Alternatively, 
one can demonstrate a certain behaviour under analytic continuation,  
which is independent of the 
dynamics of a particular theory, provided certain conditions 
are met by the exact result \cite{F97}. In this report we will not pursue 
this very interesting but still uncertain subject. 

Finally, we emphasize that renormalons can be discussed only in the context 
of processes for which a hard scale, say $Q\gg\Lambda$, exists and a 
(possibly only partial) perturbative treatment and power expansion 
is possible. For $Q\sim \Lambda$ this framework breaks down (i.e.~the 
OPE would have to be summed) and there 
is nothing we have to say about this region in this report. The 
non-perturbative regime where all scales are of order $\Lambda$ is 
inaccessible with the methods reviewed here. 

\subsection{The Borel plane}
\label{borelplane}
\begin{figure}[t]
   \vspace{-4.5cm}
   \epsfysize=27cm
   \epsfxsize=18cm
   \centerline{\epsffile{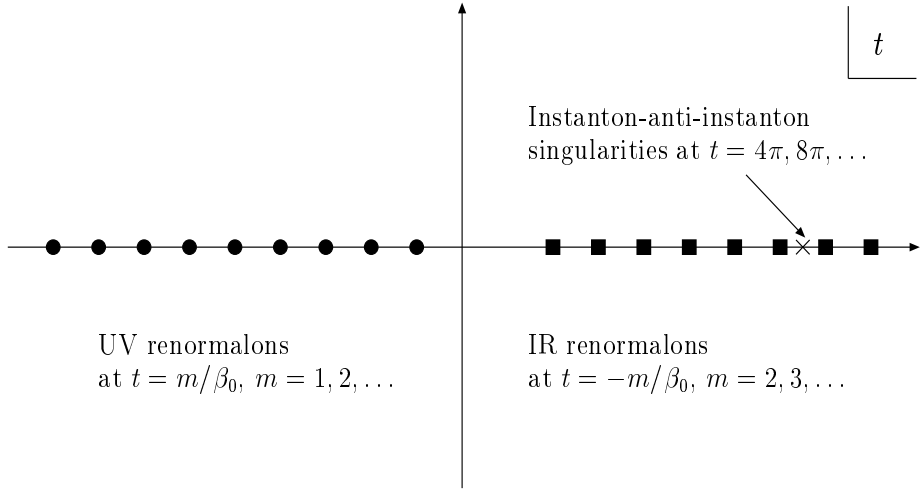}}
   \vspace*{-15.3cm}
\caption[dummy]{\small Singularities in the Borel plane of 
$\Pi(Q^2)$, the current-current correlation function in QCD. 
Shown are the singular points, but not the cuts attached to each 
of them. Recall that $\beta_0 < 0$ according to (\ref{beta0}).
\label{fig3}}
\end{figure}

We summarize what is known about singularities in the Borel 
plane. Recalling the definition of the Borel 
transform (\ref{defborelt}), the Borel plane for the Adler function 
(current-current correlation functions) is portrayed in 
Fig.~\ref{fig3}. Please note that the figure does not show what 
is {\em not} known. We distinguish three sets of singularities:

{\em Ultraviolet renormalons} are located at $t=m/\beta_0$, with positive 
integer $m$, i.e. $u=-1,-2,\ldots.$ The first UV renormalon is the 
singularity closest to the origin of the Borel plane and hence 
governs the large-order behaviour of the series expansion of the 
Adler function. According to 
(\ref{bestapprox}) the minimal term is of order $\Lambda^2/Q^2$, 
using (\ref{onelooprun}). A more precise analysis 
\cite{BZ92} shows that it is of order
\begin{equation}
\label{realuv}
\delta D_{\rm UV} \sim \frac{Q^2\Lambda^2}{\mu^4}\,\times\,
\mbox{logarithms}.
\end{equation}
However, since UV renormalons produce sign-alternating factorial divergence 
in QCD, we do not take them as an indication that extra terms should 
be added to the perturbative expansion. Eq.~(\ref{realuv}) supports this 
interpretation: since the coupling renormalization scale $\mu$ is 
arbitrary, one can make the minimal term small by increasing $\mu$. In 
this way, one systematically cancels (approximately) factorially 
large constants against powers of $\ln(Q^2/\mu^2)$. Note that 
$\delta D_{\rm UV}$ is polynomial in $Q$ (up to logarithms) 
and therefore cannot be 
confused with an infrared $1/Q^2$ power correction.

For the current-current correlation function all UV renormalons are double 
poles, if one restricts oneself to the set of bubble graphs in 
Fig.~\ref{fig1}. Beyond this approximation, only the first singularity 
at $u=-1$ has been analysed in detail \cite{BBK97}. This analysis uses 
renormalization group methods suggested by \cite{P78} and developed 
further in \cite{VZ94,dCP95,Ben95,BS96}. These will be the subject of 
Section~\ref{uvren}. The result is a complicated branch point 
structure attached to the point $u=-1$.

UV renormalons are theory-specific, but 
process-independent.\footnote{Read: The process dependence factorizes 
and is calculable, see Section~\ref{uvren}.} In theories with 
four-dimensional rotational invariance, UV renormalons are always 
located at positive integer multiples of $1/\beta_0$, provided the 
theory contains no power divergences. If it does, the semi-infinite 
series of UV renormalons begins at some negative integer multiple 
of $1/\beta_0$. If $O(4)$ invariance is broken, UV renormalons 
can also occur at half-integer $u$. An example of this kind is 
heavy quark effective theory, because it contains the heavy quark 
velocity four-vector \cite{BB94a}. 

{\em IR renormalons} are located at $t=-m/\beta_0$, with $m=2,3,\ldots$, 
i.e. $u=2,3,\ldots$. As discussed in Section~\ref{fope} the minimal 
term associated with the subseries due to the first 
IR renormalon is of order $(\Lambda/Q)^4$. Contrary to the situation for 
UV renormalons, the minimal term is $\mu$-independent and cannot be decreased 
\cite{BZ92}. (We are using dimensional regularization, see the 
remarks in Section~\ref{fope}.) This suggests that the ambiguities 
caused by IR renormalons have physical significance. For current-current 
correlation functions one can associate them with condensates. The 
singularity at $u=1$ is absent, because there is no dimension-2 
condensate in the OPE \cite{P79}. The set of diagrams of Fig.~\ref{fig1} 
leads to double poles for all IR renormalons expect for 
$u=2$, which is a single pole \cite{Ben93a}. Beyond this approximation, 
only the first singularity has been analysed in detail 
\cite{Mue85,Z92}, making use of the renormalization properties of the 
gluon condensate. This will be discussed in Section~\ref{irren}. 
The result is that the simple pole is turned into a branch cut, 
but the structure is simpler than for the first UV renormalon.

IR renormalons are process-dependent and the absence of an 
IR renormalon at $u=1$ is specific to processes without identified 
hadrons in the initial and final state, for which vacuum matrix 
elements are relevant. For example, there exists 
a leading singularity at $u=1$ in deep inelastic scattering,  
that is naturally connected 
with $1/Q^2$ twist-4 corrections \cite{Mue93}. In general, 
observables that can be related to off-shell Green functions 
(non-exceptional external momentum configurations) have IR renormalons 
at positive integer $u$. For time-like processes and on-shell Green 
functions, singularities at half-integer $u$ are quite common, often 
beginning at $u=1/2$, which leads to power corrections suppressed 
only as $\Lambda/Q$ \cite{KS95a,DW95,AZ95a}. 
For time-like processes one can also construct 
physical quantities, 
which are IR finite, but arbitrarily IR sensitive \cite{MW95,BBM97}. 
Such quantities 
have IR renormalon poles at $u=\gamma$ with $\gamma$ positive and 
arbitrarily close to zero.

Note that theories without self-interactions of massless particles 
such as `real' QED with massive leptons are not expected to have 
IR renormalons.

In addition to renormalon singularities, {\em instantons} are known 
to produce factorially divergent series \cite{Lip77}. In QCD instantons 
carry topological charge and hence they cannot be related to the 
perturbative expansion. However, configurations of $n$ instantons 
and $n$ anti-instantons with topological charge zero produce 
singularities at $t=4\pi n$ \cite{BF77}, the position of the singularity 
being related to the action of the field configuration. Instanton 
singularities are not associated with either large or small momenta,  
but with the number of diagrams, which increases rapidly with order 
of perturbation theory. Because of their semi-classical origin, 
instanton singularities are under better control than renormalon 
singularities. For example, not only the form of the singularity, 
but also the residue can be calculated. For the current-current 
correlation function this calculation is carried out in \cite{Bal91}. 
However, in QCD, and in fact most other interesting renormalizable 
theories, instanton singularities are far away from the origin of the 
Borel plane. Hence we do not expect them to play a role in the 
large-order behaviour of perturbative expansions in QCD. Nor do they 
represent a dominant source of power corrections. Instanton-induced 
factorial divergence is reviewed in \cite{LeGZ90}.

What do we {\em really} know? What we have said appears to be 
compelling on physics grounds 
and is (probably) correct, but mathematical proofs are rare. 
Although the rules are set by specifying the Lagrangian, results 
on global properties of series expansions are difficult to obtain 
in renormalizable field theories. For example, in the above discussion 
we have implicitly assumed that one has not applied arbitrary 
subtractions in defining the coupling. Otherwise any singularity 
could be obtained. Provided that only minimal subtractions 
are applied, it was shown in \cite{BS96} for off-shell Green functions 
that to any finite order 
in an expansion in $1/N_f$ of (massless) QED and QCD, where 
$N_f$ is the number of flavours, the Borel transform is analytic, except 
for UV and IR renormalon singularities at the expected positions. 
But this may 
tell us more about deficiencies of the $1/N_f$ expansion 
than anything else: 
instanton singularities are absent, because they are exponentially 
small effects in $1/N_f$. To the knowledge of the author, the 
strongest result has been obtained by \cite{DFR88}, although 
for the scalar $\Phi^4$ theory. There it was shown that the Borel 
transform is analytic in a disc around the origin of the Borel plane 
of radius at least 
as large as the distance of the first UV renormalon from the origin. 
The existence of the first UV renormalon singularity was almost 
established and could be avoided only through improbable cancellations. 

\cite{tH77} has shown that, even if the Borel transform of Green functions 
in QCD had no singularities on the positive real axis, it could not 
reconstruct the Green function non-perturbatively, because the 
analyticity domain in $\alpha_s$ of the Borel sum would be in 
conflict with the horn-shaped analyticity region that follows 
from the (assumed) {\em non-perturbative} analyticity properties 
of Green functions in momentum space.\footnote{By non-perturbative we mean 
that the existence of resonances is crucial. It is not enough 
that the cut in the Adler function is generated by perturbative 
logarithms $\ln(-q^2/\mu^2)$. See \cite{Khu81} for 
a discussion of this point.} This is often interpreted to the 
effect that the Borel integral must diverge at positive infinity. However, 
once we give up the idea that Green functions should be reconstructible 
from their Borel integrals, this conclusion does not follow. 
IR renormalons signal that further contributions should be added to 
perturbative expansions. 
In fact only after one sums not only perturbative series, but the 
entire OPE, can one hope to recover the correct analyticity 
properties. As far as the Borel transform defined by (\ref{defborelt}) 
is concerned, one usually finds that the Borel integral (defined 
as principal value) converges, provided that all kinematic 
invariants ($Q$ for the Adler function) are larger than $c\Lambda$, 
where $\Lambda$ is the QCD scale and $c$ a constant of order 1.


\section{Renormalons from Feynman diagrams}
\setcounter{equation}{0}
\label{sectrenfeyn}

This section deals solely with properties of perturbative expansions, 
and phenomenological applications do not concern us here. We will try to 
learn as much as possible about renormalons from Feynman diagrams. 
Our basic tool to look at diagrams is an expansion in the number of 
massless fermions, although, of course, we are mainly interested in 
statement that are true beyond this expansion. After setting up the rules 
of the $1/N_f$ expansion, we consider UV renormalons in Section~\ref{uvren}, 
first to next-to-leading order in $1/N_f$. The purpose of this exercise is 
to motivate the subsequent, general, renormalization group analysis. 
In Sect.~\ref{irren} we discuss IR renormalons. Our treatment will 
be more qualitative for these, mainly because a general process 
independent factorization theorem for IR renormalons does not hold. 
The subsequent two subsections address the question of scheme-dependence 
of large-order behaviour and methods to calculate or represent bubble 
diagrams, which we will need in Section~\ref{pheno}.

More precisely, let us anticipate that the 
asymptotic behaviour due to UV and IR renormalons takes the form
\begin{equation}
\label{exl}
r_n = \sum_i K_i\,(a_i\beta_0)^n\,n!\,n^{b_i}\left(1+\frac{c_{i1}}{n} 
+\ldots\right).
\end{equation}
We will try to calculate the parameters $K_i$, $a_i$, etc., and to 
understand why $\beta_0$ enters. We emphasize the diagrammatic point 
of view, although we shall 
then see that every positive result can be obtained more elegantly 
by solving renormalization group equations. However, we believe that 
the diagrammatic analysis is useful to understand why some quantities 
in (\ref{exl}) can be calculated and others cannot.

\subsection{The flavour expansion}
\label{flavourexp}

We begin the analysis with the set-up of the 
{\em flavour expansion}. That is, we consider 
QCD (or any SU($N_c$) gauge theory) or QED with $N_f$ massless 
fermion flavours and we expand in $1/N_f$. Because we are interested 
in properties of (classes of) Feynman diagrams, we use the 
flavour expansion also for QCD, even though one loses asymptotic 
freedom and everything that is crucial for the QCD {\em vacuum}. In the 
flavour expansion (the large-$N_f$ limit) the gluon 
self-couplings are perturbations, generally speaking.

Even if the flavour expansion could be summed, it would not reproduce 
QCD. We have already mentioned instanton effects as 
exponentially small, and hence non-perturbative effects in $1/N_f$. 
Besides there is evidence from lattice QCD \cite{IKK97}
and supersymmetric QCD \cite{Sei94} for a phase structure in $N_f$, 
so that the large-$N_f$ (IR free) region and small-$N_f$ 
(asymptotically free) region are not analytically connected. This 
being said, we shall nevertheless see that the flavour expansion is 
quite instructive also in QCD.

The flavour expansion is obtained in the limit $N_f\to \infty$, where 
$N_f$ is the number of massless fermion flavours in QED or QCD, keeping 
$a_s=-\beta_{0f} \alpha_s\propto N_f \alpha_s$ fixed. In this limit, 
fermion loops with two gluon legs are special, because they count as 
$N_f \alpha_s\propto a_s = {\cal O}(1)$. In leading order one is led to 
the set of diagrams with a single chain of fermion bubbles, 
such as in Fig.~\ref{fig1} for 
the current-current two-point functions. The flavour expansion as an 
organizing principle is implicit in the works of \cite{L77,tH77}. It 
was used in \cite{Coq81,EPPT82,PP84,KKO91} 
to obtain renormalization group functions in QED in 
the $\overline{\rm MS}$ and on-shell schemes and then in 
\cite{Ben93a,Bro93} for the photon propagator. Since then it has been 
applied to a variety of processes in QCD, for which we refer to 
Section~\ref{pheno}.

More precisely, we call a gluon propagator with any number of fermion 
bubbles inserted and summed over a {\em chain}. The effective propagator 
for a chain in covariant gauge is
\begin{equation}
\label{chainprop} 
D_{\mu\nu}(k) = \frac{(-i)}{k^2} 
\left(g_{\mu\nu}-\frac{k_\mu k_\nu}{k^2}\right)\,
\frac{1}{1+\Pi_0(k^2)} + (-i)\,\xi\,\frac{k_\mu k_\nu}{k^4},
\end{equation}
where $\Pi_0$ is given by (\ref{simpleloop}) and $\xi$ is the gauge-fixing 
parameter. The counterterms for the fermion loops are included and 
we have taken the limit $\eps=(4-d)/2 \to 0$ in dimensional 
regularization. In Landau gauge, $\xi=0$, the propagator is particularly 
simple.\footnote{For gauge-invariant quantities, and in QED, one can 
neglect all $k_\mu k_\nu$-terms, as long as one is interested only 
in large-order behaviour.} When Feynman diagrams are written in terms 
of chains, all other interactions are suppressed by powers of $N_f$. 
Let $\gamma$ be a diagram consisting of $n_c$ chains, $f$ fermion loops with 
more than two gluon legs (i.e. fermion loops 
other than those absorbed into chains), 
and $v_{3,4}$ three-gluon (four-gluon) vertices. Then the diagram 
contributes to the flavour expansion at order $N_f^{-d(\gamma)}$, 
where
\begin{equation}
d(\gamma) = n_c-f-v_3-v_4.
\end{equation}
Examples of diagrams at leading and next-to-leading order to pair creation  
from an external current are shown in Fig.~\ref{fig4}.
\begin{figure}[t]
   \vspace{-5.7cm}
   \epsfysize=33cm
   \epsfxsize=22cm
   \centerline{\epsffile{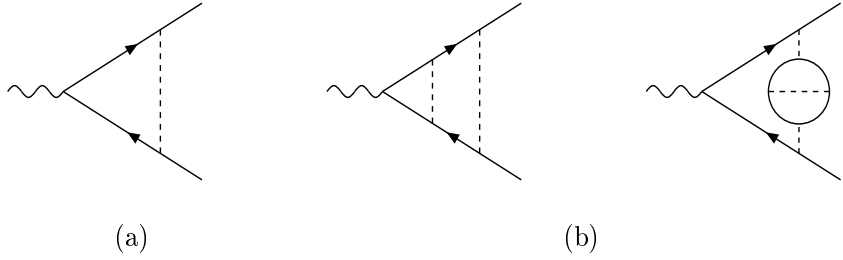}}
   \vspace*{-22.8cm}
\caption[dummy]{\small Pair creation of quarks by an external current: 
(a) Leading order in the flavour expansion; (b) representatives at 
next-to-leading order. Chains are displayed as dashed lines. \label{fig4}}
\end{figure}

The chain propagator becomes particularly useful after applying Borel 
transformation \cite{Ben93a}. Using the definition (\ref{defborelt}), 
one has
\begin{equation}
\label{effprop}
B[\alpha_s D_{\mu\nu}](u) = 
\frac{(-i)}{k^2} \left(g_{\mu\nu}-\frac{k_\mu k_\nu}{k^2}\right)
\left(-\frac{\mu^2}{k^2} e^{-C}\right)^{\!u} +  
(-i)\,\xi\,\frac{k_\mu k_\nu}{k^4},
\end{equation}
where $u=-\beta_{0f} t$. Hence Borel transformation of a chain results 
in an analytically regularized gluon propagator, except for the gauge 
parameter dependent piece. Note that inserting 
the {\em renormalized} chain propagator after taking $\eps\to 0$ is correct 
only if the diagram into which it is inserted does not require further 
subtractions. For the moment, we postpone the 
issue of subtractions. 

The Borel transform of diagrams with one chain is obtained by 
Borel-transforming the chain as in the previous paragraph. If the 
number of chains $n_c>1$, one uses the fact that the Borel transform of a 
product of series is a convolution. Suppressing the Lorentz indices, 
the relevant identity is
\begin{equation}
\label{convolution}
B\!\left[\prod_{j=1}^{n_c} \alpha_s D(k_j)\right](u) =
\frac{1}{(-\beta_{0f})^{n_c-1}}\intl_0^u \left[\prod_{j=1}^{n_c} d u_j
\right]
\delta\!\left(\!u-\sum_{j=1}^{n_c} u_j\right)\,\prod_{j=1}^{n_c}
B\!\left[\alpha_s D(k_j)\right]\!(u_j).
\end{equation}
If both ends of a chain attach to fermion or ghost lines, 
one obtains $D_{\mu\nu}(k)$ always in conjunction with a factor 
of $\alpha_s$. On the other hand if a chain attaches to a 
three-gluon or four-gluon vertex, factors of $1/\alpha_s$ will be 
left over after use of (\ref{convolution}). These factors can be 
dealt with by applying an appropriate number of derivatives in 
$u$ at the end. Thus, the Borel transform can be obtained 
by first calculating the skeleton diagram with all gluon propagators 
analytically regularized with regularization parameters $u_j$. Then 
for a given value of the Borel parameter $u$ one integrates over 
all regularization parameters $u_j$ with the delta-function 
constraint of (\ref{convolution}). This completes the set-up of the 
flavour expansion. 

Our goal is to find the factorially divergent contributions 
from diagrams with an arbitrary number of chains, i.e. the 
singularities in $u$ of their Borel transforms. 
As far as singularity structure is concerned, the second step above 
-- integrating over the $u_j$ -- is trivial, 
once the singularities in the space of 
variables $u_j$ are known. Finally, we will see in the 
flavour expansion regularities that allow us to sum partial contributions 
to all orders in $1/N_f$. 

To prepare the subsequent discussion, consider expanding 
(\ref{exl}) in $1/N_f$. For simplification, we assume that there is 
only one component (no sum over $i$). Write 
\begin{equation}
a\beta_0 = a \beta_{0f}\left(1+\frac{\beta_0-\beta_{0f}}{\beta_{0f}}
\right),
\end{equation}
the second term being $O(1/N_f)$. Furthermore, we expand
$K=K^{[0]}\,(1+K^{[1]}/N_f+\ldots)$ and likewise for $b$ and $c_1$. 
We assume that $c_1^{[0]}=0$. This is always true, if the leading order 
contribution to the flavour expansion is a one-loop skeleton diagram, 
because one-loop diagrams can result only in a simple pole in the 
Borel transform. Then
\begin{eqnarray}
\label{rnnfexp}
r_n &=& K^{[0]} (a\beta_{0f})^n\,n!\,n^{b^{[0]}}\Bigg(
1+\frac{1}{N_f}\left[
N_f\frac{\beta_0-\beta_{0f}}{\beta_{0f}}\,n + 
b^{[1]}\ln n+K^{[1]}+\frac{c_1^{[1]}}{n} + 
O\!\left(\frac{1}{n^2}\right)\right]
\nonumber\\
&&+ \,O\!\left(\frac{1}{N_f^2}\right)\Bigg).
\end{eqnarray}
In the following we will identify the origin of the 
various terms in this equation.

\subsection{Ultraviolet renormalons}
\label{uvren}

In this section, we discuss in detail the first UV renormalon, 
located at $u=-1$, in QED and QCD. 

\subsubsection{QED}

Most explicit calculations of renormalon behaviour have foucssed on 
diagrams with one chain.
Large-order behaviour due to UV renormalons from 
diagrams with two chains (other than chains inserted into chains) 
was first considered in \cite{VZ94}. Further 
work is due to \cite{BS96,PdR97}.  
The characterization of singularities 
of the Borel transform for an arbitrary number of chains 
below follows \cite{BS96}. Sometimes, instead of 
being general we take pair creation of quarks from a vector current 
and the two-point function of two vector currents as illustrative 
examples.

Let $\Gamma$ represent a diagram with $n_c$ chains. Such a diagram 
is expressed as a series in $\alpha_s$, whose Borel transform is 
denoted by $B_\Gamma(u)$. Let 
$G_\Gamma(\underline{u})\equiv G_\Gamma(u_1,\ldots,u_{n_c})$ be the 
Feynman integral that is obtained by replacing each chain/dressed gluon by 
its Borel transform of form $1/(k_i^2)^{1+u_i}$ (cf.~(\ref{chainprop})), 
where $k_i$ is the 
momentum of the $i$th dressed gluon line. The two latter quantities 
are related by
\begin{equation}
\label{relbg}
B_\Gamma(u) =
\frac{1}{(-\beta_{0f})^{n_c-1}}\intl_0^u \prod_{i=1}^{n_c} d u_i\,
\delta\!\left(u-\sum_{i=1}^{n_c} u_i\right)\,G_\Gamma(\underline{u}).
\end{equation}
The singularity structure of $G_\Gamma(\underline{u})$ 
follows straightforwardly 
from earlier results on analytic regularization 
\cite{Spe68,Poh74,BM77} in the context of renormalization of 
field theories. 

Consider one-particle irreducible (1PI) subgraphs $\gamma$ of $\Gamma$ 
and let $\omega(\gamma)$ be the (naive) degree of UV divergence 
of $\gamma$ obtained in the standard way from UV power counting of lines 
and vertices in $\gamma$. For a given point 
$\underline{u}_0=(u_{01},\ldots,u_{0 n_c})$ in the space of 
(complex) regularization parameters $u_i$ define the modified degree 
of divergence
\begin{equation} 
\label{uvdegree}
\omega_{\underline{u}_0}(\gamma) = \omega(\gamma)  - 2 u_0(\gamma),
\end{equation}
where $u_0(\gamma)=\sum_{l\in\gamma} {\rm Re\,}(u_{0l})$ is the sum over 
the real parts of the analytic regularization parameters of all lines of 
$\gamma$. With this definition the subgraph has no over-all UV divergence 
if $\omega_{\underline{u}_0}(\gamma)<0$. One then finds that 
$G_\Gamma(\underline{u})$ has poles of ultraviolet origin at 
those points $\underline{u}_0$ for 
which there exists a 1PI subgraph $\gamma$ of $\Gamma$ such that
$u_0(\gamma)$ is integer and $\omega_{\underline{u}_0}(\gamma)\geq 0$.

For example, the vertex graph in Fig.~\ref{fig4}a has $\omega(\gamma)=0$ 
and hence leads to singularities at $u_1=0,-1,-2,\ldots$, where $u_1$ is the 
single regularization parameter. The box subgraph in the two-chain 
vertex graph in Fig.~\ref{fig4}b is ultraviolet convergent, 
$\omega(\gamma)=-2$ and leads to singularities at 
$u_1+u_2=-1,-2,\ldots$, where $u_{1,2}$ are the two 
analytic regularization parameters for the two gluon propagators.

A forest is a set of non-overlapping subgraphs. In the present context 
we can restrict these subgraphs to be 1PI. Let ${\cal F}$ be a maximal 
forest, i.e. a forest such that for any $\gamma$ not in 
${\cal F}$ the union ${\cal F}\cup\gamma$ is no longer a forest. Then 
the singularities of $G_\Gamma(\underline{u})$ are characterized by 
\begin{equation}
\label{singofg}
G_{\Gamma}(\underline{u}) = \sum_{\cal F} \prod_{
{\scriptsize \begin{array}{c} \gamma\in 
{\cal F}:\omega_{\underline{u}_0}(\gamma)\geq 0
\\[0.0cm] u_0(\gamma)\,\,{\rm integer}
\end{array}}
}\frac{g_{\cal F} (\underline{u})}{u_0(\gamma)-u(\gamma)},
\end{equation}
where the functions $g_{\cal F} $ are analytic in a
vicinity of the point $\underline{u}_0$, the sum extends over all 
maximal forests, and $u(\gamma)$ is defined analogously to $u_0(\gamma)$. 
Barring cancellations between 
different forests, (\ref{singofg}) allows us to obtain the nature 
of UV renormalon singularities for any diagram in the flavour expansion. 
Note that a maximal forest of an $n$-loop skeleton diagram can have at most 
$n$ elements. Hence, an $n$-loop skeleton diagram 
can have at most $n$ singular factors in 
(\ref{singofg}). 

Let us illustrate (\ref{singofg}) by examples:

{\em One chain.} The single regularization parameter $u$ coincides 
with the Borel parameter. The diagram of Fig.~\ref{fig4}a gives rise 
to simple poles\footnote{Note that we consider ultraviolet renormalon 
poles only.} at $u=0,-1,\ldots$. Any single chain one-loop diagram 
must result in simple poles. The pole at $u=0$ corresponds to an 
explicit logarithmic ultraviolet divergence. It is cancelled by the 
self-energy diagrams, so that the pair creation amplitude is UV finite. 
The pole at $u=-1$ gives rise to the first UV renormalon singularity 
at lowest order in the flavour expansion. Its residue gives $K^{[0]}$ 
in (\ref{rnnfexp}). Furthermore $b^{[0]}=0$ and, since there is 
no subleading singularity at leading order in $1/N_f$, $c_1^{[0]}=0$, 
as assumed for (\ref{rnnfexp}). The explicit expression is 
\begin{equation}
\label{vertexoneloop}
B_{\Gamma_{4a}}(u) = \frac{e^C}{6\pi\mu^2}\,\frac{1}{1+u}\,
(q^2\gamma_\mu-\not\!q q_\mu) + \ldots,
\end{equation}
where $C$ comes from the fermion loop (\ref{simpleloop}) and the 
dots denote terms that vanish when the external `quarks' are 
on-shell.

The residue of the pole at 
$u=-1$ follows from the coefficient of the $d^4k/k^6$-term in the 
expansion of the Feynman integrand for $k\gg q$, where $q$ stands for 
the external momentum and we have in mind the integrand of the 
skeleton diagram with all $u_i$ set to zero. 
Likewise, the residue of the pole at $u=-2$ follows 
from the $d^4k/k^8$-term and so on. When the gluon propagator is 
Borel transformed, $d^4k/k^6$ becomes $d^4k/k^{6+2 u}$, and it is seen 
that the pole occurs when $u$ is such that the integral is 
logarithmic by power counting. The fact that the pole follows from the 
{\em expansion} of the Feynman integrand is very important, because 
it implies that the residue is polynomial in the external momentum 
$q$. On dimensional grounds alone, the residue of a pole at $u=-n$ 
can be written as the insertion of an operator of dimension $4+2 n$. 
From this point of view there is not much difference between ordinary 
UV divergences and UV renormalon singularities. 
The former produce poles at $u=0$ in the Borel transform. 
They can be compensated 
by counterterms, that is, insertions of operators of dimension 4 
with appropriately chosen coefficients. The latter can be compensated 
by insertions of higher-dimension operators\footnote{Power ultraviolet 
divergences regulated dimensionally also cause UV renormalons, but 
at {\em positive} $u$ with the definition of $u$ chosen here. These 
are evidently related to counterterms of dimension smaller 
than 4.} \cite{P78}. In particular, the leading UV renormalon at $u=-1$ 
leads to considering 
dimension-6 operators. From the structure $q^2\gamma_\mu-\not\!q q_\mu$ 
in (\ref{vertexoneloop}) it can be deduced that 
the first UV renormalon is proportional to the 
zero-momentum insertion of 
the operator \cite{VZ94,dCP95}
\begin{equation}
\label{o6}
{\cal O}_6=\frac{1}{g_s^2}\,(\bar{\psi}\gamma_\mu\psi)\,\partial_\nu 
F^{\mu\nu}
\end{equation}
into the three-point function. At this order, the three-point function 
with insertion of ${\cal O}_6$ is needed only at tree level and the 
coefficient of ${\cal O}_6$ is adjusted to reproduce the normalization 
of (\ref{vertexoneloop}). In 
(\ref{o6}) $F^{\mu\nu}$ is the field strength of an external abelian 
gauge field that relates to the external vector 
current through $\partial_\mu 
F^{\mu\nu} = j_V^\nu$. The factor $1/g_s^2$ is convention.

Note that there is only a limited number of 1PI one-loop graphs $\Gamma$, 
which can have a UV renormalon pole at $u=-1$. The condition is 
$\omega(\Gamma)\geq -2$. This generalizes to all loops: a diagram 
$\Gamma$ with $\omega(\Gamma)< -2$ can have a singularity at $u=-1$ 
only from {\em subgraphs} with a larger degree of divergence.

\begin{figure}[t]
   \vspace{-4cm}
   \epsfysize=33cm
   \epsfxsize=22cm
   \centerline{\epsffile{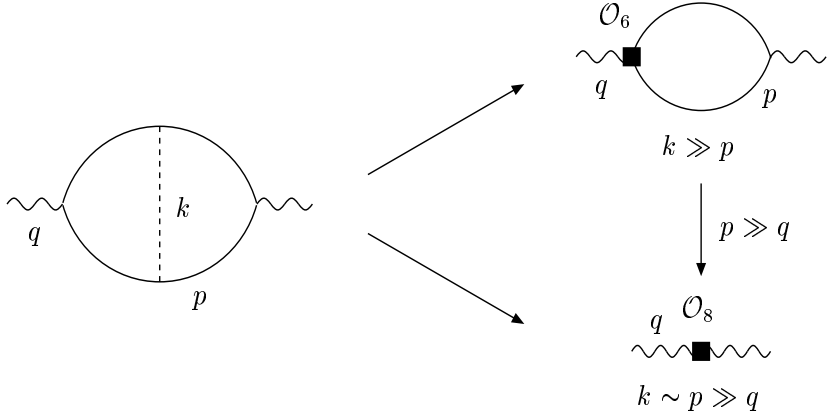}}
   \vspace*{-22.6cm}
\caption[dummy]{\small Leading order contribution 
in the flavour expansion 
to the two point function of vector currents $j_V$ (left) and its 
reduced diagrams with operator insertions (right). The momentum 
$p$ is the loop momentum for the fermion loop. \label{fig5}}
\end{figure}
For the vector current two-point function (see Fig.~\ref{fig5}) a 
maximal forest contains two elements, for example the left one-loop 
vertex subgraph and the two-loop (skeleton) diagram itself. Each of the 
two gives one singular factor $1/(1+u)$. This explains why all 
UV renormalons in the Adler function turned out to be double poles 
as discussed in Section~\ref{sectbascon}.

The double pole arises from the loop momentum region, where both 
loop momenta are large but ordered: $k\gg p\gg q$. In this case one 
can contract the vertex subgraph to a point, as shown in the upper diagram 
of Fig.~\ref{fig5}. This amounts to inserting the operator 
${\cal O}_6$ with exactly the coefficient that we found 
from the analysis of the vertex graph above. 
The region where both loop momenta are large but of the same order, 
$k\sim p\gg q$, contributes a simple pole $1/(1+u)$. Because 
both loop momenta are much larger than the external momentum, this 
region can again be compensated by a local counterterm. The relevant 
operator is
\begin{equation}
{\cal O}_8 = \frac{1}{g_s^4}\,\partial_\nu F^{\nu\mu}\partial^\rho
F_{\rho\mu},
\end{equation}
as shown in the lower diagram of Fig.~\ref{fig5}. This gives 
rise to a $1/n$-correction to the leading asymptotic behaviour 
of perturbative coefficients in order $\alpha_s^{n+1}$, 
cf.~(\ref{dex}). In the upper 
diagram of Fig.~\ref{fig5}, after contraction of the vertex subgraph, 
one can have $p\gg q$ or $p\sim q$. 
In the first case, we get the double pole as already mentioned. 
The loop integral over $p$ in the upper diagram of Fig.~\ref{fig5}  
can also be contracted and one obtains another 
contribution to the coefficient of ${\cal O}_8$ as indicated by the 
vertical arrow in the figure. The important point 
to note is that the second factor $1/(1+u)$ that comes from 
the loop integration over $p$ is related to 
the {\em logarithmic} contribution $d^4p/p^4$ in the upper 
diagram of Fig.~\ref{fig5} and hence it is 
related to the entry in the anomalous 
dimension matrix of dimension-6 operators that describes mixing of 
${\cal O}_6$ into ${\cal O}_8$. Thus, this 
contribution to the coefficient of ${\cal O}_8$ is the product of 
the coefficient of ${\cal O}_6$ and an entry of the mixing matrix. 
In the second case, $p\sim q$, the integration over $p$ does not produce 
further singularities in $u$ and the net result is $1/(1+u)$ from the 
insertion of ${\cal O}_6$. The residue of this pole is determined by the 
coefficient of ${\cal O}_6$ times the value of the 
one-loop $p$-integral. Because $p\sim q$, the residue is 
non-polynomial in $q$. (It contains a logarithm of $q^2$.)

To summarize, the singularity at $u=-1$ of the two-point function of 
two currents at leading order in the flavour expansion is described 
by two universal constants, one from the one-loop vertex subgraph and 
the other from the region $k\sim p\gg q$. They are associated with 
the operators ${\cal O}_6$ and ${\cal O}_8$ respectively. 

Let us draw an analogy with counterterms that arise in the ordinary 
renormalization process in dimensional regularization, for instance. 
A two-loop diagram, in general, has a double pole in $\epsilon$. The 
double pole arises from large and ordered loop momenta and can be 
expressed recursively in one-loop subgraphs. The coefficient of the double 
pole is already determined by one-loop renormalization group functions. 
The single pole in $\eps$ is in general 
non-local in external momenta. The non-locality comes from the region 
where only one-loop momentum is large and the non-local 
contribution to the single pole is determined in terms of 
the UV divergence of a one-loop subgraph. The genuine two-loop 
contribution to the single pole (and two-loop anomalous dimension 
functions) arises from the region where both loop momenta are of the same 
order and large. The analogy with the discussion of the singularity 
at $u=-1$ is clear.

{\em Two chains.} The case of two chains is only slightly more 
involved. Consider as an example the two-chain vertex diagram in 
Fig.~\ref{fig6}. Call the regularization parameter of the left chain 
$u_1$ and the other one $u_2$. Let the loop momentum $k_1$ run through 
the `inner' vertex subgraph and $k_2$ through the box subgraph. 
There are two maximal forests. (Others lead to vanishing scaleless 
integrals.) The first, ${\cal F}_1$, consists of the inner vertex 
subgraph and the 
diagram itself, the second, ${\cal F}_2$, of the box subgraph 
and the diagram itself.
According to (\ref{singofg}) the leading singularities are
\begin{eqnarray}
\label{vertsub}
{\cal F}_1:&& \frac{1}{1+u_1+u_2}\,\frac{1}{1+u_1} \qquad\quad\,\,
\longrightarrow\quad
\frac{\ln(1+u)}{1+u},
\\
\label{boxsub}
{\cal F}_2:&& \frac{1}{1+u_1+u_2}\,\frac{1}{1+u_1+u_2} \quad
\longrightarrow\quad\frac{1}{(1+u)^2}.
\end{eqnarray}
The arrows indicate the resulting singularity of the Borel transform after 
integration over $u_{1,2}$ according to (\ref{relbg}). The second 
forest, containing the box subgraph, results in a double pole, to be 
compared with the single pole at leading order in the flavour expansion 
(Fig.~\ref{fig4}a). This translates into an enhancement of the 
large-order behaviour of perturbative coefficients by a factor of 
$n$, which was first noted in \cite{VZ94}. This enhancement can also 
be obtained by counting logarithms of loop momentum $\ln k^2$. 
It should also be taken into account that there are of the order of $n$ 
ways to distribute $n$ fermion loops over the two photon lines of the 
box subgraph. Viewed this way, the enhancement is combinatorial in 
origin.
\begin{figure}[t]
   \vspace{-6.1cm}
   \epsfysize=35cm
   \epsfxsize=23.33cm
   \centerline{\epsffile{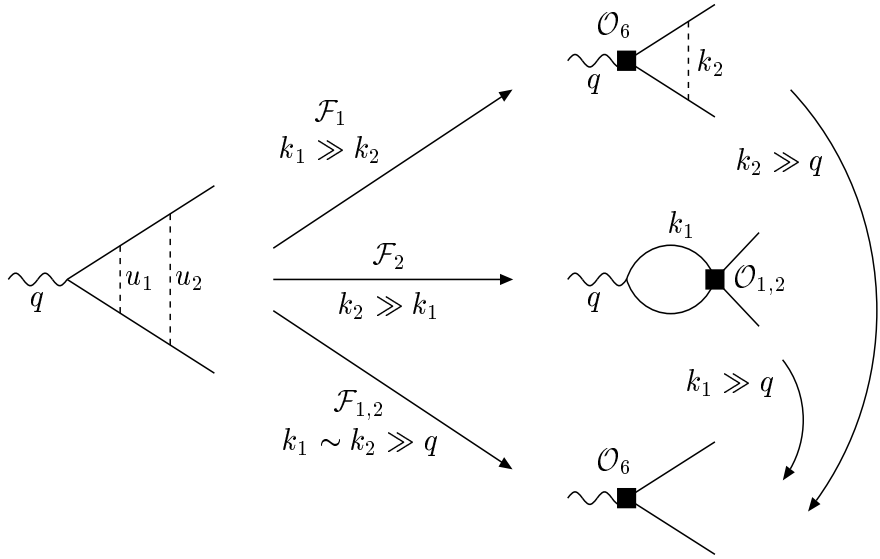}}
   \vspace*{-20.7cm}
\caption[dummy]{\small A two-chain vertex integral and the 
contributions to its UV renormalon singularity. The straight arrows 
indicate the contractions that lead to insertions of dimension-6 
operators. The arrows to the right indicate contractions that 
correspond to logarithmic operator mixing among the dimension-6 
operators. \label{fig6}}
\end{figure}

Let us analyse again in more detail the relation between 
singular terms near $u=-1$ and loop momentum regions. A pictorial 
representation of this relation is shown in Fig.~\ref{fig6}. 

We begin with the forest ${\cal F}_1$. When $k_1\gg k_2$, the inner 
vertex can be contracted. Because it contains only the chain with 
parameter $u_1$, the result is a singular factor $1/(1+u_1)$. Its residue 
can be described by an insertion of ${\cal O}_6$ with the coefficient 
already determined from the singularity of the one-loop  
vertex function at leading order in the flavour expansion. When 
$k_2\gg q$, in addition to $k_1\gg k_2$, one obtains a factor $1/(1+u)$, 
in addition to $1/(1+u_1)$ from the first contraction, 
because the contracted graph contains both $u_i$. The residue of this 
pole is proportional to the logarithmic mixing of ${\cal O}_6$ into 
itself. The result (\ref{vertsub}) then follows. 
Compared to leading order in the flavour 
expansion, there is an additional factor $\ln(1+u)$, which 
translates into a $\ln n$ in the large-order behaviour. The 
logarithm is due to (part of the) anomalous dimension 
of ${\cal O}_6$.\footnote{After summation of all diagrams,  
the anomalous dimension of ${\cal O}_6$ is found to vanish.} 
We can therefore 
identify (part of) $b^{[1]}$ in (\ref{rnnfexp}) with this anomalous 
dimension. When $k_2\sim q$, 
there is no further singular factor and we end up with $\ln(1+u)$ 
in the Borel transform. Using (\ref{bpolelog}), this corresponds to 
a $1/n$-suppression in large orders relative to leading order in the 
flavour expansion. It can be obtained from the order-$\alpha_s$ 
correction to the vertex function with one insertion of 
${\cal O}_6$. Hence (part of) $c_1^{[1]}$ follows from a 
first-order perturbative 
calculation. Finally, when $k_1\sim k_2\gg q$, the entire two-loop graph 
is contracted as indicated by the lowest arrow in Fig.~\ref{fig6}. The 
result is a single singular factor $1/(1+u)$ and one obtains a new 
contribution to the coefficient function of ${\cal O}_6$, which corrects 
the leading order coefficient function by an amount suppressed by 
$1/N_f$. This is a contribution to $K^{[1]}$ in (\ref{rnnfexp}). 

Turning to ${\cal F}_2$, the discussion can be essentially repeated. 
Note only that the box subgraph leads us to introduce two four-fermion 
operators
\begin{eqnarray}
{\cal O}_1 &=& (\bar{\psi}\gamma_\mu\psi)(\bar{\psi}\gamma^\mu\psi),
\\
{\cal O}_2 &=& (\bar{\psi}\gamma_\mu\gamma_5\psi)
(\bar{\psi}\gamma^\mu\gamma_5\psi).
\end{eqnarray}
Furthermore, one obtains an enhancement by a factor of $n$ rather 
than $\ln n$ from mixing of ${\cal O}_{1,2}$ into ${\cal O}_6$, because 
the box subgraph contains two chains so that $u({\rm box})=u_1+u_2$ 
in (\ref{singofg}). This results in (\ref{boxsub}). 
Since in (\ref{rnnfexp}) we assumed that 
the large-order behaviour has only one component, we do not 
identify the contributions from ${\cal F}_2$ with the parameters 
of (\ref{rnnfexp}).

It is clear from this example 
how the interpretation of singularities extends to 
diagrams with any number of chains and that the combinatorial 
structure is identical to the one that arises in ordinary 
renormalization of Feynman integrals. As far as the singular point 
$u=-1$ is concerned, there can be an insertion of exactly one 
dimension-6 counterterm and then logarithmic operator mixing. 
An important point to note is that the region 
$k_1\sim k_2\sim\ldots\sim k_m\gg q$ in an $m$-chain contribution 
to the vertex function results only in a simple pole $1/(1+u)$ whose 
residue is not related to that of lower-order subgraphs. Hence 
it corrects the coefficient function of ${\cal O}_6$ at some order 
in the $1/N_f$-expansion, but with a numerical coefficient of order 
unity otherwise. {\em Beyond} the flavour expansion, the coefficients 
of the dimension-6 operators must therefore be considered as 
non-perturbative constants in the sense that they receive 
unsuppressed contributions from classes of diagrams with any number 
of chains. The fact that the over-all normalization $K$ 
cannot be calculated has been emphasized in 
\cite{Gru93a,Ben93b,VZ94}. 

Consider now the second diagram in Fig.~\ref{fig4}b. 
This diagram can be thought of as a chain inserted in one of the bubbles 
of a chain. It is a correction to the effective propagator 
(\ref{effprop}) and in this sense `universal'. This diagram is 
special for the following reason: up to now we have only considered 
the possibility that a forest of large-momentum subgraphs 
gives rise to a dimension-6 operator insertion from the largest 
loop momentum followed by logarithmic mixing among these operators. 
However, in general it is possible that the smallest subgraph 
in a forest has a logarithmic ultraviolet divergence, which 
gives $1/u_1$, proportional to a dimension-4 counterterm. One can 
then pick up the $d^4k/k^6$ piece from the reduced diagram, in which 
the smallest subgraph is contracted, and obtain $1/((1+u)\,u_1)$ 
in total. In other words, the logarithmic mixing of dimension-4 
operators is followed by one insertion of a dimension-6 operator. 
In individual graphs of the type shown in the left half of 
Fig.~\ref{fig4}b, such contributions to the singularity at 
$u=-1$ exist. However, the Ward identity of QED implies that 
all such contributions cancel and we therefore ignored them. 
The only non-cancelling renormalization parts of the electromagnetic 
vertex function reside in the photon vacuum polarization. They 
first appear in the second diagram of 
Fig.~\ref{fig4}b.\footnote{For the following discussion we 
imply that the self-energy type contributions are added inside the 
vacuum polarization insertion in Fig.~\ref{fig4}b.} 

Call the Borel parameter of the chain in the bubble $u_3$ and those of 
the chains that connect to the bubble $u_{1,2}$. The singularities from the 
two-loop/one-chain vacuum polarization have already been discussed in 
part. From $u_3\to -1$, one obtains $1/((1+u)\,(1+u_3)^2)$, which 
after integration over the $u_i$ results in $\ln(1+u)/(1+u)$. 
In the sum of all diagrams, this contribution is always cancelled 
\cite{BS96}. But the vacuum polarization is also UV divergent and this 
results in a single pole $1/u_3$ for the two-loop vacuum polarization 
subdiagram with coefficient proportional to 
$\beta_{1}=N_f/(4\pi)^2$, the two-loop coefficient of the QED 
$\beta$-function. After adding the diagram with the charge renormalization 
counterterm, the singularity structure is
\begin{equation}
\label{singandcount}
\frac{K_{\rm vert}^{[0]}}{1+u_1+u_2+u_3}\,\frac{\beta_1}{u_3} - 
\frac{K_{\rm vert}^{[0]}}{1+u_1+u_2}\,\left[\frac{\beta_1}{u_3} - 
\,\mbox{finite terms}\right],
\end{equation}
where the second term comes from the counterterm and $K_{\rm vert}^{[0]}$ 
is the 
residue of the simple pole in the one-chain vertex graph 
(Fig.~\ref{fig4}a). Note that in the counterterm the first factor 
has no $u_3$. This can be seen as follows: let $k$ be the momentum 
of the two photon lines with indices $u_{1,2}$ that join to the 
vacuum polarization insertion. On dimensional grounds the 
vacuum polarization insertion is proportional to $(-\mu^2/k^2)^{u_3}$;  
this factor combines with the other two chain propagators 
to $u_1+u_2+u_3$. On the other hand the counterterm insertion has 
no momentum dependence and the two chain propagators combine 
with index $u_1+u_2$ only. The UV divergence at $u_3=0$ cancels 
in the difference (\ref{singandcount}) and one obtains, 
after integration over the $u_j$, 
\begin{equation}
\label{leadbeta1}
-\frac{K_{\rm vert}^{[0]}}{1+u}\,\beta_1\ln(1+u)
\end{equation}
for the most singular term. It yields a $\ln n$ enhancement in the 
large-order behaviour relative to the leading order vertex in 
the flavour-expansion and gives another contribution to $b^{[1]}$ in 
(\ref{rnnfexp}). The finite terms in (\ref{singandcount}) 
are renormalization scheme dependent. If we assume that the subtractions 
do not themselves introduce factorial divergence -- as is true in 
$\overline{\rm MS}$-like schemes --, the subtraction dependent 
singular terms are $\ln(1+u)$, and hence are suppressed by 
one power of $n$ relative to the leading order vertex in 
the flavour expansion. This scheme-dependence affects only 
$c_1^{[1]}$ of (\ref{rnnfexp}). 
The $\ln n$-enhancement from inserting a chain 
into a chain has been noted in \cite{Z92}. This paper also 
demonstrates diagrammatically how these logarithms exponentiate 
to $n^{\beta_1/\beta_0^2}$, when one iterates the process of inserting 
chains into chains. We will see later how this exponentiation 
follows from renormalization group equations. 

In addition to the leading singularity (\ref{leadbeta1}), one also 
obtains $1/(1+u)$ with a residue that does not factorize into 
a one chain residue and an anomalous dimension (such as $\beta_1$).  
It is to be interpreted as a $1/N_f$ correction to the coefficient 
function of ${\cal O}_6$. This correction was noted  
by \cite{Gru93a,BZ93} and provided the first diagrammatic 
evidence that the over-all renormalization of renormalon divergence 
cannot be computed without resorting to the flavour expansion. 
Its value was calculated in \cite{Ben95} and  
can also be inferred from \cite{Bro93}.

To summarize: the parameter $b$ in the asymptotic estimate of 
(\ref{exl}) follows from the anomalous dimension matrix of dimension-6  
operators and the $\beta$-function. A rather straightforward extension 
of the above analysis leads to the conclusion that only one-loop 
anomalous dimensions and the two-loop $\beta$-function are required. 
Higher coefficients contribute to pre-asymptotic corrections 
parametrized by $c_1$ etc.. These pre-asymptotic corrections 
are also accessible through calculations involving a {\em finite} 
number of loops. In particular, in addition to two-loop anomalous dimensions 
and the three-loop $\beta$-function, the one-loop 
corrections to Green functions with one zero-momentum insertion 
of a dimension-6 operator are requiresd. {\em Only} the normalization 
$K$ cannot 
be computed in a finite number of loops. But its scheme-dependence 
is trivial and arises only through the counterterm for the 
simple fermion loop ($C$ in (\ref{simpleloop})). All other 
scheme-dependence is $1/n$-suppressed, see Section~\ref{schemedep} for 
a further discussion of scheme dependence.

Finally, we mention that the leading large-$n$ behaviour can also 
be found by counting logarithms of loop momentum. (Recall the 
remarks in Section~\ref{sectbascon}.) The logarithms arise from 
the running coupling and logarithmically divergent loop integrals.
For diagrams with two chains 
one has to take care of the correct argument of the coupling and the 
hierarchy of loop momenta. The contribution from the forest 
${\cal F}_2$ is treated by this method in \cite{VZ94}. 

\subsubsection{QCD}
\label{qcd}
There exists no complete analysis of non-abelian diagrams at the time 
of writing. 
The analysis of multi-chain diagrams in QED showed that higher order 
corrections in $1/N_f$ do not modify the location of UV renormalon 
singularities, but only their `strength', specified by $b$ in 
(\ref{exl}). Not so in QCD, where one expects that higher order 
contributions move the singularity from $m/\beta_{0f}$ to $m/\beta_0$ 
after a partial resummation of the flavour expansion. 
As seen from (\ref{rnnfexp}) this shift should be visible in the 
flavour expansion as systematically enhanced corrections of the form
\begin{equation}
\label{enhanced}
\left(\frac{\beta_0-\beta_{0f}}{\beta_{0f}}\right)^k\,n^k\equiv 
\left(\frac{\delta\beta_0}{\beta_{0f}}\right)^k\,n^k
\end{equation}
at order $1/N_f^k$. The goal of this section is to identify the 
new elements in the non-abelian theory that lead to precisely this 
factor for $k=1$. The general pattern for arbitrary $k$ should then 
be transparent also.
\begin{figure}[t]
   \vspace{-5.3cm}
   \epsfysize=30cm
   \epsfxsize=20cm
   \centerline{\epsffile{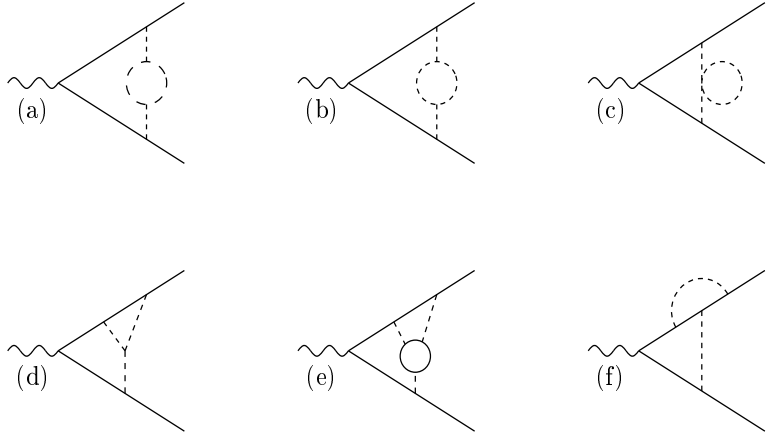}}
   \vspace*{-17.9cm}
\caption[dummy]{\small Some non-abelian vertex diagrams (a-e) 
at next-to-leading order in the flavour expansion. The long-dashed 
circle denotes a ghost loop. \label{fig7}}
\end{figure}

Consider again pair creation from an external abelian vector current 
and the first UV renormalon singularity at $u=-1$. 
The additional non-abelian diagrams are shown in Fig.~\ref{fig7}a-e. 
In QCD gauge cancellations are more complicated than in QED and,  
in a general covariant gauge,  the longitudinal piece in the 
chain propagator (\ref{chainprop}) 
has to be kept. It is convenient to perform the 
analysis in Landau gauge, $\xi=0$. If one chooses another gauge, 
one encounters UV divergent subgraphs, which are not regulated 
by one of the regularization parameters $u_i$, because the 
longitudinal part of the chain propagator carries no $u_i$. It is 
then necessary to choose another intermediate regularization for these 
subgraphs. For a complete treatment, this is also 
necessary for fermion loops with three and four gluon legs.

According to (\ref{enhanced}) we should find contributions to the 
large-order behaviour which are enhanced by one power of $n$. Some of 
the contributions of this type are clearly not related to the 
$\beta$-function, for example the contribution from the 
box subgraph/four-fermion operator insertions to the diagram of 
Fig.~\ref{fig6}. It is not difficult to keep  
these contributions apart from those relevant to restoring the non-abelian 
$\beta$-function. Consider the non-abelian vertex subgraph $\gamma$ 
of diagram \ref{fig7}d. It is logarithmically UV divergent. When the loop 
momentum of $\gamma$ is large compared to all other momenta, the 
subgraph can be contracted to a point. The two leading contributions 
in its UV behaviour correspond to a dimension-4 counterterm 
($u_0(\gamma)=0$ in (\ref{singofg})) and a dimension-6 counterterm 
($u_0(\gamma)=-1$ in (\ref{singofg})). The second contribution, where 
one picks up the $d^4k/k^6$ term from the first loop, is of the 
same type as discussed for QED. The first contribution, however, has 
an obvious connection with the $\beta$-function and we follow 
only this type of contribution in this section.

The first coefficient of the $\beta$-function follows from the 
one-loop pole part of the charge renormalization constant 
\begin{equation}
\label{renconst}
Z_g=Z_1 Z_3^{-1/2} Z_2^{-1},
\end{equation}
where $Z_2$ is the quark wave function renormalization constant, $Z_3$ the 
gluon wave function renormalization constant, and $Z_1$ the renormalization 
constant for the quark-gluon vertex. Note that at the one-loop order a pole 
in $1/\eps$ in dimensional regularization is in one-to-one 
correspondence with a pole at $1/u(\gamma)$ in a logarithmically 
UV divergent subgraph $\gamma$. 
In QED $Z_1=Z_2$ and the only non-cancelling 
logarithmically divergent subgraphs occur in the photon vacuum 
polarization. We have already seen that these subgraphs give 
rise to a logarithmically enhanced contribution to the large-order 
behaviour proportional to the second coefficient of the 
$\beta$-function. In QCD one has to keep track of all other 
logarithmically UV divergent subgraphs and their counterterms. 

There is a potential difficulty in QCD, because the gluon self-energy 
is quadratically divergent by power counting. A quadratic divergence 
gives rise to a UV renormalon singularity at $u=+1$ on top of an IR 
renormalon singularity at the same position. Consider for example 
the tadpole diagram in Fig.~\ref{fig7}c. It contains
\begin{equation}
\int\frac{d^dk}{(k^2)^{1+u_1}},
\end{equation}
which is zero in dimensional regularization, but should be interpreted 
as a UV and IR renormalon pole at $u_1=1$ with opposite signs. A 
UV renormalon at $u=1$ would complicate the discussion, because a 
singularity at $u=-1$ could in principle be obtained by inserting 
a dimension-2 counterterm first and then a dimension-8 operator. However, 
gauge invariance requires the gluon self-energy to have the tensor 
structure $g_{\mu\nu}-k_\mu k_\nu/k^2$ in the external gluon momentum $k$, 
while a non-cancelling quadratic divergence 
would have the tensor structure of the metric tensor $g_{\mu\nu}$. 
Consequently, the pole at $u=1$ should be interpreted as purely infrared. 
Another way to say this is that there is no gauge-invariant 
dimension-2 operator in QCD that could serve as a counterterm. 

Consider the diagrams of Fig.~\ref{fig7} explicitly. Diagram \ref{fig7}a 
is obtained by substituting one fermion loop by a ghost loop,
\begin{equation}
\beta_{0f}\left[\ln(-k^2/\mu^2)+C\right] 
\to
-\frac{N_c}{48\pi} \left[\ln(-k^2/\mu^2)+C^\prime\right], 
\end{equation}
where $N_c$ is the number of colours. The enhancement of diagram \ref{fig7}a 
by a factor of $n$ in the large-order behaviour is combinatorial: 
at order $n+1$ in perturbation theory the chain in the leading order 
diagram of Fig.~\ref{fig4} has $n$ loops and there are $n$ ways to 
replace one fermion loop by a ghost 
loop. In terms of the Borel transform, since one factor of $n$ 
is equivalent to one factor $1/(1+u)$, the ghost loop diagram results 
in
\begin{equation}
\label{ghostcon}
\frac{K_{\rm vert}^{[0]}}{1+u}\,\left(-\frac{N_c}{48\pi\beta_{0f}}\right)
\frac{1}{1+u}.
\end{equation}
For the gluon loop (diagram \ref{fig7}b) the same argument leads to 
a contribution  
\begin{equation}
\label{gluoncon}
\frac{K_{\rm vert}^{[0]}}{1+u}\,\left(-\frac{25 N_c}{48\pi\beta_{0f}}\right)
\frac{1}{1+u}.
\end{equation}
The full contribution from the gluon loop is more involved, 
because the gluon loop itself consists of chains. This results in 
further singular terms at $u=-1$, but they are not related to the 
$\beta$-function. Turning to diagram \ref{fig7}d, 
we pick up the logarithmic UV divergence of the vertex subgraph as 
discussed above. Together 
with the counterterm, the relevant singularity structure is 
\begin{equation}
\frac{1}{u_2+u_3}\left(\frac{1}{1+u_1+u_2+u_3}-\frac{1}{1+u_1}\right),
\end{equation}
where $u_1$ is the parameter of the lower chain in Fig.~\ref{fig7}d 
and $u_{2,3}$ are the parameters in the vertex subgraph. 
Integrating over the $u_i$ one 
obtains $1/(1+u)$ as the leading singularity at $u=-1$. However, 
when using (\ref{convolution}) for the diagram with three chains, 
we assumed three powers of $\alpha_s$ 
from the vertices while diagram \ref{fig7}d has only two. To 
compensate for the factor $1/\alpha_s$, one has to take one derivative 
in $u$. The result, putting in the correct constants and taking into 
account that there is an identical contribution from a symmetric diagram, 
reads
\begin{equation}
\label{threegluoncon}
\frac{K_{\rm vert}^{[0]}}{1+u}\,\left(-\frac{3 N_c}{8\pi\beta_{0f}}\right)
\frac{1}{1+u}.
\end{equation}
The factor $1/\beta_{0f}$ arises from the prefactor in 
(\ref{convolution}). 
The tadpole diagram \ref{fig7}c vanishes. The only logarithmically 
divergent subgraph of diagram \ref{fig7}e is the fermion loop. 
However, since the fermion loop itself produces no singularities 
in any $u_i$, this region can be considered as an order-$\alpha_s$ 
renormalization of the three-gluon vertex. Hence, for the present 
discussion, this diagram can be considered $1/n$-suppressed relative 
to diagram \ref{fig7}d. The diagram \ref{fig7}f has to be 
reconsidered, because its colour factor in the non-abelian case 
is $C_F (C_F-N_c/2)$. The $C_F^2$-part of the logarithmic UV divergence 
cancels with a self-energy insertion as in the abelian case. The 
non-abelian part contributes to $Z_1/Z_2$ in general. But in 
Landau gauge the vertex subgraph is in fact UV finite and diagram 
\ref{fig7}f does not contribute to $Z_1/Z_2$. Adding together the three 
non-vanishing contributions (\ref{ghostcon}), (\ref{gluoncon}) and 
(\ref{threegluoncon}), one obtains
\begin{equation}
\label{sumofnonab}
\frac{K_{\rm vert}^{[0]}}{1+u}\,\left(-\frac{11 N_c}{12\pi\beta_{0f}}\right)
\frac{1}{1+u}
\end{equation}
or, since $\delta\beta_0=-(11 N_c)/(12\pi)$, 
\begin{equation}
\label{recna}
r_n \sim K_{\rm vert}^{[0]}\,\beta_{0f}^n\,n!\left[\frac{\delta\beta_0}
{\beta_{0f}}\,n\right].
\end{equation}
Comparison with (\ref{rnnfexp}) shows that this is exactly what is needed 
at sub-leading order in the flavour expansion to restore the non-abelian
$\beta$-function.

As already mentioned, (\ref{sumofnonab}) represents only a 
fraction of all contributions to the singularity at $u=-1$ 
from the non-abelian diagrams. The ones not discussed should be associated 
with insertions of dimension-6 operators for some of the 
subgraphs of the diagrams and their interpretation parallels the QED 
case. A complete analysis of these contributions remains to be done.

Expanding (\ref{exl}) to yet higher order in $1/N_f$, one obtains terms 
of the form $(\delta\beta_0)^2$, $\delta\beta_0 b^{[1]}$, $(b^{[1]})^2$ 
enhanced by $n^2$, $n\ln n$ and $\ln^2 n$, respectively. The origin 
of these terms is roughly as follows: consider a forest of 
nested subgraphs $\gamma_1\subset\gamma_2\subset\Gamma$ of a three-loop 
diagram at next-to-next-to-leading order in the flavour expansion. 
Assume that $\gamma_{1,2}$ are both logarithmically UV  
divergent. Then (\ref{singofg}) permits three contributions to the 
singularity at $u=-1$ from such a forest:
\begin{eqnarray}
\mbox{(i)}\,:&& u_0(\gamma_1)=0\quad u_0(\gamma_2)=0\quad u_0(\Gamma)=1,
\nonumber\\
\mbox{(ii)}\,:&& u_0(\gamma_1)=0\quad u_0(\gamma_2)=1\quad u_0(\Gamma)=1,
\\
\mbox{(iii)}\,:&& u_0(\gamma_1)=1\quad u_0(\gamma_2)=1\quad u_0(\Gamma)=1.
\nonumber
\end{eqnarray}
The first line amounts to picking up the logarithmic UV divergences 
in the first two subgraphs. This is a contribution to the terms of the  
form $K_{\rm vert}^{[0]}(\delta\beta_0)^2$.\footnote{The required enhancement 
by $n^2$ is most easily seen in the contribution from two ghost loops 
in one chain. In this particular case it can be obtained again from 
a simple counting argument.} The third line amounts 
to picking up the $d^4k/k^6$ term in $\gamma_1$. The subsequent two 
contractions can then be associated with logarithmic operator mixing 
among dimension-6 operators. This is a contribution to terms of the 
form $(b^{[1]})^2$. The second line represents the obvious intermediate 
case. This discussion has been crudely simplified in that we ignored 
again that four-fermion operators also lead to an enhancement by 
a factor of $n$. However, this enhancement occurs only once. The effect 
of the anomalous dimension of these operators is then $\ln n$ for 
every loop in the flavour expansion. 

Despite the somewhat sketchy treatment of the QCD case, the 
general pattern that leads to the restoration of the non-abelian 
$\beta_0$ seems to be simple. It confirms the heuristic argument that 
$\beta_0$ has to appear, because this coefficient is tied to the 
leading ultraviolet logarithms.

It would be nice to recover the full QCD $\beta_0$ already from 
vacuum polarization subgraphs in order to preserve the association 
of renormalons with the running coupling at each vertex, which is suggested 
by abelian theories. This can indeed be done \cite{Wat97}, at least 
at one loop, by a diagrammatic rearrangement (the `pinch technique') 
that absorbs parts of the vertex graphs into an `effective charge'. 
As far as large-order behaviour is concerned, one then has to 
demonstrate that after this rearrangement no contributions enhanced 
by a factor of $n$ (and not related to dimension-6 insertions) 
are left over.

\subsubsection{Renormalization group analysis}
\label{rganalysis}

We have treated the diagrammatic approach at length in order to familiarize 
the reader with the idea that UV factorization can be applied 
to the problem of UV renormalons. Diagrammatically in the flavour 
expansion a recursive construction of operator insertions emerges, which 
is completely analogous to the recursive structure of renormalization 
in an expansion in the coupling, except that  
higher-dimension operators are implied in the case of renormalons. 
This paves the ground to 
introducing the renormalization group treatment, originally suggested 
by \cite{P78} and exemplified in the scalar $\phi^4$-theory. The idea 
was worked out for QCD in \cite{BBK97}, on which this section is based. 

The renormalization group equations are formulated most easily for 
ambiguities or, equivalently, imaginary parts of Borel-type 
integrals introduced in 
Section~\ref{divseries}. In QCD UV renormalons lie on the negative 
Borel axis and do not lead to ambiguities. It is technically convenient 
to consider the integral
\begin{equation}
\label{irr}
I[R](\alpha_s) = \int\limits_{0+i\epsilon}^{-t_c+i\epsilon} d t\,
e^{-t/\alpha_s}\,B[R](t) \qquad 
-\frac{2}{\beta_0}>t_c>-\frac{1}{\beta_0}>0,
\end{equation}
given a series expansion $R$ and its Borel transform as defined 
in (\ref{defborelt}). The integral is 
complex and its imaginary part is unambiguously related to the 
first UV renormalon singularity at $t=1/\beta_0$ ($u=-1$) or large-order 
behaviour (compare (\ref{div}) and (\ref{im})). 

The statement of factorization is that the imaginary part of 
$I[R]$ can be represented as 
\begin{equation}
\label{parisi}
\mbox{Im}\,I[R](\alpha_s,p_k) = \frac{1}{\mu^2}\sum_i
C_i(\alpha_s)\,R_{{\cal O}_i}(\alpha_s,p_k).
\end{equation} 
In this equation ${\cal O}_i$ denote dimension-6 operators and 
$R_{{\cal O}_i}$ the Green function from which $R$ is derived with a 
single zero-momentum insertion of ${\cal O}_i$. $C_i(\alpha_s)$ 
are the coefficient functions, which are independent of any external 
momentum $p_k$ of $R$ and in fact independent of the quantity $R$. 
They play the same role as the universal renormalization constants 
in ordinary renormalization. 
The coefficient function being universal, 
the dependence of the UV renormalon 
divergence on the observable $R$ is contained in the factors 
$R_{{\cal O}_i}$. These factors can be computed order by order 
in $\alpha_s$ by conventional methods. 
The dimension-6 operators 
may be thought of as an additional term,  
\begin{equation}
\Delta {\cal L} = -\frac{i}{\mu^2}\sum_i
C_i(\alpha_s)\,{\cal O}_i, 
\end{equation}
in the QCD Lagrangian with coefficients such that for {\em any} $R$ the 
imaginary part of $I[R]$ is compensated by the additional 
contribution to $R$ from $\Delta {\cal L}$. From the requirement 
that $\Delta {\cal L}$ be independent of the renormalization scale $\mu$ 
or from a comparison of the renormalization group equations satisfied by  
$I[R]$ and $R_{{\cal O}_i}$ it can be derived that 
\begin{equation}
\label{rge}
\left[\left(\beta(\alpha_s)\frac{d}{d\alpha_s} - 1\right)\delta_{i j} - 
\frac{1}{2}\gamma_{ij}(\alpha_s)\right] C_j(\alpha_s) = 0,
\end{equation}
where $\gamma(\alpha_s)$ is the anomalous dimension matrix of the 
dimension-6 operators 
defined such that the renormalized operators satisfy
\begin{equation}
\label{defandim}
\left(\delta_{ij}\,\mu\frac{d}{d\mu}+\gamma_{ij}\right)
{\cal O}_j = 0.
\end{equation}
The unusual `$-1$' in (\ref{rge}) originates from the factor 
$1/\mu^2$ in (\ref{parisi}). 
The solution to the differential equation (\ref{rge}) can be written as
\begin{equation}
\label{solcoeff}
C_i(\alpha_s) =
e^{-1/(\beta_0\alpha_s)}\alpha_s^{-\beta_1/\beta_0^2}\,F(\alpha_s)\,
E_i(\alpha_s),
\end{equation}
where 
\begin{equation}
\label{fi}
F(\alpha_s) = \exp\left(\,\int\limits_0^{\alpha_s} dx\left[
\frac{1}{\beta_0 x^2}-\frac{\beta_1}{\beta_0^2 x}-\frac{1}{\beta(x)}
\right]\right)
\end{equation}
has a regular series expansion in $\alpha_s$ and incorporates the 
effect of terms of higher order than $\beta_1$ in the $\beta$-function 
and 
\begin{equation}
\label{ei}
E_i(\alpha_s) = \exp\left(\,\int\limits_{\alpha_0}^{\alpha_s} dx\,
\frac{\gamma_{ij}^T(x)}{2 \beta(x)}\right) \hat{C}_j
\end{equation}
takes into account the anomalous dimension matrix. Thus, the 
coefficient functions are determined 
up to $\alpha_s$-independent 
integration constants $\hat{C}_i$.\footnote{The lower limit 
$\alpha_0$ in (\ref{ei}) is arbitrary. A change of $\alpha_0$ can be 
compensated by adjusting the integration constants.} Because 
the $\alpha_s$-dependence in (\ref{irr}) translates into $n$-dependence 
of large-order behaviour, we deduce that this 
$n$-dependence is completely determined. Only over-all normalization factors 
related to the integration constants do not follow from the renormalization 
group equation. However, these integration constants are 
process-independent numbers; they depend only on the Lagrangian that 
specifies the theory. It is in this precise sense that ultraviolet 
renormalon divergence is universal.

When (\ref{parisi}), together with the solution 
for the coefficient functions, 
is translated into large-order behaviour and expanded 
formally in $1/N_f$, one can verify that it is consistent with the 
diagrammatic analysis. The unspecified integration constants are 
related to the over-all normalization of renormalon singularities. 
We have already seen that 
its calculation requires more input than renormalization group 
properties. Since in fact we concluded that it cannot be calculated 
at finite $N_f$, it follows that the renormalization group treatment 
already gives everything one can hope to obtain for UV 
renormalons without approximations. 

To proceed we specify a basis of dimension-6 operators. In general, one 
is also interested in processes induced by external currents. For simplicity, 
we consider only vector and axial-vector currents and we let them be 
flavour singlets. Thus, 
in expressions like $(\bar{\psi} M\psi)$, a sum over flavour, colour and 
spinor indices is implied, and $M$ is a matrix in colour and spinor space, 
but unity in flavour space. The generalization to broken flavour symmetry 
will be  indicated below.
To account for the external currents, two (abelian) 
background fields $v_\mu$ and 
$a_\mu$, which couple to the vector and axial-vector current, are introduced. 
Their fields 
strengths $F_{\mu\nu}=\partial_\mu v_\nu-\partial_\nu v_\mu$ and 
$H_{\mu\nu}=\partial_\mu a_\nu-\partial_\nu a_\mu$ satisfy 
$\partial_\mu F^{\mu\nu}=j_V^\nu$ and 
$\partial_\mu H^{\mu\nu}=j_A^\nu$. A basis of dimension-6 
operators is then given by 
\begin{eqnarray}
{\cal O}_1 &=& (\bar{\psi}\gamma_\mu\psi) (\bar{\psi}\gamma^\mu\psi)
\qquad\qquad\,\,\,
{\cal O}_2 \,=\, (\bar{\psi}\gamma_\mu\gamma_5\psi) (\bar{\psi}
\gamma^\mu\gamma_5\psi)
\nonumber\\[0.2cm]
{\cal O}_3 &=& (\bar{\psi}\gamma_\mu T^A\psi) (\bar{\psi}\gamma^\mu 
T^A \psi)
\qquad
{\cal O}_4 \,=\, (\bar{\psi}\gamma_\mu\gamma_5 T^A\psi) (\bar{\psi}
\gamma^\mu\gamma_5 T^A\psi)
\nonumber
\end{eqnarray}
\vspace*{-0.6cm}
\begin{eqnarray}
\label{basis}
{\cal O}_5 &=& \frac{1}{g_s}\,f_{ABC} \,G_{\mu\nu}^A G_\rho^{\nu\,B} 
G^{\rho\mu\,C} 
\end{eqnarray}
\vspace*{-0.6cm}
\begin{eqnarray}
{\cal O}_6 &=& \frac{1}{g_s^2}\,(\bar{\psi}\gamma_\mu\psi)\,
\partial_\nu F^{\nu\mu}
\qquad
{\cal O}_7 \,=\, \frac{1}{g_s^2}\,(\bar{\psi}\gamma_\mu\gamma_5\psi)\,
\partial_\nu H^{\nu\mu}
\nonumber\\
{\cal O}_8 &=& \frac{1}{g_s^4}\,\partial_\nu F^{\nu\mu}\,
\partial^\rho F_{\rho\mu}
\qquad\quad\!\!
{\cal O}_9 \,=\, \frac{1}{g_s^4}\,\partial_\nu H^{\nu\mu}\,
\partial^\rho H_{\rho\mu},
\nonumber
\end{eqnarray}
where the over-all factors $1/g_s^k$ have been inserted for convenience. We  
neglected gauge-variant operators and operators that vanish by 
the equations of motion. We also assume that all $N_f$ quarks are 
massless. Chirality then allows us to omit four-fermion operators of 
scalar, pseudo-scalar or tensor 
type. Diagrammatically, they cannot be generated in 
massless QCD, because 
the number of Dirac matrices on any fermion line that connects 
to an external fermion in a four-point function is always odd. 
The coefficients $C_i$ corresponding to these operators therefore 
vanish exactly.

The leading-order anomalous dimension matrix is easily obtained. 
The mixing of four-fermion operators was obtained in 
\cite{SVZ79} and the mixing of ${\cal O}_5$ into itself can 
be inferred from \cite{NT83,Mor84}. Writing 
$\gamma=\gamma^{(1)}\alpha_s/(4\pi)+\ldots$ and 
\begin{equation}
\label{matrix}
\gamma^{(1)} = \left(
\begin{array}{ccc}
A & 0 &B \\
0 & \gamma_{55} & 0\\
0 & 0 & C
\end{array}\right), 
\end{equation}
the mixing of four-fermion operators is described by 
\begin{equation}
A=\left(
\begin{array}{cccc}
0 & 0 & \frac{8}{3} & 12 \\[0.1cm]
0 & 0 & \frac{44}{3} & 0 \\[0.1cm]
0 & \frac{6 C_F}{N_c} & -\frac{9 N_c^2+4}{3 N_c}+\frac{8 N_f}{3} & 
\frac{3 (N_c^2-4)}{N_c} \\[0.1cm]
\frac{6 C_F}{N_c} & 0 &\frac{3 (N_c^2-4)}{N_c} -\frac{4}{3 N_c} & 
-3 N_c
\end{array}\right),
\end{equation} 
with $C_F=(N_c^2-1)/(2 N_c)$, $N_c$ the number of colours. 
The non-zero entries of the $4\times4$ sub-matrices $B,C$ are: 
\begin{eqnarray}
&&B_{11}=B_{22}=8 (2 N_c N_f+1)/3,\nonumber\\
&&B_{12}=B_{21}=8/3,\nonumber\\
&&B_{31}=B_{32}=B_{41}=B_{42}=8 C_F/3,\\
&&C_{11}=C_{22}=-2 b,\nonumber\\
&&C_{33}=C_{44}=-4 b,\nonumber\\
&&C_{13}=C_{24}=8 N_c N_f/3.\nonumber
\end{eqnarray}
The mixing of 
${\cal O}_5$ into itself is given by $\gamma_{55}=-8(N_c-N_f)/3$. 
Note that due to a cancellation of different 
diagrams the entry $\gamma_{53}$ vanishes. As a 
consequence ${\cal O}_5$ decouples from the mixing at leading order 
\cite{NT83}.

To solve (\ref{ei}) with $\gamma(\alpha_s)$ and $\beta(\alpha_s)$ 
evaluated at leading order, let $b=-4\pi\beta_0$, and let $2 b\lambda_i$, 
$i=1,\ldots,4$, be the eigenvalues of $A$ and $\lambda_5=\gamma_{55}/(2 b)$. 
Let $U$ be the matrix that diagonalizes $A$. Since the integration 
constants $\hat{C}_i$ cannot be calculated and can be considered as 
non-perturbative, we do not keep track of factors multiplying 
these constants in the following, unless they 
are exactly zero. Thus we only note that no element of $U$ vanishes 
for values of $N_f$ of interest. Since $C$ is triangular, 
one obtains
\begin{eqnarray}
E_i(\alpha_s) &=& \sum_{k=1}^4 C^{[1]}_{ik}\alpha_s^{-\lambda_k}
\qquad i=1,\ldots, 4
\nonumber\\
E_5(\alpha_s) &=& C^{[1]}_5\alpha_s^{-\lambda_5} 
\\
E_i(\alpha_s) &=& C^{[2]}_i\alpha_s+ 
\sum_{k=1}^4 C^{[1]}_{ik}\alpha_s^{-\lambda_k}
\qquad i=6,7
\nonumber\\
E_i(\alpha_s) &=& C^{[2]}_i\alpha_s+ C^{[3]}_i\alpha_s^2 + 
\sum_{k=1}^4 C^{[1]}_{ik}\alpha_s^{-\lambda_k}
\qquad i=8,9
\nonumber
\end{eqnarray}
with $\alpha_s$-independent non-vanishing constants $C^{[l]}$ that depend 
on the nine integration constants $\hat{C}_i$ 
and the elements of $\gamma^{(1)}$. The 
exponents $\lambda_k$ are reported in Table~\ref{tab1}. At leading 
order it is consistent to set $F(\alpha_s)=1$. This completes the 
evaluation of the coefficient functions in (\ref{solcoeff}). 
\begin{table}[t]
\addtolength{\arraycolsep}{0.2cm}
\renewcommand{\arraystretch}{1.25}
$$
\begin{array}{c|ccccc}
\hline\hline
N_f & \lambda_1 & \lambda_2 &\lambda_3 &\lambda_4 & \lambda_5  \\ 
\hline
3 & 0.379 & 0.126 & -0.332 & -0.753 & 0  \\
4 & 0.487 & 0.140 & -0.302 & -0.791 & 4/25 \\
5 & 0.630 & 0.155 & -0.275 & -0.843 & 8/23 \\
6 & 0.817 & 0.172 & -0.254 & -0.910 & 4/7    \\ 
\hline\hline
\end{array}
$$
\caption{\label{tab1}\small 
Numerical values of $\lambda_i$ ($N_c=3$).}
\end{table}

As an example, we consider the Adler function defined in (\ref{adlerdef}). 
In addition to the coefficient functions we need the $R_{{\cal O}_i}$, 
the current-current correlation function with a single insertion of 
${\cal O}_i$. Since we do not follow over-all constants, it is 
sufficient to know that
$R_{{\cal O}_i}(\alpha_s,q)\propto \alpha_s^0$, 
$i=1,\ldots, 4$, $R_{{\cal O}_5}(\alpha_s,q)\propto \alpha_s$, 
$R_{{\cal O}_i}(\alpha_s,q)\propto \alpha_s^{-1}$, $i=6,7$, and  
$R_{{\cal O}_i}(\alpha_s,q)\propto \alpha_s^{-2}$ for $i=8,9$. 
Having determined the $\alpha_s$-dependence of (\ref{parisi}), 
we use (\ref{div}) and (\ref{im}), and find 
\begin{equation}
\label{asfinal}
r_n \stackrel{n\to \infty}{=} \beta_0^n\,n!\,n^{\beta_1/\beta_0^2}
\left[\,\sum_{i=1}^4 K_i\,n^{2+\lambda_i} + K_5\,n^{-1+\lambda_5} +
K_6 + K_8\, n\right]\left(1+O(1/n)\right)
\end{equation}
for the coefficient at order $\alpha_s^{n+1}$. Because we 
consider vector currents, the operators ${\cal O}_{7,9}$ are not needed. 
The normalization constants $K_i$ are undetermined. The leading 
asymptotic behaviour is 
\begin{equation}
\label{adleras}
r_n \stackrel{n\to \infty}{=} K_1\,\beta_0^n\,n!\,
n^{2+\beta_1/\beta_0^2+\lambda_1} = K_1\,\beta_0^n\,n!\,
n^{\{1.59,1.75,1.97\}},
\end{equation}
for $N_f=\{3,4,5\}$. Note that the leading-order result in the 
flavour expansion corresponds to the term $K_8 n$ in (\ref{asfinal}) 
because in the large-$N_f$ limit $K_1$ is suppressed by one power 
of $N_f$ compared to $K_8$. For the Adler function,  
UV renormalons dominate the 
large-order behaviour and hence (\ref{adleras}) represents the 
strongest divergent behaviour at large $n$.

We assumed that the external vector current is flavour-symmetric. 
In reality, the current is $j_\mu=\bar{\psi}\gamma_\mu Q\psi$,  
with $Q_{ij}=
\mbox{diag}(e_u,e_d,\ldots)$ a matrix in flavour space and 
flavour indices are summed over. Since flavour 
symmetry is broken only by the external current (all quarks are 
still considered as massless), the `QCD operators' ${\cal O}_{1-5}$ 
remain unaltered. The basis of `current operators' ${\cal O}_{6-9}$ 
has to be modified to include the operators $(\mbox{tr} Q)
\,\bar{\psi}\gamma^\mu
\psi\,\partial_\nu F^{\nu\mu}$ and $\bar{\psi}\gamma^\mu Q 
\psi\,\partial_\nu F^{\nu\mu}$ instead of ${\cal O}_6$. 
This ensures that 
mixing of four-fermion operators into the current operators contributes 
proportionally to $\mbox{tr} Q^2=\sum_f e_f^2$ and 
$(\mbox{tr} Q)^2=(\sum_f e_f)^2$, as 
required by the existence of `flavour non-singlet' and `light-by-light 
scattering' terms. The matrices $B$ and $C$ in (\ref{matrix}) change, 
but their pattern of non-zero entries does not. Thus, as we are not  
interested in over-all constants, (\ref{asfinal}) carries over to 
the present case. 

Eq.~(\ref{asfinal}) holds 
when the series is expressed in terms of the $\overline{\mbox{MS}}$ 
renormalized coupling $\alpha_s$. If a different coupling is 
employed that is related 
to the $\overline{\mbox{MS}}$ coupling by a factorially divergent 
series, the coefficients $r_n$ change accordingly and 
(\ref{asfinal}) may not be valid. We return to the problem of 
scheme-dependence in Section~\ref{schemedep}.

It is interesting to note that sub-leading corrections 
to the asymptotic behaviour can be computed without introducing 
further `non-perturbative' parameters in addition to the constants 
$\hat{C}_i$ already present at leading order. As a rule, to obtain 
the coefficient of the $1/n^k$ correction, one needs the 
$\beta$-function coefficients $\beta_0,\ldots,\beta_{k+1}$, the 
$(k+1)$-loop anomalous dimension matrix and the $k$-loop correction 
to Green functions with operator insertions. For simplicity, 
suppose there is only a single operator ${\cal O}$ and 
$R_{\cal O} = 1 + e_1\alpha_s+\ldots$. 
Then, using (\ref{solcoeff}-\ref{ei}), one finds
\begin{equation}
\label{imnlo}
\mbox{Im}\,I[R](\alpha_s,p_k) = {\rm const}\cdot 
e^{-1/(\beta_0\alpha_s)}\,(-\beta_0\alpha_s)^{-\beta_1/\beta_0^2+
\gamma_0/(2 \beta_0)}\,\left(1+s_1\alpha_s+\ldots\right),
\end{equation}
where
\begin{equation}
\label{s1nlo}
s_1 = e_1+\frac{\gamma_1}{2\beta_0}-\frac{\gamma_0\beta_1}{2\beta_0^2}-
\frac{\beta_2}{\beta_0^2}+\frac{\beta_1^2}{\beta_0^3}.
\end{equation}
The corresponding large-order behaviour is
\begin{equation}
\label{asnlo}
r_n \stackrel{n\to \infty}{=} K\,\beta_0^n\,
\Gamma\!\left(n+1+\frac{\beta_1}{\beta_0^2}-\frac{\gamma_0}{2\beta_0}\right)
\left[1+\left(-\frac{1}{\beta_0}\right)\,\frac{s_1}{n}+ O\!
\left(\frac{1}{n^2}\right)\right].
\end{equation}
The extension to higher terms in $1/n$ is straightforward.

The renormalization group treatment can in principle be extended to the 
next singularity in the Borel plane at $u=-2$. One has to consider 
single insertions of dimension-8 operators and double insertions 
of dimension-6 operators. In practice, this is probably already too 
complicated to be useful. 

Note that the idea of compensating UV renormalons (to be precise, 
the imaginary part of the Borel integral due to UV renormalons) by 
adding higher-dimension operators has much in common with the idea 
of reducing the cut-off dependence of lattice actions by adding 
higher-dimension operators, known as Symanzik improvement \cite{Sym83}. 
This analogy has been taken up by \cite{BD84}. The fact that the 
normalization of UV renormalons cannot be calculated is reflected  
in the statement that the coefficients of higher-dimension 
operators in Symanzik-improved actions 
have to be tuned non-perturbatively in order that a certain power 
behaviour in the lattice spacing is eliminated completely.

\subsection{Infrared renormalons}
\label{irren}

Infrared renormalons are more interesting than ultraviolet renormalons 
from the phenomenological point of view. Despite this fact, there has been 
less work on diagrammatic aspects beyond diagrams with a single 
chain. A general classification of IR renormalon singularities for 
an arbitrary Green function comparable to the classification of 
UV renormalons presented above is not known at this time. This is 
probably due to the fact that IR properties of Green functions depend 
crucially on external momentum configurations, while UV properties 
depend on external momenta trivially, through diagrams with counterterm 
insertions. The structure of UV renormalization is also simpler 
than IR factorization, which deals with collinear and soft divergences 
on a process-by-process basis. The same increase in complexity may be 
expected when dealing with IR renormalons. Nevertheless, this is an 
area where progress can be made and should be expected in the nearer 
future. 

In the following we restrict ourselves to a qualitative discussion 
of diagrammatic aspects of IR renormalons. This discussion divides 
into off-shell and on-shell processes.
More details on the connection of IR renormalons and 
non-perturbative power corrections can be found in 
Section~\ref{nonp} and many explicit cases will be reviewed in 
Section~\ref{pheno}. 

\subsubsection{Off-shell processes}

In QCD off-shell, euclidian Green functions of external (electromagnetic or 
weak) currents are of interest. They are related to physical processes 
such as the total cross section in $e^+ e^-\to\,\,$hadrons or moments 
of deep inelastic scattering structure functions through dispersion 
relations. 

In the flavour expansion the Borel transform of a diagram with 
chains is represented by the integral (\ref{relbg}). We suppose that 
there are no power-like infrared divergences. Then for off-shell 
Green functions at euclidian momenta it follows from 
properties of analytic regularization that the Borel transform has 
IR renormalon singularities at non-negative integer $u$.\footnote{Recall 
that in the flavour expansion $u=-\beta_{0f}t$, where $t$ is the Borel 
parameter. In QCD $u=-\beta_0 t$, so that in both cases, QED and QCD, 
IR renormalons are located at positive $u$.} However, the structure of the 
singularity in terms of subgraphs is different from (\ref{singofg}) 
as different notions of irreducibility apply to ultraviolet and infrared 
properties. The methods used in \cite{BS96} could be extended to 
this situation.

Consider the two-point function of two quark currents, defined in 
(\ref{currentcorr}), with external momentum $q$. 
IR renormalons arise from regions of small loop momentum $k\ll q$, 
where the integrand becomes IR sensitive. For massless, off-shell 
Green functions, the IR sensitive points are those where a collection 
of internal lines has zero momentum. There has to be a connected path 
of large external momentum from one external vertex to the other. 
Hence, a general graph can be divided into a sum of contributions 
of the form shown in Fig.~\ref{fig8}a: A `hard' subgraph to which both 
external vertices connect and a `soft' subgraph of small momentum 
lines, which connects to the hard subgraph through an arbitrary number 
of soft lines. In terms of the operator expansion (OPE), the soft subgraph 
corresponds to the matrix element of an operator and the hard part 
to the coefficient function. An analysis of the leading IR renormalon 
contribution ($t=-2/\beta_0$) to the current-current correlation function 
based on factorization of hard and soft subgraphs can be found in  
\cite{Mue85}. 
\begin{figure}[t]
   \vspace{-6.5cm}
   \epsfysize=36cm
   \epsfxsize=24cm
   \centerline{\epsffile{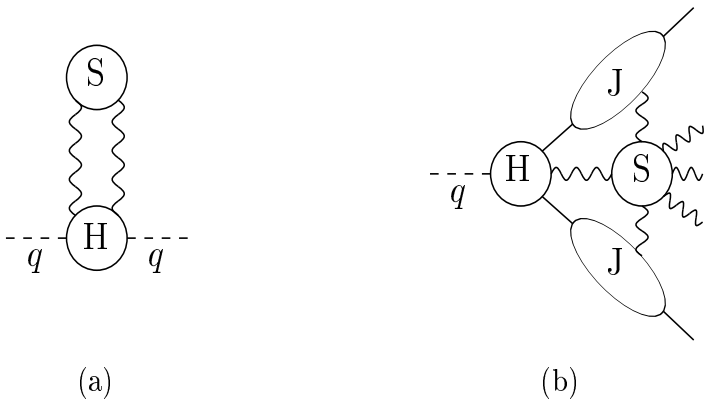}}
   \vspace*{-23.6cm}
\caption[dummy]{\small Infrared regions that give rise to infrared 
renormalons. (a) For a current-current two-point function at euclidian 
momentum. The external currents are shown as dashed lines. 
(b) For an event-shape variable in $e^+ e^-$ annihilation 
near the two-jet limit. Wavy lines represent collections of soft lines. 
\label{fig8}}
\end{figure}

In Section~\ref{renormalonexample} we 
considered the leading-order diagrams of Fig.~\ref{fig1} in the 
loop momentum region, where the soft part consisted of a single 
gluon line (or chain). The general classification would also allow 
a quark line or more than one line in the soft part. These parts 
are associated with condensates in the OPE containing quark fields. 
For the analysis of IR renormalons soft quark lines alone play no 
role, because they cannot be `dressed' with bubbles, which is 
necessary in order to turn IR sensitivity in a skeleton diagram 
into a factorially divergent series expansion. 

An immediate consequence of the factorization expressed by 
Fig.~\ref{fig8}a is that in order for the diagram to contribute 
to an IR renormalon at $t=-m/\beta_0$, the soft part must connect 
to the hard part by not more than $2 m$ gluon lines. This follows 
from the fact that each additional such line adds one hard 
propagator to the hard part, which counts as $1/q$. On dimensional 
grounds this factor must be compensated by a power of one of the 
small momenta $k_i$. Such factors result in a suppression of 
the large-order behaviour which is related to integrals that 
generalize
\begin{equation}
\int\limits_0^q dk^2\,k^{m-2} \left[\beta_0\ln(k^2/q^2)\right]^n 
\sim \left(-\frac{2\beta_0}{m}\right)^n n!.
\end{equation}
In general, the location of IR renormalons and the possible contributions 
to a singularity at a particular point follow from such IR {\em power} 
counting arguments.

The leading-order diagrams in the flavour expansion, 
Fig.~\ref{fig1}, result in $d^4k/k^2$ for small $k$. This 
leads to a singularity 
at $t=-1/\beta_0$ for each diagram, which can be associated with 
the operator $A_\mu^A A^{\mu,A}$. Gauge invariance of the 
current-current two-point function requires that these leading 
contributions cancel in the sum of diagrams. 
After this cancellation the leading term 
is $d^4k$, associated with a singularity at $t=-2/\beta_0$ and 
the operator $G_{\mu\nu}^A G^{\mu\nu,A}$ 
as discussed in Section~\ref{sectbascon}. Consider now the 
diagram with two chains shown in Fig.~\ref{fig9}a. If both 
gluon momenta are small, power counting gives 
$d^4k_1/k_1^2\,d^4k_2/k_2^2$ which can contribute to 
the singularity at $t=-2/\beta_0$. This contribution  
must be associated with 
the $(A_\mu)^4$ term in the operator $G_{\mu\nu}^A G^{\mu\nu,A}$ 
and it is hence related to the leading order in the flavour expansion 
by gauge invariance. Except for this trivial contribution, the region 
when both gluon momenta are small contributes only to subleading 
renormalon singularities at $t>-2/\beta_0$. When one of the gluon 
lines is hard and only one is soft, a contribution to the 
order $\alpha_s$ correction of the coefficient function of 
$G_{\mu\nu}^A G^{\mu\nu,A}$ is obtained. 
Because one loses one power of $\alpha_s$, 
this contribution is $1/n$-suppressed in large orders relative to 
the leading order in the flavour expansion. We conclude that 
the leading IR renormalon at $u=2$ is determined by diagrams 
with only a single soft chain, up to contributions constrained 
by gauge invariance and up to a calculable multiplicative factor that 
follows from the coefficient function of $G_{\mu\nu}^A G^{\mu\nu,A}$. 
These diagrams are shown in Fig.~\ref{fig9}b, 
where the shaded circle denotes an arbitrary 
collection of soft lines. Note the difference with the corresponding 
analysis for UV renormalon singularities, in which case 
diagram~\ref{fig9}a was found to be enhanced relative to the leading 
order in the flavour expansion rather than suppressed. 
The diagrams of type \ref{fig9}b 
have been considered further in \cite{Z92,Gru93a,BZ93}. It was 
found that the residue of the IR renormalon singularity receives 
contributions from arbitrarily complicated graphs in the shaded circle 
and remains uncalculable \cite{Gru93a,BZ93} despite the simpler 
over-all diagram 
structure compared to the UV renormalon case. A graph-by-graph comparison 
of some contributions to 
the first IR and first UV renormalon is summarized in 
Table~\ref{tab2}.
\begin{table}[b]
\addtolength{\arraycolsep}{0.2cm}
\renewcommand{\arraystretch}{1.25}
$$
\begin{array}{c|ccc}
\hline\hline
\mbox{Diagram} & \mbox{Fig.~\ref{fig1}} & \mbox{Fig.~\ref{fig9}a} & 
\mbox{Fig.~\ref{fig9}b}  \\ 
\hline
\mbox{UV} & n & n^2 & n\ln n  \\
\mbox{IR} & 1 & 1/n^\dagger & \ln n  \\
\hline\hline
\multicolumn{4}{l}{\mbox{\footnotesize ${}^\dagger$Ignoring $O(1)$ 
fixed by gauge invariance.}}
\end{array}
$$
\caption{\label{tab2}\small 
Comparison of contributions of various diagrams to the leading UV and 
IR renormalon behaviour. For the UV renormalon the displayed factor 
multiplies $\beta_0^n n!$, for the IR renormalon $(-\beta_0/2)^n n!$. In 
the case of Fig.~\ref{fig9}b we refer to the diagram with a chain inserted 
into a chain analogous to Fig.~\ref{fig4}b.}
\end{table}
\begin{figure}[t]
   \vspace{-3.2cm}
   \epsfysize=27cm
   \epsfxsize=18cm
   \centerline{\epsffile{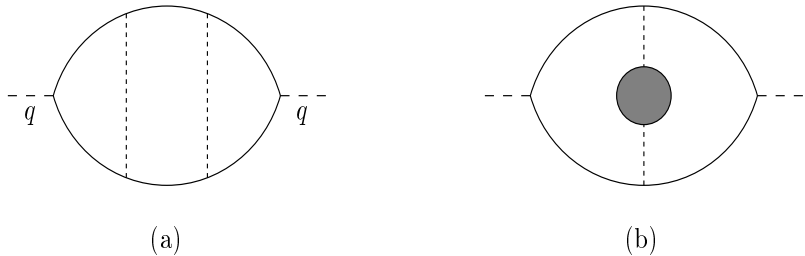}}
   \vspace*{-20cm}
\caption[dummy]{\small Two diagrams at higher order in the flavour 
expansion. \label{fig9}}
\end{figure}
 
A complete characterization of IR renormalon singularities must account 
not only for powers of small momenta but also for logarithms of $k/q$.
The soft subgraphs contains renormalization parts, when some soft 
momenta are larger than others: $k_1\ll k_2$. These renormalization 
parts lead to logarithms whose coefficients are given by renormalization 
group functions and introduce the effect of higher order coefficients 
in the $\beta$-function and operator anomalous dimensions into the 
large-order behaviour. Technically, in the flavour expansion, 
this occurs in a way similar to the UV renormalon case. 
In particular, there 
is no difference between UV and IR renormalons as far as the 
mechanism that restores the non-abelian $\beta$-function 
coefficient $\beta_0$ is concerned (see Section~\ref{qcd}).

Once factorization is established, the most elegant characterization 
of IR renormalon singularities follows from first identifying 
the `operator content' of the soft subgraph and then from 
deriving an evolution (renormalization group) equation for it. 
Consider a physical quantity such as the Adler function (\ref{adlerdef}) 
or its discontinuity and its series expansion $\sum r_n\alpha_s^{n+1}(Q)$ 
in $\alpha_s$ normalized at $Q$. The IR renormalon behaviour of the 
coefficients $r_n$ leads to an ambiguity in the Borel integral with 
a certain scaling behaviour in $Q$. This scaling behaviour must be 
matched exactly by higher-dimension terms in the OPE. For 
simplicity, we assume that there is only one operator ${\cal O}$ of 
dimension $d$ with anomalous dimension $\gamma$ as defined 
in (\ref{defandim}) and coefficient function 
$C(1,\alpha_s(Q))=c_0+c_1\alpha_s(Q)+\ldots$. The scaling behaviour 
is given by
\begin{eqnarray}
\frac{1}{Q^d}\,C\left(Q^2/\mu^2,\alpha_s\right)\,
\langle 0|{\cal O}|0\rangle(\mu) &=&
\mbox{const}\,\cdot
e^{d/(2\beta_0\alpha_s(Q))}\,(-\beta_0\alpha_s(Q))^
{d\beta_1/(2\beta_0^2)}
\nonumber\\
&&\hspace*{-4cm}
\cdot\,F(\alpha_s(Q))^{d/2}\,\exp\left(-\int\limits_{\alpha_0}^{\alpha_s(Q)} 
\!\!d x\,\frac{\gamma(x)}{2\beta(x)}\right) C(1,\alpha_s(Q)),
\end{eqnarray}
where $F$ is defined in (\ref{fi}). Using (\ref{div}) and (\ref{im}), 
the large-order behaviour
\begin{equation}
\label{iras}
r_n \stackrel{n\to \infty}{=} K\,\left(\frac{2\beta_0}{d}\right)^n\,
\Gamma\!\left(n+1-\frac{d\beta_1}{2\beta_0^2}+\frac{\gamma_0}{2\beta_0}
\right)\left[1+\left(-\frac{d}{2\beta_0}\right) \frac{s_1}{n}+
O\!\left(\frac{1}{n^2}\right)\right]
\end{equation}
with
\begin{equation}
\label{irasnlo}
s_1=\frac{c_1}{c_0}-\frac{\gamma_1}{2\beta_0}+
\frac{\gamma_0\beta_1}{2\beta_0^2}+
\frac{d \beta_2}{2\beta_0^2}-\frac{d\beta_1^2}{2\beta_0^3}
\end{equation}
follows. Note the different signs of the anomalous dimension terms 
compared to (\ref{asnlo}). (Otherwise the first UV renormalon can 
formally be obtained from setting $d=-2$.) The global normalization $K$ is 
not determined. This equation is valid provided the renormalization 
counterterms do not absorb factorial divergence into the definition 
of renormalized parameters \cite{Mue85,Ben93b}; see also 
Section~\ref{schemedep}.

For current-current correlation functions the leading IR renormalon 
corresponds to $d=4$ and ${\cal O}=\alpha_s G_{\mu\nu}^A G^{\mu\nu,A}$. 
Taking into account that for this operator $\gamma_0=0$ and 
$\gamma_1=2\beta_1$, one reproduces the leading asymptotic 
behaviour and the $1/n$ correction, obtained in \cite{Mue85} and 
\cite{Ben93b}, respectively. The $1/n^2$ correction could be computed 
also, if the two-loop correction to the coefficient function 
of the gluon condensate were known. 

An important point is that the unknown constant $K$ is a universal 
property of the soft part in Fig.~\ref{fig8}a, that is a property 
of the operator ${\cal O}$. Hence for correlation functions with 
different currents, which differ only in their hard part, the 
{\em difference} in the leading IR renormalon behaviour is 
calculable. We refer to this property as {\em universality} of the 
leading IR renormalon or $1/Q^4$ power correction. Note, however, 
that universality is more restricted for IR renormalons than for 
UV renormalons, because it refers to a specific class of processes, 
in the present case given by various current-current correlation 
functions. Let us also note that for certain operators $K$ can be exactly 
zero. These are operators like $\bar{q} q$, which are protected 
from perturbative contributions to all orders in perturbation theory.

Our discussion has focused on the current-current correlation 
functions. The generalization to other off-shell quantities is 
straightforward.

\subsubsection{On-shell processes}

For on-shell, minkowskian processes the classification of IR sensitive 
regions of a Feynman integral is more complicated than for 
off-shell quantities. As is well known, in addition to soft, 
zero-momentum lines, collinear configurations 
of massless lines (`jets') have to be considered. 
Furthermore, on-shell propagators 
do not give power suppression, even if the line momentum is of 
order $q$. As a consequence, soft subgraphs, which connect to on-shell 
propagators, 
cannot be parametrized by local operators. As an example, the infrared 
regions that contribute to power corrections to two-jet-like 
observables in $e^+ e^-$ annihilation are shown in Fig.~\ref{fig8}b. 
Non-local operators that parametrize power corrections to a 
class of jet observables were first analysed in \cite{KS95a}. 
 
It is characteristic of off-shell processes that IR renormalons 
occur only at positive integer $u$, which implies power corrections 
as powers of $1/Q^2$ and not powers of $1/Q$, where $Q$ is the 
`hard' scale of the process. For on-shell quantities the generic 
situation leads to IR renormalons at positive half-integers and integers 
and a series of power corrections in $1/Q$. To illustrate this point, 
we consider a simpler case than Fig.~\ref{fig8}b, a system 
with one heavy quark. More precisely, we consider the mass shift 
$\delta m=m-m_{\overline{\rm MS}}(m_{\overline{\rm MS}})$, the 
difference between the pole mass and the $\overline{\rm MS}$ mass 
of a heavy quark. This is in fact the quantity where IR renormalons 
leading to linear suppression in the hard scale, here $m$, have 
been found first \cite{BB94a,BSUV94}.

\begin{figure}[t]
   \vspace{-6.2cm}
   \epsfysize=30cm
   \epsfxsize=20cm
   \centerline{\epsffile{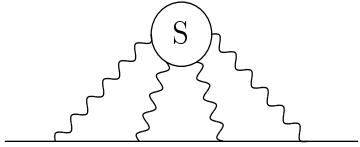}}
   \vspace*{-21.2cm}
\caption[dummy]{\small Infrared regions that contribute to the 
first IR renormalon in the mass shift $\delta m$.\label{fig10}}
\end{figure}
It is a trivial consequence of IR power counting to see that the 
IR contribution to the mass shift is suppressed only linearly in $m$. 
The one-loop contribution to the heavy quark self-energy 
$\Sigma(p^2)$ evaluated at $p^2=m^2$ is
\begin{equation}
\Sigma(m^2)\propto m \int\frac{d^4 k}{k^2 (2p\cdot k+k^2)} 
\sim \int dk.
\end{equation} 
When the one-loop diagram is dressed with vacuum polarization (`bubble') 
insertions one obtains $(-2\beta_0)^n n! \alpha_s^{n+1}$ in large 
orders, i.e. an IR renormalon singularity at $u=1/2$. The IR 
sensitive regions in an arbitrary diagram are shown in 
Fig.~\ref{fig10}. The important difference to Fig.~\ref{fig8}a is 
that one obtains a contribution to the singularity at $u=1/2$ 
for an arbitrary number of gluon couplings to the heavy quark line, 
because the heavy quark propagators are nearly on-shell. The 
IR renormalon singularity cannot be associated with a {\em local} 
operator as in the case of off-shell correlation functions. The situation 
is still simple, though. As far as the leading IR renormalon 
is concerned, the numerator of the heavy quark propagator can be 
approximated by $m\!\!\not\!v+m$, where $p=m v$ is the heavy quark momentum 
and $v^2=1$. Hence, using also the on-shell condition, 
gluons couple only through the combination $v\cdot A$. 
In a temporal axial gauge with $v\cdot A=0$, they 
decouple and the leading IR renormalon can be seen to 
correspond to an operator bilinear in the quark field with 
fields at non-coincident positions. In a general 
gauge a phase factor
\begin{equation}
\mbox{P}\,e^{ig\int_C ds\,v\cdot A(s)}
\end{equation}
accounts for the non-vanishing temporal soft gluon couplings \cite{BSUV94} 
and makes the non-local operator gauge-invariant. 

In high-energy processes involving massless quarks there are in addition 
collinear-sensitive regions such as `J' in Fig.~\ref{fig8}b. 
However, it seems that power corrections from hard-collinear regions 
(energy $\omega$ much larger than transverse momentum $k_\perp$) 
are always suppressed by powers of $Q^2$ rather than $Q$. There is 
no proof to all orders of this statement yet, but the following heuristic 
argument may illustrate the point: let $p$ be the momentum of 
a fast on-shell particle, $p\sim Q$, after emission of a hard-collinear 
on-shell particle with $\omega\sim Q$, $k_\perp\ll Q$, where 
$k_\perp$ is the transverse momentum relative to $p$. Then the propagator 
\begin{equation}
\frac{1}{(p+k)^2} = \frac{1}{p\left(\omega-\sqrt{\omega^2+k_\perp^2}
\right)}
\end{equation}
is expanded in $k_\perp^2/\omega^2\sim k_\perp^2/Q^2$ and $Q$ enters 
only quadratically. Since the same is true of the hard-collinear 
phase space, it may be argued that the transverse momentum, and hence $Q$, 
always enters quadratically as long as energies are large.

As a consequence, if $1/Q$ power corrections exist and if one is interested 
only in those, the diagram of Fig.~\ref{fig8}b can be somewhat 
simplified. The jet parts J can be replaced by Wilson line operators 
to which soft gluons couple through a phase factor. In general, 
the leading-order eikonal approximation may not be sufficient and the 
first-order correction to it must be kept. However, many hadronic 
event shape observables, which are particularly interesting 
with respect to $1/Q$ power corrections, 
have a linear suppression of soft regions built into their 
definitions. For such observables, the analysis simplifies further, since 
the conventional eikonal approximation can be used. Systematic 
investigations of power corrections to such quantities beyond one 
gluon emission have been started in \cite{KOS97,DLMS97}.\footnote{A 
detailed analysis of (the cancellation of) $\Lambda/m$ power corrections 
at two loops for inclusive heavy quark decay can be found in 
\cite{SAZ98}.} We will 
return to this topic in Section~\ref{eventshapes} in connection with 
the phenomenology of power corrections to hadronic event 
shape variables.

\subsection{Renormalization scheme dependence}
\label{schemedep}

The answer to the following question is overdue: Since the perturbative 
coefficients can be altered arbitrarily by changing the renormalization 
convention, which convention has been implicit in the 
derivation of large-order behaviour? The short answer is that a 
renormalization prescription must be used in which the subtraction 
constants are not factorially divergent. This ensures that bare and 
renormalized parameters are related by convergent series (although 
every coefficient of the series 
diverges when the the cut-off is removed) and that 
no factorial divergence is `hidden' in the formal definition of 
the renormalized parameters. Such schemes have been called regular in 
\cite{Ben93b}. 

Before addressing scheme transformations in general, let us consider 
the issue of subtractions in the flavour-expansion. Suppose $R$ is 
a renormalization scheme invariant quantity, which depends only 
on the strong coupling, for example the Adler function (\ref{adlerdef}). 
We calculate its Borel transform in leading order in the 
flavour expansion in four dimensions by 
inserting {\em renormalized} fermion loops (\ref{simpleloop}) into 
a gluon line. Since by assumption the quantity is scheme-invariant, 
no further subtractions, except for $C$ in (\ref{simpleloop}) 
are needed and the calculation in four dimensions is justified. 
The result has the form
\begin{equation}
\label{sch1}
B[R](u) = \left(\frac{Q^2}{\mu^2} e^C\right)^{-u} F(u).
\end{equation}
The function $F$ is scheme and scale independent, but the Borel transform 
is not, because it is defined as a Borel transform with respect to the 
scheme and scale dependent coupling $\alpha_s$. The prefactor 
in (\ref{sch1}) can be combined with the exponent in the 
(formal) Borel integral 
\begin{equation}
\int\limits_0^\infty dt\,e^{-t\left(\frac{1}{\alpha_s}-\beta_{0f}
\left[\ln\frac{Q^2}{\mu^2}+C\right]\right)}\, F(u)
\end{equation}
such that the exponent is manifestly scheme and scale invariant 
at leading order in the flavour expansion. The 
definition of the Borel transform can be modified 
in such a way as to preserve manifest 
scale and scheme independence beyond the leading order of the 
flavour expansion \cite{Ben93a,Gru93a}, but the definition in terms of 
perturbative coefficients becomes complicated.

Suppose now that $R$ is not in itself physical, but requires additional 
subtractions beyond the renormalization of the fermion 
loops in the chain. For example $R$ may be the gluon/photon vacuum 
polarization or the quark mass shift discussed in Section~\ref{irren}. 
In this case the result of the calculation takes the form 
\begin{equation}
\label{sch2}
B[R](u) = \left(\frac{Q^2}{\mu^2} e^C\right)^{-u} F(u) - S(u),
\end{equation}
where $F(u)$ has a pole at $u=0$. This pole is cancelled by the 
scheme-dependent but momentum-independent subtraction 
function $S(u)$, which is arbitrary 
otherwise. UV and IR power counting relates the UV and IR renormalon 
poles of $F(u)$  to the behaviour of loop 
diagrams at large and small momentum. 
In order that these relations remain valid, the function 
$S(u)$ must not introduce singularities in $u$ other than at $u=0$. 
At this leading 
order in the flavour expansion, the subtraction function can be 
expressed in terms of renormalization group functions 
\cite{EPPT82,PP84,BB94a}, the $\beta$-function in the case of the 
vacuum polarization, and the anomalous dimension of the quark mass 
in the case of the mass shift. The requirement that $S(u)$ be analytic 
except at $u=0$ 
results in the requirement that the renormalization group functions 
have {\em convergent} series expansions in $\alpha_s$, or at least 
they should not diverge as fast as factorials. This is indeed 
true, at least to leading order in the flavour expansion, for 
the $\overline{\rm MS}$ definition of the coupling and the quark mass, 
but it is obviously not true for `physical' definitions of the 
coupling, because the perturbative expansions of 
physical quantities do have renormalons. Of course, 
once the large-order behaviour of two physical quantities expressed 
as series in a coupling, defined in a regular scheme, is known, 
the two physical quantities can always be related directly to each other, 
and the large-order behaviour of this relation can be found.

Once $S(u)$ is specified at leading order in the flavour expansion, 
it appears as a counterterm in higher orders, for example as a 
vacuum polarization insertion in the second diagram of 
Fig.~\ref{fig4}b. In this case $S(u)-\beta_1/u=\sum_{k=0} s_k u^k$ 
appears as `finite terms' in (\ref{singandcount}) and contributes 
to the singularity at $u=1$ as
\begin{equation}
\frac{K_{\rm vert}^{[0]}}{1+u_1+u_2}\,s_k u_3^k\quad\longrightarrow
\quad s_k\,(1+u)^k\ln(1+u).
\end{equation}
Since $s_k$ is proportional to the $\beta_{k+2}$ in the large-$N_f$ 
limit, it follows that the scheme-dependent $\beta$-function coefficient 
$\beta_{k+2}$ ($k>0$) enters as a $1/n^{k+1}$ correction to the 
large-order behaviour, provided $S(u)$ is analytic in the complex plane 
with the origin removed. This is 
in accordance with the general results (\ref{s1nlo}, \ref{asnlo}) and 
(\ref{iras}, \ref{irasnlo}). We emphasize that these general results 
are valid only in regular renormalization schemes. It is reasonable 
to conjecture (and true to leading order in the flavour expansion) that 
the $\overline {\rm MS}$ scheme is regular, but since the 
$\overline {\rm MS}$ scheme is defined only order by order in 
$\alpha_s$ or $1/N_f$, there is no proof of this conjecture. Above and 
below when we state(d) that a certain large-order behaviour is valid 
in the $\overline {\rm MS}$ scheme, it is always tacitly assumed 
that the $\overline {\rm MS}$ anomalous dimensions are 
convergent series in $\alpha_s$ or, at least, do not diverge as 
fast as factorials. In this context it is interesting to note 
that the series expansion of the $\beta$-function up to the highest 
order known today \cite{RVL97} is indeed much better behaved than physical 
quantities, which are expected to have divergent series expansions. 

The transformation properties of the large-order behaviour under 
changes of the series expansion parameter $\alpha_s$ are as follows: 
suppose $R=\sum_{n=0}^\infty r_n\alpha_s^{n+1}$, with the large-order 
behaviour\footnote{The subsequent equations are valid not only for 
the dominant large-order behaviour but also for subleading components 
from the second UV or IR renormalon etc.. Keeping $b$ in the denominator 
of the $1/n$ correction term in (\ref{scheme1}) proves convenient, when 
one goes to yet higher order in $1/n$.}
\begin{equation}
\label{scheme1}
r_n = K\,\left(\frac{1}{S}\right)^n\,\Gamma(n+1+b)\left[1+S\,
\frac{c_1}{n+b}+\ldots\right]
\end{equation}
and suppose that $\alpha_s$ is related to $\bar{\alpha}_s$, the coupling 
in the new scheme, by
\begin{equation}
\label{couprel}
\alpha_s = \bar{\alpha}_s +\delta_1\bar{\alpha}_s^2+\delta_2
\bar{\alpha}_s^3+\ldots.
\end{equation}
Then the parameters of the expression analogous to (\ref{scheme1}) 
in the new scheme are given by
\begin{eqnarray}
\label{trafp}
\bar{K} &=& K\,e^{\delta_1 S},
\\
\bar{S} &=& S,
\\
\bar{b} &=& b,
\\
\bar{c}_1 &=& c_1-S (\delta_1^2-\delta_2) - b\delta_1.
\end{eqnarray}
For these relations to be valid one can allow that the couplings are 
related by divergent series, provided the divergence is slower than for 
the $r_n$. The easiest way to obtain these transformation properties is 
to examine the transformation of the ambiguity of the Borel integral 
or the variant (\ref{irr}). Recall that $b$ and $c_1$ are calculable, 
but $K$ is not. However, the scheme dependence of the normalization 
is known and involves only the relation of the couplings at one loop. 
This is analogous to the transformation property of the QCD scale 
parameter $\Lambda$.

The case, where the scheme is fixed, but the renormalization scale 
of $\alpha_s$ is changed, is covered as a special case of 
(\ref{couprel}). With $\bar{\alpha}_s=\alpha_s(\mu')$ and 
$\delta_1=-\beta_0\ln({\mu'}^2/\mu^2)$, this 
leads to a trivial scale-dependence of $K$,
\begin{equation}
K(\mu') = \left(\frac{{\mu'}^2}{\mu^2}\right)^{-\beta_0 S} K(\mu).
\end{equation}
For UV renormalons in QCD $(-\beta_0 S)$ is a negative integer 
and the over-all 
normalization decreases when the renormalization scale is increased 
\cite{BZ92}. For IR renormalons it is exactly opposite.

The transformation properties can be generalized to the case, 
where (\ref{couprel}) is allowed to be arbitrary. In this case, 
the large-order behaviour of $R$ may end up being dominated 
by the large-order behaviour of (\ref{couprel}). From the point of 
view of analysing power corrections (IR and UV behaviour) to $R$ 
via renormalons, expressing $R$ through a non-regular coupling 
seems unnatural, since the coupling parameter `imports' power 
corrections not related to the physical process $R$ itself.

As in the case of low orders in perturbation theory \cite{Ste81}, 
one can find certain scheme independent combinations of the parameters 
that characterize the large-order behaviour \cite{Ben93b}. Restricting 
attention to physical quantities that depend on only one scale 
(`effective charges'), these parameters can be read off 
from the large-order behaviour of the 
effective charge $\beta$-functions just as in low orders 
of perturbation theory \cite{Gru80}. One finds that $S$, $b$ and 
\begin{eqnarray}
K_{\rm eff} &=& \beta_0 K\,e^{-r_1 S},
\\
c_{1\rm eff} &=& c_1-\frac{b+2}{S}+b r_1+S (r_1^2-r_2)+
\frac{\beta_1}{\beta_0}
\end{eqnarray}
are scheme and scale independent, provided the relation 
(\ref{couprel}) does not diverge too fast.

One may also wonder about the situation when a quark has intermediate 
mass $m\gg \Lambda$ but $m\ll Q$, where $Q$ is the scale of the hard 
process. A physical $\beta$-function would continuously interpolate from 
the $N_f+1$ to the $N_f$ flavour theory. In massless subtraction 
schemes one may ask whether $\beta_0^{[N_f+1]}$ or $\beta_0^{[N_f]}$ 
determines the factorial growth of perturbative coefficients. The 
answer depends on whether one considers UV or IR renormalons. 
For UV renormalons, $\beta_0^{[N_f+1]}$ is relevant. For IR renormalons, 
the typical loop momentum falls below $m$ beyond a certain order, 
in which case the massive quark effectively decouples. In large 
orders the perturbative coefficients become close to those of 
the $N_f$ flavour theory even though $Q$ is much larger than $m$, 
provided the coupling constants in the 
$N_f+1$ and $N_f$ flavour theory are matched as usual. The decoupling 
of intermediate mass quarks has been studied in \cite{BBB95}.

\subsection{Calculating `bubble' diagrams}
\label{bubbles}

Many of the applications reviewed in Section~\ref{pheno} are based 
on the analysis of diagrams with a single chain of fermion loops. 
In this section we summarize various methods to represent or calculate 
this class of diagrams and the relations between these methods.

We begin with some definitions. We consider observables $R$ and 
subtract the tree contribution. The radiative corrections take the form 
$\sum_{n=0}^\infty r_n\alpha_s^{n+1}$. We assume that $R$ is gauge-invariant 
and does not 
involve external gluon legs at tree level, so that the first-order 
correction $r_0$ comes from diagrams with a single gluon line. 
The coefficients $r_n$ are polynomials in $N_f$:
\begin{equation}
\label{flavourseries}
r_n=r_{n0}+r_{n1} N_f+\ldots+r_{nn} N_f^n.
\end{equation}
The set of fermion loop diagrams (`bubble diagrams') is gauge-invariant 
and gives the coefficient $r_{nn}$ with the largest power of $N_f$, the 
number of light flavours. In the following we do not consider the 
other terms in (\ref{flavourseries}). 

\begin{figure}[t]
   \vspace{-3cm}
   \epsfysize=30cm
   \epsfxsize=20cm
   \centerline{\epsffile{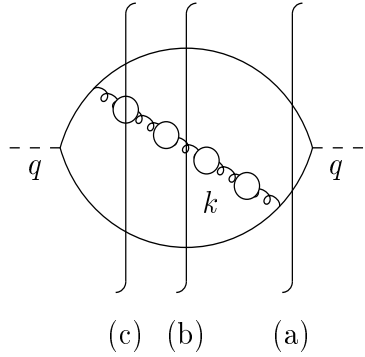}}
   \vspace*{-22cm}
\caption[dummy]{\small The three different types of cuts relevant 
for bubble graphs, here for $e^+ e^-\to$ hadrons. The cuts may be 
weighted to give an event shape variable.\label{fig11}}
\end{figure}
In general, the first-order correction to $R$ may be the sum of a 
one-loop virtual and a one-gluon real emission contribution. The fermion 
bubble corrections are (Fig.~\ref{fig11}): 
Fermion loops inserted into the virtual 
gluon line [cut (a)] or fermion loops inserted into the `real' gluon line, 
which can be either part of the final state [cut (b)] or 
split into a fermion pair (`cut bubble') [cut (c)]. In case (c), the 
gluon is not real anymore. In case (b) the fermion loops are scaleless 
integrals, which vanish in dimensional regularization. The virtual 
corrections of type (a) can be represented as 
\begin{equation}
\label{rvirt}
R_{\rm virt} = \int dk^2\,F_{\rm virt}(k,Q)\,\frac{1}{k^2}
\frac{\alpha_s}{1+\Pi(k^2)} = 
\int dk^2\,F_{\rm virt}(k,Q)\,\frac{\alpha_s(k\,\exp[C/2])}{k^2},
\end{equation}
where $\Pi(k^2)$ is given by (\ref{simpleloop}), $k$ is the momentum 
of the gluon line, and $Q$ stands collectively for external momenta. 
The fermion loop insertions are summed to all orders into $1/(1+\Pi(k^2))$. 
The real corrections (c) can be represented as
\begin{equation}
\label{rreal}
R_{\rm real} = \int dk^2\,F_{\rm real}(k,Q)\,\frac{1}{k^2}
\frac{\beta_{0f}\alpha_s^2}{|1+\Pi(k^2)|^2},
\end{equation}
where the virtuality of the gluon line, $k^2$, is now the invariant mass 
of the fermion pair into which the gluon splits. In writing (\ref{rreal}) 
we have separated the two-particle phase space over $k_{1,2}$ 
for the cut bubble by introducing a factor 
$d^4k\,\delta^{(4)}(k-k_1-k_2)$. Note that all 
dependence on $\alpha_s$ in 
(\ref{rvirt}, \ref{rreal}) is either explicit or in $\Pi(k^2)$.
If $R$ requires subtractions in addition to those for the fermion 
loops, the above integrals have divergences. Even if $R$ is finite 
after coupling renormalization, the integrals are ill-defined, because 
the Landau pole lies in the integration domain. However, their 
perturbative expansions are defined (but divergent). 
The integral (\ref{rreal}), understood as an expansion in $\alpha_s$, 
does not include the first-order correction with no gluon splitting, 
as seen from the fact that its expansion starts at order $\alpha_s^2$. 
It turns out that in the {\em summed} expression (\ref{rreal}) 
-- appropriately defined -- the 
first-order real correction is contained as an 
`end-point' contribution of order $1/\alpha_s$ from the lower 
limit $k^2=0$ and that (\ref{rreal}) gives the correct 
result for (b) and (c) together.

\subsubsection{The Borel transform method}
\label{btm}
The Borel transform $B[R](u) = \sum_{n} r_n/n! \,(-\beta_{0f})^{-n} 
u^n$ can be used as a generating function for the perturbative 
coefficients:
\begin{equation}
\label{genfunction}
 r_n = (-\beta_{0f})^n \frac{d^n}{du^n} B[R](u)_{|_{u=0}}\,.
\end{equation}
The Borel transform of bubble graphs is obtained using the relations
\begin{eqnarray}
\label{bvac}
B\!\left[\frac{\alpha_s}{1+\Pi(k^2)}\right] & =& 
\left(-\frac{k^2}{\mu^2} e^C\right)^{-u},
\\
\label{bimvac}
B\!\left[\frac{\beta_{0f}\alpha_s^2}{|1+\Pi(k^2)|}\right] &=& 
-\frac{\sin(\pi u)}{\pi}\left(-\frac{k^2}{\mu^2} e^C\right)^{-u}\!,
\end{eqnarray}
on (\ref{rvirt}, \ref{rreal}) to obtain $B[R_{\rm virt}](u)$ and 
$B[R_{\rm real}](u)$. The integrals for $B[R_{\rm virt}](u)$ 
then look like those that appear in evaluating the lowest-order 
correction $r_0$, except that the gluon propagator is raised to the power 
$1+u$ \cite{Ben93a}. However, the integral over $k^2$ obtained for 
$B[R_{\rm real}](u)$ does not converge in the vicinity of $u=0$ 
and cannot be used in (\ref{genfunction}). Constructing the analytic 
continuation of the integral in the usual way by integrating by parts
and defining $\xi=-k^2/\mu^2\,e^C$, we obtain
\begin{eqnarray}
\label{bvirt}
B[R_{\rm virt}](u) &=& \int\limits_0^\infty\frac{d\xi}{\xi}\,\xi^{-u}\,
F_{\rm virt}(\xi,Q/\mu),
\\
\label{breal}
B[R_{\rm real}](u) &=& -\frac{\sin(\pi u)}{\pi u}\int
\limits_0^{\xi_{\rm max}} 
d\xi\,\xi^{-u}\,\frac{d}{d\xi}F_{\rm real}(\xi,Q/\mu),
\end{eqnarray}
with a kinematic upper limit $\xi_{\rm max}$. 
The virtual and real corrections have infrared divergences separately. 
These result in singularities at $u=0$, which cancel in the sum of 
virtual and real corrections. With this pole subtracted 
$B[R_{\rm real}](u)$ approaches a constant at $u=0$ and hence gives 
rise to a contribution to $r_0$, see (\ref{genfunction}). 
It can be shown that this contribution 
is exactly the order $\alpha_s$ contribution from real gluon emission 
despite the fact that this contribution belongs to the cuts (b) 
in Fig.~\ref{fig11} while (\ref{rreal}) followed from the cuts (c).

The resolution to the paradox lies in the unconventional IR regularization 
implied in calculating the Borel transforms \cite{BB95b}. If we keep 
dimensional regularization, the cuts (b) vanish, except for the one with 
no fermion loop. However, we also have to take into account the 
counterterms for the fermion loops that do not lead to vanishing 
scaleless integrals. The Borel transform of the one-gluon emission 
together with the counterterm contributions is proportional to 
$\exp(-u/\epsilon)$, which should be set to zero in the limit 
$\epsilon\to 0$. Thus the one-gluon emission contribution disappears 
together with all other contributions of type (b). It reappears as 
part of (c) in (\ref{breal}). 

If $R$ requires ultraviolet renormalization in addition to 
coupling constant renormalization, (\ref{bvirt}) has to be amended 
by a subtraction function as discussed in Section~\ref{schemedep}. 
The calculation of the subtraction function is described in detail 
in \cite{EPPT82,PP84,BB94a,BBB95}. If $R$ needs infrared 
subtractions and receives only virtual corrections, the procedure 
is essentially identical. The case when $R$ requires IR subtractions 
and receives real and virtual corrections has not been worked out 
in detail so far.

\subsubsection{The dispersive method}
\label{dm}

The bubble diagrams can also be calculated by using the dispersion 
relation 
\begin{eqnarray}
\label{dispersiverep}
\frac{1}{1+\Pi(k^2)} &=& \frac{1}{\pi}\int_0^\infty d\lambda^2\,
\frac{1}{k^2-\lambda^2} \frac{{\rm Im}\Pi(\lambda^2)}{|1+\Pi(\lambda^2)|^2}
\nonumber\\
&&\hspace*{-1cm}
+\,\int_{-\infty}^\infty d\lambda^2\,\frac{1}{k^2-\lambda^2}
\frac{\lambda_L^2}{(-\beta_{0f}\alpha_s)}\,\delta(\lambda^2-\lambda_L^2)
\end{eqnarray}
in (\ref{rvirt}) \cite{BB95a}. Here 
\begin{equation}
\label{landau}
\lambda_L^2 =-\mu^2\exp[-1/(-\beta_{0f}\alpha_s)-C]
\end{equation}
is the position of the Landau pole. This leads to a very 
intuitive characterization of IR renormalon singularities 
\cite{BBZ94,BB95a,BBB95,DMW96}. Note that, since 
$\mbox{Im}\Pi(\lambda^2) = \pi\beta_{0f}\alpha_s$, the first term 
on the right hand side has the same $\alpha_s$ dependence as the 
real term (\ref{rreal}). Moreover, the integral over $k$ left after 
inserting (\ref{dispersiverep}) in (\ref{rvirt}), 
\begin{equation}
r_{0,\rm virt}(\lambda^2,Q)= \int dk^2\, F_{\rm virt}(k,Q)
\frac{1}{k^2-\lambda^2},
\end{equation}
coincides with the first-order virtual correction calculated with 
a massive gluon.\footnote{Since we assumed that the observable is 
gauge-invariant and does not involve the three-gluon coupling 
in the order $\alpha_s$ correction, this identification is meaningful. 
Because of this the present method is often called the `massive gluon' 
method. In general, the identification holds only for virtual 
corrections.} Because the $\alpha_s$ dependence for virtual and real 
corrections is the same after application of the dispersive representation 
(\ref{dispersiverep}), the Borel transform can be represented in 
the particularly simple form \cite{BB95a,BBB95}
\begin{eqnarray}
\label{bxi}
B[R](u) &=& -\frac{\sin(\pi u)}{\pi u}\int
\limits_0^\infty 
d\xi\,\xi^{-u}\,\frac{d}{d\xi}T(\xi,Q/\mu),
\\
T(\xi,Q/\mu) &=& r_{0,\rm virt}(\xi,Q/\mu) + F_{\rm real}(\xi,Q/\mu)\,
\theta(\xi_{\rm max}-\xi)
\end{eqnarray}
where we set $\xi=\lambda^2/\mu^2\,e^C$ in the virtual contribution. 
If the observable $R$ is sufficiently inclusive, one finds that 
\begin{equation}
F_{\rm real}(\xi,Q/\mu) =  r_{0,\rm real}(\xi,Q/\mu),
\end{equation}
where $r_{0,\rm real}$ denotes the correction from emission of a 
virtual gluon with mass $\Lambda$. That is, 
the set of bubble diagrams can be evaluated by taking an integral 
over the first-order virtual and real correction evaluated with 
a finite gluon mass. `Sufficiently inclusive' means that the cuts (c) 
in Fig.~\ref{fig11} are not weighted. Total cross sections and total  
decay widths are 
sufficiently inclusive, but event shape observables in $e^+ e^-$ 
annihilation are not \cite{NS95,BB95b}. It follows from 
(\ref{genfunction}) and (\ref{bxi}) that the coefficients $r_n$ 
can be computed in terms of logarithmic moments of the 
function $T(\xi,Q/\mu)$. 

The series given by the bubble graphs are divergent because of IR and 
UV renormalons. One may still {\em define} the sum of the series 
by defining the Borel integral (\ref{borelint}) as a principal value 
or in the upper/lower complex plane. 
Let $a_s=-\beta_{0}\alpha_s$ and $u=-\beta_0 t$.\footnote{For the set 
of bubble graphs
$\beta_0=\beta_{0f}=N_f T/(3\pi)$. In order that $a_s$ be positive for 
positive $\alpha_s$, we formally consider negative $N_f$. In practical 
applications of the following equation one usually departs from the 
literal evaluation of fermion bubble graphs and uses the full QCD 
$\beta_0$. Since it is negative, $a_s$ is then positive.}
Then 
\begin{eqnarray}
\label{sumrep}
R &\equiv& \int\limits_0^{\infty+i\epsilon}d t\,e^{-t/\alpha_s}\,
B[R](t)
\nonumber\\
&=&\int_0^\infty d\xi\, \Phi(\xi)\,\frac{d}{d\xi}T(\xi,Q/\mu)
+[T(\xi_L-i\eps,Q/\mu)-T(0,Q/\mu)],
\end{eqnarray}
where the {\em effective coupling} $\Phi$ is given by 
\begin{equation}
\label{phi}
\Phi(\xi)= -\frac{1}{\pi}\arctan\left[\frac{a_s\pi}
{1+a_s\ln(\xi)}\right] -\,\theta(-\xi_L-\xi)
\end{equation}
and $\xi_L=-\exp(-1/a_s)$ is related to the position of the Landau 
pole, cf.~(\ref{landau}). The derivation of (\ref{sumrep}, \ref{phi}) 
requires some care and can be found in 
\cite{BB95a,BBB95}.\footnote{The representation 
(\ref{sumrep}, \ref{phi}) of bubble graphs has been 
derived in a slightly different way by \cite{DMW96}. There, the effective 
coupling $\Phi$ is called $\alpha_{\rm eff}$ and the 
distribution function $T$ `characteristic function', 
denoted by ${\cal F}$. \cite{DMW96} do not include the Landau pole 
contribution in the dispersion relation (\ref{dispersiverep}), because 
they have in mind a physical coupling rather than the $\overline{\rm MS}$ 
coupling. As  a consequence  
the term $T(\xi_L-i\eps,Q/\mu)-T(0,Q/\mu)$ is absent from their result. 
This difference is irrelevant for the study of power corrections 
induced by IR renormalons, because one needs to know only the 
function $T(\xi)$ for this purpose, and not the Borel integral. 
As shown in \cite{BBB95} leaving out 
the Landau pole contribution in (\ref{dispersiverep}) implies 
a redefinition of the strong coupling, which differs from the 
standard one by $\Lambda^2/Q^2$ power corrections {\em not} related to 
renormalons and infrared properties. The possible implications 
of such additional power corrections are also discussed 
in \cite{Gru97,AZ97}.} Despite the $\theta$-function the 
effective coupling is continuous 
at $\xi=-\xi_L$ and approaches a finite value as $\xi\to 0$. 

The attractiveness of the dispersive method results from the fact 
that renormalon properties follow directly from the distribution 
function $T(\xi)$ (we omit the second argument for brevity) without 
the integration over $\xi$ having to be done. $R$, defined by 
(\ref{sumrep}), has an imaginary part due to the term $T(\xi_L-i\eps)$. 
This imaginary part persists as $\epsilon\to 0$, because 
$\xi_L<0$ and $T(\xi)$ has a cut for $\xi<0$.
The imaginary part of the Borel integral is directly related 
to renormalon singularities, cf.~(\ref{bpole}, \ref{im}) in 
Section~\ref{divseries}. Because $\xi_L=-\exp(1/a_s)\ll 1$, 
one can expand
\begin{equation}
\label{smallxi}
T(\xi) = \sum_{k,l} c_{kl} \left(\sqrt{\xi}\,\right)^k\ln^l\xi,
\end{equation}
where we anticipated that the expansion goes in powers of $\sqrt{\xi}$ 
and logarithms of $\xi$. Since only the imaginary part for negative 
$\xi$ is related to IR renormalons, it follows that IR renormalon 
singularities are characterized by {\em non-analytic} terms in the 
small-$\xi$ expansion of the distribution $T$ \cite{BBZ94}. Taking into 
account the value of $\xi_L$, the following correspondences are found 
between non-analytic terms in $\xi$, renormalon singularities and 
power corrections ($n,m$ non-negative integer):
\begin{eqnarray}
\label{correspondences}
&&\hspace*{-0.4cm}\xi^n \ln^{m+1} \xi\qquad\!\!
\longleftrightarrow\quad \frac{1}{(n-u)^{m}}
\qquad\,\,\,\,\,\,\,\,\,\,
\longleftrightarrow\quad\left(\frac{\Lambda^2}{Q^2}\right)^n 
\ln^m(\Lambda^2/Q^2),\\
&&\hspace*{-0.4cm}
\xi^{1/2+n} \ln^m \xi \quad\longleftrightarrow\quad \frac{1}{(1/2+n-u)^m}
\quad\longleftrightarrow\quad \left(\frac{\Lambda^2}{Q^2}\right)^{1/2+n} 
\ln^m(\Lambda^2/Q^2).
\end{eqnarray}
These relations provide a direct implementation of the 
correspondence between perturbative infrared behaviour and power 
corrections. 

It is clear that analytic terms in (\ref{smallxi}) are not related 
to IR renormalons, because analytic terms arise from large and small 
momenta.\footnote{For large $k$ the propagator 
$1/(k^2-\lambda^2)$ can be Taylor-expanded and gives rise to (only) 
analytic terms in $\lambda$.} 
Note, however, that analytic terms in  $T(\xi_L-i\eps)$ in 
(\ref{sumrep}) are important for the real part of (\ref{sumrep}) 
to coincide with the principal value of the Borel integral. Although 
the relevance of the principal value is far from obvious, the term 
$T(\xi_L-i\eps)-T(0)$, which is exponentially small in $\alpha_s$ 
(`non-perturbative'), should still be kept for the following reason. 
One would like the sum of the bubble diagrams to equal roughly the 
sum of the perturbative series truncated at its minimal term. There are 
cases \cite{BBB95} for which the real part of $T(\xi_L-i\eps)-T(0)$ 
is parametrically larger in $Q^2$ than the minimal term. In these
cases, (\ref{sumrep}) without the Landau pole contribution comes 
nowhere close to the sum of the perturbative expansion truncated at 
its minimal term. There may of course be non-perturbative corrections 
parametrically larger than the minimal term. However, without any 
positive evidence for them, one would like to avoid introducing them 
by hand.

If one takes $a_s$ negative, ambiguities in the Borel integral arise 
from ultraviolet renormalons. In this case one finds a correspondence 
between UV renormalon singularities and non-analytic terms in 
the expansion of the distribution 
function $T(\xi)$ at {\em large} $\xi$.

If $R$ requires renormalization beyond coupling renormalization, this 
manifests itself as $T(\xi)\sim \ln\xi$ at large $\xi$. Then the 
integral over $\xi$ in (\ref{sumrep}) does not converge. The 
renormalized $R$ includes subtractions, after which the integral 
becomes convergent. The modifications of (\ref{bxi}) and (\ref{sumrep}) 
relevant to quantities requiring additional renormalization can be 
found in \cite{BBB95}. The subtraction function analogous to $S(u)$ 
in (\ref{sch2}) can in fact be determined entirely from the asymptotic 
behaviour of the first-order virtual corrections in the limit of 
large gluon mass. In the $\overline{\rm MS}$ scheme, the subtractions 
do not introduce factorial divergence. As a consequence the non-analytic 
terms in the small-$\xi$ expansion of $T(\xi)$ remain 
unaffected.\footnote{If one uses $\overline{\rm MS}$ subtractions 
for infrared divergences, one cancels a $\ln\xi$ term in the 
small-$\xi$ expansion of the distribution $T$ of the corresponding 
hard scattering coefficient, but all other non-analytic terms 
remain unmodified.} 

\subsubsection{The loop momentum distribution function}
\label{lmd}

The fact that for euclidian quantities renormalons can be characterized
in terms of the loop momentum 
distribution function $F_{\rm virt}(k^2/Q^2)$ of (\ref{rvirt}) in a 
transparent way has been emphasized by \cite{N95}. We have already 
exploited in Section~\ref{renormalonexample} the fact that the 
small and large momentum expansion of $F_{\rm virt}(k^2/Q^2)$ suffices 
for this purpose. In addition to this, the loop momentum distribution 
function provides an easily visualized answer to the question of  
which momentum scales contribute most to a given perturbative 
coefficient. 

From this perspective, the summation of bubble 
graphs can be considered as the extension of Brodsky-Lepage-Mackenzie 
scale-setting \cite{BLM83} envisaged in \cite{LM93}. Thus extended, the 
BLM scale $Q^*$ is given by 
\begin{equation}
r_0\alpha_s(Q^*) = \, \mbox{Eq.}~(\ref{sumrep}).
\end{equation}
Note that the BLM scale is small compared to $Q$ if the cumulative 
effect of higher order perturbative corrections is large. But 
a small BLM scale need not be indicative of a large intrinsic perturbative 
uncertainty, as renormalon ambiguities can still be small. 

For minkowskian quantities a loop momentum distribution function 
that generalizes (\ref{rvirt}) does not exist \cite{N95b} and the 
distribution function $T(\xi,Q/\mu)$ is more useful. For 
euclidian quantities 
the relation between the loop momentum distribution function and 
the distribution function $T(\xi,Q/\mu)$ is given by 
\cite{BBB95b,N95b}
\begin{equation}
T(\xi,1) = \int\limits_0^\infty ds \frac{s}{s+\xi}\,F_{\rm virt}(s),
\end{equation}
where $T(\xi,1)=r_{0,\rm virt}(\xi)$ is only from virtual corrections. 
In turn it follows from (\ref{bvirt}), that the loop momentum 
distribution function can be obtained from the Borel transform 
by an inverse Mellin transformation.  


\section{Renormalons and non-perturbative effects}
\setcounter{equation}{0}
\label{nonp}

In the previous section we have emphasis on the diagrammatic analysis 
of renormalon divergence. In QCD, IR renormalons are taken as an indication 
that a perturbative treatment is not complete and that further terms 
in a power expansion in $\Lambda/Q$, where $\Lambda$ is the QCD scale and 
$Q$ is a `hard' scale, should be added. The perturbative expansion itself is 
ambiguous to the accuracy of such terms 
unless it is given a definite summation 
prescription. In this section we address the question in what 
sense IR renormalons are related to non-perturbative, power-like 
corrections and how perturbative and non-perturbative contributions 
combine to an unambiguous result. In order to examine non-perturbative 
corrections, one has to resort to a solvable model. In 
Section~\ref{sigmamodel} we consider 
the non-linear $O(N)$ $\sigma$-model in the $1/N$ expansion 
as a toy model. 
After general remarks regarding QCD, we consider explicitly 
the matching of IR contributions to twist-2 coefficient functions 
in deep-inelastic scattering and UV contributions to the 
matrix elements of twist-4 operators.

\subsection{The $O(N)$ $\sigma$-model}
\label{sigmamodel}

The euclidian action of the non-linear $O(N)$ $\sigma$-model is given by 
\begin{equation}
S=\frac{1}{2} \int d^d x\,\partial_\mu\sigma^a\partial_\mu\sigma^a,
\end{equation}
where $d=2-\epsilon$ and the fields are subject to the constraint
$\sigma^a\sigma^a =N/g$.
The index `$a$' is summed from 1 to $N$. The `length' of the $\sigma$ field 
is chosen such that a $1/N$ expansion can be obtained. Solving the 
constraint locally for $\sigma^N$, an interacting theory 
for the remaining $N-1$ components is obtained, 
which can be treated perturbatively 
in $g$. Perturbation theory is rather complicated in this theory, because 
the $\sigma$ field is dimensionless and the Lagrangian contains an 
infinite number of interaction vertices after elimination of $\sigma^N$. 
In perturbation theory the fields $\sigma^a$, $a=1,\ldots, N-1$, 
are massless and the perturbative expansion is plagued by severe 
IR divergences. Despite this fact, $O(N)$ invariant Green functions 
are IR finite \cite{Eli83,Dav81} and a sensible perturbation expansion 
is obtained for them.

The non-linear $O(N)$ $\sigma$-model can be solved 
non-perturbatively (in $g$) in an expansion 
in $1/N$ \cite{BLS76}. The $1/N$ expansion follows from introducing 
a Lagrange multiplier field $\alpha(x)$, which makes the generating 
functional 
\begin{equation}
Z[J] = \int {\cal D}[\sigma] {\cal D}[\alpha]\,\exp\left(-S[\sigma,
\alpha] + \int d^d x\,J^a(x)\sigma^a(x)\right),
\end{equation}
with
\begin{equation}
S[\sigma,\alpha] = \frac{1}{2}\int d^d x\,\left\{
\partial_\mu\sigma^a\partial_\mu\sigma^a + \frac{\alpha}{\sqrt{N}} 
\left(\sigma^a\sigma^a-\frac{N}{g}\right)\right\}
\end{equation}
quadratic in the $\sigma$ field. One then integrates over $\sigma$ and 
performs a saddle point expansion of the $\alpha$ integral. There is a 
non-trivial saddle point at 
\begin{equation}
\bar{\alpha}_0 = \sqrt{N}\left(g_0\mu^\epsilon
\Gamma\!\left(\frac{\epsilon}{2}\right)
(4\pi)^{\frac{\epsilon-2}{2}}\right)^{2/\epsilon},
\end{equation}
where $g_0$ denotes the bare coupling and $\mu$ is the renormalization scale 
of dimensional regularization. Defining the renormalized 
coupling $g(\mu)$ by $g_0^{-1}=Z g^{-1}$ with 
\begin{equation}
Z = 1+g(\mu) \Gamma\!\left(\frac{\epsilon}{2}\right)
(4\pi)^{\frac{\epsilon-2}{2}},
\end{equation}
the saddle point approaches
\begin{equation}
\bar{\alpha} \equiv \sqrt{N} m^2 = \sqrt{N}\mu^2\,e^{-4\pi/g(\mu)}
\end{equation}
as $\epsilon\to 0$. As a consequence the $\sigma$ field acquires a 
mass $m$, which is non-perturbative in $g$. 
Furthermore, at leading order in the $1/N$ expansion, 
\begin{equation}
\beta(g) = \mu^2 \frac{\partial g}{\partial \mu^2} = \beta_0 g^2, 
\qquad \beta_0=-\frac{1}{4\pi}
\end{equation}
is exact and the model is asymptotically free. 
The Feynman diagrams of the $1/N$ expansion are constructed from 
the $\sigma$ propagator $\delta^{ab}/(p^2+m^2)$, the propagator 
for $\alpha-\bar{\alpha}$, 
\begin{equation}
\label{alphaprop}
D_\alpha(p) = 4\pi\sqrt{p^2 (p^2+4 m^2)}\left[\ln\frac{
\sqrt{p^2+4 m^2}+\sqrt{p^2}}{\sqrt{p^2+4 m^2}-\sqrt{p^2}}\right]^{-1},
\end{equation}
and the $\sigma^2\alpha$ vertex $\delta^{ab}/\sqrt{N}$. By definition 
bubble graphs of $\sigma$ fields are already 
summed into the $\alpha$ propagator and are to be omitted.

The non-linear $O(N)$ $\sigma$-model has often been used as a toy 
field theory, because it has some interesting features in 
common with QCD. It has only massless 
particles in perturbation theory, but exhibits dynamical mass generation 
non-perturbatively and a mass gap in the spectrum. It is asymptotically 
free, as is QCD, and $m$ is the analogue of the QCD scale $\Lambda$. In 
the following we consider the structure of the short-distance/operator 
product expansion (OPE) of euclidian correlation functions in the 
$\sigma$-model as a toy model for the OPE in QCD. 
The $\sigma$-model has been analysed from this 
perspective in the papers \cite{D82,D84,NSVZ84,NSVZ85,Ter87,BBK98}, 
on which this section is based. 

Because the $\sigma$ field is dimensionless, there exist an infinite 
number of operators of any given dimension that can appear in the OPE. 
In leading order of the $1/N$ expansion, the matrix elements 
factorize and, using the constraint $\sigma^a\sigma^a=N/g$, the number 
of independent matrix elements is greatly reduced. 
In the following it will 
be sufficient to consider the (vacuum) matrix elements of the operators
\begin{equation}
{\cal O}_0=1,\quad 
{\cal O}_2=g \partial_\mu\sigma^a\partial_\mu\sigma^a,\quad
{\cal O}_4=g^2 \partial_\mu\sigma^a\partial_\mu\sigma^a
\partial_\nu\sigma^b\partial_\nu\sigma^b
\end{equation}
to illustrate the point. Note that the equations of motion yield 
\begin{equation}
\label{sigmaeom}
\alpha = -\frac{g}{\sqrt{N}}\partial_\mu\sigma^a\partial_\mu\sigma^a,
\end{equation} 
so that $\langle {\cal O}_2\rangle=-m^2$ at leading order in $1/N$. 
Because of factorization one has $\langle {\cal O}_4\rangle=m^4$ 
at this order.

One can consider as examples the OPE of the amputated two-point function 
$\Gamma(p)$ of the $\sigma$ field and of the two-point function of the 
$\alpha$ field. Because of (\ref{sigmaeom}) the second quantity 
can also be interpreted as the 
two-point correlation function of the scale invariant current 
$j=(-g)/\sqrt{N}\,\partial_\mu\sigma^a\partial_\mu\sigma^a$. 
Introducing a factorization scale $\mu$ satisfying 
$m\ll\mu\ll p$, the OPE of $\Gamma(p)$ reads
\begin{equation}
\Gamma(p) = \sum_n C^\Gamma_n(p^2,\mu)\,\langle {\cal O}_n\rangle(\mu,m)
= p^2+m^2+ O(1/N)
\end{equation}
and realizes an expansion in $m^2/p^2$.
From the second equality one deduces $C^\Gamma_0=p^2$ and 
$C^\Gamma_2=-1$. All other coefficient functions vanish in leading 
order in $1/N$. The OPE of the 
current-current correlation function reads
\begin{eqnarray}
S(p^2,m) &\equiv& i\int d^d x\,e^{ipx}\,\langle 0| T\left(j(x) j(0)\right)
|0\rangle = \sum_n C^S(p^2,\mu)\,\langle {\cal O}_n\rangle(\mu,m)
\nonumber\\
&=& (2\pi)^2\delta^{(2)}(p)\langle\alpha\rangle^2 + D_\alpha(p) + 
O(1/N).
\end{eqnarray}
In the following we drop the disconnected term proportional to 
$\langle\alpha\rangle^2$.  At leading order in $1/N$, the expansion 
\begin{equation}
\label{alphaexpa}
\frac{1}{4\pi}\frac{D_\alpha(p)}{p^2} = \hat{g}(p) + 
\frac{m^2}{p^2} \left(2 \hat{g}(p)-2\hat{g}(p)^2\right) 
+ \frac{m^4}{p^4} \left(-2 \hat{g}(p)-\hat{g}(p)^2+
4\hat{g}(p)^3\right) + \ldots
\end{equation}
follows from (\ref{alphaprop}). (We introduced 
$\hat{g}(\mu)\equiv-\beta_0 g(\mu) = 1/\ln(\mu^2/m^2)$.) Each power 
correction is multiplied by a finite series in $g(p)$. At leading order 
in $1/N$ there are no renormalons and there is no factorization scale 
dependence. The power corrections in $m^2/p^2$ follow from the 
factorizable part of matrix elements of $\sigma$ fields \cite{NSVZ84}. 
Note that the truncated expansion in $m^2/p^2$ and $\hat{g}(p)$ 
has a Landau pole at $p^2=m^2$ due to the IR behaviour of $\hat{g}(p)$.  
The correct analyticity properties of $S(p^2,m)$ are restored only  
after the OPE (the expansion in $m^2/p^2$) is summed.

\begin{figure}[t]
   \vspace{-4.8cm}
   \epsfysize=26cm
   \epsfxsize=18cm
   \centerline{\epsffile{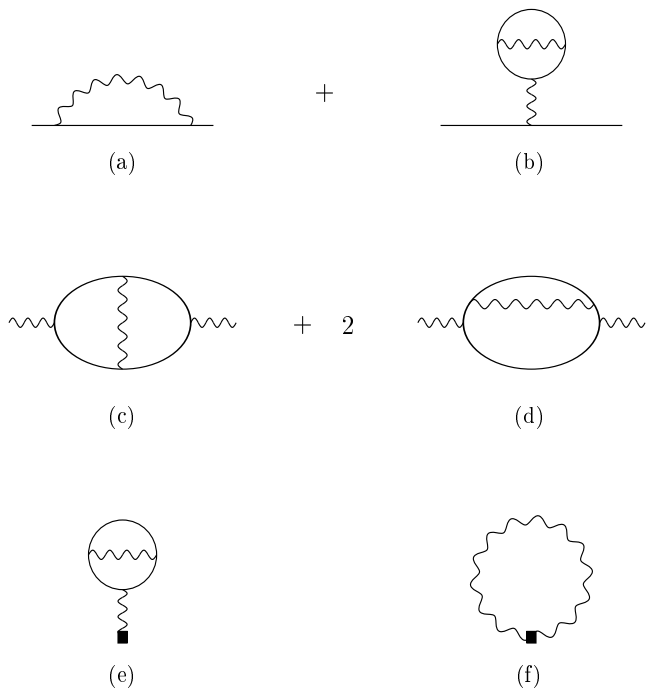}}
   \vspace*{-11.6cm}
\caption[dummy]{\small (a, b) $\sigma$-self-energy diagrams at 
order $1/N$. (c, d) Connected contributions to the 
$\alpha$ propagator at order $1/N$. (e) Non-factorizable 
contribution to the vacuum expectation value of ${\cal O}_2 \propto 
\alpha$. (f) Non-factorizable contribution to the vacuum expectation 
value of ${\cal O}_4\propto \alpha^2$. The solid lines represent  
the $\sigma$ propagator, the wavy lines the leading order 
$\alpha$ propagator (\ref{alphaprop}).\label{fig12}}
\end{figure}
To see the interplay of IR renormalons and operator matrix elements, 
one has to go to the first subleading order in $1/N$. The relevant 
Feynman diagrams in the $1/N$ expansion are shown in 
Fig.~\ref{fig12}. At this order one has to specify a factorization 
prescription in the OPE. If one uses dimensional regularization 
\cite{D82,D84} one is led to the usual situation that the 
coefficient functions have IR renormalons and to the problem how the 
corresponding ambiguities are cancelled. One can also use 
an explicit factorization scale in loop momentum integrals 
\cite{NSVZ84,NSVZ85}. In this case the coefficient functions 
contain only integrations over loop momenta $k>\mu$ and therefore 
have no IR renormalon divergence. The IR renormalon divergence 
appears as a perturbative contribution to the vacuum expectation 
values, if one attempts to separate such a perturbative part from the 
whole.

It is somewhat easier to begin with cut-off factorization, since 
it suffices to calculate the operator matrix elements. The leading 
non-factorizable contributions to the matrix elements of 
${\cal O}_2$ and ${\cal O}_4$ are shown in Fig.~\ref{fig12}e 
and \ref{fig12}f, respectively. The OPE of the 
self-energy diagram \ref{fig12}b is trivially given by the 
first correction to $\langle {\cal O}_2 \rangle$. The non-factorizable 
contribution to $\langle {\cal O}_4 \rangle$ appears as part of 
diagrams~\ref{fig12}a, \ref{fig12}c, \ref{fig12}d, when the 
$\alpha$ line is soft and the $\sigma$ lines are hard. The contribution 
to the vacuum expectation value of the operator $\alpha^2$ (which 
is proportional to ${\cal O}_4$) from Fig.~\ref{fig12}f 
is given by
\begin{equation}
\label{alphasquared}
\langle \alpha^2\rangle(\mu,m) = \int\limits_{p^2<\mu^2} 
\frac{d^2p}{(2\pi)^2}\,D_\alpha(p).
\end{equation}
Note that the restriction $p^2<\mu^2$ {\em defines} the otherwise 
singular operator product $\alpha^2$. The integral can be evaluated 
\cite{NSVZ84} with the result
\begin{equation}
\label{vev1}
\langle \alpha^2\rangle(\mu,m) = m^4\left[\mbox{Ei}(\ln A)+
\mbox{Ei}(-\ln A)-\ln\ln A -\ln(-\ln A) - 2\gamma_E\right],
\end{equation}
where $\gamma_E=0.5772\ldots$ is Euler's constant, $\mbox{Ei}(-x)=
-\int_{x}^{\infty} dt\, e^{-t}/t$ the exponential integral function and 
\begin{equation}
A = \left(\sqrt{1+\frac{\mu^2}{4 m^2}}+\sqrt{\frac{\mu^2}{4 m^2}}
\,\right)^4.
\end{equation}
Note that $F(x)=\mbox{Ei}(-x)-\ln x$ has an essential singularity at $x=0$ 
but no discontinuity. By assumption $\mu\gg m$, hence $\ln A\to 
2/\hat{g}(\mu)\gg 1$. To expand (\ref{vev1}) in this limit, up to 
terms that vanish as $\mu\to \infty$, one needs the asymptotic expansion 
of $F(x)$ at large $x$. For positive argument the asymptotic expansion 
is
\begin{equation}
F(x) = -\ln x + e^{-x} \sum_{n=0}^\infty (-1)^{n+1}\,\frac{n!}{x^{n+1}}.
\end{equation}
If the divergent series is understood as its Borel sum, the right hand 
side equals $F$. For negative, real argument, one obtains the 
asymptotic expansion
\begin{equation}
\label{aspos}
F(-x) = e^x \left[\sum_{n=0}^\infty \frac{n!}{x^{n+1}} - e^{-x}
\left(\ln x \mp i\pi\right)\right].
\end{equation}
Note the `ambiguous' imaginary part in the exponentially small term. 
The interpretation of (\ref{aspos}) is as follows (compare the discussion 
at the end of Section~\ref{divseries}): the upper (lower) sign is 
to be taken, if the (non-Borel-summable!) 
divergent series is interpreted as the Borel 
integral in the upper (lower) complex plane. With this interpretation,  
(\ref{aspos}) is {\em exact} and unambiguous. Inserting these expansions, 
the condensate is given by
\begin{eqnarray}
\label{vev2}
\langle \alpha^2\rangle(\mu,m) &=& \mu^4\sum_{n=0}^\infty \left(
\frac{\hat{g}(\mu)}{2}\right)^{n+1} \!\!n! \,+ \,2\hat{g}(\mu)\,\mu^2 m^2 
\nonumber\\ 
&&\hspace*{-1cm} + \,m^4
\left[-2\ln\frac{2}{\hat{g}(\mu)} \pm i\pi-2\gamma_E-4\hat{g}(\mu)+
\frac{\hat{g}(\mu)^2}{2}\right] + O\!\left(\frac{m^2}{\mu^2}\right).
\end{eqnarray}
The expansion for large $\mu$ has quartic and quadratic terms in 
$\mu$, parametrically larger than the `natural magnitude' of the 
condensate of order $m^4$. The power terms in $\mu$ arise from the 
quartic and quadratic divergence of the Feynman integral 
(\ref{alphasquared}), i.e. from loop momentum $p\sim \mu$. The 
$\mu$ dependence cancels with the $\mu$-dependence of the coefficient 
functions in the OPE. In particular the $\mu^4$-term cancels with the 
coefficient function of the unit operator. The important point to note 
is that the condensate is unambiguous, but separating the 
`perturbative part' of order $\mu^4$ is not, since the asymptotic 
expansion for $\mu/m\gg 1$ leads to divergent, non-sign-alternating 
series expansions, which require a summation prescription. The 
`non-perturbative part' of order $m^4$ depends on this prescription 
(via $\pm i\pi$ in (\ref{vev2})). In a purely perturbative calculation,  
one would only obtain the divergent series expansion. The infrared 
renormalon ambiguity of this expansion would lead us to correctly 
infer the existence of a non-perturbative power correction of  
order $m^4$. However, it does not allow us to say much about the 
magnitude of the power correction which is determined by other 
terms, such as $\ln(2/\hat{g})$ in (\ref{vev2}).\footnote{At first 
sight there seems to be a problem with the argument, because of 
the term proportional to $\mu^2 m^2$. However, this is a pure 
cut-off term, which cancels in physical quantities when the condensates 
are combined with coefficient functions. In dimensional regularization 
such terms are absent.}

In dimensional regularization power dependence on the factorization 
scale $\mu$ is absent and  IR renormalon divergence is 
part of the coefficient function. If the power terms 
in $\mu$ in (\ref{vev2}) are deleted, 
it seems that the remainder has a twofold 
ambiguity. This should be taken as an indication that the definition 
of a renormalized condensate in dimensional regularization 
requires some care, because the summation prescription for  
the coefficient functions depends on it.\footnote{The fact that 
coefficient functions depend on the definition of the condensates 
is of course true in any factorization scheme. However, in 
some schemes the subtleties in handling divergent 
series expansions may be avoided.} This point has been studied in detail by 
\cite{D82,D84}.

Consider as in (\ref{alphasquared}) 
the condensate of $\alpha^2$, but defined 
in dimensional regularization instead of a momentum cut-off:
\begin{equation}
\label{alphasquareddim}
\langle \alpha^2\rangle(\mu,m) = \mu^\epsilon\int\limits 
\frac{d^dp}{(2\pi)^d}\,D_\alpha(p,d) = 
\frac{m^2/(4\pi)}{\Gamma(1-\epsilon/2)}\,\left(\frac{m^2}{4\pi\mu^2}
\right)^{\!-\epsilon} 
\!\int\limits_0^\infty d\!\left(\frac{p^2}{m^2}\right) 
\left(\frac{p^2}{m^2}\right)^{\!-\epsilon}\!\!D_\alpha(p,d). 
\end{equation}
Since the integral contains no scale other than $m$, it must be 
proportional to $m^4$. $D_\alpha(p,d)$ denotes the $\alpha$ propagator 
(\ref{alphaprop}) before the limit $\epsilon\to 0$ is taken. 
However, for the following 
short-cut of the detailed analysis of \cite{D82,D84} it is 
sufficient to set $d=2$ in the $\alpha$ propagator. From the 
treatment of the integral in cut-off regularization we learn that 
we should focus on the UV behaviour of the integral. 
Hence expanding (cf.~(\ref{alphaexpa}))
\begin{equation}
D_\alpha(p) = 4\pi m^2 u \sum_{k, l} \frac{c_{kl}}{u^k}\,\frac{1}{
\ln^l u}
\end{equation}
with $u=p^2/m^2$, we obtain
\begin{equation}
\label{uvint}
\langle \alpha^2\rangle(\mu,m) = \frac{m^4}{\Gamma(1-\epsilon/2)}\,
\left(\frac{m^2}{4\pi\mu^2}\right)^{\!-\epsilon} 
\!\sum_{k, l} c_{kl} \int\limits_{u_0}^
\infty du \,\frac{u^{1-\epsilon-k}}{\ln^l u}, 
\end{equation}
with an (arbitrary and irrelevant) IR cut-off $u_0>1$. Now write
\begin{equation}
\frac{1}{\ln^l u} = \int\limits_0^\infty d v\,v^{l-1} u^v.
\end{equation}
The $u$-integration leads to UV poles of the form $1/(-2+\epsilon+k+v)$. 
Keeping only those, 
\begin{equation}
\label{uvint2}
\langle \alpha^2\rangle(\mu,m) \sim  \frac{m^4}{\Gamma(1-\epsilon/2)}\,
\left(\frac{m^2}{4\pi\mu^2}\right)^{\!-\epsilon} 
\!\sum_{k, l} c_{kl} \int\limits_{0}^
\infty dv \,\frac{v^{l-1}}{-2+\epsilon+k+v} 
\end{equation}
follows. To define the $v$-integral it is necessary to take complex 
$\epsilon$. (We also take $\mbox{Re}(\epsilon)>0$, because the 
$\sigma$-model is super-renormalizable in $d<2$.) For $k=2$, and 
only for $k=2$, one obtains a pole in $\epsilon$ which can be 
subtracted as usual. This pole arises from a logarithmically 
UV divergent 
integral. The terms with $k=0$ ($k=1$) correspond to the quartically 
(quadratically) divergent terms in (\ref{alphasquareddim}). For these 
terms the limit $\epsilon\to 0$ is finite, but depends on whether 
it is taken from the upper or the lower complex plane, because of the 
pole at $v=2-\epsilon-k$ in (\ref{uvint2}). The difference between the 
two definitions of the dimensionally renormalized condensate is
\begin{equation}
\left[\lim_{\epsilon\to +i0} - \lim_{\epsilon\to -i0}\right]
\langle \alpha^2\rangle(\mu,m) 
= 2\pi i\,m^4 \sum_{k=0}^1 \sum_{l=1} c_{kl}\,(2-k)^{l-1}.
\end{equation}
From (\ref{alphaexpa}) only $c_{01}=1$, $c_{11}=-c_{12}=2$ are 
non-zero for $k<2$ and the result is
\begin{equation}
\label{alphaamb}
\left[\lim_{\epsilon\to +i0} - \lim_{\epsilon\to -i0}\right]
\langle \alpha^2\rangle(\mu,m) 
= 2\pi i\,m^4.
\end{equation}
The approximations made do not allow us to calculate the condensate 
itself. However, comparison of (\ref{alphaamb}) with (\ref{vev2}) 
demonstrates that the difference between the two limits coincides 
with the difference in the $m^4$ terms in (\ref{vev2}), when the 
perturbative parts are subtracted. It is interesting to note that 
although power divergences do not give rise to counterterms 
in dimensional regularization, they have not completely disappeared 
in the limit $\epsilon \to 0$ in the sense that they render the 
limit non-unique.

A more precise analysis also demonstrates that the 
summation prescription for the divergent series expansions of the 
coefficient functions depends on how the limit $\epsilon\to 0$ is 
taken. The OPE of Green functions is 
unambiguous, if the limit is taken in the same way as for the 
condensates. To this end the works of \cite{D82,D84} begin 
with an analysis 
of the OPE of bare Green functions in $d$ dimensions. The 
OPE exists also in the regularized theory ($\epsilon$ finite),
\begin{equation}
\Gamma(p,\epsilon) = \sum_n C^\Gamma_n(p^2,\epsilon)\,
\langle {\cal O}_n\rangle(m,\epsilon),
\end{equation}
taking the self-energy as an example. In the regularized theory 
the separation in coefficient functions and matrix elements is unique 
and well-defined without further prescriptions. However, 
the analytic structure in $\epsilon$ of the individual terms on 
the right hand side is different from that of the unexpanded 
self-energy. The latter has a straightforward limit as 
$\epsilon\to 0$ (we assume that counterterms have been included), 
but a condensate of dimension $d$ on the right hand side has 
poles at $\epsilon=2 k/l$ ($k<d$, 
$l$ positive integer) related to the power 
divergences of the operator. The poles accumulate at $\epsilon=0$ 
and hence the limit $\epsilon\to 0$ has to be taken from complex 
$\epsilon$. But the limit $\epsilon\to 0$ has to be taken in the same way 
for all terms in the OPE and this is how the definitions of 
renormalized condensates and coefficient functions are related to each 
other. At finite $\epsilon$ there are no renormalon singularities 
in the coefficient functions and the Borel integrals are defined. 
When $\epsilon$ approaches zero, singularities develop in the 
Borel transform but the limit also entails a prescription of how the 
contour is to be chosen in the Borel integral to avoid the 
singularities. 

To see this in more detail, it may be helpful to consider the 
integrals
\begin{equation}
\label{btone}
\sum_n g^{n+1} \int\limits_0^\lambda d k^2\,\beta_0^n\ln^n k^2 
\quad\longrightarrow\quad
\int\limits_0^\lambda dk^2\,(k^2)^{\beta_0 t}.
\end{equation}
To the right of the arrow the Borel transform of the series is 
indicated, which has a single IR renormalon pole at $\beta_0 t=-1$. 
In the regularized theory, the corresponding series is 
\begin{equation}
\sum_n g_0^{n+1}\int\limits_0^\lambda d k^2 
\left(-\frac{\beta_0(\epsilon)}{\epsilon}\right)^n (k^2)^{-n\epsilon},
\end{equation}
where $\beta_0(\epsilon)$ is a function that approaches $\beta_0$ 
as $\epsilon\to 0$. These integrals do not lead to divergent series,  
contrary to the logarithmic integrals for vanishing $\epsilon$. For any 
given $n$, $\epsilon$ can be made small enough for the integral to  
converge. However, for any given $\epsilon$, there always exists 
an $n$ beyond which the integrals diverge and have to be defined in the 
sense of an analytic continuation in $n$. This is the reason why the 
limit $\epsilon\to 0$ and the 
large-order behaviour $n\to \infty$ do not commute. Taking the integrals, 
one finds accumulating poles at $\epsilon=1/n$ in the sum of the series. 
The Borel transform with respect to $g$ is given by
\begin{equation}
\int\limits_0^\lambda d k^2\,\exp\left(-\frac{\beta_0(\epsilon)t}{\epsilon} 
\left[(k^2)^{-\epsilon}-1\right]\right).
\end{equation}
The `$-1$' in the exponent takes into account the coupling 
renormalization counterterms. As $\epsilon\to 0$ one recovers 
the Borel transform in (\ref{btone}). But for any finite $\epsilon$ 
the behaviour of the $k^2$ integrals at small $k^2$ is very different. 
In fact for $\epsilon>0$ the integral diverges, so we define 
the integral as the analytic continuation from negative $\epsilon$. As a 
result one finds that the pole at $\beta_0 t=-1$, which arises 
in the limit $\epsilon\to 0$, should be interpreted as 
$1/(1\pm i0 + \beta_0 t)$ depending on whether the limit is taken from 
the upper or lower right half plane. The corresponding difference 
in the Borel integrals cancels exactly the difference in the 
condensates. 

The OPE of the self-energy can be obtained exactly at order $1/N$ 
\cite{BBK98}, and the result confirms what one would expect from the 
above discussion. The expansion in $m^2/p^2$ of diagram (a) of 
Fig.~\ref{fig12}\footnote{The self-energy in \cite{BBK98} is subtracted 
at zero momentum, in which case diagram (b) is subtracted completely.} 
can be expressed in the form 
\begin{equation}
\Sigma(p) = \frac{p^2}{N} \int_0^\infty\! dt\,
\sum_{n=0}^\infty \left(-\frac{m^2}{p^2}\right)^n
\Bigg\{e^{-t/g(p)}\Bigg[F_{\rm p}^{(n)}[t]\frac{1}{g(p)}
+G_{\rm p}^{(n)}[t]\Bigg] - H_{\rm np}^{(n)}[t]\Bigg\}. 
\label{sesigex}
\end{equation}
The explicit expressions for the functions $F_{\rm p}^{(n)}[t]$, 
$G_{\rm p}^{(n)}[t]$, $H_{\rm np}^{(n)}[t]$ can be found in \cite{BBK98}, 
but only the structure of the result is of importance. The two terms 
that are multiplied by $e^{-t/g(p)}$ can be interpreted as Borel transforms 
of perturbative expansions of coefficient functions. The function 
$H_{\rm np}^{(n)}[t]$ originates from the loop momentum region, where 
the momentum of the $\alpha$ propagator in diagram (a) is of order $m$, 
and hence probes its long-distance behaviour. This term corresponds to 
condensates of $\alpha^2$. 

The functions $F_{\rm p}^{(n)}[t]$, $G_{\rm p}^{(n)}[t]$, 
$H_{\rm np}^{(n)}[t]$  
have singularities in the complex $t$ plane 
at integer values $t=\pm k$, $k=1,2,\ldots$. These are just the 
UV and IR renormalon singularities. All IR renormalon singularities at 
positive values of $t$ cancel in the integrand, so that the integral, 
and hence the OPE, is well defined. 
The cancellation of a particular singularity at $t= t_0$ occurs 
between $G_{\rm p}^{(n)}[t]$  and $H_{\rm np}^{(n+t_0)}[t]$ and thus 
involves a cancellation between a short-distance coefficient 
and an operator matrix element over different orders in the power 
expansion. As a consequence of the singularities 
in individual terms of the sum over $n$, the summation and 
the integration over $t$ cannot be interchanged, 
unless the integration 
contour is shifted slightly above (or below) the real axis. 
This amounts to a simultaneous prescription for summing the 
divergent series expansions of coefficient functions as well as a 
definition of the renormalized condensates.
Only after such a definition can the OPE be truncated at a given 
order in $m^2/p^2$.

In Section~\ref{divseries} we asked why the Borel integral should 
play a privileged role in defining divergent series and whether the 
association of IR renormalons with power corrections does not rely 
too much on this idea. The $O(N)$ $\sigma$ model in the $1/N$ 
expansion provides an example which confirms the picture assumed 
there. The Borel integral emerges as the natural way to define 
the divergent series that arise in the limit $\epsilon \to 0$. In 
particular, the Borel representation (\ref{sesigex}) emerges naturally 
in the exact OPE of the self-energy.

The $\sigma$ model is still special because the leading, 
factorizable contributions in $1/N$ to the condensates are unambiguous 
or factorization scale independent. As a consequence the 
power-like ambiguities in defining perturbative expansions are 
parametrically smaller in $1/N$ than the actual condensates. This 
tells us that some caution is necessary in identifying the 
magnitude of the `renormalon ambiguity' with the magnitude of power 
corrections. It is probably more appropriate to say that power 
corrections are expected to be {\em at least} as large as 
perturbative ambiguities. However, a similar parametric suppression 
of perturbative ambiguities 
does not seem to take place in neither the large-$N_c$ nor the 
large-$N_f$ limit of QCD.

\subsection{IR renormalons and power corrections}

We have shown above how IR renormalons arise in asymptotically 
free theories, when one performs an asymptotic expansion in 
$\Lambda/Q$, where $\Lambda$ is the intrinsic scale of the 
theory and $Q$ a large external scale. In the following we 
summarize the conclusions from the $\sigma$-model with respect 
to applications in QCD, recollecting in part the remarks of 
Section~\ref{fope}. The tacit assumption 
is that the structure of short-distance expansions in QCD is 
as in the $\sigma$-model. We then check perturbatively in a QCD example 
that the power divergences of matrix elements match with 
IR renormalons.

\subsubsection{Summary}
\label{sumy}

First, let us emphasize that IR renormalon ambiguities are not a 
problem of QCD, but a problem of doing perturbative calculations 
in QCD, which implicitly or explicitly require some kind of 
factorization, and an expansion in a ratio like $\Lambda/Q$. If 
we could do non-perturbative calculations, IR renormalons would 
just be artefacts that appear in the expansion of the exact 
(and well-defined) result. 

Let us imagine that an observable $R$, which depends on at least 
the two scales $Q$ and $\Lambda$, can be written as an 
expansion\footnote{We assume that $R$ has been made dimensionless.}
\begin{equation}
R(Q,\Lambda) = \sum_n \frac{C_n(Q,\mu)}{Q^{d_n}} \otimes 
\langle\langle {\cal O}_n\rangle\rangle(\mu,\Lambda).
\end{equation}
The product may be a normal product or a convolution and the 
operator ${\cal O}_n$ of dimension $d_n$ may be local or non-local. 
The matrix element may be a vacuum matrix element or a matrix element 
between hadron states. (We use the double bracket to indicate that 
the external state may be complicated.) 
We assume that the $C_n(Q/\mu)$ can be 
calculated as a series in $\alpha_s$. It is not obvious that such 
an expansion in powers and logarithms of $\Lambda/Q$ always exists. 
Or one may know only the form of the first term, but not the form 
of power corrections to it. This is the most interesting situation from 
the point of view of IR renormalons.

We assume that factorization is done in dimensional regularization. 
If one uses another factorization scheme, the wording of the following 
changes but the conclusions do not. In dimensional regularization 
the coefficient functions $C_n(Q,\mu)$ have IR renormalons from 
integrating Feynman integrals over loop momenta much smaller than 
$Q$. With regard to power corrections we note:

(i) Renormalon ambiguities in $C_n(Q,\mu)$ are power-suppressed. 
Non-perturbatively they are cancelled by ambiguities in defining the 
(renormalized) matrix elements 
$\langle\langle {\cal O}_m\rangle\rangle$ with $d_m>d_n$. 
Contrary to the $\sigma$-model one cannot trace this cancellation 
non-per\-tur\-bative\-ly in QCD. However, if QCD is a consistent 
theory and if $R$ is physical, this cancellation must occur. 
In this way, IR renormalons in $C_n$ lead us to introduce 
parameters for power corrections with a dependence on $Q$ (given 
by $C_m/Q^{d_m} \langle\langle {\cal O}_m\rangle\rangle$) that 
matches the scaling behaviour of the renormalon ambiguity. This 
is the minimalistic, but also most rigorous and most universally
applicable use of IR renormalons. 

(ii) The analysis of Feynman diagrams gives some information on the 
form of the operator ${\cal O}_m$. But IR renormalons provide no 
information on the magnitude of $\langle\langle {\cal O}_m\rangle\rangle$. 
It is natural to think of $\langle\langle {\cal O}_m\rangle\rangle$ 
as at least as large as the renormalon ambiguity. The $\sigma$-model 
is an example where the matrix elements are parametrically larger than 
their ambiguities, both at large $N$ and at $g\ll 1$.

(iii) The IR renormalon approach to power corrections does not provide 
a `non-perturbative method'. Viewed from the low-energy side, IR renormalons 
are related to {\em ultraviolet} properties of {\em operators} and 
not to matrix elements. The analysis of the $\sigma$-model shows that 
the IR renormalon in $C_n$ is related to a power divergence of degree 
$d_m-d_n$ of ${\cal O}_m$. In a cut-off factorization scheme 
with factorization scale $\mu$, divergent 
series appear in the expansion of matrix elements 
for $\mu/\Lambda\gg 1$, and the same 
statement holds. In Section~\ref{disex} we demonstrate this for 
deep inelastic scattering in QCD by evaluating the ultraviolet 
contributions to twist-4 operator matrix elements perturbatively. 
As a consequence of being UV with respect to the scale $\Lambda$, 
IR renormalons do not distinguish 
matrix elements of the same operator but taken between different states.

(iv) Renormalon factorial divergence is closely connected with logarithms 
of loop momentum, which in turn are related to the running coupling. 
This leads to the universal appearance of $\beta_0$, $\beta_1$, etc., in 
the large-order behaviour. On the other hand, power corrections inferred 
from IR renormalons and power corrections in general 
have {\em nothing} to do with the low-energy 
properties of the running coupling. They are process-dependent and, 
generally speaking, non-universal.

IR renormalons can be universal for a restricted set of observables, 
if the same operator appears in their short-distance 
expansion.\footnote{If the operator is non-local and is multiplied 
with the coefficient function in the sense of a convolution, the situation 
is more complicated, because one also has to unfold the convolution.} 
However, universality of IR renormalons does not imply universality 
of non-perturbative effects. This is true only if the operator is not only 
the same, but is also taken between the same external states. 

(v) If this strong form of universality holds for a set of 
observables, one can relate power 
corrections to them on the basis of knowing only the IR renormalon 
behaviour of coefficient functions. In particular, one can relate 
the leading power correction on the basis of the perturbative 
expansion at leading power. For simplicity, consider two observables 
\begin{eqnarray}
\label{obs1}
R &=& C_0 + \frac{C_1}{Q^d}\,\langle\langle {\cal O}\rangle\rangle,
\\
\overline{R} &=& \overline{C}_0 + \frac{\overline{C}_1}{Q^d}\,
\langle\langle {\cal O}\rangle\rangle,
\end{eqnarray}
and denote by $\delta C_{0 |t=-d/(2\beta_0)}$ the renormalon ambiguity 
in $C_0$ of order $1/Q^d$, which is related directly to the large-order 
behaviour. Then it follows that
\begin{equation}
\label{renandcoeff}
\frac{\delta\overline{C}_0}{\delta C_0}{}_{\big|t=-d/(2\beta_0)} = 
\frac{\overline{C}_1}{C_1},
\end{equation}
and this ratio can be expanded in $\alpha_s(Q)$. In particular, the 
ratio of the uncalculable normalizations of IR renormalon behaviour 
is given by the ratio of the coefficient functions $C_1$, 
$\overline{C}_1$ evaluated to lowest order in $\alpha_s$. Conversely, 
knowing the left-hand side of (\ref{renandcoeff}), the 
relative magnitude of $1/Q^d$ power corrections of the two observables 
can be predicted systematically as an expansion in $\alpha_s(Q)$. 
One observable has to 
be used to fix the absolute normalization, i.e. to determine 
$C_1\langle\langle {\cal O}\rangle\rangle$ from data. The procedure 
described parallels the phenomenological use of the OPE in 
standard situations such as QCD sum rules or deep inelastic 
scattering.

Note that in practice, in connection with renormalons, universality 
often takes the status of an assumption. This is so, because to 
establish universality, one needs to know enough of the operator 
structure of power corrections that it may be possible to 
compute $C_1$ and $\overline{C}_1$ directly, thus by-passing 
(\ref{renandcoeff}) and the IR renormalon argument.

(vi) There is the problem of consistently combining (divergent) perturbative 
expansions in dimensional renormalization with phenomenological 
parametrizations of power corrections. For the purpose of discussion,  
let us consider the simplified structure of (\ref{obs1}) 
with only one parameter and a single, corresponding IR renormalon 
singularity in $C_0$ at $t=-d/(2\beta_0)$. If we knew the singularity 
exactly, we could subtract it from the series. Recalling that 
the $\mu$-dependence of the IR renormalon singularity is an 
over-all factor $(\mu/Q)^d$ (up to logarithms), we write (\ref{obs1})
as 
\begin{equation}
\label{rewrite}
R = \left[C_0-C_0^{\rm as}\left(\frac{\mu}{Q}\right)^{\! d}\,\right] + 
\frac{1}{Q^d}\,\left[C_0^{\rm as}\,\mu^d+C_1
\,\langle\langle {\cal O}\rangle\rangle\right],
\end{equation}
where $\Lambda<\mu<Q$ and $C_0^{\rm as}$ denotes the exact asymptotic 
behaviour. Both square brackets are now separately 
well-defined. Note that this rewriting results in exactly the 
same representation as would be obtained with cut-off factorization. 
In reality the subtraction can be carried out at best approximately. 
Moreover, $C_0$ is known only in the first few orders. 

Suppose we choose $\mu$ as close to $\Lambda$ as possible for $\alpha_s(\mu)$ 
to be perturbative. In this case the subtraction 
is effective only as one gets close to the minimal term of the 
series expansion of $C_0$. It may turn out that the phenomenological 
determination of the power suppressed term is large compared 
to the last term kept in the expansion of $C_0$. In this case 
$C_0^{\rm as}\,\mu^d$ is small and IR renormalons are not an issue. 
It is sometimes argued \cite{NSVZ85} that such a 
numerical fact is at the basis of the success of QCD sum rules.

It may also turn out that the phenomenological 
determination of the power suppressed term is not large, but 
of the order of the last known term in the truncated series. 
In this case the phenomenological power correction may 
parametrize the effect of higher order perturbative corrections rather 
than a truly non-perturbative effect. It is still reasonable 
to use such an effective parametrization, because, as illustrated 
by (\ref{rewrite}), the dominant contribution to perturbative 
coefficients in sufficiently large orders can be combined 
with $\langle\langle {\cal O}\rangle\rangle$. Moreover, if the 
minimal term in $C_0$ is reached at not very high orders, 
the sum of higher order corrections parametrized in this way, 
may indeed scale approximately like a power correction. 

The important conclusion is that combining power corrections 
with truncated perturbative series is meaningful in the sense 
that the error incurred is never larger and most likely smaller 
than the error one would obtain without using information on power 
corrections. The improvement comes from the fact that the error 
is now determined by the degree to which the perturbative correction 
is non-universal in intermediate orders 
rather than by the size of the perturbative correction itself. 
For a related discussion see \cite{D84,MS96}.

\subsubsection{Example: DIS structure functions}
\label{disex}

In this section\footnote{This section is based on unpublished notes 
worked out in collaboration with V.M.~Braun and L.~Magnea. It is somewhat 
more technical and can be 
omitted for first reading. The reader may return to it in connection 
with Section~\ref{dismod}, where we discuss the renormalon model of 
twist-4 corrections to deep inelastic scattering structure functions.} 
we demonstrate the 
cancellation of IR renormalons in coefficient functions with ultraviolet 
contributions to matrix elements at the one-loop order and to 
twist-4 accuracy for the longitudinal structure function in deep 
inelastic scattering. This example serves to illustrate the operator 
interpretation of IR renormalons in a more involved situation than 
the OPE of current-current correlation functions discussed in 
Section~\ref{fope}. The motivation for choosing this more complicated 
example is that it is of interest in context with the phenomenological 
modelling of twist-4 corrections discussed in Section~\ref{dismod}.

\begin{figure}[t]
   \vspace{-3.9cm}
   \epsfysize=30cm
   \epsfxsize=20cm
   \centerline{\epsffile{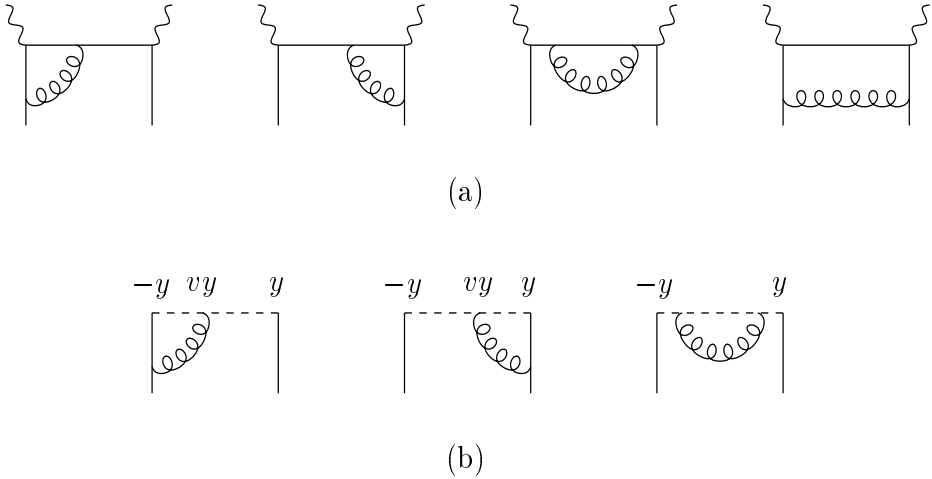}}
   \vspace*{-19.1cm}
\caption[dummy]{\small (a) Diagrams that contribute to the twist-2 
coefficient function in the operator product expansion 
of the hadronic tensor. (Wave-function 
renormalization on the external quark legs is not shown.) The 
wavy lines denote the external current with momentum $q$. When the 
gluons are dressed with fermion loops, these diagrams contribute 
at leading order in the flavour expansion. (b) Diagrams 
that give contributions to the matrix elements of twist-4 operators 
to leading order in the flavour expansion. The third diagram is 
scaleless and vanishes. \label{fig13}}
\end{figure}
We begin with some notation. The (spin-averaged) deep-inelastic scattering 
cross section of a virtual photon with momentum $q$ from a
nucleon with momentum $p$ is obtained from the hadronic tensor
\begin{eqnarray}
W_{\mu\nu} &=& \frac{1}{4\pi}\sum_\sigma\int d^4z\,e^{iq z}\,
\langle N(p,\sigma)|j_\mu(z) j_\nu(0)|N(p,\sigma)\rangle
\\
&=& \left(g_{\mu\nu}-\frac{q_\mu q_\nu}{q^2}\right)\frac{F_L}{2 x} 
- \left(g_{\mu\nu}+\frac{q^2}{(p\cdot q)^2}\,p_\mu p_\nu -\frac{p_\mu
q_\nu+p_\nu q_\mu}{p\cdot q}\right)\frac{F_2}{2 x},
\nonumber
\end{eqnarray}
where $x=Q^2/(2 p\cdot q)$ and $Q^2=-q^2$ and $j_\mu$ is the 
electromagnetic current $\bar{\psi}\gamma_\mu \psi$. In the following, 
the spin average over $\sigma$ is always implicitly understood. 
At leading order in the expansion in $1/Q^2$, the longitudinal 
structure function can be written as a convolution
\begin{equation}
\label{conv2}
F_L(x,Q^2)_{|\rm twist-2} = 
\int\limits_x^1\frac{d\xi}{\xi}\,C_{2,L}(\xi,\alpha_s(Q),Q^2/\mu^2)\,
F(x/\xi,\mu) + \,\mbox{gluon contribution}.
\end{equation}
Here $F$ is the usual quark distribution, defined through 
the matrix element
\begin{equation}
\label{twist2}
\langle N(p)|\bar{\psi}(y)\!\not\!y \psi(-y)|N(p)\rangle(\mu) = 
2 \,p\cdot y\int\limits_{-1}^1 d\xi\,e^{2 i\xi p\cdot y}\, 
F(\xi,\mu),
\end{equation}
where $y$ is the light-like projection of $z$, 
$y_\mu=z_\mu-(z^2 p_\mu)/(2p\cdot z)$ for $p^2=0$. 
The quark fields at positions 
$y$ and $-y$ are joined by a path-ordered exponential that makes the 
operator product gauge-invariant. We do not write out the path-ordered 
exponential explicitly. 
We will check the matching of IR renormalons and UV contributions 
to twist-4 operators only to leading order in the flavour expansion. 
The $N_f$ massless quarks are assumed to have identical electric 
charges and 
in (\ref{twist2}) a sum over the $N_f$ quark flavours 
is assumed. The leading order in the flavour expansion is equivalent 
to the analysis of IR regions of the one-loop diagrams (see 
Fig.~\ref{fig13}a) with an 
important exception: there is also a gluon contribution to $F_L$ at 
one loop (not shown in the Figure), 
but it does not have an internal gluon line. Consequently, 
there are no contributions of order $\alpha_s^{n+1}$ (with $n>0$) 
to the gluon matrix elements in leading order of the flavour expansion, 
and hence we will not consider them here. However, it is important 
to keep in mind that there are $1/Q^2$ power corrections from gluon 
matrix elements as well and that they are not suppressed in any way. 
The flavour expansion does not treat soft quark lines and renormalons 
appear in the gluon matrix elements only at next-to-leading order 
in the flavour expansion. In leading order of the flavour expansion 
there is a contribution from diagrams where a quark (or anti-quark) in a 
cut quark loop connects to the external hadron state, which we do not 
consider here. This contribution is not relevant for pure 
non-singlet quantities. 
With these restrictions in mind, we continue to analyse 
the non-singlet contribution to the quark matrix elements as shown 
in Fig.~\ref{fig13}.

The coefficient function $C_{2,L}$ vanishes at order $\alpha_s^0$. 
To obtain it at leading order in the flavour expansion, it is therefore 
sufficient to evaluate (\ref{twist2}) between quark states 
at tree level, which gives $F(\xi)=\delta(1-\xi)$. One then 
finds $C_{2,L}$ from the quark deep-inelastic scattering cross 
section according to (\ref{conv2}). The hadronic tensor at leading 
order in the flavour expansion requires the one-loop diagrams of 
Fig.~\ref{fig13}a dressed with fermion loops. It has 
been calculated in \cite{BB95b,DMW96,SMMS96,DW96a}.   
For the Borel transform of the longitudinal structure function 
close to the leading IR renormalon pole at $u=1$, we obtain 
\begin{equation}
\label{irtwist2}
B\!\left[\frac{F_L}{2 x}\right](u,x) = \frac{K}{Q^2}\,
\left\{-8\xi^2+4\delta(1-\xi)
\right\} * F(x/\xi),
\end{equation}
where `$*$' denotes the convolution product as in (\ref{conv2}) 
and 
\begin{equation}
\label{Kdef}
K = \frac{C_F}{4\pi}\,\frac{\mu^2\,e^{-C}}{1-u}.
\end{equation}
($C$ is the subtraction constant for the fermion loop, $C=-5/3$ 
in the $\overline{\rm MS}$ scheme.) The IR renormalon pole at $u=1$ 
corresponds to a twist-4 $1/Q^2$ power correction to (\ref{conv2}).
In the remainder of this section, we reproduce the leading IR 
renormalon in the twist-2 coefficient function $C_{2,L}$ from 
the analysis of twist-4 matrix elements. 

A complete analysis of twist-4 operators and their coefficient 
functions has been performed in \cite{JS81,EFP83,J83}. Here we 
follow the treatment of
\cite{BB88}, who work directly with non-local twist-4 operators 
rather than their expansion into local operators. The twist-4  
contributions to the longitudinal structure function can be 
written as 
\begin{equation}
\label{conv4}
F_L(x,Q^2)_{|\rm twist-4} = \frac{1}{Q^2}\sum_i\int d\{\xi\} 
\,C_{4,L}^i(x,\{\xi\},\alpha_s(Q),Q^2/\mu^2)\,T_i(\{\xi\},\mu)
\end{equation}
with multi-parton correlations $T_i$ defined below. At tree level 
the relevant part of the light-cone expansion of the current product 
is
\begin{equation}
\label{curprod}
iT\left(j_\mu(z)j_\nu(-z)\right)_{|\rm twist-4} = \frac{1}{128\pi^2}\,
\frac{4g_{\mu\nu}}{z^2-i0}\int\limits_0^1 d\tau \,\tau(1+\ln\tau)\,
Q_1(\tau z) + \ldots,
\end{equation}
where
\begin{equation}
Q_1(y) = \int\limits_{-1}^1 dv\,\left[4\,{\cal O}_3(v,y)-2 i (1-v^2)\, 
{\cal O}_7(v,y) + \ldots\right] + (y\leftrightarrow -y)
\end{equation}
and the three-particle operators ${\cal O}_{3,7}$ are defined as
\begin{eqnarray}
\label{twist4op1}
{\cal O}_3(v,y) &=& \frac{1}{2}\epsilon_{\alpha\beta\gamma\delta}\,
\bar{\psi}(y) y^\alpha\gamma^\beta\gamma_5 g_s G^{\gamma\delta}(v y)
\psi(-y),
\\
\label{twist4op2}
{\cal O}_7(v,y) &=& 
\bar{\psi}(y) \!\not\!y y^\beta D^\alpha g_s G_{\alpha\beta}(v y)\psi(-y).
\end{eqnarray}
Path-ordered exponentials that connect fields at different 
points are again understood. 
The dots denote contributions that can be found in \cite{BB88}, but which
are needed neither for the longitudinal part of the structure function 
nor in leading order of the flavour expansion. Two-gluon 
operators and four-fermion operators not related to a $\bar{\psi}G\psi$ 
operator through the equation of motion are not relevant at leading 
order. For the two-gluon operators, this follows from the fact that 
the third diagram in Fig.~\ref{fig13}b vanishes because it contains 
no scale. The nucleon matrix elements of the three-particle operators 
${\cal O}_{3,7}$ are parametrized as
\begin{eqnarray}
\label{tis1}
&&\hspace*{-1cm}\langle N(p)|{\cal O}_3(v,y)|N(p)\rangle(\mu) =
2\,p\cdot y \int d\xi_1 d\xi_2\,e^{ip\cdot y [\xi_1(1-v)+\xi_2 (1+v)]}\,
T_3(\xi_1,\xi_2,\mu), 
\\
\label{tis2}
&&\hspace*{-1cm}\langle N(p)|{\cal O}_7(v,y)|N(p)\rangle(\mu) =
2\,(p\cdot y)^2 \int d\xi_1 d\xi_2\,e^{ip\cdot y [\xi_1(1-v)+\xi_2 (1+v)]}\,
T_7(\xi_1,\xi_2,\mu). 
\end{eqnarray}
The dependence on the 
renormalization scale will be suppressed in the following. It is now 
straightforward to take the Fourier transform and discontinuity of 
(\ref{curprod}) to obtain the longitudinal structure function in the form
of (\ref{conv4}):
\begin{equation}
\label{t37}
\frac{F_L(x,Q^2)}{2 x}{}_{|\rm twist-4} =  
\frac{1}{Q^2} \int d\xi_1 d\xi_2\,\left[C_{4,L}^3(x,\xi_1,\xi_2)\,
T_3(\xi_1,\xi_2) + C_{4,L}^7(x,\xi_1,\xi_2)\,
T_7(\xi_1,\xi_2)\right],
\end{equation}
where
\begin{eqnarray}
\label{c43}
C_{4,L}^3(x,\xi_1,\xi_2) &=& \frac{4 x}{\xi_2-\xi_1}\left\{
\frac{x}{\xi_2^2} \left(1+\ln\frac{x}{\xi_2}\right) \theta(\xi_2-x) 
- (\xi_2\leftrightarrow \xi_1)\right\},
\\
\label{c47}
C_{4,L}^7(x,\xi_1,\xi_2) &=& -\frac{4 x^2}{(\xi_2-\xi_1)^2}
\Bigg\{\left(\frac{1}{\xi_2^2} \left(1+\ln\frac{x}{\xi_2}\right)+
\frac{2}{\xi_2 (\xi_2-\xi_1)}\,\ln\frac{x}{\xi_2}\right) \theta(\xi_2-x) 
\nonumber\\
&&\hspace*{-2cm}+ \,(\xi_2\leftrightarrow \xi_1)\Bigg\}.
\end{eqnarray}
Since we are interested in UV contributions to the matrix 
elements of multi-parton operators, the coefficient functions at tree level 
as quoted suffice.

Up to this point we have been rather general. 
Let us now consider the UV renormalization of the three-particle 
operators ${\cal O}_{3,7}$. There are logarithmic UV divergences, which 
lead to logarithmic scaling violations. However, power counting tells us 
that the operators also have quadratic divergences, which can appear 
in quark matrix elements through the diagrams shown in 
Fig.~\ref{fig13}b. Since the quadratic divergences depend on 
the factorization scheme, one has to compute the quark matrix elements 
of ${\cal O}_{3,7}$ in the same way as the twist-2 coefficient function, 
which means that we consider their Borel transform in leading order of the 
flavour expansion. Then the (Borel transform of the) 
matrix element of ${\cal O}_7$ 
between quark states of momentum $p$ is given by
\begin{eqnarray}
\langle p|{\cal O}_7(v,y)|p\rangle &=& 
(-i) 4\pi C_F\,(-\mu^2 e^{-C})^{-u}\,e^{2 i p\cdot y}
\int\frac{d^4 k}{(2\pi)^4}\,\frac{k^2 y_\mu-(k\cdot y)\,k_\mu}
{(k^2)^{1+u} (p-k)^2}
\nonumber\\
&&\hspace*{-2cm}
\bar{u}(p)\left\{e^{-ik\cdot y (1+v)} \!\not\!y (\!\not\!p-\!\not\!k) 
\gamma_\mu + e^{-ik\cdot y (1-v)} \gamma_\mu (\!\not\!p-\!\not\!k) 
\!\not\!y \right\} u(p)
\end{eqnarray}
and a similar result holds for ${\cal O}_3$. Strictly speaking, the 
integral vanishes for $p^2=0$, because it does not contain a scale. 
This is the usual fact that matrix elements vanish perturbatively 
in factorization schemes that do not introduce an explicit factorization 
scale. One can isolate the quadratic divergence by keeping 
$p^2\!\not= 0$, since the quadratic divergence is independent of 
$p^2$. Power divergences lead to non-Borel summable UV renormalon 
singularities in QCD and the quadratic divergence is seen as a pole 
at $u=1$ in the integral above. The integral can be done exactly. 
It is crucial for the singularity structure in $u$ that $y$ is 
exactly light-like. 
Close to $u=1$, we find for the Borel transforms\footnote{To avoid 
cumbersome notation, we do not write $B[\ldots]$ in what follows, 
but the Borel transform is understood.}
\begin{eqnarray}
\label{quaddiv3}
&&\hspace*{-1.5cm}
{\cal O}_{3}(v,y){}_{|\rm q.\,div.} =
(-K)\int\limits_0^1 d\alpha\,(2-\alpha)
\left\{\bar{\psi}(y)\!\!\not\!y \psi(y 
[\alpha v-\bar{\alpha}]) +\bar{\psi}(y[\alpha v+\bar{\alpha}] )\!\!\not\!y 
\psi(-y)\right\},
\\
\label{quaddiv7}
&&\hspace*{-1.5cm}
{\cal O}_{7}(v,y){}_{|\rm q.\,div.} =
2\,p\cdot y\,K\int\limits_0^1 d\alpha\,\bar{\alpha}
\left\{\bar{\psi}(y)\!\!\not\!y \psi(y 
[\alpha v-\bar{\alpha}]) +\bar{\psi}(y[\alpha v+\bar{\alpha}] )\!\!\not\!y 
\psi(-y)\right\},
\end{eqnarray}
where $\bar{\alpha}=1-\alpha$ and $K$ is as defined in (\ref{Kdef}). Note 
that $K$ is proportional to $\mu^2$, so these equations take the 
form expected for a quadratic divergence. The quadratic divergence is 
independent of the external states and (\ref{quaddiv3}, \ref{quaddiv7}) 
are written as operator relations. The power divergent 
part takes the form of an integral over the leading-twist operator 
(\ref{twist2}). This is exactly what one needs to match the 
IR renormalon singularity in the coefficient function at leading twist. 
Taking the nucleon matrix elements, the quadratically divergent 
part of $T_{3,7}$ is expressed in terms of the twist-2 quark distribution 
as follows:
\begin{eqnarray}
&&\hspace*{-1.5cm}
T_3(\xi_1,\xi_2){}_{|\rm q.\,div.} = (-K)\left\{\frac{1}{\xi_1}
\left(1+\frac{\xi_2}{\xi_1}\right) F(\xi_1)\,\theta(\xi_1-\xi_2) + 
(\xi_1\leftrightarrow\xi_2)\right\}\theta(\xi_1)\theta(\xi_2),
\\
&&\hspace*{-1.5cm}
T_7(\xi_1,\xi_2){}_{|\rm q.\,div.} = 2 K\left\{
\frac{\xi_2}{\xi_1}\,F(\xi_1)\,\theta(\xi_1-\xi_2) + 
(\xi_1\leftrightarrow\xi_2)\right\}\theta(\xi_1)\theta(\xi_2).
\end{eqnarray}
Inserting these expressions into (\ref{t37}), using (\ref{c43}, 
\ref{c47}), and taking the remaining integrals except for one convolution, 
one finds that the result takes the following form:
\begin{equation}
\label{uvtwist4}
B\!\left[\frac{F_L}{2 x}{}_{|\rm twist-4}\right](u,x) 
\stackrel{\rm q.\, div.}{=} \frac{K}{Q^2}\,
\left\{G_3(\xi)+G_7(\xi)
\right\} * F(x/\xi),
\end{equation}
where
\begin{eqnarray}
G_3(\xi) &=& 4\xi^2\left[1+2(1+\ln\xi)\ln\frac{1-\xi}{\xi} + \ln^2\xi 
+2 \mbox{Li}_2(1-\xi)\right]
\\ 
G_7(\xi) &=& 4\xi^2\left[1-2(1+\ln\xi)\ln\frac{1-\xi}{\xi} - \ln^2\xi 
-2 \mbox{Li}_2(1-\xi) - \delta(1-\xi)\right]
\end{eqnarray}
with $\mbox{Li}_2$ the dilogarithm function. There is a remarkable 
cancellation (for which we do not have an explanation) 
between the two contributions from the two operators 
and the sum
\begin{equation}
G_3(\xi)+G_7(\xi)=8\xi^2-4\delta(1-\xi)
\end{equation}
leads to the coincidence of (\ref{uvtwist4}) and (\ref{irtwist2}), 
except for the over-all sign. Hence we have shown that the first IR 
renormalon singularity in $C_{2,L}$ cancels the UV renormalon 
singularity at the same position in a perturbative evaluation of 
UV contributions to twist-4 matrix elements. 

The matching of IR renormalons in coefficient functions and UV 
contributions to matrix elements exhibited here and in the $\sigma$-model 
is a  general feature of perturbative factorization and 
short-distance expansions, or asymptotic expansions in ratios of 
mass scales in general, in quantum field theories. QCD has a mass 
gap and is supposed to be well defined in the infrared. 
The complicated structure of the short-distance expansion, including 
renormalons, reflects the fact that quantum fluctuations are distributed 
over all distance scales. However, if 
care is taken of defining all terms in the expansion consistently, 
the unambiguous expansion that is obtained may be hoped to be asymptotic 
to the exact, non-perturbative result.


\section{Phenomenological applications of renormalon divergence}
\setcounter{equation}{0}
\label{pheno}

In this section we turn to manifestations of renormalon divergence 
in particular physical processes. During the past few years the number 
of processes considered has been rapidly expanding as has been the 
number of next-to-leading and next-to-next-to-leading perturbative 
calculations. The interest in renormalons stems from the fact that 
they provide a link between perturbative and non-perturbative physics, 
because, on the one hand, renormalons account for a large part of the  
higher order perturbative coefficients and, on the other hand, in still 
higher orders they merge with the treatment of non-perturbative 
power corrections. 

Following this general idea, three main strains of applications with 
more or less emphasis on the perturbative or power correction aspect 
have developed. We briefly summarize the questions, methods and problems 
associated with each of the three in Section~\ref{direct} before 
turning to the details of process-specific applications. These 
applications deal exclusively with QCD processes. 

In order that there be renormalons, there must be a perturbative 
expansion. Hence the processes analysed in what follows satisfy two 
requirements: they contain at least one scale, which is large compared 
with $\Lambda$, the scale where QCD becomes non-perturbative, and 
one can isolate a part of the process that depends 
{\em only} on large scales, such that it can be expanded perturbatively 
in $\alpha_s$. The large scale may be provided by large energy 
transfer in the high-energy collision of massless particles or 
by the mass of a quark much heavier than $\Lambda$. Applications 
of renormalons to hard reactions of massless particles are reviewed 
in Sections~\ref{hard1} and \ref{hard2}. The first section 
concentrates on processes that admit an operator product expansion (OPE)
or are related to an OPE by dispersion relations. The second section 
deals with genuinely time-like processes. Finally, observables 
involving heavy quarks are discussed in Section~\ref{heavyQ}.

\subsection{Directions}
\label{direct}

We summarize the main uses of renormalons. The starting point is  
series expansions in $\alpha_s=\alpha_s(\mu)$,
\begin{equation}
\label{serdef}
R(\{q\},\alpha_s,\mu) = \sum_{n=0}^\infty r_n(\{q\},\mu)\,
\alpha_s^{n+1},
\end{equation}
where $\{q\}$ denotes a set of kinematic variables which must all be 
large compared to $\Lambda$, and $\mu$ denotes the renormalization and, 
if present, factorization scale. $R$ may be either a physical quantity 
or a short-distance coefficient in a factorization formula for a 
physical quantity. Without loss of generality the series starts at 
order $\alpha_s$.

\subsubsection{Large perturbative corrections}
\label{lpc}

Since renormalons dominate the large-order behaviour of the 
perturbative coefficients $r_n$, the question of whether they can be 
used to improve truncated perturbative series suggests itself. 
In an ideal situation we would compute the asymptotic behaviour 
and combine it with exact results in low orders so as to approximate 
the Borel integral, as was done using the instanton-induced divergence 
for improving perturbative calculations of 
critical exponents \cite{LeGZ77}. 
Even if the series were not Borel-summable, we would 
be able to improve the perturbative prediction to an accuracy 
limited only by the leading power correction. 

The large-order behaviour due to renormalons cannot be used in this 
way, because the over-all normalization cannot be computed. As a 
consequence only ratios of coefficients can be computed. If 
\begin{equation}
\label{lo}
r_n = K\,(a\beta_0)^n\,\Gamma(n+1+b)\left(1+\frac{c_1}{n+b} +
\frac{c_2}{(n+b) (n+b+1)} + \ldots\right),
\end{equation}
the ratio of consecutive coefficients is given by 
\begin{equation}
\label{rat}
\frac{r_n}{r_{n-1}} = a\beta_0\,(n+b)\left(1-\frac{c_1}{(n+b) (n+b-1)} + 
\frac{c_1^2-2 c_2}{(n+b)^3} +\ldots\right)
\end{equation}
and the parameters $a,b,c_i$ are calculable as discussed in 
Section~\ref{sectrenfeyn}. 
An attempt to use this observation for the cross section in 
$e^+ e^-$ into hadrons was made in \cite{Ben93b}. There exist a few 
observables for which the first correction term in brackets is known 
and one -- the difference between the pole mass and the $\overline{\rm 
MS}$ mass of a heavy quark -- for which even the $1/n^3$ correction 
can be obtained (see Section~\ref{pole}). 
In practice it is often difficult to carry out 
this idea, because the large-order behaviour is not as simple as 
(\ref{lo}). There may be several components with the same value 
of $a$, but with different normalization constants. This is the 
case with ultraviolet (UV) renormalon divergence as discussed 
in Section~\ref{rganalysis}. Then the ratio of asymptotic coefficients 
depends on ratios of normalization constants. For UV renormalons 
these ratios are process-independent and in principle one may 
think of determining them from a set of observables. In practice, 
this does not seem feasible, given in particular that the available 
exact series are not very long and reach $n=2$ at best. The 
application of the strategy outlined here is therefore restricted to 
observables whose large-order behaviour is dominated by infrared 
(IR) renormalons and which in addition exhibit a relatively simple 
IR renormalon structure. 

\begin{table}[t]
\addtolength{\arraycolsep}{0.2cm}
\renewcommand{\arraystretch}{1.7}
$$
\begin{array}{c|c|c}
\hline\hline
\mbox{Observable} & \mbox{Series} & \mbox{LPC} \\ 
\hline
4\pi^2 Q^2\frac{d\Pi}{d Q^2} & 
1+a_s\left(1+1.6 a_s+6.4 a_s^2+\ldots\right) & 1/Q^4  \\
\frac{1}{6}\int_0^1 dx F_3(x,Q) & 
1-a_s\left(1+3.6 a_s+19 a_s^2+\ldots\right) & 1/Q^2  \\
\langle 1-T\rangle(Q) & 
\hspace*{-1.3cm}1.05 a_s\left (1+9.6 a_s+\ldots\right)& 1/Q  \\
\hline\hline
\end{array}
$$
\caption[dummy]{\label{tab3}\small 
Comparison of perturbative series in $a_s=\alpha_s(Q)/\pi$ 
($\overline{\rm MS}$ scheme) and leading power corrections (LPC). 
Results are taken from 
\cite{GKL91,SS91} (Adler function of vector currents, 1st line, 
$N_f=3$), \cite{LV91} (Gross-Llewellyn-Smith sum rule, 2nd line, 
$N_f=3$), \cite{KNMW89} (average $1-T$, 3rd line, $N_f=5$).}
\end{table}

For all these reasons one resorts to either qualitative or less 
rigorous approaches. There are indeed interesting patterns in low 
order perturbative coefficients. Referring to Table~\ref{tab3}, 
which compares the perturbative expansions of three observables, we 
observe that the series in brackets (i) all have positive 
coefficients and (ii) the larger the coefficients are the larger the 
leading power correction is.

All this may well be accidental. But remembering that larger power 
corrections are associated with faster growth of perturbative 
coefficients due to IR renormalons, one may also speculate whether 
the observed pattern may be a manifestation of IR renormalon behaviour 
down to very low orders. This raises obvious questions: Why then is 
there no trace of sign-alternating UV renormalon behaviour, which 
should dominate the asymptotic behaviour for the Adler function 
(first line) and perhaps also deep inelastic scattering sum rules 
(second line)?\footnote{We recall that the leading UV renormalon 
leads to a minimal term of the series of order $1/Q^2$ and hence 
dominates all observables with IR power corrections smaller than 
$1/Q^2$.} The coefficients in the table are scheme-dependent, and 
the comparison may look completely different in another scheme. 
Why should the $\overline{\rm MS}$ scheme be special?

There is a simple `approximation' to the perturbative coefficients 
that allows us to study part of these questions. Write 
\begin{equation}
\label{flavourseries2}
r_n=r_{n0}+r_{n1} N_f+\ldots+r_{nn} N_f^n = r_0
\left[d_n (-\beta_0)^n+\delta_n
\right],
\end{equation}
where $d_n=(-6\pi)^n r_{nn}/r_0$, $\beta_0=-(11-2N_f/3)/(4 \pi)$ 
and $N_f$ is the number of massless quarks. We then obtain the 
coefficients $d_n$ from a calculation of fermion bubble graphs 
(see Section~\ref{bubbles}) and neglect the remainder $\delta_n$. 
The `model' for the series constructed in this way has UV and IR 
renormalons at the correct positions, although the nature of the 
singularity and the over-all normalization are not reproduced 
correctly. It has been suggested 
in \cite{BB95a,N95,BBB95} to use this approximation\footnote{ 
Note that this is not a systematic approximation, because it has no 
tunable small parameter. Formally, however, it can be obtained  
as a `large-$\beta_0$' limit or a large (and negative!) $N_f$ 
limit. Instead of `Naive Non-Abelianization' we will refer to the 
approximation as the `large-$\beta_0$ approximation' or 
`large-$\beta_0$ limit'.} 
quantitatively and the procedure is often referred to as 
`Naive Non-Abelianization' (NNA). The term was coined by Broadhurst 
\cite{BG95} who observed empirically that the remainder 
$\delta_1$ at second order $\alpha_s^2$ is typically rather small 
compared to $d_1(-\beta_0)$ 
in the $\overline{\rm MS}$ scheme. This empirical fact, together 
with the fact that the method can be viewed as an extension 
of Brodsky-Lepage-Mackenzie scale setting \cite{BLM83}, still provides 
the principal motivation for considering fermion loop diagrams. 
An important point is that one should expect the NNA or 
large-$\beta_0$ approximation to work 
quantitatively only, {\em if} the contribution associated with the 
(one-loop) running coupling is large in higher orders. If it turns 
out to be small, there is no reason to expect that the NNA approximation 
is a good approximation to the exact higher order coefficient.

There are very few calculations that go beyond the calculation of 
fermion loop diagrams, and much of what follows relies on this class 
of diagrams. 
An interesting observation in the context of fermion bubble calculations 
is that {\em in the $\overline{MS}$ 
scheme} the (scheme-dependent, see Section~\ref{schemedep}) 
normalization of UV renormalons is suppressed compared 
to the normalization of IR renormalons, and hence the onset of 
UV renormalon behaviour is delayed \cite{Ben93b}. This suggests 
that UV renormalons are irrelevant to intermediate orders 
in perturbation theory in that scheme;  
it also suggests an explanation for why the series in 
Table~\ref{tab3} exhibit a fixed-sign pattern. We will return 
to estimates of perturbative coefficients from NNA in 
Sections~\ref{hard1} and \ref{heavyQ}.

\subsubsection{The power of power corrections}

Aside from their obvious connection with perturbation theory, renormalons 
are primarily discussed in connection with power corrections. If 
$a<0$ in (\ref{lo}), the attempt to sum the series with the help of 
Borel summation leads to ambiguities of order
\begin{equation}
\label{poweragain}
\delta R \sim \left(\frac{\Lambda^2}{q^2}\right)^{1/a} \times\,
\mbox{logarithms of\,}\,q/\Lambda
\end{equation}
in defining the perturbative contribution. In QCD these ambiguities 
arise from long distances and are interpreted as the ambiguity in 
defining what one means by `perturbative' and `non-perturbative'. 
As a consequence one identifies the {\em scaling} with $q$ of 
some power corrections through the value of $a$. Additional logarithmic 
variations of (\ref{poweragain}) can also be determined, but not 
the absolute magnitude of the power suppressed contribution. Note the 
analogy with the standard formalism for deep-inelastic scattering: 
scaling violations, logarithmic only at leading power, can be 
computed in perturbation theory, but not the parton densities. 

Early phenomenologically oriented 
discussions of IR renormalons concentrated on the 
question of whether or not there could be a $1/Q^2$ power correction to 
current-current correlation functions \cite{BY92,Z92,Ben93a} 
which would imply larger non-perturbative corrections than the 
$1/Q^4$ correction incorporated through the OPE. 
From the present perspective this discussion appears historical. If 
there is an OPE the IR renormalon structure is consistent with it 
by construction.

At the same time one has to be aware of the fact that IR renormalons 
imply power corrections, but that the converse is not true. There may be 
power corrections larger than those indicated by renormalons, especially 
for time-like processes, and next to nothing is known theoretically 
about them. These may be power corrections to coefficient functions from 
short distances, power corrections from long-distances that do not 
`mix' with perturbation theory, or, for time-like processes, 
power corrections related to 
violations of parton-hadron duality, i.e. the possibility that the 
power expansion is not an asymptotic one after continuation to the 
time-like region. 
Our attitude towards this problem is that if power 
corrections indicated by renormalons are large, there is a good chance 
that one has found the dominant ones.

Identifying power corrections through IR renormalons is especially 
interesting for processes that do not admit an OPE and 
for which the result is not obvious. Along this 
line the heavy quark mass was considered 
in \cite{BB94a,BSUV94}. The first investigations of hard 
QCD processes, in particular event shape observables in 
$e^+ e^-$ annihilation into hadrons, from this perspective appeared 
in \cite{CS94,MW95,Web94}. Most often power corrections to these 
observables are large, being suppressed only by one power of 
the large momentum scale. For event shape observables and other 
hadronic quantities the understanding of even the leading power 
corrections is still not complete, although significant progress 
has been made over the past four years. At this point it seems 
that the analysis of IR renormalons would merge with the general 
problem of classifying IR sensitive regions in Feynman integrals 
beyond leading power.

\subsubsection{Models}
\label{modelssect}

The absolute magnitude of power corrections cannot be calculated 
with perturbative methods. Additional assumptions are needed, which may 
be difficult to justify. The result is a model for power 
corrections. Such models have the advantage that they are consistent 
with short-distance properties of QCD -- exactly the point that is 
most problematic for other models of low-energy QCD --, although they 
cannot be derived from QCD. Two models, for different purposes, have 
been developed.

In the Dokshitzer-Webber-Akhoury-Zakharov (DWAZ) model\footnote{The 
models proposed by \cite{DW95} and \cite{AZ95a} do not coincide 
exactly. They share however the crucial assumption that power 
corrections are universal. See Section~\ref{eventshapes} for 
a more discriminative discussion.} for 
event shape variables 
\cite{DW95,AZ95a} it is assumed that $1/Q$ (where $Q$ is the 
centre-of-mass energy) power corrections in the fragmentation 
of quarks and gluons in $e^+ e^-\to\,$hadrons
can be accounted for by one parameter only. Hence, for all 
(averaged) event shape variables we may write schematically
\begin{equation}
S_{|1/Q} = K_S\cdot\frac{\langle \mu_{had}\rangle}{Q}.
\end{equation}
It follows from the universality assumption that the relative 
magnitudes of $1/Q$ power corrections to different observables 
are predicted (see the discussion in Section~\ref{sumy}). The 
simplicity of the model is appealing and has led to numerous 
comparisons with experimental data on the energy dependence of 
averaged event shapes. We follow this in more detail 
in Section~\ref{eventshapes}.

The second model, proposed in \cite{DMW96,SMMS96}, concerns  
the dependence of twist-4 corrections to deep-inelastic scattering 
cross sections on the scaling variable $x$. The OPE constrains these 
to be of the form of (\ref{conv4}) with unknown multi-parton 
correlations. The model assumes that the $x$-dependence can be 
approximated by the $x$-dependence of the IR renormalon 
contribution to the twist-2 coefficient function folded with the 
ordinary parton densities. The structure functions are then 
expressed as 
\begin{equation}
\label{fl2}
F_P(x,Q)/(2 x) 
= \sum_i\int\limits_x^1\frac{d\xi}{\xi}\,f_i(x/\xi,\mu)\,
\left[C^i_{2,P}(\xi,Q,\mu)+A^i_P(\xi)\frac{\Lambda^2}{Q^2}\right]
+ \ldots,
\end{equation}
with calculable functions $A^i_P(\xi)$. Usually only quarks are taken 
into account in the sum over $i$. It is clear that the 
target dependence at twist-4 is the same as at twist-2 in this model 
and a prerequisite for it to work is that the genuine target dependence
at twist-4 is small compared to the twist-4 correction as a whole.
In \cite{DMW96} the model has been motivated by the assumption 
that the bulk of the twist-4 correction can be accounted for as 
an integral over a universal, IR-finite coupling constant. 
In \cite{BBM97} it was argued that the model could be justified, if 
the twist-4 matrix elements normalized at $\mu$ 
were dominated by their UV contributions from $\Lambda < k < \mu$ 
rather than by fluctuations with $k\sim \Lambda$. This 
interpretation follows indeed directly from the matching calculation 
performed in Section~\ref{disex}. Since the UV  
contributions to twist-4 matrix elements are equivalent to 
IR contributions to twist-2 coefficient functions, the 
`ultraviolet dominance' suggestion amounts to stating  
that the model provides an effective parametrization of perturbative 
contributions not taken into account in the truncated series 
expansion of $C_{2,P}^i$. We discuss this model further 
in Section~\ref{dismod} and, for fragmentation functions 
in $e^+ e^-$ annihilation, in Section~\ref{fragmod}.
The advantage of both models introduced here is simplicity. In both cases,  
success or failure in comparison with data leads to interesting hints 
on the nature of power corrections.

\subsection{Hard QCD processes I}
\label{hard1}

In this section we summarize results on renormalons for inclusive 
hadronic observables in $e^+ e-$ annihilation, $\tau$ decay and 
deep inelastic scattering (DIS). Since the scaling of power corrections 
is known from OPEs, the emphasis is on 
potentially large higher order perturbative corrections and, 
in Section~\ref{dismod}, on modelling the $x$-dependence of 
twist-4 corrections to DIS structure functions. 

\subsubsection{Inclusive hadroproduction in $e^+ e^-$ annihilation}
\label{inclepem}

The inclusive cross section $e^+ e^-\to\gamma^*\to\,$hadrons is given by 
the vector current spectral function:
\begin{equation}
R_{e^+ e^-}(q^2) = \frac{\sigma_{e^+ e^-\to\,\rm hadrons}}
{\sigma_{e^+ e^-\to \mu^+\mu^-}} = 12\pi\left(\,\sum_q e_q^2\,\right)
\mbox{Im}\,\Pi(q^2+i0),
\end{equation}
where the vacuum polarization $\Pi(q^2)$ is defined by 
(\ref{currentcorr}).\footnote{A different quark electric charge factor 
is understood for the `light-by-light' contributions. However, in the 
following `light-by-light' terms do not play an important role.} 
The Adler function (see (\ref{adlerdef})) 
is expanded as
\begin{equation}
\label{larged}
D(Q^2) = 4\pi^2\,Q^2\,\frac{d\Pi(Q^2)}{dQ^2} = 1+\frac{\alpha_s}{\pi}
\sum_{n=0} \alpha_s^{n}\,
\left[d_n (-\beta_0)^n+\delta_n\right],
\end{equation}
see (\ref{flavourseries2}). The normalization is such that $d_0=1$, 
$\delta_0=0$. As mentioned after (\ref{flavourseries2}) the coefficients 
$d_n$ are computed in terms of fermion bubble diagrams, in the present 
case the diagrams shown in Fig.~\ref{fig1}. 

The exact result for these diagrams was obtained in \cite{Ben93a,Bro93} 
in the context of QED. Adjusting the colour factors and over-all 
normalization, the Borel transform is found to be
\begin{equation}
\label{borelpolaroper} 
B[D](u)=\sum_{n=0}\frac{d_n}{n!}\,u^n = \frac{32}{3} 
\left(\frac{Q^2}{\mu^2} e^C\right)^{-u} \frac{u}{1-(1-u)^2} \,
\sum_{k=2}^\infty \frac{(-1)^k k}{(k^2-(1-u)^2)^2}.
\end{equation}
The representation in terms of a single sum is due to \cite{Bro93}. 
In the $\overline{\rm MS}$ scheme $C=-5/3$. 
The coefficients $d_n$ are presented in Table~\ref{tab4} for 
$\mu=Q$. With reference 
to (\ref{larged}), we call the approximation of neglecting the 
$\delta_n$ the `large-$\beta_0$' approximation. For comparison we 
show $\delta_{1,2}$ obtained from the exact perturbative coefficients 
\cite{GKL91,SS91} and $\delta_3$ obtained from the estimate 
of \cite{KS95z}. The `large-$\beta_0$' approximation is quite good 
at order $\alpha_s^{2,4}$ but overestimates the coefficient at 
order $\alpha_s^3$ considerably. It should be noted that the comparison 
depends on the choice $\mu=Q$ and the approximation cannot be expected  
to work well for arbitrary choices 
of scale or scheme \cite{BB95a,BBB95}. This has been a 
point of criticism of the `large-$\beta_0$' 
approximation \cite{Chy95}. We discuss this point further in 
the context of $\tau$ decay below. 
\begin{table}[t]
\addtolength{\arraycolsep}{0.2cm}
$$
\begin{array}{c|c|c|c}
\hline\hline
n & d_n & d_n (-\beta_0)^n & \delta_n \\ 
\hline
0 & 1       & 1      & 0 \\
1 & 0.6918  & 0.4955 & 0.0265 \\
2 & 3.1035  & 1.5919 & -0.9464 \\
3 & 2.1800  & 0.8009 & 0.0860 \\
4 & 30.740  & -      & - \\
5 & -34.534 & -      & - \\
6 & 759.74  & -      & - \\
7 & -3691.4 & -      & - \\
8 & 42251   & -      & - \\
\hline\hline
\end{array}
$$
\caption[dummy]{\label{tab4}\small 
Perturbative corrections to the Adler function in the 
$\overline{\rm MS}$ scheme: the `large-$\beta_0$ limit' in comparison 
with the remainder, $\delta_{1,2}$, to the exact result and an 
estimate thereof for $\delta_3$. Results for $N_f=3$.}
\end{table}

The renormalon singularities of the Adler function have already been 
discussed in Section~\ref{borelplane}. The UV renormalon poles at 
$u=-1,-2,\ldots$ are double poles. The IR renormalon poles at 
$u=2,3,\ldots$ are also double poles, with the exception of $u=2$. 
{}In the large-$N_f$ limit one expects an IR renormalon pole at $n$ to 
take the form $1/(n-u)^{1+\gamma_0/(2\beta_{0f})}$, where 
$\gamma_0$ is the $N_f$-part of the 
one-loop anomalous dimension of an operator of 
dimension $2 n$, see~(\ref{iras}). It follows that the singularity 
at $n=2$ has to be a simple pole, because the operator $\alpha_s G G$ 
has no anomalous dimension in the large-$N_f$ limit. It has been 
checked in \cite{BenDiss} that there is a dimension-6 operator 
with $\gamma_0=2\beta_{0f}$, which leads to a double pole at $n=3$. 
Since there is no operator of dimension 2 in the OPE of the Adler 
function, there is no IR renormalon pole at $u=1$. 

The Borel transform of the vacuum polarization is obtained by dividing 
$B[D](u)$ by $(-u)$. One then notes \cite{Ben93a,LM94} the symmetry 
$B[\Pi](1+u)=B[\Pi](1-u)$, which interchanges UV and IR renormalon poles. 
This symmetry implies that the small and large momentum behaviours of 
the diagrams of Fig.~\ref{fig1} are related \cite{BenDiss}. 
Note that this symmetry relates 
the IR renormalon pole at $u=2$ which corresponds to the gluon operator 
$\alpha_s GG$ to the pole at $u=0$, which corresponds to (external) charge 
renormalization. Likewise the IR renormalon pole at $u=3$ and the UV 
renormalon pole at $u=-1$ are related, and both are described in terms of 
dimension-6 operators. It is not known whether this symmetry persists
in higher orders of the flavour expansion.

It is interesting to break down the $d_n$ into contributions from 
the leading renormalon pole in order to check how fast the asymptotic 
regime is reached. To this end we decompose $B[D](u)$ into the sum of 
the leading poles according to
\begin{eqnarray}
B[D](u) &=& e^{-5/3}\left\{\frac{4}{9}\frac{1}{(1+u)^2}+\frac{10}{9} 
\frac{1}{1+u}\right\} + e^{10/3}\,\frac{2}{2-u} 
\nonumber\\
&&\hspace*{-1.5cm}+\,e^{-10/3}\left\{-\frac{2}{9}\frac{1}{(2+u)^2}-
\frac{1}{2} \frac{1}{2+u}\right\} + \ldots.
\end{eqnarray}
This breakdown is given in Table~\ref{tab5}. One can see that the 
asymptotic behaviour sets in late and the low-order coefficients 
$n\sim 1$-$5$ are not dominated by a single renormalon pole. The 
irregularities in low orders are due to cancellations between 
IR and UV renormalons in every second order. The sum over 
contributions from IR renormalon poles does not converge, because 
of the over-all factors $e^{5 n/3}$ for an IR renormalon pole at 
$u=n$. If one chooses the scheme with $C=0$, the asymptotic regime sets in 
earlier. In this case the series is dominated 
by sign-alternating behaviour 
from UV renormalons starting at low order. 
\begin{table}[t]
\addtolength{\arraycolsep}{0.2cm}
$$
\begin{array}{c|r|r|r|r|r|r }
\hline\hline
n & d_n\hspace*{0.5cm} & \mbox{UV}(-1) & \mbox{IR}(2) & \mbox{UV}(-2) & 
\mbox{IR}(3) & \mbox{IR}(4) \\ 
\hline
0 & 1       & 0.294  & 28.03 & -0.011 & -11.0 & -50.9 \\
1 & 0.6918  & -0.378 & 14.02 & 0.006  & -11.0 & -7.28 \\
2 & 3.1035  & 0.923  & 14.02 & -0.007 & -12.2 & -0.91 \\
3 & 2.1800  & -3.27  & 21.02 & 0.013  & -17.1 & 1.36  \\
4 & 30.740  & 15.1   & 42.05 & -0.028 & -29.3 & 3.41  \\
5 & -34.534 & -85.6  & 105.1 & 0.078  & -59.7 & 6.82  \\
6 & 759.74  & 574    & 315.4 & -0.256 & -141  & 14.1  \\
7 & -3691.4 & -4442  & 1104  & 0.975  & -380  & 31.3  \\
8 & 42251   & 38923  & 4415  & -4.214 & -1149 & 76.1  \\
\hline\hline
\end{array}
$$
\caption[dummy]{\label{tab5}\small 
Breakdown of $d_n$ into contributions from the leading IR and UV 
renormalon poles. The integer in brackets denotes the position of the 
pole.}
\end{table}
 
In \cite{Ben93b} a result for 
the ratio of asymptotic coefficients due to the first IR renormalon 
was obtained that does not rely on the large-$\beta_0$ limit. This 
uses the known anomalous dimension of the operator $\alpha_s GG$ and 
the second-order Wilson coefficient \cite{CGS85,ST90} to obtain 
$b$ and $c_1$ in (\ref{rat}). The result can only be useful in 
intermediate orders, before the asymptotically dominant UV renormalon 
behaviour takes over. However, Table~\ref{tab5} suggests that higher IR 
renormalons are very important at low orders because of their 
enhanced over-all normalization in the $\overline{\rm MS}$ scheme. 
Hence the method outlined 
in Section~\ref{lpc} is not expected to be useful for the Adler 
function, at least in the $\overline{\rm MS}$ scheme.

Taking the large-$\beta_0$ approximation as a model for the entire 
series, we can also estimate the ambiguity in summing the series. 
We estimate this by dividing the absolute value of the 
imaginary part of the Borel integral (\ref{im}) by 
$\pi$, an estimate that comes close to the minimal term of the 
series. Restricting the attention to the first IR renormalon pole, 
we find\footnote{The factor $1/\pi$ comes from the $1/\pi$ in 
(\ref{larged}). We determine $\Lambda_{\overline{\rm MS}}$ from 
$\alpha_s(m_\tau)=0.33$, using the one-loop relation (to be consistent 
with the large-$\beta_0$ approximation) and $N_f=3$. This gives 
$\Lambda_{\overline{\rm MS}}=215\,$MeV.}
\begin{equation}
\delta D(Q^2) = \left(-\frac{2}{\beta_0}\right)
\frac{e^{10/3}}{\pi}\,\frac{\Lambda^4_{\overline{\rm MS}}}
{Q^4} \approx \frac{0.06\,\mbox{GeV}^4}{Q^4}.
\end{equation}
This should be compared with the contribution from the gluon condensate 
\begin{equation}
\label{phengl}
\frac{2\pi^2}{3}\langle\frac{\alpha_s}{\pi} GG\rangle \,\frac{1}{Q^4} 
\approx \frac{0.08\,\mbox{GeV}^4}{Q^4},
\end{equation}
which is marginally larger than the perturbative ambiguity. (The present 
estimate agrees with \cite{N95}.) Note that 
the phenomenological value of the gluon condensate \cite{SVZ79} may in 
part parametrize higher order perturbative corrections, because it is 
extracted from comparison of data with a theoretical prediction that 
includes only a first-order radiative correction.

In the large-$\beta_0$ approximation there is a simple relation 
between the Borel transform of the Adler function and that of 
the inclusive cross section $e^+ e^-\to\,$hadrons, 
because the $\beta$-function has 
exactly one term $\beta_0 \alpha_s^2$. Writing 
\begin{equation}
\label{larger}
R_{e^+ e^-} = N_c\left(1+\frac{\alpha_s}{\pi}
\sum_{n=0} \alpha_s^{n}\,
\left[d_n^R (-\beta_0)^n+\delta_n^R\right]\right),
\end{equation}
and neglecting $\delta_n^R$, we have \cite{BY92}
\begin{equation}
B[R](u) = \sum_{n=0}\frac{d_n^R}{n!}\,u^n = \frac{\sin(\pi u)}{\pi u}\, 
B[D](u).
\end{equation}
This follows directly from the fact that the $Q^2$-dependence 
factorizes in (\ref{borelpolaroper}) in the large-$\beta_0$ approximation 
\cite{Ben93a}. The sin attenuates the renormalon singularities. 
In particular, the first IR renormalon pole at $u=2$ is eliminated. 
This is an artefact of the large-$\beta_0$ approximation. Beyond 
this approximation the renormalon singularities are branch cuts, which 
are suppressed  but not eliminated by analytic continuation to Minkowski 
space. In large orders, $d_n/d_n^R\sim n$.

More on numerical aspects of the Adler function in 
the large-$\beta_0$ approximation can be found in 
\cite{N95,BBB95,LM95}. The distribution function $T(\xi)$ that enters 
the integral representation (\ref{sumrep}) of the 
(principal value) Borel integral is given in 
\cite{BBB95} (for $R_{e^+ e^-}$) and \cite{N95b} (for $D$). 

\subsubsection{Inclusive $\tau$ decay into hadrons}

The inclusive $\tau$ decay rate into hadrons yields one of the 
most accurate determinations of the strong coupling $\alpha_s$. 
Subsequent to the detailed analysis of \cite{BNP92} in the framework 
of the OPE \cite{SVZ79}, a lot of effort 
has gone into controlling and understanding the uncertainties in the 
perturbative series that enters the prediction and into the 
question of whether there could be other non-perturbative corrections 
than those incorporated in the OPE, in particular power corrections suppressed 
only by $\Lambda^2/m_\tau^2$. The latter question touches 
also the issue of 
parton-hadron duality, although from the point of view of duality there 
is no reason that violations of it should scale as $1/m_\tau^2$. 
Since renormalons have nothing to say about 
this and since experimental evidence does not support 
`non-standard' non-perturbative corrections (such as small-size 
instanton corrections \cite{NP94,BBB93,NP95}), we focus on the 
accuracy of the perturbative prediction in this section. 
Its renormalon structure was analysed in \cite{BenDiss}. Numerical 
investigations of the large-$\beta_0$ limit were performed 
by \cite{BBB95,N95b} and by \cite{LM95,MT96} for the total decay width 
and for weighted spectral functions by \cite{Neu96}. \cite{ANR95} 
investigated the uncertainties due to UV renormalons specifically. 

The total hadronic width is very well known experimentally, and we quote  
the result from \cite{Ale98}:
\begin{equation}
R_\tau = \frac{\Gamma(\tau^-\to\nu_\tau+\,\rm hadrons)}{\Gamma(\tau^-\to
\nu_\tau e^-\bar{\nu}_e)} = 3.647\pm 0.014.
\end{equation}
The error in $\alpha_s(m_\tau)$ obtained from this measurement 
is largely theoretical. The theoretical prediction follows from the 
correlation functions of the charged vector and axial-vector 
currents, which are decomposed as 
\begin{equation}
\Pi^{\mu\nu}_{V/A}(q) = \left(q_\mu q_\nu-g_{\mu\nu} q^2\right) 
\Pi^{(1)}_{V/A}(q^2) +q_\mu q_\nu \Pi^{(0)}_{V/A}(q^2).
\end{equation}
Making use of the exact, non-perturbative analyticity properties of the 
correlation functions, one obtains
\begin{equation}
\label{rep2}
R_\tau= 6\pi i \oint\limits_{|s|=m_\tau^2} \frac{d s}{m_\tau^2} \left(
1-\frac{s}{m_\tau^2}\right)^2\left[\left(1+2\frac{s}{m_\tau^2}
\right) \Pi^{(1)}(s) + \Pi^{(0)}(s),
\right]\,
\end{equation}
where the integral extends over a circle of radius $m_\tau^2$ 
in the $s=q^2$ plane 
and $\Pi^{(i)}(s)=\Pi^{(i)}_V(s)+\Pi^{(i)}_A(s)$. 
This equation includes decays into strange quarks. 
Small electroweak corrections have to be applied. Eq.~(\ref{rep2}) 
has a meaningful perturbative expansion, because the smallest 
scale involved is $m_\tau$.

We treat quark mass terms as power corrections in $m_{d,s}^2/m_\tau^2$ 
and refer to the perturbative 
expansion of $R_\tau$ in $\alpha_s$ in the 
limit $m_{d,s}=0$ as the perturbative contribution. As before, we write 
\begin{equation}
\label{largetau}
R_\tau = N_c (|V_{ud}|^2+|V_{us}|^2) \left(1+\frac{\alpha_s}{\pi}
\sum_{n=0} \alpha_s^{n}\,
\left[d_n^\tau (-\beta_0)^n+\delta_n^\tau\right]\right),
\end{equation}
and obtain an exact result in the approximation where the 
remainders $\delta_n^\tau$ are neglected. The Borel transform follows 
from inserting (\ref{borelpolaroper}) into (\ref{rep2}). Taking 
advantage of the factorized dependence on $s=-Q^2$ in 
(\ref{borelpolaroper}), the result is \cite{BenDiss}
\begin{equation}
\label{taubtexact}
B\big[R_\tau\big](u) = \sum_{n=0}\frac{d_n^\tau}{n!}\,u^n
=B[D](u)\,\sin(\pi u)
\left[\frac{1}{\pi u}+\frac{2}{\pi(1-u)}-\frac{2}{\pi(3-u)}+
\frac{1}{\pi(4-u)}\right].
\end{equation}
The sin attenuates all renormalon poles except those 
at $u=3,4$. The point $u=1$ is regular, but we note that if a  
power correction of order $\Lambda^2/m_\tau^2$ to $D$ existed, it 
would not be suppressed by a factor of $\alpha_s$ after taking the 
integral in (\ref{rep2}). 

\begin{table}[t]
\addtolength{\arraycolsep}{0.2cm}
$$
\begin{array}{c|c|c|c|c|c|c}
\hline\hline
n & d_n^{\tau,\overline{\rm MS}} & d_n^{\tau,V} & 
M_n^{\tau,\overline{\rm MS}} &
d_n^{\tau,\overline{\rm MS}} (-\beta_0)^n & \delta_n^\tau &
M_{n,\rm exact}^{\tau,\overline{\rm MS}}\\ 
\hline
0 & 1       & 1       & 1     & 1     & 0      & 1     \\
1 & 2.2751  & 0.6084  & 1.521 & 1.629 & 0.027  & 1.530 \\
2 & 5.6848  & 0.8788  & 1.819 & 2.916 & -0.245  & 1.803 \\
3 & 13.754  & -0.3395 & 1.984 & 5.053 & -1.650 & 1.915 \\
4 & 35.147  & 3.7796  & 2.081 & -  & -  & -  \\
5 & 84.407  & -14.680 & 2.134 & -  & -  & -  \\
6 & 248.83  & 99.483  & 2.170 & -  & -  & -  \\
7 & 525.38  & -664.00 & 2.187 & -  & -  & -  \\
8 & 3036.0  & 5400.1  & 2.210 & -  & -  & -  \\
\hline\hline
\end{array}
$$
\caption[dummy]{\label{tab6}\small 
Perturbative corrections to $R_\tau$ in the 
$\overline{\rm MS}$  and $V$ scheme. For the partial sums we take 
$\alpha_s(m_\tau)=0.32$ in the $\overline{\rm MS}$ scheme. The 
last three columns compare the `large-$\beta_0$ limit' with the 
remainder, $\delta_{1,2}^\tau$, to the exact result and an 
estimate thereof for $\delta_3^\tau$. 
$M_{n,\rm exact}^{\tau,\overline{\rm MS}}$ gives partial sums with 
$\delta_n^\tau$ taken into account.}
\end{table}
In Table~\ref{tab6} we show the coefficients 
$d_n^\tau$ in the $\overline{\rm MS}$ scheme and in the scheme with $C=0$, 
with $\mu=m_\tau$ in both cases. 
In the present approximation the second scheme coincides with the 
$V$ scheme, where the coupling is defined through the static 
heavy quark potential. 
The table also shows the partial sums
\begin{equation}
M_N(\alpha_s) = 1+\sum_{n=1}^N d_n (-\beta_0\alpha_s)^n, 
\end{equation}
which quantify how much the first-order radiative correction is 
modified by higher order corrections. Compared to the Adler function, 
see Table~\ref{tab4}, the onset of the 
sign-alternating UV renormalon divergence 
is delayed, because the integration in (\ref{rep2}) enhances the over-all 
normalization of IR renormalons relative to UV renormalons. (This effect 
holds beyond the large-$\beta_0$ limit.) 
In the $\overline{\rm MS}$ scheme the low orders are dominated 
by fixed-sign behaviour and the series can be summed to a parametric 
accuracy of order $\Lambda^4/m_\tau^4$ without interference of UV 
renormalons. The situation is different in the $V$ scheme, where 
UV renormalon residues are larger and IR renormalon residues are 
smaller. Comparison with exact results shows that the large-$\beta_0$ 
approximation is very good at order $\alpha_s^{2,3}$, but seems to 
overestimate the next order, if we trust the estimate of \cite{KS95z} 
more than the large-$\beta_0$ estimate. In Table~\ref{tab7} we show 
the contributions to $d_n^{\tau,\overline{\rm MS}}$ 
from the leading renormalon poles, 
to be compared with Table~\ref{tab5} for the Adler function. The 
relevant decomposition of the Borel transform is now
\begin{equation}
B\left[R_\tau\right] = e^{-5/3}\,\frac{2}{15}\frac{1}{1+u} + 
 e^{-10/3}\,\frac{2}{135}\frac{1}{2+u} + 
 e^5\left\{\frac{8}{3}\,\frac{1}{(3-u)^2}-\frac{8}{9}\,\frac{1}{3-u}
\right\} +\ldots,
\end{equation}
which shows explicitly the suppression of residues of the leading UV 
renormalon poles. However, Table~\ref{tab7} illustrates that the 
coefficients $d_n^{\tau,\overline{\rm MS}}$ are only approximately
dominated by the IR renormalon pole at $u=3$. On the other hand, in the 
$V$ scheme (not shown in the Table) 
the leading UV renormalon pole describes the coefficients 
well for $n>5$. 
\begin{table}[t]
\addtolength{\arraycolsep}{0.2cm}
$$
\begin{array}{c|r|r|r|r}
\hline\hline
n & d_n^{\tau,\overline{\rm MS}}\hspace*{0.3cm} & 
\mbox{UV}(-1) & \mbox{IR}(3) & \mbox{IR}(4) \\ 
\hline
0 & 1       & 0.025  & 0     & -87.31 \\
1 & 2.2751  & -0.025 & 14.66 & -27.28 \\
2 & 5.6848  & 0.050  & 19.54 & -16.37 \\
3 & 13.754  & -0.151 & 29.32 & -14.32 \\
4 & 35.147  & 0.604  & 52.12 & -16.37 \\
5 & 84.407  & -3.022 & 108.6 & -23.02 \\
6 & 248.83  & 18.13  & 260.6 & -38.37 \\
7 & 525.38  & -126.9 & 709.4 & -73.86 \\
8 & 3036.0  & 1015   & 2162  & -161.1 \\
\hline\hline
\end{array}
$$
\caption[dummy]{\label{tab7}\small 
Breakdown of $d_n^{\tau,\overline{\rm MS}}$ 
into contributions from the leading IR and UV 
renormalon poles. The integer in brackets denotes the position of the 
pole.}
\end{table}

The $\alpha_s^3$ correction ($n=2$) adds about 0.3 to the partial 
sums in Table~\ref{tab6}. If we truncate the series at its minimal 
term ($n=7$) the cumulative effect of higher order corrections 
amounts to 0.4, slightly larger than the third-order 
correction.\footnote{One may object that the large-$\beta_0$ 
approximation overestimates this number, because it may overestimate  
already the coefficient for $n=3$. However, if the actual growth of 
coefficients were slower than in the large-$\beta_0$ approximation, we 
would be able to add more terms.} This amounts to a reduction of 
$\alpha_s(m_\tau)$ needed to reproduce the data. To make this 
more precise \cite{BBB95} (see also \cite{N95b,LM95}) computed the 
principal value of the Borel integral as a function of $\alpha_s$. 
For $\alpha_s(m_\tau)=0.32$, they find $M^\tau_\infty=2.23$, close to the 
value $M_7^\tau=2.19$ that would have been obtained from truncating the 
series expansion (see Table~\ref{tab6}). Note that $M^\tau_\infty$ is 
scheme-dependent, but $\alpha_s M^\tau_\infty$ is not, provided schemes 
are consistently related in the large-$\beta_0$ 
approximation \cite{BB95a}.\footnote{However, the corrections to the 
large-$\beta_0$ approximation may be different in different schemes.}  
Accounting for electroweak 
and power corrections, $R_\tau$ is given by
\begin{equation}
R_\tau = 3\,(|V_{ud}|^2+|V_{us}|^2)\, S_{EW} \left\{
1+\delta^{({\rm pt})}+\delta_{EW}+\delta_{\rm power}\right\}\,,
\end{equation}
Making use of the analysis of power corrections in \cite{BNP92} 
and their approximate $\alpha_s$-independence, the experimental 
measurement quoted above translates into
\begin{equation}
\delta^{({\rm pt})}_{\rm exp} = 0.211\pm 0.005.
\end{equation}
The error is purely experimental and no theoretical error 
has been assigned to 
$\delta_{\rm power}$. (The analysis of 
power corrections in \cite{Ale98} leads to 
$\delta^{({\rm pt})}_{\rm exp} = 0.20$.)
The theoretical prediction, based on 
the series in the large-$\beta_0$ approximation, is
\begin{equation}
\label{largebres}
\delta^{({\rm pt})} = \frac{\alpha_s(m_\tau)}{\pi}\left[M_\infty^
\tau(\alpha_s(m_\tau)) + \delta_1^\tau\alpha_s(m_\tau)+
\delta_2^\tau\alpha_s(m_\tau)^2
\right],
\end{equation}
where the terms in the series known exactly are taken into account. 
This result for $\delta^{({\rm pt})}$ is shown as curve `i' in 
Fig.~\ref{fig14}. Compared to perturbation theory truncated at order 
$\alpha_s^3$ (curve `ii'), the value of $\alpha_s(m_\tau)$ is reduced 
by 15\% from about 0.35 to 0.31. This is somewhat less than the reduction 
caused by adding the $\alpha_s^3$ correction (compare curves 
`ii' and `iv'). 
\begin{figure}[t]
   \vspace{-4cm}
   \epsfysize=15cm
   \epsfxsize=10cm
   \centerline{\epsffile{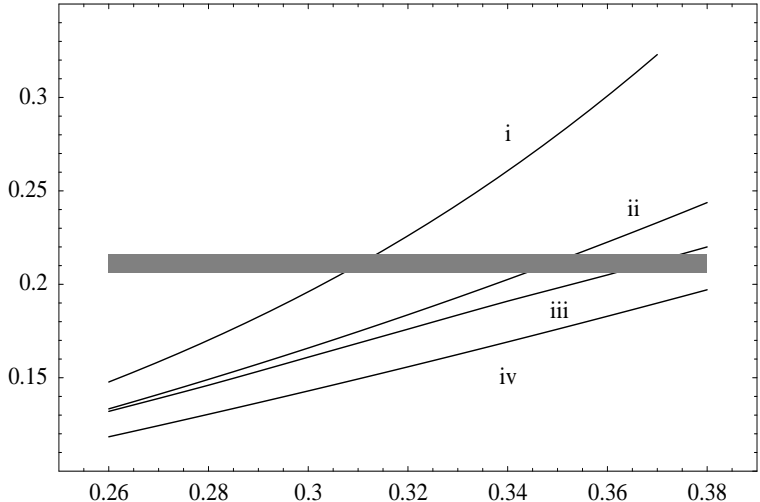}}
   \vspace*{-3.7cm}
\caption[dummy]{\small $\delta^{({\rm pt})}$ as a function of 
$\alpha_s(m_\tau)$ ($\overline{\rm MS}$ scheme) 
for various truncations/partial resummations 
of the perturbative expansion: (i) Large-$\beta_0$ resummation 
according to (\ref{largebres}). (ii) Fixed-order perturbation theory 
up to (including) $\alpha_s^3$. (iii) Resummation of running coupling 
effects from the contour integral only (see text for discussion). 
(iv) Fixed-order perturbation theory 
up to $\alpha_s^2$. The shaded bar gives the experimental measurement 
with experimental errors only. The figure is an update from 
\cite{BBB95}.\label{fig14}}
\end{figure}

How reliable is the large-$\beta_0$ approximation for the unknown 
higher order perturbative contributions? Clearly, there is no answer 
to this question. If we knew, we could do better. It seems safe to 
conclude that higher order corrections add positively and reduce  
$\alpha_s$. As a consequence, we may argue that the theoretical 
error should not be taken symmetric around the fixed order 
$\alpha_s^3$ result, but rather as the variation between 
curves `i' and `ii'. This understanding of the `systematics' of 
higher order corrections is taken into account in \cite{Ale98}, 
where the error of fixed-order perturbation theory is computed 
from a variation around an assumed positive value for the 
$\alpha_s^4$ correction. An important point is that incorporating 
systematic shifts due to higher order perturbative corrections 
in $\tau$ decay may bring us closer to the `true' value of 
$\alpha_s$, but need not improve the consistency with other measurements, 
if similar systematic effects exist there and are not taken into 
account. 

In the above discussion, renormalon ambiguities in the perturbative 
prediction play no role, because they are very small, reflecting the 
fact that the minimal term is attained at rather large $n$. In principle 
there is an error of order $\Lambda^2/m_\tau^2$ that arises when the 
series is truncated at the onset of UV renormalon divergence. The 
large-$\beta_0$ approximation suggests that the numerical coefficient 
of this term is very small, so that this uncertainty is insignificant 
in the $\overline{\rm MS}$ scheme. (Recall that the magnitude of this 
term is scheme-dependent, see Section~\ref{schemedep} and 
\cite{BZ92}.) Related to this is the observation made above that 
the coefficients do not show sign-alternation up to relatively high 
orders, see Table~\ref{tab6}. \cite{ANR95} have investigated 
UV renormalons in $\tau$ decay in great detail, using conformal 
mappings to eliminate this uncertainty. They found rather sizeable 
variations of $\pm 0.05$ in $\alpha_s(m_\tau)$, depending on the 
precise implementation of the mapping procedure. There is a problem 
in applying these mappings to series that do not yet show 
sign-alternation, because the mapping then produces amplifications of 
coefficients rather than cancellations. We therefore feel that 
the conclusion of \cite{ANR95} may be too pessimistic.

In curve `iii' of Fig.~\ref{fig14} we show the result for the 
perturbative contribution to $R_\tau$ based on the implementation 
of a partial resummation of running coupling effects suggested by 
\cite{LP92}. This resummation takes into account a series 
of `$\pi^2$-terms' that arise when integrals of powers  
of $\alpha_s(-\sqrt{s})$ 
are taken according to (\ref{rep2}). Because the largest effect comes 
from $\beta_0$, this resummation is included in the large-$\beta_0$ 
approximation which takes into account running coupling effects 
not only in the contour integral (\ref{rep2}) but also in the 
spectral functions. Comparison of `i' and `iii' with  
`ii' shows that the effect of the two resummations tends into different 
directions relative to the fixed-order result. The explanation 
suggested in \cite{BBB95} reads that the convergence of the 
partial resummation of \cite{LP92} is limited by the UV renormalon 
behaviour of the Adler function. As seen from Table~\ref{tab4} 
this limitation is more serious for $D$ than it is for $R_\tau$. 

The large-$\beta_0$ approximation is scheme and scale dependent in the 
sense that the terms dropped (the remainders $\delta_n$) are of different 
size in different schemes. Such scheme\--de\-pen\-dence is expected for 
partial resummations and the real question is in which schemes the 
approximation works best. The requirement of scheme-independence 
emphasized by \cite{Chy95,MT96} misses this point. Since empirically 
the approximation seems to work well in the $\overline{\rm MS}$ scheme, 
one cannot expect it to work well in schemes that differ from 
$\overline{\rm MS}$ by large parameter redefinitions that are formally 
of sub-leading order. \cite{MT96} proposed to implement the 
large-$\beta_0$ limit for the effective charge $\beta$-function 
that corresponds to $R_\tau$. In Fig.~\ref{fig14} this implementation 
falls below the fixed order result `ii'. This 
resummation scheme implies that the correction to be added to the 
third order result in the $\overline{\rm MS}$ scheme is negative 
despite the regular fixed-sign behaviour observed in the exact 
coefficients up to order $\alpha_s^3$.

As the spectral functions in $\tau$ decay are well measured, additional 
information can be obtained from their moments. \cite{Neu96} has analysed 
in detail the leading-$\beta_0$ resummations for the moments.

Finally, we mention that when (\ref{sumrep}) is used to compute 
the principal value Borel integral $M_\infty$, the 
`Landau pole contribution' in square brackets is very 
important.\footnote{The distribution function 
required for $\tau$ decay can be found in \cite{BBB95}.} 
Although formally of order $\Lambda^2/m_\tau^2$, leaving this term 
out results in a very small value for $M_\infty$. The omission of this 
term is equivalent to a redefinition of the coupling constant which is 
related to the $\overline{\rm MS}$ coupling by large $1/Q^2$ corrections 
not related to renormalons. This point is discussed in detail in 
\cite{BBB95}. 

\subsubsection{Deep-inelastic scattering: sum rules}

Consider the Gross-Llewellyn-Smith (GLS) and polarized Bjorken (Bj) 
sum rules,
\begin{eqnarray}
&&\int_0^1 d x\,F_3^{\nu p+\bar{\nu} p}(x,Q) = 
6\left(1-\frac{\alpha_s}{\pi}\sum_{n=0}\alpha_s^n [d_n^{\rm GLS} 
(-\beta_0)^n+\delta_n^{\rm GLS}]\right),
\\
&&\int_0^1 d x\,g_1^{e p-e n}(x,Q) = 
\frac{1}{3}\,\left|\frac{g_A}{g_V}\right|
\left(1-\frac{\alpha_s}{\pi}\sum_{n=0}\alpha_s^n [d_n^{\rm Bj} 
(-\beta_0)^n+\delta_n^{\rm Bj}]\right).
\end{eqnarray}
The nucleon structure functions $F_3$ and $g_1$ are defined in the 
standard way. In both cases the twist-4 $\Lambda^2/Q^2$ 
corrections are given by the 
matrix element of a single operator \cite{JS81,SV82,EFP83}. The 
perturbative corrections are known exactly to order 
$\alpha_s^3$ \cite{LV91}. The normalization is such that 
$d_0=1$, $\delta_0=0$.

The IR renormalon singularity at $t=-1/\beta_0$ ($u=1$) 
that corresponds to the twist-4 operator was first discussed 
by \cite{Mue93}. The strength of the leading UV renormalon 
at $t=1/\beta_0$ is determined in \cite{BBK97}. Combining 
both pieces of information, we finds \cite{BBK97}
\begin{eqnarray}
\label{dis2}
C_{\rm GLS}(\alpha_s) &\stackrel{n\to \infty}{=}&
\sum_n (-\beta_0)^n\,n!\,\Big[K^{UV}_{GLS}\,(-1)^n\,
n^{1+\beta_1/\beta_0^2+\lambda_1} 
\nonumber\\
&&\hspace*{+1.5cm}
+\,K^{IR}_{GLS}\,n^{-\beta_1/\beta_0^2-(4/3b)(N_c-1/N_c)}
\Big]\,\alpha_s^{n+1},
\end{eqnarray}
where $C_{\rm GLS}(\alpha_s)$ denotes the perturbative contribution 
to the GLS sum rule and the anomalous dimension of the 
twist-4 operator calculated by \cite{SV82} has been used. 
In this equation $\beta_{0,1}$ are the first two coefficients of the 
$\beta$-function, $b=-4\pi\beta_0$, and 
$\lambda_1$ is related to the anomalous dimension matrix of four-fermion 
operators, see Table~\ref{tab1}. For $N_f>2$, the UV renormalon behaviour 
dominates the asymptotic behaviour at very large $n$ because of its larger 
power of $n$. However, the over-all normalizations are not known. 
Since the $\overline{\rm MS}$ scheme favours large residues of IR 
renormalons, one expects fixed-sign IR renormalon behaviour in 
intermediate orders. The first three terms in the series known exactly 
are indeed of the same sign in the $\overline{\rm MS}$ scheme.

The large-$\beta_0$ approximation to the perturbative part of the 
sum rules has been investigated in \cite{Ji95,LM95}. 
The large-$\beta_0$ approximations 
to the GLS and Bj sum rules coincide, because the perturbative 
contributions to the sum rules differ only by `light-by-light'  
contributions starting at order $\alpha_s^3$. These contributions 
are subleading in the large-$\beta_0$ approximation. The Borel 
transform that is 
relevant in the large-$\beta_0$ approximation can be inferred 
from \cite{BK93} and is given by
\begin{equation}
\label{borelgls}
B[\mbox{GLS/Bj}](u) = \sum_{n=0} \frac{d_n^{\rm GLS/Bj}}{n!}\, u^n 
= \left(\frac{Q^2}{\mu^2}\,e^C\right)^{-u}\,\frac{1}{9}\left\{
\frac{8}{1-u}+\frac{4}{1+u}-\frac{5}{2-u}-\frac{1}{2+u}\right\}.
\end{equation}
It is much simpler than the Borel transform for the Adler function 
(\ref{borelpolaroper}), because the $\alpha_s$ correction comes from 
one-loop diagrams in DIS and from two-loop diagrams for the Adler 
function. In particular, there are only four renormalon poles, all 
other being suppressed at leading order. But since the leading 
singularities at $u=\pm 1,\pm 2$ are present, we may still try a 
numerical analysis. 

\begin{table}[t]
\addtolength{\arraycolsep}{0.2cm}
$$
\begin{array}{c|c|c|c|c|c}
\hline\hline
n & d_n^{\rm GLS} &  
M_n^{\rm GLS} &
d_n^{\rm GLS} (-\beta_0)^n & \delta_n^{\rm GLS} &
M_{n,\rm exact}^{\rm GLS}\\ 
\hline
0 & 1      & 1     & 1     & 0       & 1     \\
1 & 2      & 1.473 & 1.432 & -0.291  & 1.376 \\
2 & 6.389  & 1.830 & 3.277 & -1.354  & 1.586 \\
3 & 22.41  & 2.125 & 8.233 & -4.040  & 1.737 \\
4 & 103.7  & 2.449 & -  & -  & -  \\
5 & 525.9  & 2.837 & -  & -  & -  \\
6 & 3362   & 3.423 & -  & -  & -  \\
7 & 22990  &  -    & -  & -  & -  \\
8 & 1.92\cdot 10^5  & -  & -  & -  & -  \\
\hline\hline
\end{array}
$$
\caption[dummy]{\label{tab8}\small 
Perturbative corrections to the GLS (Bj) sum rules in the 
large-$\beta_0$ limit. 
All results in the $\overline{\rm MS}$ scheme and for $N_f=3$. 
To compute the partial sums 
we take $\alpha_s(Q^2=3\,{\rm GeV}^2)=0.33$. The 
last three columns compare the large-$\beta_0$ limit with the 
remainder, $\delta_{1,2}^{\rm GLS}$, to the exact result and an 
estimate thereof for $\delta_3^{\rm GLS}$. 
$M_{n,\rm exact}^{\rm GLS}$ gives partial sums with 
$\delta_n^{\rm GLS}$ taken into account.}
\end{table}
The coefficients $d_n^{\rm GLS}=d_n^{\rm Bj}$ are displayed in 
Table~\ref{tab8} and compared with the exact result and an estimate of 
the $\alpha_s^4$ correction from \cite{KS95z}. We note that while 
the large-$\beta_0$ approximation gives the higher order corrections 
with the correct sign, it generally overestimates them, a tendency 
already observed for the Adler function and $\tau$ decay. Taken at 
face value, the large-$\beta_0$ approximation implies that the 
minimal term of the series is reached at order $\alpha_s^{3,4}$ 
at $Q^2=3\,$GeV${}^2$, a momentum transfer relevant to the CCFR 
experiment. Hence it is not clear whether at $Q^2=3\,$GeV${}^2$ 
the perturbative prediction could be improved by further exact 
calculations of higher order corrections. Further improvement would 
then require the inclusion of twist-4 contributions, and in particular 
a practically realizable procedure to combine them consistently 
with the perturbative series. 

In this context it is interesting to note that the integral over 
loop momentum is dominated by $k\sim 450\,$MeV at order $\alpha_s^3$ and 
$k\sim 330\,$MeV at order $\alpha_s^4$.\footnote{These estimates can be 
obtained from converting the Borel transform into the loop 
momentum distribution \cite{N95}, see Section~\ref{lmd}.} As for 
the Adler function, we estimate the ambiguity in summing the perturbative 
expansion by the imaginary part of the Borel integral (\ref{im}) (divided 
by $\pi$) from the first IR renormalon pole alone. This gives 
($\Lambda_{\overline{\rm MS}}=215\,$MeV as above)
\begin{equation}
\frac{1}{6}\,\delta {\rm GLS}(Q^2) = \left(-\frac{1}{\beta_0}\right)
\frac{8 e^{5/3}}{9\pi}\,\frac{\Lambda^2_{\overline{\rm MS}}}
{Q^2} \approx \frac{0.10\,\mbox{GeV}^2}{Q^2}.
\end{equation}
This should be compared to the twist-4 contribution to the 
same quantity estimated 
by QCD sum rules \cite{BK87}, 
\begin{equation}
-\frac{8}{27}\,\frac{\langle\langle {\cal O}_4\rangle\rangle}{Q^2} 
\approx -\frac{0.1\,\mbox{GeV}^2}{Q^2},
\end{equation}
where $\langle\langle {\cal O}_4\rangle\rangle$ is the 
reduced nucleon matrix element of a certain local twist-4 operator. 
The two are comparable, which suggests that the treatment of 
perturbative corrections beyond those known exactly is 
as important for a determination of $\alpha_s$ as the twist-4 
correction.

\cite{SMMS96} and \cite{MMS97} have considered moments of the longitudinal 
structure function $F_L$ and the non-singlet contribution to $F_2$, 
respectively, in the large-$\beta_0$ approximation. The second case 
is more difficult, because it requires collinear factorization to be 
carried out in the large-$\beta_0$ limit, while this is not 
necessary for $F_L$ in leading order. The approximation is found to be  
quite good for larger moments ($N>4$), typically overestimating 
the exact result by some amount, but fails completely for the 
lower moments of $F_2$. This may be due to the fact that smaller moments 
are more sensitive to the small-$x$ region in which other effects 
not incorporated in the large-$\beta_0$ limit are important \cite{SMMS96}.

\subsubsection{Twist-4 corrections to DIS structure functions}
\label{dismod}

In this section we discuss applications of the `renormalon model' for 
twist-4 corrections to deep-inelastic scattering (DIS) quantities 
suggested in \cite{DMW96,SMMS96}. The 
basic aspects of the model, its virtues and limitations, have already 
been outlined in Section~\ref{modelssect}, see (\ref{fl2}). 

To make the idea more explicit, we consider the structure functions 
$F_2$ and $F_L$ as examples. One first computes the dependence of the 
first IR renormalon residue (related to twist-4 operators, see 
Section~\ref{disex}) on the scaling variable $x=-q^2/(2 p\cdot q)$. 
At present all such calculations have been done only for one-loop 
diagrams dressed by vacuum polarization insertions, i.e. in the 
formal large-$\beta_0$ limit. It is usually most convenient to extract 
the residue from the expansion of the distribution function $T(\xi)$ 
introduced in section~\ref{dm}. The result is\footnote{A common 
over-all normalization is omitted, because it plays no role 
in what follows. See Section~\ref{disex} for definitions and the 
derivation of the result for $F_L$ in terms of UV properties of 
twist-4 distributions.} \cite{BB95b,DMW96,SMMS96,DW96a}
\begin{eqnarray}
A^2_L(x) &=& 8 x^2-4\delta(1-x), 
\\
A^2_2(x) &=& -\frac{4}{[1-x]_+}+4+2 x+12 x^2-9\delta(1-x)-\delta'(1-x) 
\end{eqnarray}
for $F_L/(2 x)$ and $F_2/(2 x)$. The `+' prescription is defined as usual 
by $\int_0^1d x \,[f(x)]_+ t(x) = \int_0^1 d x \,f(x) \,(t(x)-t(1))$ 
for test functions 
$t(x)$. The result is then represented as
\begin{equation}
\label{par1}
F_P(x,Q) = F_P^{\rm tw-2}(x,Q)\left(1+\frac{D_P(x,Q)}{Q^2}+ O(1/Q^4)
\right),
\end{equation}
where $F_P^{\rm tw-2}(x,Q)$ is the leading-twist result for the 
structure function $F_P$ and 
\begin{equation}
\label{par}
D_P(x,Q) = \frac{1}{F_P^{\rm tw-2}(x,Q)} 
\sum_i \int\limits_x^1\frac{d\xi}{\xi}\,f_i(x/\xi,\mu)\,\Lambda_i^2 
A_P^{2,i}(\xi)
\end{equation}
is the model parametrization of the (relative) twist-4 correction. 
Here $f_i(x/\xi,\mu)$ are standard (leading-twist) parton densities, 
$i$ sums over quarks and gluons, and $\Lambda_i$ are scales of 
order $\Lambda$ which provide the over-all normalization. We recall 
(Section~\ref{disex}) 
that twist-4 corrections take the form (\ref{par}) 
if the twist-4 matrix elements are substituted by their power 
divergence \cite{BBM97}.

The over-all normalization has been treated differently in the 
literature. In the approach of 
\cite{DMW96}, it is suggested to parametrize 
the normalization of all $1/Q^2$ power corrections by a single 
process-independent number, to be extracted from the data once.
\cite{SMMS96} originally suggested to fix the over-all normalization 
parameter-free by the normalization of the renormalon ambiguity. 
This turned out to fit the data poorly and the authors 
subsequently also treated the over-all normalization as a free parameter 
\cite{MSSM97}. In \cite{BBM97} it is suggested that the normalization 
should be adjusted in a process-dependent way and only the 
shape of the $x$-distribution taken as a prediction of the model. 
Because of difficulties in constructing the gluon contribution 
in the model, one may think of adjusting the normalization of 
quark and gluon contributions separately. 

\begin{figure}[p]
   \vspace{-3.8cm}
   \epsfysize=16cm
   \epsfxsize=12cm
   \centerline{\epsffile{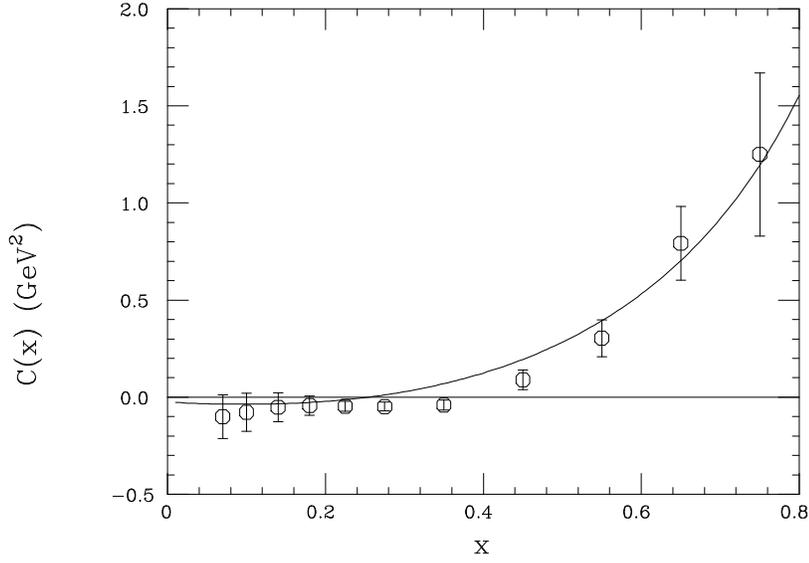}}
   \vspace*{-4.5cm}
\caption[dummy]{\small Relative twist-4 contribution $D_2(x)$ 
(called $C(x)$ here)  defined by 
(\ref{par}) to the structure function $F_2$ in the `renormalon model' 
compared with the data analysis of 
\cite{VM92}. Plot taken from \cite{DMW96}. \label{fig15}}
\end{figure}
\begin{figure}[p]
   \vspace{0.3cm}
   \epsfysize=8.3cm
   \epsfxsize=11cm
   \centerline{\epsffile{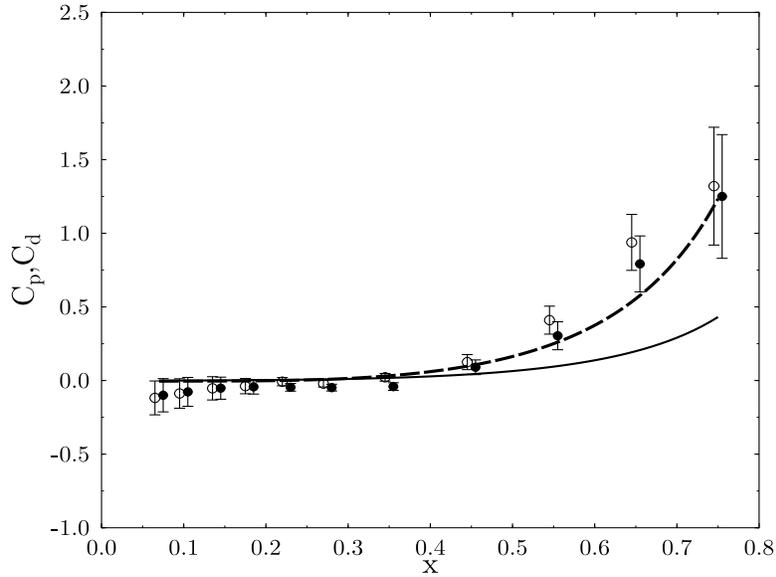}}
   \vspace*{-0.5cm}
\caption[dummy]{\small  Relative twist-4 contribution $D_2(x)$ 
(called $C_{p,d}(x)$ here)  defined by (\ref{par}) 
to the proton (deuteron) structure function $F_2$ in the `renormalon model' 
(dashed line) compared with 
proton (filled circles) and deuteron (empty circles) data 
\cite{VM92}. Plot taken from \cite{MSSM97}. The solid 
curve shows the literal estimate of the renormalon ambiguity.
\label{fig16}}
\end{figure}
The `renormalon model' of twist-4 corrections has drawn much of its 
inspiration from Fig.~\ref{fig15} first shown by \cite{DMW96} 
(see also \cite{DW96a,MSSM97}). The shape of the twist-4 correction 
to the structure function $F_2$ calculated from the model reproduces the 
shape required to fit experimental data very well. Note that 
the renormalon model contains only the non-singlet contribution 
to $F_2$, which is expected to dominate except for small values of $x$. 

Encouraged by this observation, \cite{SMMS96,DW96a} 
considered the longitudinal structure function 
$F_L$, while \cite{DW96a,MSSM97} considered the structure function 
$F_3$. The polarized structure function 
$g_1$ has been analysed by \cite{DW96a,MMMSS96}. Other polarized 
structure functions were examined by \cite{LS98} and the 
transversity distribution $h_1$ by \cite{MS97}.\footnote{Note, 
however, that \cite{MS97} did not consider the correlation functions 
of physical currents and therefore the result is not applicable to 
a measurable deep inelastic scattering process.} 
Recently, \cite{SMMS98} added a model 
prediction for the singlet contribution to $F_2$, which modifies 
Figs.~\ref{fig15} and \ref{fig16} at small $x$, below those $x$ 
for which comparison with data is possible. It is interesting 
to compare this prediction with other model parametrizations of 
twist-4 corrections at small $x$. The treatment of singlet contributions 
is more difficult and ambiguous in the renormalon model than non-singlet 
contributions.\footnote{See Section~\ref{fragmod} for a discussion of 
this point in the context of fragmentation.} 
The calculation relies on singlet quark contributions, 
which are then reinterpreted as gluon contributions according 
to the procedure suggested by \cite{BBM97}. In any case, the renormalon 
model cannot be applied at $x$ so small that logarithms of $x$ 
need to be resummed.

One may naturally wonder whether there is an explanation for why the 
model seems to work in cases where it can be compared with 
measurements. Several hints are provided by the comparisons 
shown in Figs.~\ref{fig15}-\ref{fig17}.
\begin{figure}[p]
   \vspace{-1.6cm}
   \epsfysize=10cm
   \epsfxsize=7cm
   \centerline{\epsffile{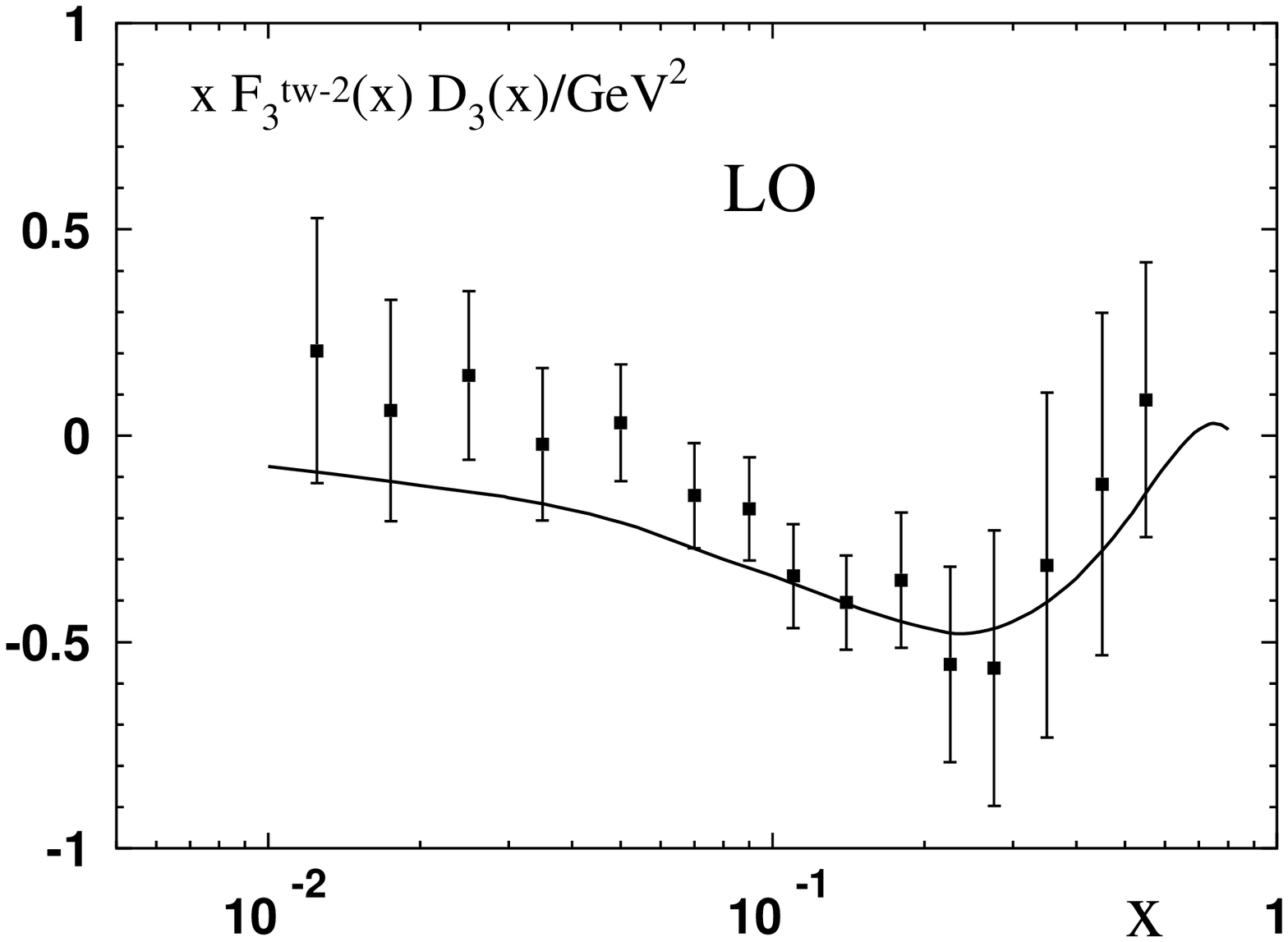}}
   \vspace*{-4cm}
   \epsfysize=10cm
   \epsfxsize=7cm
   \centerline{\epsffile{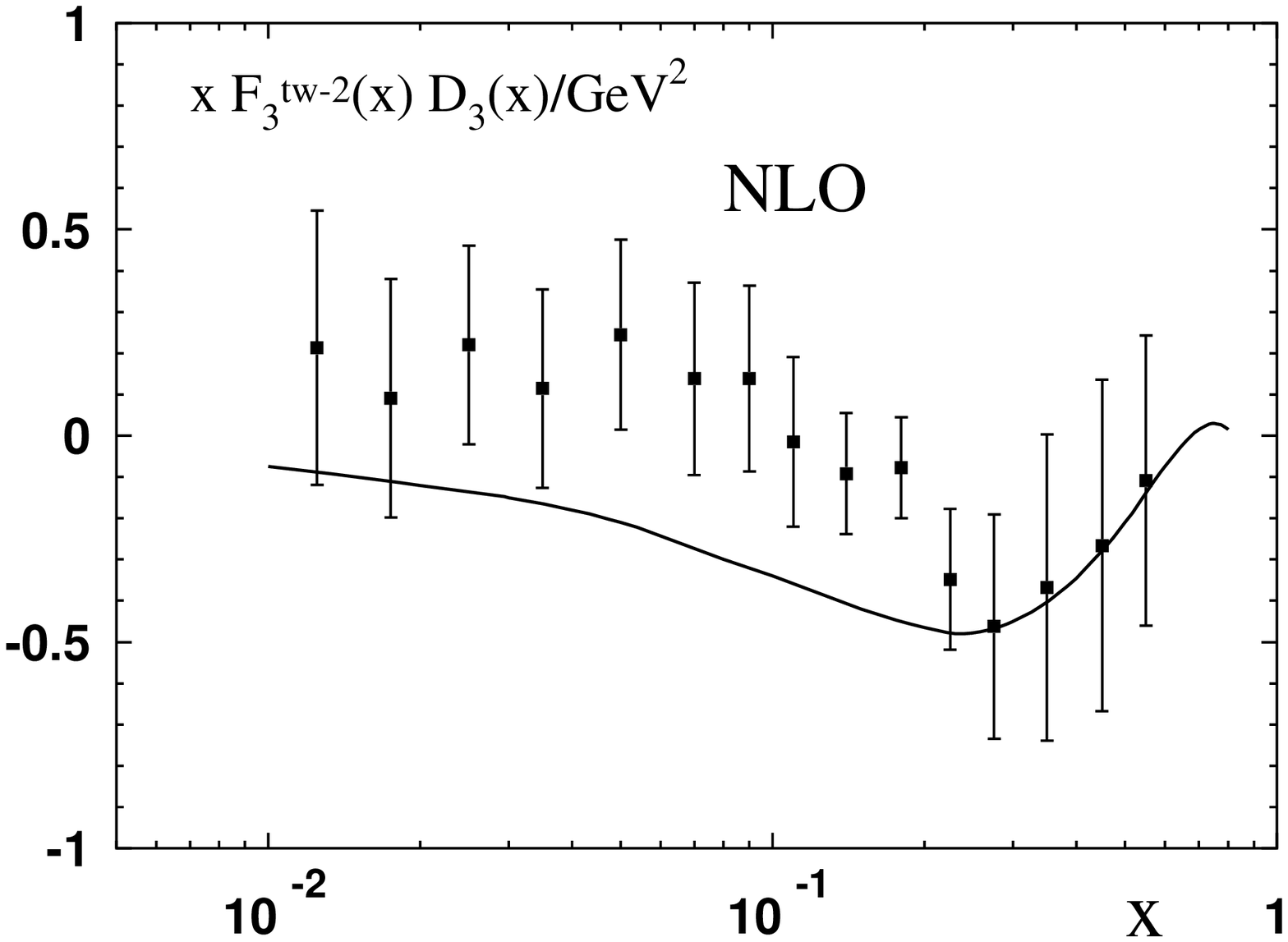}}
   \vspace*{-4cm}
   \epsfysize=10cm
   \epsfxsize=7cm
   \centerline{\epsffile{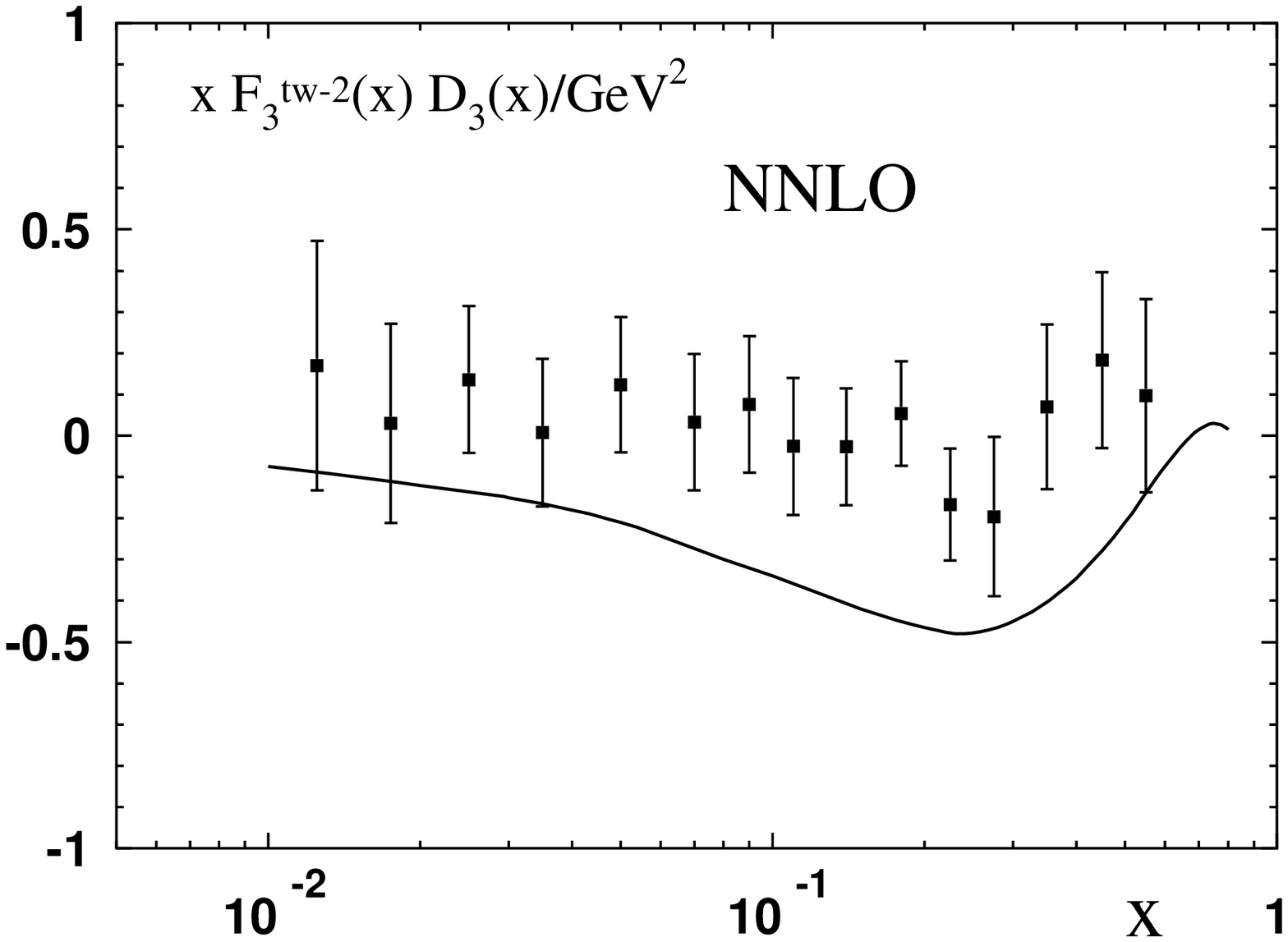}}
   \vspace*{-2cm}
\caption[dummy]{\small Twist-4 correction to $x F_3$ as extracted from 
the (revised) CCFR data. The three plots show the effect of including 
leading order (LO), next-to-leading order (NLO) and 
next-to-next-to-leading order (NNLO) QCD corrections in the 
twist-2 term. The data points are quoted from the analysis 
of \cite{KKPS97}. Overlaid is the shape obtained from the 
`renormalon model' for the $1/Q^2$ power correction. \label{fig17}}
\end{figure}

We recall that the model for twist-4 corrections is target-independent 
in the sense that all target-dependence enters trivially through the 
target dependence of the twist-2 distribution functions. In terms of 
moments $M_n$, (\ref{par}) implies
\begin{equation}
\frac{M_n^{\rm tw-4}}{M_n^{\rm tw-2}}_{|\rm hadron 1} = \,\,\,
\frac{M_n^{\rm tw-4}}{M_n^{\rm tw-2}}_{|\rm hadron 2} 
\end{equation}
exactly. Hence the model is useful only if the genuine twist-4 target 
dependence is small compared to the magnitude of the 
twist-4 correction itself. Fig.~\ref{fig16} shows that this 
is indeed the case for $F_2$ of protons against deuterons, in particular 
in the region of large $x$. 

It is known that higher-twist corrections (as well as higher order 
perturbative corrections) are enhanced as $x\to 1$ 
(see for example \cite{BBL89}). This is in part an effect of kinematic 
restrictions near the exclusive region and the renormalon model 
reproduces such enhancements.\footnote{This is seen most easily in the 
dispersive approach discussed in Section~\ref{dm}, in which the 
radiated gluons acquire an invariant mass that modifies the phase 
space boundaries.} For the structure functions it is found that 
power corrections related to renormalons are of order 
\begin{equation}
\label{scaledis}
\left[\frac{\Lambda^2}{Q^2 (1-x)}\right]^{\!n},
\end{equation}
at least those related to diagrams with a single gluon line 
\cite{BB95b}. This provides some insight into the kinematic region 
in which the twist expansion breaks down.\footnote{The possibility 
to use renormalons for this purpose was first noted by \cite{Agl95}. 
However, the result of this paper was not confirmed by 
\cite{BB95b,DMW96}.} It also tells us that the increase of the 
twist-4 correction towards larger $x$ seen in the model and the data 
in Figs.~\ref{fig15} and \ref{fig16} may to a large extent be 
the correct parametrization of such a kinematic effect. Note 
that (\ref{scaledis}) can be understood as following from the fact that 
the hard scale in DIS is $Q\sqrt{1-x}$ at large 
(but not too large) $x$. 

It is also possible that both the experimental parametrization of 
higher-twist corrections and the model provide effectively a 
parametrization of higher order perturbative corrections to 
twist-2 coefficient functions. As far as data are concerned, it should be  
kept in mind that it is obtained from subtracting from the measurement 
a twist-2 contribution obtained from a truncated perturbative 
expansion. As far as the renormalon model is concerned, it is best 
justified by the `ultraviolet dominance hypothesis' \cite{BBM97} 
(see Section~\ref{modelssect}). Since UV contributions to twist-4 
contributions can also be interpreted as contributions to twist-2 
coefficient functions, a `perturbative' interpretation of the 
model prediction suggests itself. Note that higher order corrections 
in $\alpha_s(Q)$ vary more rapidly with $Q$ than lower order ones, 
and may not be easily distinguished from a $1/Q^2$ behaviour, 
if the $Q^2$-coverage of the data is not rather large. An interesting 
hint in this direction is provided by the analysis of CCFR data 
on $F_3$ of \cite{KKPS97}, reproduced in Fig.~\ref{fig17}. The 
figure shows how the experimentally fitted twist-4 correction 
gradually disappears as NLO and NNLO perturbative corrections 
to the twist-2 coefficient functions are 
included. At the same time, the renormalon model for the 
twist-4 corrections reproduces well\footnote{Compared to \cite{KKPS97} 
we have rescaled the renormalon model prediction (solid curve) by a 
factor 1.5. As mentioned above 
we treat the over-all normalization as an adjustable parameter.} 
the shape of data at leading order, and hence parametrizes successfully 
the effect of NLO and (approximate) 
NNLO corrections. This is an important 
piece of information, relevant to quantities for which an NNLO or 
even NLO analysis is not yet available.

Note that whether the model is interpreted as a model for twist-4 
corrections or higher order perturbative corrections is insignificant 
inasmuch 
as renormalons are precisely related to the fact that 
the two cannot be separated unambiguously. The model clearly cannot be 
expected to reproduce fine structures of twist-4 corrections. Its  
appeal draws from the fact that it provides a simple way to 
incorporate some contributions beyond LO or NLO in perturbation 
theory, which may be the dominant source of discrepancy with data 
at accuracies presently achievable.

\subsection{Hard QCD processes II}
\label{hard2}

In this section we summarize results on hard processes that do not 
admit an OPE. We do not follow the historical 
development and begin with fragmentation functions in $e^+e^-$ 
annihilation, which provide a continuation of Section~\ref{dismod}. 
We then turn to hadronic event shape observables in $e^+e^-$ 
annihilation and deep-inelastic scattering. These are the 
simplest observables with $1/Q$ power corrections and 
renormalon-inspired phenomenology has progressed furthest in this 
area. Soft gluons play an important role for $1/Q$ power corrections. 
The issue of soft gluon resummation near the boundary of partonic 
phase space and power corrections is taken up in Section~\ref{dy}, 
where the Drell-Yan process is studied from this perspective. 
Finally, in Section~\ref{otherhard} we summarize work related to 
renormalons on other hard processes not covered so far.

\subsubsection{Fragmentation in $e^+ e^-$ annihilation}
\label{fragmod}

Inclusive single particle production in $e^+ e^-$ annihilation, 
$e^+ e^-\to\gamma^*,Z^0\to H(p)+X$, is the time-like analogue 
of DIS. The double differential cross section 
can be expressed as 
\begin{eqnarray}
\label{def}
\frac{d^2\sigma^H}{dx d\cos\theta}(e^+e^-\to HX)\!&=&\! 
\frac{3}{8}\,(1+\cos^2\theta)\,\frac{d\sigma_T^H}{dx}(x,Q^2)
+\frac{3}{4}\sin^2\theta\, \frac{d\sigma_L^H}{dx}(x,Q^2)
\nonumber\\
&&\hspace*{-3cm}+\,\frac{3}{4}\cos\theta\, \frac{d\sigma_A^H}{dx}(x,Q^2). 
\end{eqnarray}
We defined the scaling variable 
$x=2 p\cdot q/q^2$, where $p$ is the momentum of $H$, and $q$ 
the intermediate gauge boson momentum; $Q^2=q^2$ denotes the centre-of-mass 
energy squared and $\theta$ the angle between the 
hadron and the beam axis. 
In the following, we will not be concerned with the asymmetric 
contribution and 
with quark mass effects. Neglecting quark masses, $(1/\sigma_0)\, d\sigma^
H_{T/L}/dx$ (where $\sigma_0$ is the Born total annihilation cross section) 
is independent of electroweak couplings and the longitudinal cross 
section is suppressed by $\alpha_s$. We drop the superscript `$H$' 
in the following and imply a sum over all hadron species $H$. 

At leading power in $\Lambda/Q$, the formalism that describes the 
fragmentation structure functions $d\sigma_P^H/dx$ is analogous 
to that for DIS. The structure functions are convolutions of 
perturbative coefficient functions and process-independent 
parton fragmentation functions defined for example in the 
$\overline{\rm MS}$ scheme. The formalism treats logarithmic scaling 
violations in $Q$. In addition, there exists power-like scaling 
violations (`power corrections') due to multi-parton correlations 
\cite{BB91}. However, contrary to DIS, the moments of these 
multi-parton correlations are not related to matrix elements of 
local operators and the OPE cannot 
be applied to fragmentation. This provides the motivation for the 
renormalon analysis.

\begin{figure}[t]
   \vspace{-1cm}
   \epsfysize=20cm
   \epsfxsize=14cm
   \centerline{\hspace*{-6.6cm}\epsffile{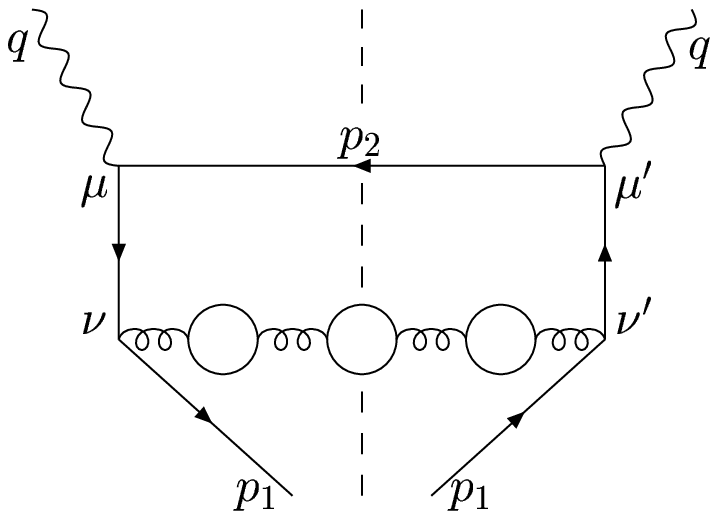}}
   \vspace*{-21cm}
   \epsfysize=20cm
   \epsfxsize=14cm
   \centerline{\hspace*{6.6cm}\epsffile{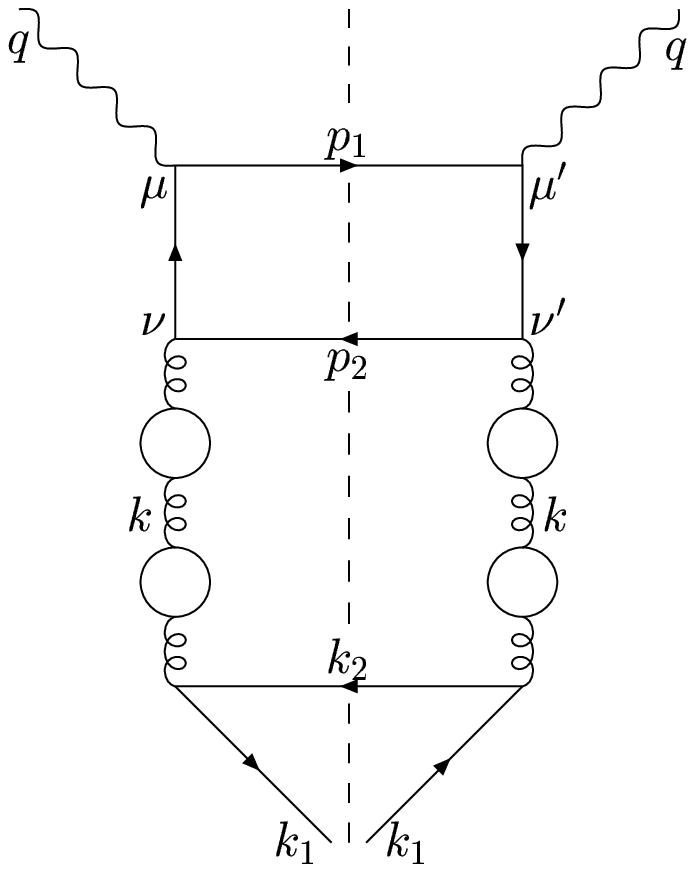}}
   \vspace*{-11cm}
\caption[dummy]{\small Primary (left) and secondary (right) quark 
fragmentation diagrams (in cut diagram representation) in the 
large-$\beta_0$ approximation or the approximation of single 
gluon emission. Note that the figure to the right appears to have two 
chains of fermion loops, but should nonetheless 
be interpreted as a single chain diagram. \label{fig18}}
\end{figure}
In the standard leading order analysis of diagrams with a single 
chain of vacuum polarizations (formally, the `large-$\beta_0$' 
approximation) there are two contributions to the fragmentation 
process, shown in Fig.~\ref{fig18}. We 
refer to the left diagram as `primary quark fragmentation' and to the 
right diagram as `secondary quark fragmentation', because in the 
first case the fragmenting quark is connected to the primary 
hard interaction vertex, while in the second case the fragmenting 
quark arises from gluon splitting $g\to q\bar{q}$. The gluon 
contributions are pure counterterms, except at order $\alpha_s$, and 
therefore are of no relevance to power corrections in the present 
approximation. The secondary quark contribution is not inclusive 
over the cut quark bubble, because it is one of those quarks that 
fragments into the registered hadron $H$.  As a consequence, when 
one uses the dispersive method described in Section~\ref{dm} to compute 
the diagrams, the calculation is not the same as a one-loop calculation 
with finite gluon mass.\footnote{For deep-inelastic 
scattering this has to be taken into account, too, for 
singlet, as opposed to non-singlet, quantities. See \cite{SMMS98} 
for a calculation of singlet contributions to DIS.} 
(They do coincide for the primary quark contribution). 
Renormalons in fragmentation were considered 
in \cite{DW97a,BBM97} for longitudinal and transverse 
components separately. In the first paper a simplified prescription 
was adopted in which all contributions were calculated with 
a finite gluon mass. In the second paper the diagrams of 
Fig.~\ref{fig18} were evaluated exactly. While the finite 
gluon mass prescription is certainly unsatisfactory, because it does 
not account for gluon splitting, it is not clear whether 
the exact evaluation is more realistic, because it accounts only 
for $g\to q\bar{q}$, but not for $g\to gg$, which is more 
important. The problem is connected with the fact that one computes 
fermion loops, but usually argues that they trace contributions 
that should naturally be written in terms of the full QCD 
$\beta$-function coefficient $\beta_0$. This argument is difficult 
to justify for a non-inclusive process such as secondary quark 
fragmentation, because restoring the full $\beta_0$ does not allow us to 
extrapolate from $g\to q\bar{q}$ to $g\to gg$. The 
conclusion is that the renormalon model 
for power corrections is more ambiguous, as far as the $x$-dependence 
is concerned for non-inclusive processes. These ambiguities are 
discussed in detail in \cite{BBM97}. 

\begin{figure}[t]
   \vspace{-2.5cm}
   \epsfysize=13cm
   \epsfxsize=9cm
   \centerline{\epsffile{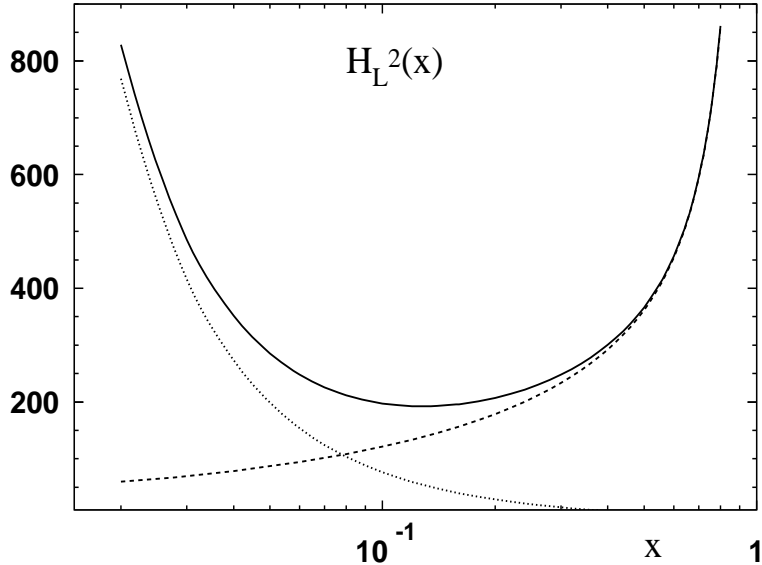}}
   \vspace*{-3cm}
\caption[dummy]{\small Shape of $\Lambda^2/Q^2$ power corrections to 
the longitudinal fragmentation cross section as a function of $x$. 
Primary quark fragmentation (dashed line), secondary quark fragmentation 
(dotted line) and their sum (solid line). 
\label{fig19}}
\end{figure}
The result for the $x$-dependence of $\Lambda^2/Q^2$ power corrections 
to the longitudinal fragmentation cross section $d\sigma_L/dx$ from 
\cite{BBM97} is shown in Fig.~\ref{fig19}. The function $H_L^2(x)$ 
is defined as in (\ref{par1}, \ref{par}) except that the scale 
$\Lambda_i^2$ in (\ref{par}) is omitted, so that $H_L^2$ is 
dimensionless, and $F_P$ is replaced by $d\sigma_L/dx$. The vertical 
scale in the figure is arbitrary and the over-all normalization 
should be adjusted to data on power corrections, once the LEP1 analysis 
becomes available. We note that the secondary quark contribution 
(which we will shortly interpret as a gluon contribution) exceeds the 
primary quark contribution at $x<0.1$, while the latter  
dominates in the region where the registered hadron takes 
away a sizeable fraction of the available energy. This is as expected. 
We also observe that the higher-twist corrections become large for 
small and large energy fraction $x$. The twist expansion breaks 
down in these regions. For large $x$ the situation is similar to DIS, 
but the behaviour at small $x$ has no analogue in DIS and will 
be discussed more below. Because the primary quark contribution 
is less ambiguous than the secondary quark contribution, we consider 
the model more reliable in the large $x$ region. However, for 
the longitudinal cross section it turns out that the small-$x$ 
region is not very different in the massive gluon model. 

The actual calculation requires the expansion of the 
distribution function that enters the dispersive 
representation (\ref{sumrep}) at small values 
of the dispersion variable $\xi$. For the secondary quark 
contribution to longitudinal fragmentation, one finds
\begin{equation}
T_L(\xi,x) \equiv \frac{1}{\sigma_0}\frac{d\sigma_L^{q,[s]}}{dx} 
=\frac{C_F\alpha_s}{2\pi}\cdot 2\cdot \left[
\frac{4}{x} -6 x+2 x^2+6 \ln x +\xi \ln \xi \,A_{2,L}^{q,[s]}(x) + 
O(\xi)\right].
\end{equation}
The coefficient $A_{2,L}^{q,[s]}(x)$ is the function that determines 
the shape of the $1/Q^2$ power correction and enters (\ref{par}). 
One then notes that
\begin{equation}
\label{forty}
2\cdot \Bigg[
\frac{4}{x} -6 x+2 x^2+6 \ln x\Bigg] = 2\cdot\frac{3}{N_f}\cdot
\left[C_L^g\ast P_{g\to q}\right](x),
\end{equation}
where $C_L^g(x)=4(1-x)/x$ is the gluon coefficient function at order 
$\alpha_s$ and $P_{g\to q}$ the gluon-to-quark splitting function. 
The asterisk denotes the convolution product. This suggests 
\cite{BBM97} that 
one can reinterpret the secondary quark contribution as a gluon 
contribution -- to be folded with the gluon fragmentation function -- 
by `deconvoluting' the gluon-to-quark splitting function. The 
power correction to the gluon contribution, 
$A^{g \leftarrow q}_{2,P}(x)$, is then defined through
\begin{equation}
\left[A^{g \leftarrow q}_{2,P}\ast P_{g\to q}\right](x) = 
A_{2,P}^{q,[s]}(x).
\end{equation}
The result can be compared with the result obtained from 
the finite gluon mass calculation \cite{DW97a}. Analysing the 
various ambiguities in restoring the gluon contributions, 
\cite{BBM97} suggested the following 
parametrization of twist-4 corrections: 
\begin{eqnarray}
\frac{d\sigma_L^{\rm tw-4}}{dx}(x,Q^2) &=& 
\frac{1 \mbox{GeV}^2}{Q^2}\int_x^1\frac{dz}{z}\,
\Bigg\{c_{q,L}\left[\delta(1-z)+\frac{2}{z}
\right]\,D_q(x/z,\mu) \nonumber\\
&&\,+ c_{g,L}\,\frac{1-z}{z^3}\,D_g(x/z,\mu)
\Bigg\}~,
\label{parL}
\\
\frac{d\sigma_{L+T}^{\rm tw-4}}{dx}(x,Q^2) &=& 
\frac{1 \mbox{GeV}^2}{Q^2}\int_x^1\frac{dz}{z}\,
\Bigg\{c_{q,L+T}\left[-\frac{2}{[1-z]_+} + 1 + 
\frac{1}{2}\delta'(1-z)\right]\,D_q(x/z,\mu) 
\nonumber\\
&&\,+ \left[c_{g,L+T}\,\frac{1-z}{z^3} + d\right]\,D_g(x/z,\mu)
\Bigg\}~,
\label{parLT}
\end{eqnarray} 
where $D_{i}$ denotes the leading-twist fragmentation function for parton 
$i$ to decay into any hadron, `$L$+$T$' the sum of longitudinal 
and transverse fragmentation cross sections and the plus distribution 
is defined as usual. The power corrections are added to the 
leading-twist cross sections as 
\begin{equation}
\frac{d\sigma_P}{dx}(x,Q^2) = \frac{d\sigma_P^{\rm tw-2}}{dx}(x,Q^2) + 
\frac{d\sigma_P^{\rm tw-4}}{dx}(x,Q^2).
\end{equation}
The constants $c_k$ and $d$ are to be fitted to data and depend 
on the order of perturbation theory and factorization scale $\mu$ 
adopted for the leading-twist prediction. The parametrization can be 
used only for $x>\Lambda/Q$, owing to strong singularities at small $x$. 
It is worth noting that the renormalon model predicts no $1/Q$ power 
corrections for the fragmentation functions at finite $x$. This is at 
variance with fragmentation models implemented in Monte Carlo 
simulations, which lead to $1/Q$ power corrections 
(see e.g. \cite{Web94b}), but consistent with \cite{BB91}.

Owing to energy conservation, the parton fragmentation functions 
disappear from the second moments
\begin{equation}
\label{moment2}
\sigma_P \equiv  \sum_H\frac{1}{2}\int_0^1\! 
dx \, x\frac{d\sigma_P^H}{dx},
\end{equation}
which can therefore be calculated in perturbation theory up to 
power corrections. (With this definition $\sigma_T+\sigma_L$ coincides 
with the total cross section $e^+ e^-\to\,\mbox{hadrons}$.) 
The power expansion of the fragmentation cross section 
has strong soft-gluon singularities and the expansion parameter 
relevant at small $x$ is $\Lambda^2/(Q^2 x^2)$. This can be related to 
the fact that in perturbation theory the hard scale relevant to 
gluon fragmentation is not $Q$, but the energy $Q x$ of the fragmenting 
gluon. \cite{DW97a,BBM97} noted that these strong singularities lead to a 
linear $\Lambda/Q$ correction to the second moment.\footnote{
A $\Lambda/Q$ correction to $\sigma_L$ was already reported in 
\cite{Web94}. However, the calculation there, which takes into account 
a gluon mass only in the phase space, is not complete.} This can be 
seen from 
\begin{equation}
\int\limits_{\Lambda/Q} dx\,\frac{1}{2}\,x\left[
\frac{\Lambda^2}{Q^2 x^2}
\right]^n \sim \frac{\Lambda}{Q}
\end{equation}
for any $n$,  
which also tells us that the correct $1/Q$ power correction is 
obtained only after resumming the power expansion at definite $x$ 
to all orders. The strong singularities at small $x$ occur only 
in the secondary quark (gluon) contribution. The result for the 
distribution function that enters (\ref{sumrep}) is 
\begin{equation}
T_L(\xi)\equiv \frac{\sigma_L}{\sigma_0} = \frac{\alpha_s}{\pi} 
\left[1-\frac{5\pi^3}{32}\,\sqrt{\xi} +\ldots\right],
\end{equation}
and, according to Section~\ref{dm}, the $\sqrt{\xi}$-term in the 
small-$\xi$ expansion indicates a $\Lambda/Q$ power 
correction.\footnote{If one evaluates the longitudinal cross section 
with a finite gluon mass, the coefficient of $\sqrt{\xi}$ 
is $2\pi^2/3$. We emphasize again that the finite gluon mass 
calculation cannot be related to renormalons for quantities 
like $\sigma_L$.} The total cross section in $e^+ e^-$ annihilation 
into hadrons is given by the sum of the transverse and longitudinal 
cross section. In $\sigma_L+\sigma_T$ all power corrections of order 
$1/Q^{1,2,3}$ cancel, compare Section~\ref{inclepem}. 

The sizeable linear 
power correction to the longitudinal (and transverse) 
cross section also leads 
to large perturbative corrections, comparable to those in other 
event shape observables. The perturbative corrections to $\sigma_L$ in the 
large-$\beta_0$ approximation can be found in \cite{BBM97}.

\cite{MW95} noted that hadronic event shape observables can have 
any power correction if one chooses an arbitrarily IR sensitive 
but IR finite weight on the phase space. The moments of 
fragmentation functions provide a simple example of a set of 
quantities that can have fractional power corrections \cite{BBM97}.
The leading power behaviour of 
\begin{equation}
\int\limits_0^1 dx\,\frac{1}{2}\,x^\gamma\,
\frac{1}{\sigma_0}\frac{d\sigma_{L,T}}{dx} 
\end{equation}
is corrected by terms of order $(\Lambda/Q)^\gamma$, where 
$\gamma$ can be arbitrarily small and positive. 
This should be compared with the 
moments of DIS structure functions, which can be described by the 
OPE, and which receive only $1/Q^2$ power 
corrections for any moment as long as the moment integral exists.

\cite{NW97} also considered heavy quark fragmentation in 
$e^+ e^-$ annihilation. Although secondary heavy quark fragmentation 
exists, it does not contribute to power corrections in $\Lambda/Q$ 
at leading order, because the gluon that splits into the heavy quark pair 
must have an invariant mass larger than $4 M^2\gg \Lambda^2$, 
where $M$ is the heavy quark mass. This eliminates the ambiguities 
for fragmentation into light hadrons mentioned above. \cite{NW97} 
find that the leading power correction is of order $\Lambda/M$. 
It can be interpreted as a power correction to the fragmentation 
function $Q\to H_Q$, which is perturbatively calculable at leading 
power. The existence of a linear power correction in $1/M$ to the 
heavy quark fragmentation function is consistent with the analysis 
based on heavy quark symmetry in \cite{JR94}. The leading power 
correction that depends on the centre-of-mass energy squared 
scales as $1/Q^2$ at finite energy fraction, consistent with what 
is found for light quarks. Note that there is an $M/Q$ power 
correction in the second moment, which comes from secondary 
heavy-quark fragmentation for the same reason as there is a $\Lambda/Q$ 
correction in case of massless quarks.

\subsubsection{Event shape observables in $e^+ e^-$ annihilation} 
\label{eventshapes}

Hadronic event shape variables in $e^+ e^-$ 
collisions can be used to measure the strong coupling, in 
particular as they are more sensitive to $\alpha_s$ than the 
total cross section. Event shape variables are computed 
theoretically in terms of quark and gluon momenta and measured 
in terms of hadron momenta. Apart from a correction for detector 
effects, the comparison of theory and data requires a correction 
for hadronization effects. It is believed that hadronization 
corrections are power suppressed in $\Lambda/Q$ (where $Q$ is the 
centre-of-mass energy) and it is known experimentally for quite 
some time that these corrections are substantial 
(see, for instance, \cite{Bar86} for an early review). Until recently, 
the traditional method to take them into account has been 
hadronization models, implemented in Monte Carlo programs 
that also simulate a parton shower. A hadronization correction  
that scales with energy as $\Lambda/Q$ provides a good description 
of the data. 

In this section we review recent developments that relate 
hadronization corrections to power corrections indicated by renormalons 
in the perturbative prediction for the event shape variable. This 
connection was suggested by \cite{MW95} for a toy model and by 
\cite{Web94} for some QCD observables, although within a simplified 
prescription that was refined later. These papers provided the 
first theoretical indications that hadronization corrections 
should scale (at least) as $\Lambda/Q$. Subsequent, more detailed 
analyses \cite{DW95,AZ95a,NS95} confirmed this conclusion. 
\cite{KS95a} also found $\Lambda/Q$ power corrections, 
potentially enhanced by inverse powers of the jet resolution 
parameter, to the 2-jet distribution in $e^+ e^-$ annihilation.

Below we consider the following set of event shape variables: 
the observable `thrust' is defined as 
\begin{equation}
\label{thr}
T = \max_{\,\vec{n}} \frac{\sum_i |\vec{p}_i\cdot\vec{n}\,|}{
\sum_i |\vec{p}_i|},
\end{equation}
where the sum is over all hadrons (partons) in the event. The thrust 
axis $\vec{n}_T$ is the direction at which the maximum is attained. 
An event is divided into two hemispheres $H_{1,2}$ by a plane 
orthogonal to the 
thrust axis. The heavier (lighter) of the two hemisphere invariant masses 
is called the heavy (light) jet mass $M_H$ ($M_L$). The jet broadening 
variables are defined through
\begin{equation}
B_k = \frac{\sum_{i\in H_k} |\vec{p}_i\times \vec{n}_T|}{2\sum_i 
|\vec{p}_i|}.
\end{equation}
In terms of these the total jet broadening is defined by 
$B_T=B_1+B_2$ and the wide jet broadening by 
$B_W=\max(B_1,B_2)$. 
Furthermore, from the eigenvalues of the tensor 
\begin{equation}
\frac{\sum_i (p_i^a p_i^b)/|\vec{p}_i|}{\sum_i |\vec{p}|} 
\end{equation}
the $C$-parameter $C=3 (\lambda_1\lambda_2+
\lambda_2\lambda_3+\lambda_3\lambda_1)$ is defined. 
All these event shape observables 
are IR safe, i.e.\ insensitive to the emission of soft or 
collinear partons at the logarithmic level. As a consequence they 
have perturbative expansions without IR divergences. 

It is relatively easy to understand that event shape observables 
are linearly sensitive to small parton momenta and are hence expected to 
receive long-distance contributions of order $\Lambda/Q$. For 
illustration we consider the average value of $1-T$ in somewhat more 
detail. At leading order, this quantity has no virtual correction, 
and we require only the matrix element for $\gamma^*\to q\bar{q} g$. 
We have seen in several instances before, that in the context of 
leading-order renormalon calculations, the gluon acquires an 
invariant mass squared, which we denote by $\xi Q^2$. To make the 
connection with hadronization, it is natural to think of this 
invariant mass as of that of a virtual gluon at the end 
of a parton cascade, before hadronization into a light hadron 
cluster with mass of order $\Lambda$ sets in. For a configuration where 
all momentum is taken by the $q\bar{q}$ pair and the virtual gluon 
is produced at rest, we have $1-T=\sqrt{\xi} \sim \Lambda/Q$, as 
compared to $1-T=0$ for the analogous configuration with a zero-energy 
massless gluon. In a more physical language, the production of a 
light hadron at rest changes the value of $1-T$ by an amount linear 
in the hadron mass over $Q$. 

For the purpose of illustration we follow \cite{Web94} and compute 
the average $\langle 1-T\rangle$ with a finite gluon mass 
$\sqrt{\xi} Q$, emphasizing 
however \cite{NS95,BB95b} that this is not equivalent to the 
computation of renormalon divergence, as the definition of thrust 
is not inclusive over gluon splitting $g\to q\bar{q}$ (see also 
Section \ref{dm} and Section \ref{fragmod} for a discussion of 
this point). The average of $1-T$ is given 
by 
\begin{equation}
\langle 1-T\rangle = \int\!\mbox{PS}[p_i]\,|{\cal M}_{q\bar{q} g}|^2\,
(1-T)[p_i].
\end{equation} 
Introducing the energy fractions $x_i=2 p_i\cdot q/q^2$, and reserving 
$x_3$ for the gluon energy fraction, we have
\begin{eqnarray}
|{\cal M}_{q\bar{q} g}|^2 &=& 
8 C_F N_c \,g_s^2\,\Bigg\{ \frac{x_1^2+x_2^2}{(1-x_1)\,(1-x_2)} + 
\xi \Bigg[\frac{2\,(x_1+x_2)}{(1-x_1)\,(1-x_2)} - \frac{1}{(1-x_1)^2}
\nonumber\\
&&
\hspace*{-1.2cm}
-\,\frac{1}{(1-x_2)^2}\Bigg]+\frac{2\xi^2}{(1-x_1)\,(1-x_2)}\Bigg\}
\,\longrightarrow \,
8 C_F N_c \,g_s^2\,\frac{2}{(1-x_1)\,(1-x_2)}.
\end{eqnarray}
For the leading correction of order $\sqrt{\xi}$, one may in fact 
set $\xi=0$ in the matrix element and $x_1=x_2=1$ in the non-singular 
terms, as done in the second line of the above expression. 
In terms of the energy fractions thrust is given by
\begin{equation}
\label{thrust1}
T = \frac{2}{2-x_3+\sqrt{x_3^2-4\xi}} \cdot \max(x_1,x_2),
\end{equation}
where we anticipated that $x_3$ is small in the region of interest. 
Note that the leading correction comes from $x_3$ of order 
$\sqrt{\xi}$ and hence $\xi$ 
cannot be dropped in this expression. The thrust variable can also be defined 
with $\sum_i |\vec{p}_i|\to Q$ in the denominator of (\ref{thr}). 
Then $T=\max(x_1,x_2)$ instead of (\ref{thrust1}). The two definitions 
agree to all orders in perturbation theory, but differ non-perturbatively 
by hadron mass effects. The phase space is 
\begin{equation}
\int\!\mbox{PS}[p_i] = \int\limits dx_1 dx_2\,\theta(x_1+x_2-(1-\xi))\,
\theta\left(\frac{1-x_2-\xi}{1-x_2}-x_2\right).
\end{equation}
We then find \cite{BB95b}
\begin{equation}
\langle 1-T\rangle = \frac{C_F \alpha_s}{\pi} 
\left(0.788 - 7.32\sqrt{\xi} +\ldots\right).
\end{equation}
If we use the alternative definition of thrust mentioned above, 
the coefficient 7.32 is replaced by 4. This value has 
been adopted in phenomenological studies initiated by 
\cite{Web94,DW95,AZ95a}. The difference constitutes an ambiguity due 
to the simplified gluon mass prescription. One may wonder how 
$\sqrt{\xi}$ enters the answer, because the phase space boundaries 
do not contain a square root of $\xi$. If we change one of the 
integration variables 
to $x_3$, we find that $x_3>2 \sqrt{\xi}$ and the linear power 
correction can be seen to arise from the fact that the integral 
over gluon energy fraction is $\int \!dx_3$ and restricted as 
indicated. The pattern of gluon radiation leads to energy integrals 
$\int\! d x_3/x_3$. IR finiteness implies that the phase space weight, 
here $1-T$, is constructed so as to eliminate the logarithmic 
divergence as $x_3\to 0$. The generic situation with event shapes 
is a linear suppression of soft gluons.

An important conclusion is that in the approximation considered 
so far the $\Lambda/Q$ power correction arises neither from 
the emission of collinear but energetic partons nor from soft quarks, 
but only from soft gluons. This is consistent with the 
analysis of fragmentation in Section~\ref{fragmod}, where the 
leading $1/Q$ power correction to the longitudinal cross section 
was seen to originate only from soft gluon fragmentation. As a consequence 
we obtain the qualitative prediction
\begin{equation}
\frac{\langle 1-T\rangle_{|1/Q, T<T_0}}{\langle 1-T\rangle_{|1/Q}} = 
\mbox{const} \times \alpha_s(Q) \qquad [\,\mbox{exp:} \,\,0.54\pm 0.16]\,.
\end{equation}
In the numerator the 2-jet region $T\approx 1$ is excluded. 
Hence a hard gluon has to be emitted, which causes an additional 
suppression in $\alpha_s(Q)$. The number in brackets 
quoted from \cite{Del97} shows some suppression, although not 
as large as expected. A slightly smaller number is obtained in \cite{Wic98}. 
However, the constant that multiplies 
$\alpha_s$ has not been estimated theoretically, and details of 
the experimental fit procedure, for which the reader should 
consult \cite{Del97}, constitute an important source of uncertainty. 
Because in $\langle (1-T)^2\rangle$ the soft gluon region is 
suppressed by two powers of $x_3$, one also expects the $1/Q$ power 
correction to this quantity to be suppressed by one power of $\alpha_s(Q)$. 
In particular, one obtains only a $1/Q^2$ power correction from the 
one gluon emission process discussed above. In both cases, however, this 
does not imply that the hadronization correction {\em relative} 
to the perturbative correction is small, because the perturbative 
coefficients at order $\alpha_s$ are also reduced in 
$\langle 1-T\rangle_{1/Q,T<T_0}$ and $\langle (1-T)^2\rangle$ 
relative to $\langle 1-T\rangle$. A recent analysis of experimental 
data at various centre-of-mass energies \cite{Wic98} reports 
that the power correction to the second moment $\langle (1-T)^2\rangle$ 
is consistent with a $1/Q^2$ behaviour. For the third moment a 
$1/Q^3$ behaviour is found, which is surprising, because 
for all $\langle (1-T)^n\rangle$ with $n\geq 2$ one expects a $1/Q^2$ 
behaviour. No matter how strong the suppression of the soft gluons, 
there should be a $1/Q^2$ power correction from hard collinear 
partons.

\cite{DW95} and \cite{AZ95a} (DWAZ) (see also \linebreak[4]
\cite{KS95b}) 
suggested that the leading power correction 
to average event shape observables may be described by a single 
(`universal') parameter multiplied by an observable-dependent, but 
calculable, coefficient. For an event shape $S$, defined such that 
its average is of order $\alpha_s$, we can write 
\begin{equation}
\label{sdwaz}
\langle S \rangle = A_S \alpha_s(\mu) + \left[B_S-A_S\beta_0
\ln\frac{\mu^2}{Q^2} \right]\alpha_s(\mu)^2 +\ldots + 
\frac{K_S(\mu)}{Q} + O(1/Q^2),
\end{equation}
see also the introductory discussion in Section~\ref{modelssect}. 
\cite{DW95} para\-met\-rize the coefficient of the power correction 
in the form 
\begin{equation}
\label{ksmu}
K_S(\mu) = \frac{4 C_F c_S}{\pi} \mu_I\left[\bar{\alpha}_0(\mu_I)-
\alpha_s(\mu)-\left(-\beta_0\ln\frac{\mu^2}{\mu_I^2} + \frac{K}{2\pi} -
2\beta_0\right)\alpha_s(\mu)^2\right],
\end{equation}
where $\mu_I$ is an IR subtraction scale (typically chosen to be $2\,$GeV), 
$\bar{\alpha}_0(\mu_I)$ is the non\--per\-tur\-ba\-tive parameter to be 
fitted and $K=(67/18-\pi^2/6) C_A-5 N_f/9$. 
The remaining terms approximately subtract the IR contributions 
contained in the perturbative coefficients $A$ and $B$ up to second order. 
The universality assumption can be tested by fitting the value of 
$\bar{\alpha}_0(\mu_I)$ or, equivalently, $K_S(\mu)$ to different 
event shape variables.  

\begin{figure}[p]
   \vspace{-2cm}
   \epsfysize=13cm
   \epsfxsize=8.6cm
   \centerline{\epsffile{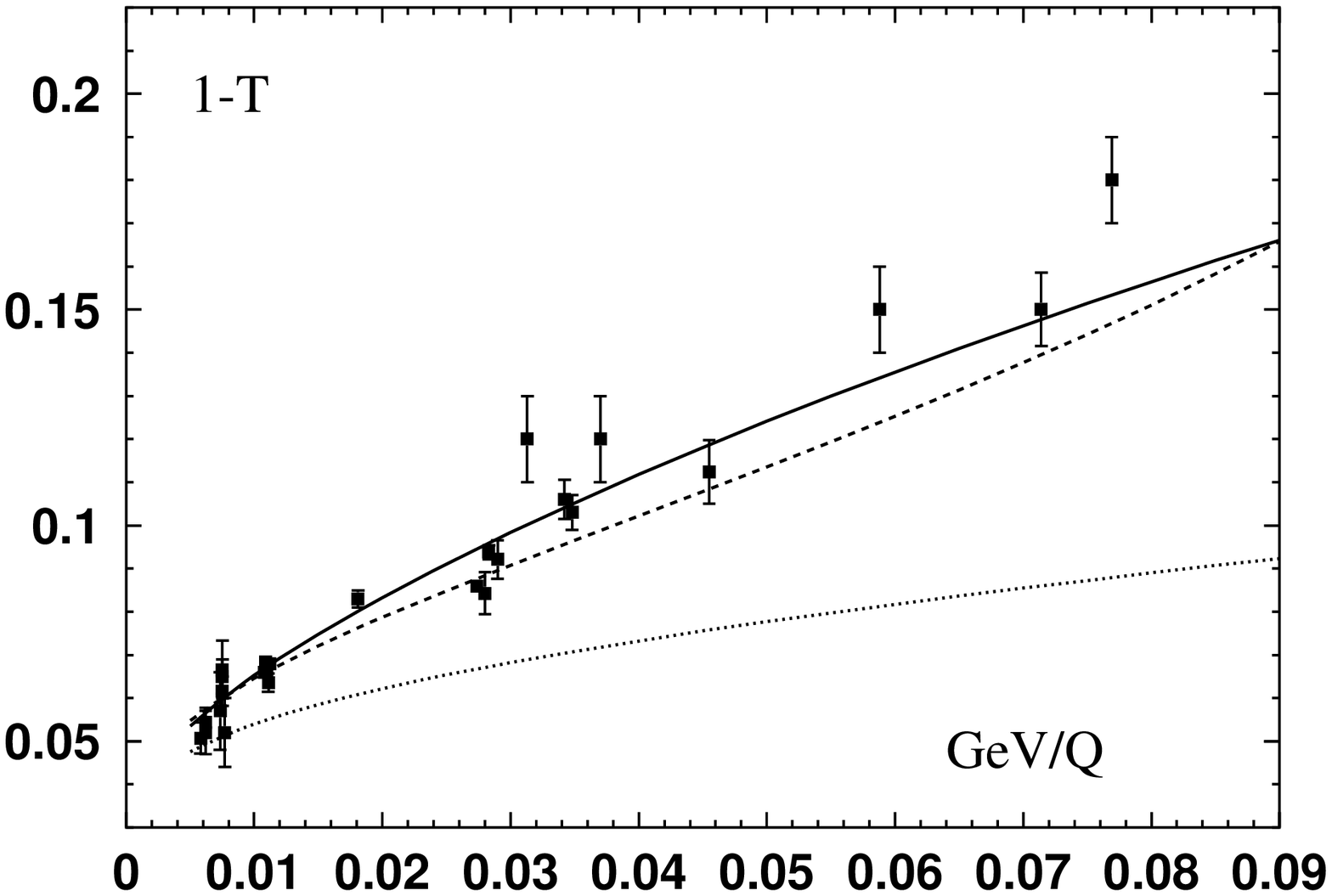}}
   \vspace*{-4.5cm}
   \epsfysize=13cm
   \epsfxsize=8.6cm
   \centerline{\epsffile{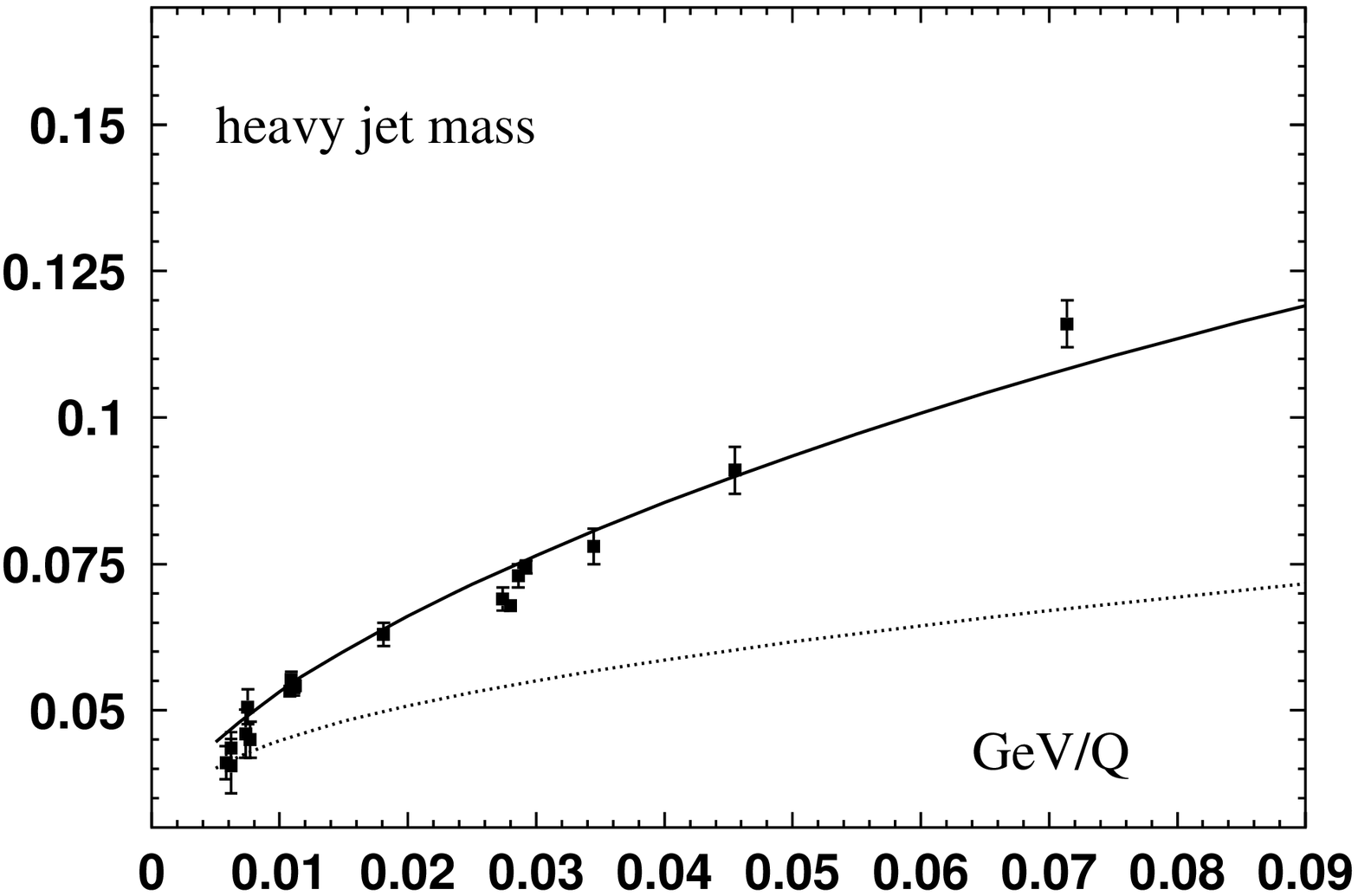}}
   \vspace*{-2.7cm}
\caption[dummy]{\small Energy dependence of $\langle 1-T\rangle$ (upper) and 
the heavy mass $\langle M_H^2/Q^2\rangle$ (lower) plotted as function 
of $1/Q$. Data compilation from 
\cite{Jad98}, see references there. Dotted line: second order perturbation 
theory with scale $\mu=Q$. Solid line: second order perturbation theory 
with power correction added according to (\ref{ksmu}) and with 
$\mu=Q$, $\mu_I=2\,$GeV. For $\bar{\alpha}_0(2\,\mbox{GeV})$ the fit values 
0.543 for thrust and 0.457 for the heavy jet mass from \cite{Jad98} 
are taken. The dashed line shows second order perturbation theory at the very 
low scale $0.07 Q$ with no power correction added. For both 
observables $\alpha_s(M_Z)$ has been fixed to 0.12. I thank O.~Biebel for 
providing me with the data points. \label{fig20}}
\end{figure}
Extensive analyses of the energy dependence of event shape variables 
and power corrections to them have been carried out by \cite{Del97} 
and members of the (former) JADE collaboration \cite{Jad98}. 
In Fig.~\ref{fig20} we compare the energy dependence of 
$\langle 1-T\rangle$ and $\langle M_H^2/Q^2\rangle$ with the prediction 
based on second order perturbation theory with and without 
a $1/Q$ power correction. It is clearly seen that (a) the second order 
perturbative result with scale $\mu=Q$ is far too small and 
(b) the difference with the data points is fitted well by a
$1/Q$ power correction. In addition to the two quantities reproduced 
here, the energy dependence of three jet fractions, the difference 
jet mass and the integrated energy-energy correlation 
can be found in \cite{Del97}. The jet broadening variables are analysed 
in \cite{Jad98}. In Table~\ref{tab9} we reproduce the fitted values 
of $\bar{\alpha}_0$ for some of these variables. For the central values 
of $\bar{\alpha}_0$ shown in the Table  
the coefficients $c_S$ in (\ref{ksmu}) are taken to be 
$c_{1-T}=1$, $c_{M_H^2/s} =1$, $c_{B_{T,W}}=1$ \cite{DW95,Web95}. 
The theoretical status of these coefficients is somewhat controversial, 
as we discuss below. Nevertheless, the measurements indicate that 
the parameter for $1/Q$ power corrections is not too different for 
the set of event shapes analysed so far. The jet broadening observables 
are special, because one expects an enhanced $(\ln Q)/Q$ power 
correction \cite{DW98,DLMS98}, which has not been taken into 
account in the experimental fits.\footnote{In their second publication 
\cite{Jad98b} performed fits to the jet broadening measures, taking 
into account the logarithmic enhancement. We refer the reader to this work, 
but do not quote their numbers in the table, since they adopt a 
normalization of the power correction different from (\ref{ksmu}), 
following \cite{DLMS98}.}
\begin{table}[t]
\addtolength{\arraycolsep}{0.2cm}
\renewcommand{\arraystretch}{1.5}
$$
\begin{array}{l|c|c}
\hline\hline
\hspace*{1.5cm}S & \bar{\alpha}_0(2\,{\rm GeV}) & \alpha_s(M_Z) \\ 
\hline
\langle 1-T\rangle \,\,\mbox{[DELPHI]} 
& 0.534\pm 0.012 &  0.118\pm 0.002 \\
\langle 1-T\rangle \,\,\mbox{[JADE]} 
& 0.543^{+0.015}_{-0.014} &  0.120^{+0.007}_{-0.006} \\
\langle M_H^2/s\rangle \,\,\mbox{[DELPHI]} 
& 0.435\pm 0.015 &  0.114\pm 0.002 \\
\langle M_H^2/s\rangle \,\,\mbox{[JADE]} 
& 0.457^{+0.212}_{-0.077} &  0.112^{+0.005}_{-0.004} \\
\langle B_T \rangle \,\,\mbox{[JADE]} 
& 0.342^{+0.064}_{-0.038} &  0.116^{+0.010}_{-0.008} \\
\langle B_W \rangle \,\,\mbox{[JADE]} 
& 0.264^{+0.048}_{-0.031} &  0.111^{+0.009}_{-0.007} \\
\hline\hline
\end{array}
$$
\caption[dummy]{\label{tab9}\small 
Fits of $\alpha_s(M_Z)$ and the power correction 
parameter $ \bar{\alpha}_0(2\,{\rm GeV})$ 
defined in (\ref{ksmu}) taken from \cite{Del97} and \cite{Jad98}. See there 
for details of the error breakdown. DELPHI does not include the LEP2 
data points.}
\end{table}

In absolute terms the power correction added to thrust and the heavy jet 
mass is about $1\,\mbox{GeV}/Q$. This is a sizeable correction of order 
$20\%$ even at the scale $M_Z$, because the perturbative contribution is 
of order $\alpha_s(M_Z)/\pi$. The fit for $\bar{\alpha}_0$ is sensitive 
to the choice of renormalization scale $\mu$ and in general to the 
treatment of higher order perturbative corrections. There is nothing 
wrong with this, because the very spirit of the renormalon approach is 
that perturbative corrections and non-perturbative hadronization 
corrections are to some extent inseparable. Hence we find it plausible 
that the $1/Q$ power correction accounts in part for large higher order 
perturbative corrections, which are large precisely because they receive 
large contributions from IR regions of parton momenta. It was noted 
in \cite{BB96} that choosing a small scale, $\mu=0.13Q$, reduces the 
second order perturbative contribution and power correction 
significantly for $\langle 1-T\rangle$. In Fig.~\ref{fig20} (dashed 
curve) we have taken a very low scale, $\mu=0.07Q$, to illustrate the 
fact that the running of the coupling at this low scale can fake 
a $1/Q$ correction rather precisely (a straight line in the figure). 
\cite{CGM98} performed an analysis of $\langle 1-T\rangle$ in the 
effective-charge scheme. This scheme selects the scale 
$\mu=0.08 Q$. \cite{CGM98} fit $\alpha_s$, a third-order perturbative 
coefficient and a $1/Q$ power correction simultaneously and find a 
reduced power correction of order $(0.3\pm 0.1)\,\mbox{GeV}/Q$ 
consistent with \cite{BB96}. 

The DWAZ model relies on the assumption of 
universality of power corrections, 
i.e. the assumption that all non-perturbative effects 
can be parametrized by one number. Different motivations  
for this assumption 
have been given in \cite{DW95}, in \linebreak[4] \cite{AZ95a}, and 
in \cite{KS95b}. 
The nature of this assumption has not been completely elucidated so far. 
In the formulation of the model of \cite{DW95} 
the $1/Q$ power correction 
to $\langle 1-T\rangle$ and $M_H^2/Q^2$ are predicted to be 
equal, but the power correction to the light jet mass  $M_L^2/Q^2$ 
is predicted to be suppressed by a factor of $\alpha_s(Q)$. 
\cite{AZ95a} argued that, in the two-jet limit, a universal 
hadronization correction is associated with each quark jet and 
hence the $1/Q$ power correction 
to $\langle 1-T\rangle$ should be twice as large as that to $M_H^2/Q^2$, 
while the $1/Q$ power correction to $M_L^2/Q^2$ should be 
as large as that to $M_H^2/Q^2$. The data reported above 
appears to favour near-equality for $\langle 1-T\rangle$ and 
$M_H^2/Q^2$. On the other hand, the very small value of the $1/Q$ 
term for the difference mass $M_d^2=M_H^2-M_L^2$ observed in 
\cite{Del97} seems to 
favour the picture of \cite{AZ95a}.\footnote{Recently, 
\cite{DLMS98} introduced a distinction of `single-jet' and 
`whole-event' properties, which revises the original formulation 
of \cite{DW95} towards the formulation of \cite{AZ95a} as far as 
thrust and jet masses are concerned. This distinction has to 
be carefully taken note of when one compares for example the fits 
of $\bar{\alpha}_0(\mu_I)$ in \cite{Jad98} with those in 
\cite{Jad98b}.}

\cite{NS95} considered the effect of gluon splitting $g\to q\bar{q}$ on 
power corrections to various event shape observables and argued 
that neither of the two answers is correct and that universality 
in the sense of the DWAZ model is unlikely to hold. They observe 
that thrust and the heavy jet mass are related by $1-T=M_H^2/Q^2$, if, 
in the two-jet limit, a soft gluon splits into two collinear quarks, 
both of which go into the same hemisphere; however, the relation 
is $1-T=2 M_H^2/Q^2$ if the quarks are emitted from the gluon 
back-to-back. As a consequence $1-T$ and $M_H^2/Q^2$ provide 
different weights on the four-parton phase space and the coefficients 
of their linearly IR sensitive contributions are not related in a 
simple way. \cite{BB95b} arrived at a similar conclusion, noting 
that event shapes resolve large angle soft gluon emission at the 
level of $1/Q$ power corrections. If collinearity of the emission 
process is not required, the association of the power correction 
to a particular jet is difficult to maintain.

The situation can be clarified either by finding an explicit 
operator parametrization of the $1/Q$ IR sensitive contribution valid to 
all orders in perturbation theory, or by explicit next-to-leading 
order calculations that take into account the emission of two gluons.

The first approach was taken by \cite{KOS97}, extending earlier 
work on jet distributions \cite{KS95a} to averaged event 
shapes. Let us define the operator 
\begin{equation}
\label{eflow}
{\cal P}(\hat{y}) = \lim_{|\vec{y}|\to\infty} 
\int\limits_0^\infty\frac{d y_0}{(2\pi)^2}\,|\vec{y}\,|^2 \hat{y}_i 
\Theta_{0i}(y^\mu),
\end{equation}
with $\Theta_{\mu\nu}$ the energy momentum tensor and 
$\hat{y}$ a unit vector, as the measure of momentum (energy) of  
soft partons (hadrons) deposited at asymptotic distances 
(for instance, in the calorimeter of the detector) in the direction 
of $\hat{y}$. Close to the two-jet limit, the soft partons are 
emitted from a pair of almost back-to-back quarks. For event shape 
weights that have (at least) a linear suppression of soft particles 
the standard eikonal approximation can be 
used for the fast quark propagators 
and the quark propagation can be described by a product of Wilson line 
operators $W_{v_1 v_2}$ with $v_1$ and $v_2$ light-like vectors 
pointing in the direction of the outgoing fast quarks. 
Squaring the matrix elements, the energy flow of soft radiation from 
the $q\bar{q}$ system is described by the distribution  
\begin{equation}
\label{es}
{\cal E}(\hat{y}) = \langle 0|W^\dagger_{v_1 v_2}\,{\cal P}(\hat{y})\,
W_{v_1 v_2}|0\rangle.
\end{equation}
In terms of these quantities, \cite{KOS97} find
\begin{equation}
\label{kos}
\langle S\rangle_{|1/Q} = \frac{1}{Q}\int\frac{d\Omega(\hat{y})}{2\pi}\,
f_S(\Omega(\hat{y}))\,{\cal E}(\hat{y}),
\end{equation}
where the integral extends over the full solid angle. The integral has 
a transparent interpretation as an observable-dependent (and calculable) 
weight of the non-perturbative energy flow distribution 
${\cal E}(\hat{y})$. There are corrections to this result from 
multi-jet configurations. These corrections are suppressed by factors 
of $\alpha_s(Q)$. Note that (\ref{kos}) embodies universality in terms 
of a universal distribution function ${\cal E}(\hat{y})$. But since 
every event shape takes a different integral of ${\cal E}(\hat{y})$, 
their $1/Q$ corrections are not related through the same non-perturbative 
parameter. The DWAZ model can be recovered, when ${\cal E}(\hat{y})$ is 
approximated by a constant. Operators similar to (\ref{eflow}) were 
also introduced by \cite{ST96,CS97}. They stress that 
event shape variables in general are most naturally defined in terms 
of calorimetric energy-momentum flow (rather than the energy-momentum of 
particles) and note that such a definition would lend 
itself more easily to an analysis of power corrections.

The second approach was followed by \cite{DLMS97,DLMS98} who presented 
a detailed analysis of IR sensitive contributions to the matrix 
elements for the emission of two partons. For event shape observables 
with a linear suppression of soft partons, the matrix elements 
can be evaluated in the soft approximation. \cite{DLMS98} find that 
for $\langle 1-T\rangle$, the jet masses, and the $C$-parameter the 
coefficient of the $1/Q$ power correction that is obtained for one gluon 
emission is rescaled by the {\em same} factor 1.8. This implies 
that these observables take the same section of the 
distribution function (\ref{es}) to leading and next-to-leading order. 
This conclusion follows from the fact that \cite{DLMS98} assume that 
the nearly back-to-back quark jets acquire an invariant mass 
that is large compared 
to $\Lambda$ (but small compared to $Q$) as a consequence of 
{\em perturbative} soft gluon radiation. In this case a soft gluon 
with energy of order $\Lambda$, which is of interest for power 
corrections, cannot determine which hemisphere becomes heavy and 
which becomes light.

It is important that the correction factor 1.8 has no parametric 
suppression, because the coupling constant in diagrams with soft 
gluon emission with momenta of order $\Lambda$ should be considered 
of order 1. In renormalon terminology this is related to the fact that 
the over-all normalization of renormalon divergence receives 
contributions from arbitrarily complicated diagrams. As a consequence 
one can expect further unsuppressed rescalings, not necessarily 
equal for the event shapes mentioned above, in still higher orders. 
\cite{DLMS97,DLMS98} argue that there are no corrections to the 
rescaling factor 1.8 from the emission of three and more partons. 
This is due to the fact that they parametrize the non-perturbative 
parameter for $1/Q$ power corrections as an integral of 
an effective coupling $\alpha_{\rm eff}$ \cite{DMW96}. In this 
language more complicated diagrams would necessitate the introduction 
of integrals of $\alpha_{\rm eff}^n$ and hence, new non-perturbative 
parameters. Since from general considerations these parameters 
cannot be expected  
to be small, these parameters presumably violate 
the simple universality hypothesis in terms of a single 
non-perturbative parameter. 

One can also consider power corrections to event shape distributions, 
rather than averaged event shapes \cite{KS95b,DW97}. Recall that at 
leading order in $\alpha_s$ the thrust distribution is 
\begin{equation}
\frac{d\sigma}{dT} = \delta(1-T).
\end{equation}
It is not difficult to see that in the approximation of one-gluon 
emission discussed earlier, the $1/Q$ power correction to the $N$th 
moment of thrust is given by
\begin{equation}
\langle T^N\rangle_{|1/Q} = -N \langle 1-T\rangle_{|1/Q} 
\equiv N\,\frac{a_T \Lambda}{Q} \quad (a_T>0),
\end{equation}
which implies
\begin{equation}
\frac{d\sigma}{dT} = \delta(1-T) + \frac{a_T\Lambda}{Q}\,
\delta^\prime(1-T) + \ldots.
\end{equation}
It is suggestive but not rigorous to interpret the correction as the 
first term in the expansion of 
\begin{equation}
\label{shift}
\delta\left(1-\left[T-\frac{a_T \Lambda}{Q}\right]\right),
\vspace*{0.2cm}
\end{equation}
so that the main effect results in a non-perturbative shift of the 
thrust value. Qualitatively, such an effect is expected on purely 
kinematic grounds from hadron mass effects. In writing 
(\ref{shift}) we have to assume that the power correction of order 
$N\Lambda/Q$ exponentiates exactly in moment space 
\cite{KS95b,DW97}. Whether exponentiation occurs in this sense 
has not yet been established. In a more general framework one would 
introduce a non-perturbative distribution function that resums the 
power corrections of order $(N\Lambda/Q)^k$ and write the 
thrust distribution as a convolution of its perturbative distribution 
with this distribution function. This is analogous to the 
introduction of shape functions in the heavy quark effective 
theory to describe 
the endpoint regions of certain energy spectra 
\cite{Neu94,BSUV94a}. The $k$th moment of this distribution function 
is related to the coefficient of $(N\Lambda/Q)^k$, which 
need not, however, be related to $a_T$. Such a distribution function 
would not be universal, i.e.\ it would be different for 
different event shapes. 

\cite{DW97} assume a distribution function of the form (\ref{shift}) and 
arrive at 
\begin{equation}
\label{ansatz}
\frac{d\sigma}{dT} = F_{\rm pert}(T-\delta T),
\end{equation}
where $F_{\rm pert}(T)$ denotes the perturbative thrust distribution and 
$\delta T$ a non-perturbative shift of order $\Lambda/Q$. They find that 
the data on thrust distributions at various energies are well 
described by the ansatz (\ref{ansatz}) down to rather small values 
of $1-T$ (see also \cite{Wic97}). 
The $C$-parameter distribution has also been successfully 
fitted with this parametrization \cite{CW98}. 

A non-perturbative distribution function in analogy with heavy quark 
decays as described above has been introduced by \cite{Kor98}, to which 
we refer for more details on the factorization of perturbative 
contributions and the evolution equations for the moments of the 
distribution function. Just like average event shapes, the distribution 
function can also be expressed in terms of the universal distribution 
${\cal E}(\hat{y})$. But again a complicated weight is taken, which 
forbids a straightforward relation of different event shape variables. 
\cite{Kor98} uses a simple three-parameter ansatz for the distribution 
function and obtains excellent agreement between the predicted and 
measured thrust distributions at all centre-of-mass energies between 
$14\,$GeV and $162\,$GeV.

\subsubsection{Event shape observables in deep-in\-ela\-stic scattering} 
\label{disevent}

Event shape variables can also be measured in  
DIS. Compared to $e^+ e^-$ annihilation, DIS offers the advantage that 
an entire range of $Q^2$ can be covered in a single experiment. Event 
shape variables in DIS are usually defined in the Breit frame, where the 
gauge boson momentum that induces the hard scattering process is 
purely space-like: $q=(0,0,0,Q)$. In leading order the target 
remnant moves into the direction opposite to $q$ and the struck parton 
moves into the direction of the virtual gauge boson. This direction 
defines the `current hemisphere', which is in many ways 
similar to one hemisphere in $e^+ e^-$ collisions. DIS event shape 
variables are then defined in close analogy to those for 
$e^+ e^-$ annihilation, but with the sum over hadrons (partons) 
restricted to the current hemisphere. 
\begin{figure}[t]
   \vspace{-3.5cm}
   \epsfysize=15cm
   \epsfxsize=11cm
   \centerline{\epsffile{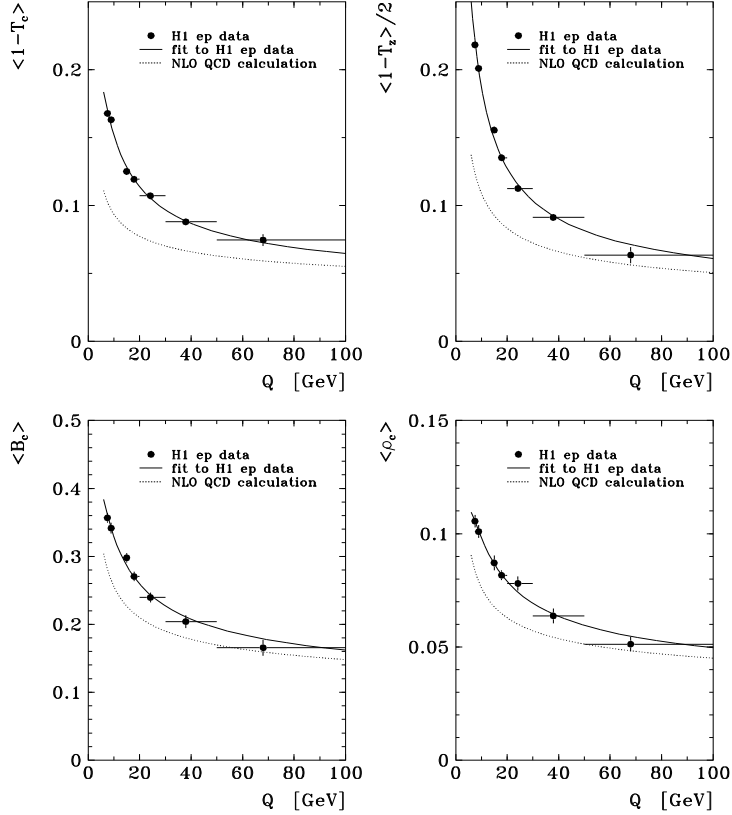}}
   \vspace*{0.2cm}
\caption[dummy]{\small Energy dependence of $\langle 1-T_c\rangle$, 
$\langle 1-T_z\rangle/2$, the current jet broadening $\langle B_c\rangle$ 
and the current jet hemisphere invariant mass $\langle \rho_c\rangle$ 
in DIS compared to NLO perturbation theory with and without $1/Q$ 
power correction. Figure taken from  \cite{H197}. \label{fig21}}
\end{figure}

As for event shape variables in $e^+ e^-$ annihilation,  
$\Lambda/Q$ power corrections are expected for 
their DIS analogues. \cite{DW98} 
computed the coefficient using the finite gluon mass prescription 
for the one gluon emission diagrams. The predicted event shape average 
is then represented in the form (\ref{sdwaz}, \ref{ksmu}).  The H1 
collaboration \cite{H197} compared the prediction to their data 
over a range of momentum transfers $Q$ from $7\,$GeV to $100\,$GeV. 
Their fit to the energy dependence using the parametrization (\ref{sdwaz}) 
is shown in Fig.~\ref{fig21} and the corresponding values of 
$\bar{\alpha}_0(2\,\mbox{GeV})$ are reproduced in Table~\ref{tab10}.
\begin{table}[t]
\addtolength{\arraycolsep}{0.2cm}
\renewcommand{\arraystretch}{1.5}
$$
\begin{array}{l|c|c}
\hline\hline
\hspace*{1cm}S & \bar{\alpha}_0(2\,{\rm GeV}) & \alpha_s(M_Z) \\ 
\hline
\langle 1-T_c\rangle  
& 0.50^{+0.07}_{-0.04} &  0.123^{+0.007}_{-0.005}\\
\langle 1-T_z\rangle/2 
& 0.51^{+0.11}_{-0.05} &  0.115^{+0.007}_{-0.005} \\
\langle \rho_c\rangle 
& 0.52^{+0.03}_{-0.02} &  0.130^{+0.007}_{-0.006} \\
\langle B_c \rangle
& 0.41^{+0.04}_{-0.02} &  0.119^{+0.007}_{-0.005} \\
\hline\hline
\end{array}
$$
\caption[dummy]{\label{tab10}\small 
Fits of $\alpha_s(M_Z)$ and the power correction 
parameter $ \bar{\alpha}_0(2\,{\rm GeV})$ (defined in (\ref{ksmu}))
to DIS event shape variables taken from \cite{H197}. See there for 
definitions of the quantities listed. The error is almost entirely 
theoretical.}
\end{table}

It is remarkable that $\bar{\alpha}_0(2\,{\rm GeV})$, the parameter 
for the $1/Q$ power correction, comes out nearly identical for the 
four event shapes shown in Fig.~\ref{fig21}, and, 
moreover, that its value is the same within 
errors as for event shapes in $e^+ e^-$ annihilation. This supports 
the idea that hadronization of the current jet in DIS is similar to 
hadronization in one hemisphere in $e^+ e^-$ annihilation. From a 
theoretical point of view the universality between DIS and $e^+ e^-$ 
annihilation is not obvious, because the factorization of the 
remnant and the current jet cannot be expected beyond leading 
power in $\Lambda/Q$, since soft gluons can connect the two. 

It is important to note that the data teach theorists an interesting 
fact, but that the numerical agreement for 
$\bar{\alpha}_0(2\,{\rm GeV})$ cannot be considered as significant 
to the accuracy at which it appears. For tests of universality it would 
be more useful to fit all event shapes with a common value for 
$\alpha_s(M_Z)$. The fact that the fitted $\alpha_s(M_Z)$ is different 
for the observables in Table~\ref{tab10}  
introduces a systematic uncertainty in 
$\bar{\alpha}_0(2\,{\rm GeV})$. In addition, the parameter that enters 
the prediction is $c_S \bar{\alpha}_0(2\,{\rm GeV})$. The value 
for $\bar{\alpha}_0(2\,{\rm GeV})$ follows once $c_S$ is computed 
in a particular prescription. The ambiguities in theoretical 
calculations of $c_S$ are large and the fact that the gluon mass 
prescription gives consistent results may also be an interesting 
coincidence. 

\cite{DSW98} have extended their calculation for event shapes in 
DIS to fragmentation processes in DIS. Data on the energy fraction 
dependence of power corrections to fragmentation functions 
would be highly interesting, as the same effects as discussed in 
Section~\ref{fragmod} for fragmentation in 
$e^+ e^-$ collisions are expected to occur in DIS. At the same time,  
an entire range in $Q^2$ can be scanned in $e p$ collisions. 
So far the theoretical calculation has been done only for 
quark-initiated DIS. At energies of the HERA collider one 
expects a large contribution from gluon-initiated DIS. In the 
leading-order renormalon model the gluon contribution can be 
reconstructed from quark-singlet contributions by the deconvolution 
method \cite{BBM97} discussed in section~\ref{fragmod}.

\subsubsection{Drell-Yan production and soft gluon resummation}
\label{dy}

We have considered power corrections to hadronic 
final states in $e^+ e^-$ annihilation and to DIS. Renormalon divergence 
also appears in the hard scattering coefficients in hadron-hadron 
collisions. The simplest hadron-hadron hard scattering process is 
Drell-Yan production of a lepton pair or a massive vector boson,
$A+B\to \{\gamma^*,W,Z\}(Q) + X$, where $X$ is any hadronic final state. 
At leading power $d\sigma/dQ^2 = \sigma_0 W(\tau,Q^2)$, where 
$\sigma_0$ is the Born cross section, $\tau=Q^2/s$, and 
\begin{equation}
\label{factformula}
W(\tau,Q^2) = \sum_{i,j} \int\limits_0^1\frac{\mbox{d}x_i}{x_i}\,
\frac{\mbox{d} x_j}{x_j} f_{i/A}(x_i,Q^2) f_{j/B}(x_j,Q^2)\,
\omega_{ij}(z,\alpha_s(Q)),
\end{equation}
with $z=Q^2/(x_1 x_2 s)$ and $s$ is the centre-of-mass energy squared of 
$A$ and $B$. In the following we are concerned with 
renormalon divergence and long-distance contributions to the 
hard scattering factor $\omega_{ij}(z,\alpha_s(Q))$. It is convenient 
to work in moment space, in which 
\begin{equation} 
\label{moments}
W(N,Q^2) \equiv \intl_0^1 d\tau\,\tau^{N-1}\,W(\tau,Q^2) 
= f_{q/A}(N,Q^2) f_{\bar{q}/B}(N,Q^2)\, 
\omega_{q\bar{q}}(N,\alpha_s(Q)),
\end{equation}
where the right hand side is expressed in terms of 
moments of the parton distributions (hard scattering factor) with 
respect to $x_i$ ($z$).

When $Q$ is large, one can consider large moments $1\ll N\ll Q/\Lambda$. 
Conventional, fixed-order perturbation theory fails for high moments, 
because one encounters corrections $\alpha_s^n \ln^m N$ with $m$ up 
to $2 n$. The physical origin of these corrections is that there exist 
three scales $Q$, $Q/\sqrt{N}$ and $Q/N$ and the logarithms are ratios of 
these scales. These scales appear because, for large $N$, the moment 
integral is dominated by $Q^2\sim s$, which leaves little phase space 
for the hadronic system $X$. In a perturbative calculation,  
the energy available 
for real emission is constrained to be of order $Q/N$ 
and the IR cancellation between virtual and real correction becomes 
numerically ineffective. 

The logarithmically enhanced contributions can be resummed systematically 
to all orders in perturbation theory \cite{Ste87,CT89}. The result 
has the exponentiated form\footnote{In the remainder of this 
section we restrict attention to the $q\bar{q}$ annihilation 
subprocess.}
\begin{equation}
\label{omega} 
\omega_{q\bar{q}}(N,\alpha_s(Q)) = H(\alpha_s(Q))\, \exp\left[
E(N,\alpha_s(Q))\right] + R(N,\alpha_s(Q)),
\end{equation}
where $R(N,\alpha_s(Q))$ vanishes as $N\to \infty$, 
$H(\alpha_s(Q))$ is independent of $N$, and the 
exponent is given by
\begin{eqnarray} 
\label{exponent} 
E(N,\alpha_s(Q)) &=&\intl_0^1 d z\,\frac{z^{N-1}-1}{1-z}\,
\Bigg\{2\!\!\intl_{Q^2 (1-z)}^{
Q^2 (1-z)^2} \!\frac{d k_t^2}{k_t^2}\,A(\alpha_s(k_t))+
B(\alpha_s(\sqrt{1-z} Q))
\nonumber\\[-0.2cm]
&&\hspace*{-1.5cm}+\,C(\alpha_s((1-z) Q))\Bigg\}.
\end{eqnarray}
The function $A$ is related to soft-collinear radiation and 
also referred to as `cusp' or `eikonal' anomalous dimension. The 
function $B$ relates to the DIS process, which enters when 
the parton densities are factorized. 
The function $C$, not needed for the resummation 
of next-to-leading logarithms, relates to the Drell-Yan process 
(see \cite{Ste87} for details). The arguments of the coupling 
constants reflect the physical scale relevant to the respective 
subprocess.

Renormalon divergence is also related to soft gluons and one may ask 
what the precise relation to soft gluon resummation is. This question 
has guided the work on renormalons in Drell-Yan production. Note 
that the integrals in (\ref{exponent}) are formal, because they 
include integration over the Landau pole of the coupling. It was 
already noted in \cite{CSS89} that this implies sensitivity to 
the large-order behaviour in perturbation theory. \cite{CS94} 
performed the first quantitative analysis and found that the ambiguity 
due to the Landau poles in (\ref{exponent}) in conventional 
leading or next-to-leading order resummations scales as 
$\Lambda/Q$. Leading order resummations of logarithms of $N$ 
need only keep the first-order term in $\alpha_s$ of 
$A(\alpha_s)=a_0\alpha_s+\ldots$. At this order  
$B$ and $C$ can be set to zero. One then finds for 
the Borel transform (defined by (\ref{defborelt}) and using 
$u=-\beta_0 t$ as usual) of the exponent 
\begin{equation} 
\label{lla}
B[E_{\rm LLA}](N,u) \stackrel{u\rightarrow 1/2}{=} 
\frac{4 (N-1)}{1-2 u}\,a_0.
\end{equation}
The pole at $u=1/2$ leads to an ambiguity of order $\Lambda/Q$ in 
defining the exponent at leading-logarithmic accuracy, which was noted by 
\cite{CS94}. The question 
arises of whether this ambiguity indicates a power correction of order 
$\Lambda/Q$ to the hard scattering factor of the Drell-Yan 
cross section or whether the ambiguity appears as the consequence 
of a particular implementation of soft gluon resummation that was 
not designed to be accurate beyond leading power.

This question has been studied by \cite{BB95b} at the level of 
one gluon virtual and real corrections with vacuum polarization 
insertions and accounting for gluon splitting into a $q\bar{q}$ 
pair. Even in this approximation the functions $A$, $B$ and $C$ 
that enter the exponent become infinite series. The large-order 
terms in these series account for highly subleading logarithms in 
$N$ and are not needed for the resummation of such logarithms 
to a given accuracy. On the other hand the Borel transform of 
the exponent becomes
\begin{equation} 
\label{bexp}
B[E](N,u) \stackrel{u\rightarrow 1/2}{=} \frac{4 (N-1)}{1-2 u}
\left[B[A](1/2)-\frac{1}{4}B[C](1/2)\right],
\end{equation}
and the residue of the pole at $u=1/2$ involves the series expansion 
of $A$ and $C$ to all orders. \cite{BB95b} found that, when all 
orders are taken into account, the expression in square brackets 
is zero, and the pole is cancelled. After this 
cancellation the leading power correction to Drell-Yan production 
turns out to be of order $N^2\Lambda^2/Q^2$, at least in the 
approximation mentioned above. Note that the function $B$, related to 
the DIS process, does not appear in (\ref{bexp}). 
This is due to the argument of the coupling, which is larger, 
$\sqrt{1-z} Q$, in this case. In general, the 
terms introduced by performing collinear factorization in the 
DIS scheme are found not to be 
relevant to the discussion of potential $\Lambda/Q$ 
corrections. This is expected, because higher-twist corrections 
scale only as $\Lambda^2/Q^2$ in DIS.

The physical origin of the cancellation becomes more transparent in 
terms of the sensitivity of the one-gluon emission amplitude to an 
IR cut-off. To this end we choose a cut-off $\mu$ and require the energy 
and transverse momentum of the emitted gluon to be larger than $\mu$. 
We are interested in terms of order $\mu$ in the cut-off. To this 
accuracy the one-gluon emission contribution in moment space can be 
written as
\begin{equation}
 \label{wsoft}
W^{[1]}_{\rm real}(N,\mu) = 2 \,\frac{C_F\alpha_s}{\pi} \int
\limits_0^{1-2 \mu/Q} 
d z\,z^{N-1}\!\intl_{\mu^2}^{Q^2 (1-z)^2/4}\!\frac{d k_t^2}{k_t^2}\,
\frac{1}{\sqrt{(1-z)^2-4 k_t^2/Q^2}}\,.
\end{equation}
The expansion at small $\mu$ of this integral starts with logarithms 
of $\mu$. They would be cancelled by adding the virtual correction and 
collinear subtractions, both of which can be seen not to be able to 
introduce a linear dependence on $\mu$. Expanding the square root 
in $k_t/Q$, one finds the following expression for the term of order 
$\mu/Q$ in the expansion at small $\mu$:
\begin{equation}
\label{muterm1}
W^{[1]}_{\rm real}(N,\mu) \ni 
4 \,\frac{C_F\alpha_s}{\pi} \,(N-1)\,\frac{\mu}{Q} 
\sum_{k=0}^\infty\frac{(-1)^k}{k!}\frac{\Gamma(1/2)}
{\Gamma(3/2-k)} = 0 \cdot \frac{\mu}{Q}.
\end{equation}
Hence there is in fact no linear sensitivity to an IR cut-off. One needs 
all terms in the expansion of the square root to obtain this 
cancellation. This means that to linear power accuracy the collinear 
approximation $k_t \ll k_0 \sim Q(1-z)/2$, where $k_t$ is the transverse 
momentum and $k_0$ the energy of the emitted gluon, is not valid. 
It is essential to consider also {\em large angle}, soft gluon emission 
with $k_t\sim k_0$. This conclusion \cite{BB95b} is general and extends 
beyond the Drell-Yan process. 

For the resummation of leading (next-to-leading etc.) logarithms of $N$ an 
expansion in $k_t/k_0$ is justified. The leading logarithms are obtained 
by neglecting $k_t$ under the square root of (\ref{wsoft}). This leads to 
the first term only in the sum of (\ref{muterm1}) and a non-vanishing 
coefficient of $\mu/Q$ in agreement with the pole at $u=1/2$ in 
(\ref{lla}) obtained in the same approximation.

The fact that the exact phase space for 
soft gluon emission is required to determine the coefficient of power 
corrections correctly relates to the fact that all terms in the 
expansion of the functions $A$ and $C$ in the exponent have 
to be kept for this purpose. 
In particular the function $C$, not related to the 
eikonal anomalous dimension, is needed and this rules out the 
possibility discussed in \cite{AZ95a} that the universal parameter 
for $1/Q$ power corrections is given by the integral over the 
eikonal anomalous dimension $A(\alpha_s(k_t))$. Another 
implication is that the angular ordering prescription, according to 
which the emission angles of subsequent emissions in a parton 
cascade decrease, and which generates the correct matrix elements to 
next-to-logarithmic accuracy in $N$ (see for example \cite{CWM91}), 
cannot be applied to power corrections. The intuitive argument 
that partons emitted at large angles can resolve only the total 
colour charge of the previous branching process does not hold true 
beyond leading power. 

This argument also resolves a paradox raised by 
\cite{KS95b}, who noted that $1/Q$ power corrections 
at large $N$ and to $1-T$ close to $T=1$ should be related, 
because the corresponding resummation formulae for logarithmically  
enhanced terms in perturbation theory are related. At present 
such a relation is known only to next-to-leading logarithmic 
accuracy \cite{CTTW93}. The fact that all orders in the exponent 
are needed for power corrections explains that it is 
consistent to expect 
$\Lambda/Q$ power corrections to thrust but not to the Drell-Yan 
process.

Is it possible to organize the resummation of leading, next-to-leading, 
etc., logarithms in $N$ without introducing undesired, because spurious, 
power corrections of order $\Lambda/Q$? \cite{CT89} noted that one 
may substitute 
\begin{equation}
\label{replace}
z^{N-1}-1 \to -\Theta\left(1-\frac{e^{-\gamma_E}}{N}-z\right)
\end{equation}
in (\ref{bexp}) to next-to-leading logarithmic accuracy. Then, 
for $N\ll Q/\Lambda$, which one must require for a short-distance 
treatment\footnote{Recall that the expansion parameter for power 
corrections is $N^2 \Lambda^2/Q^2$. For $N\sim Q/\Lambda$ 
the Drell-Yan process ceases to be a short-distance process,  
and factorization breaks down.}, the integration in 
(\ref{bexp}) does not reach the Landau pole and there are no power 
corrections to the exponent, unless the series expansions for 
$A$, $B$ and $C$ are themselves divergent. 

\cite{BB95b} addressed the above question in the fermion bubble 
approximation, which provides a useful toy model, because the functions 
$A$, $B$ and $C$ are infinite series expansions in $\alpha_s$. 
Ignoring complications from collinear subtractions, the partonic 
Drell-Yan cross section factorizes into
$\hat{\sigma}_{\rm DY}(N,Q) = H(Q,\mu)\,S(Q/N,\mu)$ up to 
corrections that vanish as $N\to\infty$, where $H$ depends 
only on the `hard' scale $Q$ and $S$ on the `soft' scale $Q/N$. 
Following \cite{KM93}, the soft part is expressed as the  
Wilson line expectation value
\begin{equation}
S(Q/N,\mu,\alpha_s) = \int\limits_0^1 d z\,z^{N-1} \,
\frac{Q}{2}\intl_{-\infty}^
\infty\frac{d y_0}{2\pi}\,e^{i y_0 Q (1-z)/2}\,
 \langle 0|\bar{T}\,U^\dagger_{\rm DY}(y) 
\,T\,U_{\rm DY}(0)|0\rangle,
\end{equation}
where 
\begin{equation}
U_{\rm DY}(x) = P\exp\left(i g_s\!\int\limits_{-\infty}^0 \!d s\,p_2^\mu 
A_\mu(p_2 s+x)\right)\, P\exp\left(-i g_s\!\int\limits_{-\infty}^0 
d s\,p_1^{\mu} A_\mu(p_1 s+x)\right),
\end{equation}
and $p_{1,2}$ denote the momenta of the annihilating quark and anti-quark. 
The `soft part' 
$S$ satisfies a renormalization group equation in $\mu$, which  
can be used to sum logarithms in $N$, because $S$ depends only 
on the single dimensionless ratio $Q/(N\mu)$. The solution 
to the RGE equation
\begin{equation}
\label{evolutionequation}
\left(\mu^2\frac{\partial}{\partial\mu^2} + \beta(\alpha_s)
\frac{\partial}{\partial \alpha_s}\right)\,\ln S(Q/N,\mu,\alpha_s(\mu)) = 
\Gamma_{\rm eik}(\alpha_s)\,\ln\frac{\mu^2 N^2}{Q^2} 
+ \Gamma_{\rm DY}(\alpha_s)
\end{equation}
reads 
\begin{equation}
\label{eff}
\hat{\sigma}_{\rm DY} = H(\alpha_s(Q))\cdot S(\alpha_s(Q/N))\cdot  
\exp\!\Bigg(\,\int\limits_{Q^2/N^2}^{Q^2}\!\!\frac{d k_t^2}{k_t^2} 
\bigg[\Gamma_{\rm eik}(\alpha_s(k_t))\ln\frac{k_t^2 N^2}{Q^2}
+\Gamma_{\rm DY}(\alpha_s(k_t))\bigg]\Bigg),
\end{equation}
where $S(\alpha_s(Q/N))$ denotes the initial condition for 
the evolution and in the end we have set $\mu=Q$. From the analysis 
in the fermion loop approximation, one can draw the following, 
more general, conclusions.

The anomalous dimensions $\Gamma_{\rm eik}(\alpha_s)$ and 
$\Gamma_{\rm DY}(\alpha_s)$ have convergent series expansions when defined 
in the $\overline{\rm MS}$ scheme. Since the integrations in 
the exponent of (\ref{eff}) exclude the Landau pole for 
all moments $N$ in the short-distance regime, it follows that 
the resummation, embodied by the exponent, can be carried 
out without ever encountering the divergent series and power corrections 
implied by them. The conclusion is then that the renormalon 
problem is a problem separate from soft gluon resummation.
Renormalons and power corrections enter in the 
hard part $H$ and the initial condition $S$. Because $S$ depends 
only on $Q/N$, the parameter for power corrections to $S$ 
is $N\Lambda/Q$. One finds that all power corrections of order 
$(N\Lambda/Q)^k$ to the Drell-Yan cross section are correctly 
reproduced in the soft part.  In the approximation considered 
in \cite{BB95b}, terms with $k=1$ do not exist. Note that if 
the exponentiated cross section is written in the `standard form' 
(\ref{omega}, \ref{exponent}), the initial condition 
$S(\alpha_s(Q/N))$ is absorbed into the exponent at the 
expense of a redefinition of $C$ ($\Gamma_{\rm DY}$). With this  
redefinition the functions in the exponent are divergent series.

As always, there is the question of whether the absence of renormalon 
divergence that would correspond to a $\Lambda/Q$ power correction 
is specific to the (essentially abelian) approximation of 
\cite{BB95b} and persists to more complicated diagrams. The answer 
to this question is still open. 

\cite{AZ96a,AZ96b,ASZ97} put the cancellation of 
$1/Q$ corrections to Drell-Yan production in the more general 
context of Kinoshita-Lee-Nauenberg (KLN) cancellations. Knowing 
that any potential $1/Q$ correction would come from soft particles, 
but not collinear particles, they consider KLN transition amplitudes,  
which include a sum over soft initial and final 
particles degenerate with the annihilating $q\bar{q}$ pair. 
The KLN transition amplitudes have no $1/k_0$ (where $k_0$ stands 
for the energies of the soft particles) contributions (collinear 
factorization is implicitly assumed). As a consequence, the amplitude 
squared, integrated unweighted over all phase space, is proportional to 
$dk_0 k_0$, which by power counting implies at most $1/Q^2$ 
power corrections. To make connection with a physical process, 
one has to demonstrate that the sum over degenerate initial states 
can actually be dispensed of. The authors above use the 
Low theorem to show this for Drell-Yan production in an abelian 
theory. For QCD this still remains an open problem. 

\cite{Kor96} argued that non-abelian diagrams (involving the 
three-gluon vertex) at two-loop order 
would give a non-vanishing contribution to a certain Wilson line 
operator introduced in \cite{KS95a} to parametrize $1/Q$ corrections 
to Drell-Yan production. It would be very interesting to carry 
out the two-loop calculation to see whether a non-zero linear 
infrared contribution is actually present in these diagrams. 
\cite{QS91} extended collinear factorization for Drell-Yan 
production to $1/Q^2$ corrections and showed that the same 
twist-4 multi-parton correlations enter as in DIS. 
The factorization is carried out at tree-level and 
hence may not be conclusive on the issue of a 
$1/Q$ power correction, which would require a demonstration 
that soft gluon interactions cancel to all orders in 
perturbation theory to the level of $1/Q^2$ accuracy. This is, 
at present, the missing element in a proof that there are 
no $1/Q$ long-distance sensitive regions in the Drell-Yan 
process to all orders in perturbation theory.

\cite{KS95a} have also considered power corrections to the transverse 
momentum (impact parameter) distributions in Drell-Yan production. 
In impact parameter space, they find that ambiguities in defining 
the perturbative contribution to the exponent require power-suppressed 
contributions of the form
\begin{equation}
(b\Lambda)^2 \,(\alpha \ln Q +\beta)
\end{equation} 
with $b$ the impact parameter. The leading correction is quadratic 
in $\Lambda$ and consistent with the parametrization of long-distance 
contributions suggested by \cite{CS81}.

\subsubsection{Other hard reactions }
\label{otherhard}

Renormalon divergence and the corresponding power corrections have been 
investigated for several other hard QCD processes:

{\em Hard-exclusive processes.} \cite{Mik98,GK97} considered the 
Brodsky-Lepage kernel that determines the evolution of hadron 
distribution amplitudes in the large-$\beta_0$ approximation. 
In the $\overline{\rm MS}$ definition, the series expansion of the 
kernel is convergent as expected for anomalous dimensions. The form 
factor for the process $\gamma^*+\gamma \to \pi^0$ was analysed 
in detail by \cite{GK97}. One finds two sources of renormalon 
divergence and power corrections. The first is power corrections in 
the hard coefficient function, which are present independently of the 
form of the hadron wave function. These correspond to higher-twist 
corrections in the hard scattering formalism. Additional power 
corrections are generated after integrating with the hadron wave 
function over the parton momentum fractions and these 
depend on the details 
of the wave function. These power corrections 
arise from the region of small parton momentum fraction and can be 
associated with power corrections due to the `soft' or `Feynman' 
mechanism for exclusive scattering. For the form factor of 
the above process, both power corrections are of order $1/Q^2$ or 
smaller. \cite{BS98} considered deeply virtual Compton scattering 
$\gamma^*+A\to \gamma+B$. For this process and the 
$\gamma^*\gamma\pi^0$ form factor there exist only two IR renormalon 
poles at $u=1,2$ in the hard coefficient functions. This is 
analogous to the GLS sum rule (\ref{borelgls}) and indeed the 
same diagrams are considered here and there, except for different 
kinematics. 

{\em Small-$x$ DIS.} Renormalons in the context of small-$x$ structure 
functions were discussed by \cite{Lev95,ARS96}. To be precise, 
renormalons are understood there as a certain prescription to 
implement the running coupling in the BFKL equation. There 
appears to be a $1/Q$ correction to the kernel, but in \cite{ARS96} 
it is argued that this correction is suppressed after convolution 
with the hadron wave function such that the correction to 
the structure function is only of order $1/Q^2$.

The next-to-leading order BFKL kernel has now been calculated 
\cite{FL98,CC98}. \cite{KM98} separate a `conformally invariant' part 
from a `running coupling' part and investigate the series expansion 
of the solution to the BFKL equation when the exact 
one-loop running coupling is kept 
in the running coupling part. Ignoring over-all 
factors, the result is a series expansion of the form
\begin{equation}
\sum_n\left(a y \alpha_s^3\right)^{n/2}\,\Gamma(n/2),
\end{equation}
where $a=42\zeta(3)\beta_0^2/\pi$ and $y$ is the (large) 
rapidity that characterizes a scattering process in the BFKL limit.
If we take the Borel transform with respect to $\alpha_s^3$, the
above series leads to a typical renormalon pole. The unusual 
feature is location of the 
renormalon pole depends on the kinematic variable $y$, and  
not only in over-all prefactor. When $a y \alpha_s^3\sim 1$ the series 
diverges from the outset and no perturbative approximation is possible. 
This leads to the interesting constraint $y<1/(a \alpha_s^3)$ for 
rapidities to which the BFKL treatment can be applied. The same 
constraint has been found independently by a different method \cite{Mue97}.

The {\em inclusive $\gamma^*\gamma^*$ cross section} into hadrons 
was analysed by \cite{Hau97}. In this case one finds a $1/Q^2$ power 
correction.

\subsection{Heavy quarks}
\label{heavyQ}

In this section we consider hard processes for which the large scale 
is given by the mass of a heavy quark. We first deal with the notion 
of the (pole) mass of a heavy quark itself and its relation 
to the heavy quark potential. We then discuss renormalons 
in heavy quark effective theory, their implications for 
exclusive and inclusive semi-leptonic $B$ decays, and close 
with brief remarks on renormalons in non-relativistic QCD.

\subsubsection{The pole mass}
\label{pole} 

The pole mass of a quark is defined, to any given order in perturbation 
theory, as the location of the pole in the quark propagator. It is 
IR finite, gauge independent and independent 
of renormalization conventions \cite{Tar81,Kro98}. Quarks are confined 
in QCD and quark masses are not directly measurable. The binding 
energy of quarks in hadrons is of order $\Lambda$ and it is 
natural to expect that the notion of a quark pole mass cannot be 
made more precise. Nevertheless for heavy quarks with 
mass $m\gg\Lambda$ the pole mass seemed to be the most natural 
mass definition. 

The pole mass is IR finite, but it is still sensitive to long 
distances. This IR sensitivity manifests itself in rapid 
IR renormalon divergence, when the pole mass is related to the 
bare mass or another mass definition insensitive to long distances 
such as the $\overline{\rm MS}$ mass\footnote{The 
$\overline{\rm MS}$ mass is related to the bare mass by 
subtraction of pure ultraviolet poles in dimensional regularization 
and contains no IR sensitivity at all.} \cite{BB94a,BSUV94}. 
Consider the one-loop self-energy diagram and insert fermion loops 
into the gluon line. The integral can be written as 
\begin{equation}
\label{polm}
m_{\rm pole}-m_{\overline{\rm MS}}(\mu) = 
(-i)C_F g_s^2\mu^{2\epsilon}
\int\!\frac{d^d k}{(2\pi)^d}\,\alpha_s(k\,e^{-5/6})\,\frac{\gamma^\mu 
(\!\not\!p+\!\not\!k+m)\,\gamma_\mu}{k^2\,((p-k)^2-m^2)}\big|_{p^2=m^2}.
\end{equation}
For $p^2=m^2$ the integral scales as $d^4 k/k^3$ for small $k$. 
This implies that the series expansion obtained from (\ref{polm}) 
leads to an IR renormalon singularity at $t=-1/(2 \beta_0)$ 
($u=1/2$) with a corresponding ambiguity of order $\Lambda$. 
The integral (\ref{polm}) can be done exactly or the leading divergent 
behaviour can be extracted from the expansion of the integrand 
at small $k$ as in Section~\ref{renormalonexample}. The 
asymptotic behaviour of the 
series expansion in $\alpha_s=\alpha_s(\mu)$ is 
\begin{equation}
\label{dif}
m_{\rm pole}-m_{\overline{\rm MS}}(\mu) = 
\frac{C_F e^{5/6}}{\pi}\,
\mu\,\sum_n (-2\beta_0)^n\,n!\,\alpha_s^{n+1}.
\end{equation}
The linear IR sensitivity of the pole mass has a transparent 
interpretation in terms of the static quark potential, discussed 
in the present context in \cite{BU94,BSUV94,Ben98}. An infinitely heavy 
quark interacts with gluons through the colour Coulomb potential 
$\tilde{V}(\vec{q}\,)=-4\pi C_F\alpha_s/\vec{q}^{\,2}$. The Fourier 
transform of the potential contains a linear IR contribution 
from integration over small $\vec{q}$. We see that the 
IR contribution to the pole mass of order $\Lambda$ represents 
a contribution to the self-mass from the Coulomb potential at 
large distances.\footnote{This statement will be made more precise 
in the following section.} At these distances the Coulomb potential is 
strongly modified by non-perturbative effects and hence the 
linear IR contribution seen in perturbation theory has no 
physical content. It can be discarded by a mass redefinition. 
Note that in QED (assuming one heavy and one massless lepton) 
the same divergence (\ref{div}) exists. However, the interpretation 
is different, because the series is sign-alternating. As a 
consequence the long-distance contribution to the self-mass 
and the notion of a pole mass are unambiguous in QED. 

It is clear that if the pole mass of the top quark is defined, 
as usual, as the real part of the pole in the top quark propagator, 
then the top quark pole mass is affected by the renormalon 
ambiguity just as the pole mass of a stable quark. The large 
width $\Gamma_t\gg \Lambda$ does not eliminate the problem, 
as emphasized by \cite{SW97}. This does not mean that the finite 
width does not simplify the perturbative treatment of top quarks,  
since it provides a natural IR cut-off. The point is 
that, because of the finite width, there exists no quantity 
for which the pole mass would ever be relevant. This is 
in contrast to bottom or charm quarks, where the pole mass 
is relevant for some quantities (such as meson masses), 
although for fewer than might have been expected, as will be 
discussed in subsequent sections.

The implication of the rapidly divergent series of corrections 
to the pole mass is the following: the large coefficients 
are associated with large finite renormalizations of 
IR origin. There are heavy quark decays, which are intrinsically 
less sensitive to long distances than the pole mass and whose 
perturbation expansions are expected to be well-behaved. 
Expressing such observables in terms of the pole mass 
introduces large corrections only because one has chosen 
a renormalization convention for the mass that does not reflect 
the short-distance properties of the decay process. We will return 
to this point in Section~\ref{incb}. 

The remainder of this section is devoted to a more detailed discussion 
of the perturbative expansion of 
\begin{eqnarray}
\delta m \equiv m_{\rm pole}-m_{\overline{\rm MS}}(m_{\overline{\rm MS}}) 
&=& m_{\overline{\rm MS}}(m_{\overline{\rm MS}})\,
\frac{C_F\alpha_s}{4\pi}\,
\sum_{n=0} r_n \alpha_s^{n+1} 
\nonumber\\
&&\hspace*{-4cm} = \,m_{\overline{\rm MS}}(m_{\overline{\rm MS}})\,
\frac{C_F\alpha_s}{4\pi}\,
\sum_{n=0} [d_n (-\beta_0)^n +\delta_n] \alpha_s^{n+1}
\label{poleqdef}
\end{eqnarray}
in the large-$\beta_0$ approximation \cite{BB95a,N95,BBB95,PS95} and beyond 
it \cite{Ben95}. The Borel transform of the mass shift, 
$B[\delta m/m](u) = \sum_{n=0} d_n u^n/n!$,  in the 
large-$\beta_0$ limit is not just given by the Borel transform of 
(\ref{polm}), but has to take into account the correct UV subtractions 
in the $\overline{\rm MS}$ scheme, see Section~\ref{schemedep}. The 
complete result is \cite{BB94a,BBB95}
\begin{equation} 
\label{polebt}
B[\delta m/m](u) = \left(\frac{m^2}{\mu^2}\right)^{\!-u}\,
e^{5 u/3}\,6 (1-u) 
\frac{\Gamma(u)\Gamma(1-2 u)}{\Gamma(3-u)} + \frac{\tilde{G}_0(u)}{u}
\end{equation}
where the expansion coefficients of $\tilde{G}_0(u)$ 
are given by $g_n/n!$ if the expansion coefficients of $G_0(u)$ 
in $u$ are given by $g_n$ and where 
\begin{equation}
\label{subfunction}
G_0(u) = -\frac{1}{3} (3+2 u)\frac{\Gamma(4+2 u)}{\Gamma(1-u) 
\Gamma(2+u)^2 \Gamma(3+u)}.
\end{equation}
The resulting series coefficients are shown in Table~\ref{tab11} 
for the scale choice $\mu=m_{\overline{\rm MS}}$, which will 
be assumed in what follows. 
For comparison we also show the contribution to $r_n$ from the 
subtraction term\footnote{The `3' is added to $\tilde{G}_0$ to 
cancel the pole in 
$u$.} $(3+\tilde{G}_0(u))/u$ and the separate contributions 
from the first IR and UV renormalon poles to $d_n$. The subtraction 
contribution is a convergent series and is practically negligible 
already at $n=2$. Furthermore, the series in the large-$\beta_0$ 
approximation is very rapidly dominated by the first IR renormalon. 
The coefficient $d_0$ reproduces the exact one-loop correction. 
One may also compare $d_1(-\beta_0) = 12.43$ [$N_f=4$] with 
the exact two-loop result \cite{GBGS90}: 
$r_1=8.81$.\footnote{It is more conventional 
to quote $r_1$ for the difference 
$m_{\rm pole}-m_{\overline{\rm MS}}(m_{\rm pole})$, in which 
case $r_1=11.41$, in better agreement with the large-$\beta_0$ 
approximation with respect to which the two scale choices are 
equivalent.} Moreover the $C_F^2$-term in the exact result, which is 
not reproduced in $d_1 (-\beta_0)$, is rather small and the non-abelian 
term $C_A C_F$ is large. These evidences together suggest that 
the relation between the pole mass and the $\overline{\rm MS}$ mass 
is dominated by the leading IR renormalon already in low 
orders and may even be well approximated in the large-$\beta_0$ 
limit. The numbers in Table~\ref{tab11} imply that this relation 
begins to diverge at order $\alpha_s^3$ for charm quarks and 
at order $\alpha_s^4$ or $\alpha_s^5$ for bottom quarks. 
\begin{table}[t]
\addtolength{\arraycolsep}{0.2cm}
\renewcommand{\arraystretch}{1.2}
$$
\begin{array}{c|c|c|c|c|c|c}
\hline\hline
n & d_n & d_n\,[\rm sub.] & \mbox{IR(1/2)} & \mbox{IR(3/2)} 
& \mbox{IR(2)} & \mbox{UV(-1)} \\ 
\hline
0 & 4        & -2.5   & 9.2039  & -6.091 & 7.008 & -0.7555 \\
1 & 18.7446  & 1.458  & 18.408  & -4.061 & 3.504 & 0.7555  \\
2 & 70.4906  & 1.251  & 73.631  & -5.414 & 3.504 & -1.511  \\
3 & 439.435  & 0.083  & 441.78  & -10.83 & 5.256 & 4.533   \\
4 & 3495.70  & -0.233 & 3534.3  & -28.88 & 10.51 & -18.13  \\
5 & 35358.7  & -0.083 & 35343   & -96.26 & 26.28 & 90.66   \\
6 & 423257   & 0.009  & 424116  & -385.0 & 78.83 & -544.0 \\
7 & 5939874  & 0.012  & 5937622 & -1796  & 275.9 & 3807.7 \\
\hline\hline
\end{array}
$$
\caption[dummy]{\label{tab11}\small 
Perturbative corrections to $\delta m$: 
the `large-$\beta_0$ limit' from \cite{BB95a} in comparison 
with the contribution from the subtraction function and a 
breakdown of $d_n$ into contributions from the first 
renormalon poles (location indicated in brackets). Renormalization 
scale $\mu=m$.}
\end{table}

One can make use of the fact that the leading IR renormalon singularity 
at $u=1/2$ is related to the linear UV divergence of the self-energy 
$\Sigma^{\rm static}$ of a static quark (see also Section~\ref{hqet}) 
to determine the singularity exactly up to an over-all 
constant \cite{Ben95}. The derivation is analogous to that in  
Section~\ref{rganalysis}: the linear UV divergence leads to a 
non-Borel summable 
UV renormalon singularity at $u=1/2$ in the Borel transform of 
$\Sigma^{\rm static}$, if the UV divergences are regulated 
dimensionally. Following \cite{P78}, the imaginary part of 
the Borel integral $I[\Sigma^{\rm static}]$ is proportional to the 
insertion of the dimension-3 operator $\bar{h} h$ (with 
$h$ a static quark field) and can be written as 
\begin{equation}
\mbox{Im}\,I[\Sigma^{\rm static}](\alpha_s,p,\mu) = 
E(\alpha_s,\mu)\,\Sigma^{\rm static}_{\bar{h} h}(\alpha_s,p,\mu),
\end{equation}
where 
$\Sigma^{\rm static}_{\bar{h} h}$ is the static self-energy with a 
zero-momentum insertion of $\bar{h} h$. The coefficient function 
satisfies a renormalization group equation, which is simplified 
by the fact that $\bar{h} h$ has vanishing anomalous dimension. One then 
shows that 
\begin{equation}
\mbox{Im}\,I[\delta m] = -E(\alpha_s,\mu) = 
{\rm const}\times\mu\,\exp\left(\int_{\alpha_s} \!
dx\frac{1}{2\beta(x)}\right) 
= {\rm const}\times \Lambda.
\end{equation}
The $\alpha_s$-dependence of the imaginary part of the Borel 
integral determines the large-order behaviour of the perturbative 
expansion of $\delta m$ according to (\ref{imnlo}, \ref{asnlo}). 
The present case is particularly simple, because the 
large-order behaviour is completely determined 
in terms of the $\beta$-function coefficients. Since 
$\beta_3$ is now known \cite{RVL97}, the 
result of \cite{Ben95} can be extended to $1/n^2$ corrections to the leading 
asymptotic behaviour. The result is 
\begin{eqnarray}
\label{poleasymp}
r_n &=& {\rm const}\,\times\, (-2\beta_0)^n\,\Gamma(n+1+b)
\Bigg[1+\frac{s_1}{n+b}+\frac{s_2}{(n+b)\,(n+b-1)}
\nonumber\\
&&+\,\frac{s_3}{(n+b)\,(n+b-1)\,(n+b-2)} + \ldots\Bigg],
\end{eqnarray}
where $b=-\beta_1/(2\beta_0)^2$ and\footnote{$s_3$ is given 
numerically below.} 
\begin{eqnarray}
s_1 &=& \left(-\frac{1}{2\beta_0}\right)\left(-\frac{\beta_1^2}{2
\beta_0^3}+\frac{\beta_2}{2\beta_0^2}\right),
\\
s_2 &=& \left(-\frac{1}{2\beta_0}\right)^2\left(-\frac{\beta_1^4}{8
\beta_0^6}+\frac{\beta_1^3}{2\beta_0^4}-\frac{\beta_1^2\beta_2}{4
\beta_0^5}-\frac{\beta_1\beta_2}{2\beta_0^3}+\frac{\beta_2^3}{8
\beta_0^4}+\frac{\beta_3}{4\beta_0^2}\right). 
\end{eqnarray}
Numerically, one has $b=\{0.395,0.370,0.329\}$ for 
$N_f=\{3,4,5\}$ and the coefficients in the square bracket in 
(\ref{poleasymp}) that correct the leading asymptotic behaviour 
are very small:
\begin{eqnarray}
s_1 &=& \{-0.065,-0.039,0.008\},
\\ 
s_2 &=& \{-0.057,-0.064,-0.072\},
\\
s_3 &=& \{0.054+0.111\beta_4,0.046+0.162\beta_4,0.034+0.246\beta_4\}.
\end{eqnarray}
The known $\beta$-function coefficients are all negative and 
between 0 and $-1$. It is natural to assume that $\beta_4$ is of 
order 1 to obtain an estimate of the $1/n^3$ term. The smallness 
of the pre-asymptotic corrections leads again to the conclusion 
that the series is close to its asymptotic behaviour already in 
low orders. Note that the present considerations do not assume 
the large-$\beta_0$ approximation. Corrections to (\ref{poleasymp}) 
are of order $1/n^4$ and $1/2^n$ from the next IR and UV renormalon 
pole.

Encouraged by this observation, we extrapolate (\ref{poleasymp}) 
to $n=1,2$. For $N_f=4$ we obtain
\begin{eqnarray}
\frac{r_2}{r_1} &=& 3.14\,\frac{1.00+0.14\beta_4}{0.70-0.51\beta_4},
\\
\frac{r_3}{r_2} &=& 4.47\,\frac{0.98+0.01\beta_4}{1.00+0.14\beta_4}.
\end{eqnarray}
The large dependence on $\beta_4$ in the first relation indicates 
that $n=1$ is too small for the asymptotic formula to apply. However, 
already at $n=2$ the asymptotic formula (\ref{poleasymp}) may become 
useful. Since only $r_1$ is known exactly at present, we cannot yet 
make use of this result. 

\subsubsection{The heavy quark potential}
\label{pot}

In this section we discuss renormalon divergence in the perturbative 
expansion of the heavy quark potential, that is the non-abelian Coulomb 
potential. It turns out that there is a close relation between the 
potential and the pole mass, as far as their leading IR renormalon 
divergence is concerned.

The static potential in coordinate space, $V(\vec{r}\,)$, 
is defined through a 
Wilson loop $W_C(\vec{r},T)$ of spatial extension $\vec{r}$ and 
temporal extension $T$ with $T\to\infty$. 
In this limit $W_C(\vec{r},T)\sim \exp(-i T V(\vec{r}\,))$. The potential 
in momentum space, $\tilde{V}(q)$, is the Fourier transform of 
$V(\vec{r}\,)$. We can compute the 
potential directly in momentum space from the on-shell quark-anti-quark 
scattering amplitude (divided by $i$) at momentum transfer $\vec{q}$ 
in the limit of static quarks, $m\to\infty$, and projected on the 
colour-singlet sector. In addition the sign of the 
$i\epsilon$-prescription in some of the 
anti-quark propagators has to be 
changed, such as in $D_1$ of Fig.~\ref{fig22}, 
so that the integration over zero-components 
of loop momentum can be done without encountering quark poles. 
(The quark poles correspond to iterations of the potential.) 
\begin{figure}[t]
   \vspace{-4.5cm}
   \epsfysize=30cm
   \epsfxsize=20cm
   \centerline{\epsffile{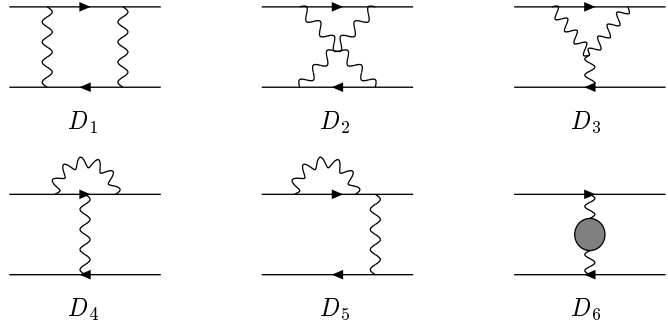}}
   \vspace*{-20.7cm}
\caption[dummy]{\label{fig22}\small 
One-loop corrections to the heavy quark potential.}
\end{figure}

We first consider renormalons for the momentum space potential 
\cite{Ben98}. The one-loop diagrams are shown in Fig.~\ref{fig22}. 
Individual diagrams have logarithmic IR divergences, which cancel 
in the combinations $D_1+2 D_5$ and $D_2+2 D_4$. Diagram $D_6$ 
represents the gluon two-point function at off-shell momentum 
$\vec{q}$. According to the general discussion of section~\ref{irren}, 
this can give rise only to power corrections suppressed at least 
as $\Lambda^2/\vec{q}^{\,2}$. Diagram $D_3$ vanishes in Feynman gauge. 
The integral relevant for $D_2$ is 
\begin{equation}
\int d^4 k \,\frac{1}{k^2 (k+q)^2 (v\cdot k)^2},
\end{equation}
where $v=(1,\vec{0}\,)$ and $v\cdot q=0$. To find the leading contribution 
from $k\sim\Lambda_{\rm QCD}\ll q$ and $k+q\sim\Lambda_{\rm QCD}\ll q$, 
which is left over after the 
IR divergence is cancelled as described above, we expand the integrand 
in $k$ (the contribution from small $k+q$ is identical). 
The integrals in each term of the expansion depend only on the 
vector $v$. Hence, in a regularization scheme that preserves Lorentz 
invariance all odd terms vanish because $v\cdot q=0$. The 
long-distance contribution is again of relative order 
$\Lambda_{\rm QCD}^2/\vec{q}^{\,2}$. A similar argument holds for all 
other one-loop diagrams.

The argument generalizes to an arbitrary diagram. Because $v\cdot q=0$ 
and because there is no other kinematic invariant linear in $q$, 
it follows from Lorentz invariance that the leading power correction 
to the potential in momentum space cannot be 
$\Lambda_{\rm QCD}/|\vec{q}\,|$, but has to be quadratic:
\begin{equation}
\label{potmom1}
\tilde{V}(\vec{q}\,) = -\frac{4\pi C_F\alpha_s(\vec{q}\,)}{\vec{q}^{\,2}}
\left(1+\ldots + \,{\rm const}\times\frac{\Lambda_{\rm QCD}^2}{
\vec{q}^{\,2}} + \ldots\right).
\end{equation}
The corresponding leading IR renormalon is located at $t=-1/\beta_0$ 
($u=1$). Let us emphasize that we are not concerned with the 
long-distance behaviour of the potential at $q\sim \Lambda_{\rm QCD}$, 
but with the leading power corrections of the form $(\Lambda_{\rm QCD}/q)^k$,  
which correct the perturbative Coulomb potential when $q$ is still 
large compared to $\Lambda_{\rm QCD}$. Renormalons cannot tell us 
anything about the potential at confining distances.

When one considers the coordinate space potential, given by the Fourier 
transform of $\tilde{V}(\vec{q}\,)$, a new situation arises. Take the 
potential generated by one-gluon exchange with vacuum polarization 
insertions. The Borel transform (in terms of $u=-\beta_0 t$) is given 
by \cite{AL95,AZ97b}
\begin{eqnarray}
B[V(\vec{r}\,)](u) &=& \left(\mu^2 e^{-C}\right)^u \int\!\frac{
d^3\vec{q}}{(2\pi)^3}\,e^{i\vec{q}\cdot \vec{r}}\,
\frac{-4\pi C_F}{(\vec{q}^{\,2})^{1+u}} 
\nonumber\\
&=&-\frac{C_F}{r}\,e^{-u C}\,(\mu r)^{2 u}\,
\frac{\Gamma(1/2+u)\Gamma(1/2-u)}{\pi\Gamma(1+2 u)}.
\label{potbt}
\end{eqnarray}
There is now a pole at $u=1/2$, which implies 
\begin{equation}
V(r) = -\frac{C_F\alpha_s(1/r)}{r}\left(
1+\ldots + \,{\rm const}\times\Lambda_{\rm QCD} r + 
\ldots\right)
\end{equation}
for the coordinate space potential. The long-distance contributions 
to the coordinate space potential are parametrically larger than 
for the momentum space potential and its series expansion diverges 
much faster.\footnote{The rapid divergence has been noticed in a 
different context by \cite{JKPST98}.} In absolute terms the 
long-distance contribution amounts to an $r$-independent constant 
of order $\Lambda$.

The leading power correction to $V(\vec{r}\,)$ originates only from 
small $\vec{q}$ in the Fourier integral and one can set 
$e^{i\vec{q}\cdot \vec{r}}$ to 1 to obtain it.
Because $\tilde{V}(\vec{q}\,)$ does not have a linear power correction, 
we can define a subtracted potential 
\begin{equation}
V(\vec{r},\mu_f) = V(\vec{r}\,)+2\delta m(\mu_f),
\end{equation}
where 
\begin{equation}
\label{deltam}
\delta m(\mu_f) = -\frac{1}{2}\int\limits_{|\vec{q}\,|<\mu_f} 
\!\!\!\frac{d^3\vec{q}}{(2\pi)^3}\,\tilde{V}(\vec{q}\,).
\end{equation}
The notation is suggestive, because the subtraction can be interpreted 
as a mass renormalization. Define the potential-subtracted (PS) mass 
\cite{Ben98}
\begin{equation}
m_{\rm PS}(\mu_f) = m_{\rm pole}-\delta m(\mu_f).
\end{equation}
Comparing the Borel transform of the pole mass (\ref{polebt}) 
with (\ref{potbt}), we note that the singularity at $u=1/2$ 
is cancelled in the difference \cite{Ben98,HSSW98}. 
Hence the PS mass definition 
is less IR sensitive in this approximation than the pole mass. As a 
consequence its relation to the $\overline{\rm MS}$ mass definition 
is better behaved (less divergent) than the corresponding 
relation for the pole mass. It can be shown that the cancellation 
extends beyond the large-$\beta_0$ 
approximation used here for illustration \cite{Ben98}. The relation 
of the PS mass to $M\equiv m_{\overline{\rm MS}}( m_{\overline{\rm MS}})$ 
at the two-loop order is given by 
\begin{eqnarray}
\label{pstoms}
m_{\rm PS}(\mu_f) &=& M\,\Bigg\{1+\frac{4\alpha_s(M)}{3\pi} 
\left[1-\frac{\mu_f}{M}\right] 
\nonumber\\
&&\hspace*{-1.2cm}+ \left(\frac{\alpha_s(M)}{\pi}
\right)^2\left[K_1-\frac{\mu_f}{3 M}\left(K_2+4\pi\beta_0\left[
\ln\frac{\mu_f^2}{M^2}-2\right]\right)\right]+\ldots\Bigg\},
\end{eqnarray}
where $K_1=13.44-1.04 n_f$ is the two-loop coefficient in the relation 
of $m_{\rm pole}$ to $M$ \cite{GBGS90}  
and $K_2=10.33-1.11 n_f$ the one-loop correction to the Coulomb 
potential in momentum space.

The PS mass has a linear dependence on the IR subtraction scale. 
Mass definitions of this kind have been advocated by 
\cite{BSUV94}, see also the review by \cite{BSU97}. These authors 
favour a subtraction based on integrals of spectral densities 
of heavy-light quark current two-point functions. 
\cite{CMU97} have computed the subtraction term for this definition 
to two-loop order. The precise form of the subtraction differs 
from the above at order $\alpha_s$ and 
higher\footnote{At order $\alpha_s$ the factor $1-\mu_f/M$ 
in (\ref{pstoms}) is replaced by $1-4\mu_f/(3 M)$.}, because 
the definitions of the subtracted masses are different. 

We can use the PS mass and subtracted potential instead of the 
pole mass and the Coulomb potential to perform Coulomb resummations 
for threshold problems. The benefit of using an unconventional 
mass definition is that large perturbative corrections related to 
strong renormalon divergence associated with the coordinate 
space potential are obviated. Physically, the crucial point is 
that, contrary to intuition, heavy quark cross sections near threshold 
are in fact less long-distance sensitive than the pole mass 
and the coordinate space potential. The cancellation is made 
explicit by using a less long-distance sensitive mass definition.  

\subsubsection{Heavy quark effective theory and exclusive $B$ decays}
\label{hqet} 

We now turn to heavy quark effective theory (HQET) in the context 
of which renormalons in heavy quark decays and the pole mass 
have been discussed first \cite{BB94a,BSUV94}. HQET is based on 
the idea that heavy hadron decays involve the large 
scale $m\gg \Lambda$ and the scale $\Lambda$ related to the 
extension of a heavy hadron. HQET formalizes the factorization 
of the two scales into perturbative coefficient functions, to which 
momenta of order $m$ contribute, and non-perturbative matrix 
elements that capture the physics on the scale $\Lambda$. New 
spin and flavour symmetries emerge below the scale $m$, which 
relate the matrix elements for different decays. This is what 
makes HQET useful (see the review \cite{Neu94a} for references 
to the original literature).

In a purely perturbative context, HQET can be viewed as the expansion 
of Green functions with heavy quark legs around the mass shell. We begin 
with the expansion of the heavy quark propagator in this perturbative 
context \cite{BB94a}, since it provides a nice example of how 
factorization introduces new renormalon poles and how they are cancelled 
over different orders in the expansion in the sum of all 
terms.\footnote{Another perturbative example of this kind is given 
by \cite{LMS95}, who consider a toy effective Lagrangian with 
four-fermion interactions and higher-dimension derivative operators 
obtained from integrating out a heavy particle.} We define 
$p=m v+k$, with $m$ the parameter of the heavy mass expansion,  
$k$ the small residual momentum $k\ll m$, and $v^2=1$. We then consider 
the inverse heavy quark propagator in full QCD, $S^{-1}$, projected as 
\begin{equation}
\frac{1+\!\not\! v}{2}\,S^{-1}_P(v k,m_Q) = \frac{1+\!\not\! v}{2}\,
S^{-1}(p,m)\,\frac{1+\!\not\! v}{2}.
\end{equation}
The Borel transform of the inverse propagator can be calculated 
exactly in the approximation of one-gluon exchange with vacuum 
polarization insertions. The expansion of the result can be cast 
into the general form
\begin{eqnarray}
\label{matchs}
B[S^{-1}_P](k,m;u) &=& m\,\delta(u)-B[m_{\rm pole}](m/\mu;u) 
\nonumber\\
&&\hspace*{-2.8cm}
+ \,B[C](m/\mu;u) \star \left(vk\delta(u)-B[\Sigma_{\rm eff}](vk/\mu;u)
\right) + O((v k)^2/m,k^2/m).
\end{eqnarray}
The asterisk denotes the convolution product of the Borel 
transforms. The second term on the right hand side is the Borel transform 
of the series that relates the pole mass to $m$. It is given by 
(see~(\ref{polebt}))\footnote{Here we take the Borel transform of the 
series including the term of order $\alpha_s^0$. This gives rise to 
$\delta(u)$.}
\begin{eqnarray}
\label{poleagain}
B[m_{\rm pole}](m/\mu;u) &=& m\,\Bigg(\delta(u) + 
\frac{C_F}{4\pi}\,\left(\frac{m^2}{\mu^2}\right)^{\!-u}\,
e^{-u C}\,6 (1-u) 
\frac{\Gamma(u)\Gamma(1-2 u)}{\Gamma(3-u)}
\nonumber\\
&&\hspace*{-1.5cm}+\,\mbox{subtractions}
\Bigg).
\end{eqnarray}
The subtraction function may be different from 
(\ref{subfunction}), if $m$ is not the $\overline{\rm MS}$ 
mass. The second line is the convolution of a coefficient 
function that depends only on the scale $m$ and the inverse 
(static) propagator of HQET that depends only on $v k$. The Borel 
transform of the latter is given by 
\begin{equation}
\label{partren} 
B[\Sigma_{\rm eff}](v k;u) = \frac{C_F}{4\pi}\,v k \left(
-\frac{2 vk}{\mu}\right)^{-2 u} e^{-u C}\,(-6)\,\frac{\Gamma(-1+2 u)
\Gamma(1-u)}{\Gamma(2+u)} \,+ \,\mbox{subtractions}.
\end{equation}
The subtraction function has no poles in $u$ and can be omitted for the 
present discussion. The Borel transform of the unexpanded 
inverse propagator $S^{-1}_P$ has IR renormalon poles only at 
positive integer $u$. Compared to this, every term in the expansion 
around the mass shell has new renormalon poles at half-integer 
$u$, and in particular at $u=1/2$. Close to $u=1$, the Borel 
transform of $S^{-1}_P$ is
\begin{equation}
B[S^{-1}_P](k,m;u) \propto \frac{\mu^2}{v\cdot k+k^2/(2 m)}\,
\frac{1}{1-u},
\end{equation}
implying an ambiguity of order $\Lambda^2/v\cdot k$. 
The residue of the pole at $u=1$ becomes divergent on-shell 
($k=0$), which causes the singularity structure of the Borel 
transform to change, when one expands in the residual 
momentum $k$. 

Because there is no singularity at $u=1/2$ in the unexpanded 
inverse propagator, the singularity has to 
cancel in the sum of all terms in the 
expansion. Inspection of (\ref{poleagain}) and (\ref{partren}) 
shows that the singularity at $u=1/2$ in the pole mass 
cancels with a singularity at the same position in the self-energy 
of a static quark. It is of conceptual importance that the 
pole at $u=1/2$ in the static self-energy is an ultraviolet pole, 
which comes from the fact that the self-energy of an infinitely heavy 
quark is linearly divergent. Similar cancellations take place 
for other poles (e.g. at $u=3/2$) over different orders in the 
heavy quark expansion. This is just a particularly simple example 
of how IR poles in coefficient functions (depending only on $m/\mu$ 
and not on $k$) cancel with UV poles in Green functions 
(depending only on $k$ and not on $m$) with 
operator insertions at zero momentum in HQET. The general 
nature of such cancellations has already been emphasized, see also 
the more complicated example in Section~\ref{disex}.

What happens if one chooses the pole mass as the renormalized 
quark mass parameter? Then the first term on the right hand side 
of (\ref{matchs}) vanishes identically and one is left with an 
apparently uncancelled pole at $u=1/2$ on the right hand side. 
This simply tells us that one has to be careful not to absorb 
long-distance sensitivity into input parameters, if one wants 
to have a manifest cancellation of IR renormalons. 

Up to now we considered the limit $\Lambda \ll k\ll m$, in which 
HQET amounts to the expansion of Green functions around the mass shell. 
For a physical heavy-light meson system, the residual momentum $k$ 
is of order $\Lambda$ and the long-distance parts of the 
factorized expressions for heavy hadron matrix elements in QCD 
are non-perturbative matrix 
elements in HQET. Then we have the usual situation that IR renormalon 
ambiguities in defining the coefficient functions must correspond 
to UV ambiguities in defining matrix elements in HQET. 
Since several processes may involve the same matrix elements, this 
leads to consistency relations on the IR renormalon behaviour 
in the coefficient functions of different processes.

In the remainder of this section we briefly consider several 
implications and applications, the latter mainly in the large-$\beta_0$ 
approximation, of renormalons in HQET.

{\em The binding energy of a heavy meson} \cite{BU94,BB94a,BSUV94}.
In HQET the mass of a meson can be expanded as 
\begin{equation}
\label{energy}
m_B=m_{b,\rm pole} + \bar{\Lambda}-\frac{\lambda_1+3\lambda_2}
{2 m_{b,\rm pole}} + O(1/m_b^2).
\end{equation}
To be specific, we have taken the pseudoscalar $B$ meson. The 
parameters $\bar{\Lambda}$ and $\lambda_{1,2}$ are the same for 
all members of a spin-flavour multiplet of HQET. $\bar{\Lambda}$ 
is interpreted as the binding energy of the meson in the 
limit $m\to \infty$. $\lambda_1$ denotes the contribution to the binding 
energy from the heavy quark kinetic energy and $\lambda_2$ 
the contribution from the spin interaction between the heavy and the light 
quark. (We follow the conventions of \cite{Neu94a}.) The fact that 
the pole mass expressed in terms of, say, the $\overline{\rm MS}$ mass 
(or another mass related to the bare mass by pure {\em ultraviolet} 
subtractions) has a divergent series expansion with an ambiguity 
of order $\Lambda$, leads to the conclusion that the 
binding energy $\bar{\Lambda}$, which is also of order $\Lambda$, 
is not a physical concept. Since $\bar{\Lambda}=m_B-m_{b,\rm pole}
+\ldots$, it depends on how the pole mass is 
defined beyond its perturbative expansion and different definitions 
can differ by an amount of order $\Lambda$, that is by as much as 
the expected magnitude of $\bar{\Lambda}$ itself. Physically, this 
is not unexpected: because quarks are confined, the meson cannot be separated 
into a free heavy and a light quark relative to which the 
binding energy could be measured.

In decay rates $\bar{\Lambda}$ appears at subleading order 
in HQET, while the quark mass 
does not appear at leading order. 
We may ask whether this would allow us to determine 
$\bar{\Lambda}$, by-passing the argument above. This is in fact not 
possible, because a term of order $\bar{\Lambda}/m$ appears 
always in conjunction with a coefficient function 
at order $(\Lambda/m)^0$ with a renormalon ambiguity of order 
$\Lambda/m$. 

One can rewrite (\ref{energy}) as 
\begin{equation}
\label{energy2}
m_B=\left[m_{b,\rm pole} - \delta m(\mu)\right] + \left[
\bar{\Lambda}+ \delta m(\mu)\right]+\ldots
\end{equation}
with a residual mass $\delta m(\mu)$ that subtracts the (leading) 
divergent behaviour of the series expansion for the pole mass. 
In order to obtain a decent heavy quark limit, the residual mass 
term should stay finite in the infinite mass limit. At the same 
time, it must be perturbative to subtract the perturbative 
expansion of $m_{\rm pole}$. This leads to a linear subtraction 
proportional to a factorization scale $\mu\gg\Lambda$, similar to the 
subtraction discussed in Section~\ref{pot}, and suggested in 
the present context by \cite{BSUV94}. If HQET is formulated 
with a residual mass of this form, the binding energy has 
a perturbative contribution of order $\mu \alpha_s(\mu)$. One 
possible definition of $\bar{\Lambda}(\mu)=\bar{\Lambda}+ \delta m(\mu)$ 
is computed to two-loop order in \cite{CMU97}. A similar strategy can 
be employed to define the binding energy on the lattice \cite{MS95}. 
In this case the inverse lattice spacing takes the place of $\mu$ 
(see section~\ref{mhq}). 

There is also the argument that the renormalon problem of 
$\bar{\Lambda}$ is actually of no relevance in practice, when we 
work only to a given finite order in perturbation theory. One 
{\em defines}, say,  a `one-loop pole mass' $m_{\rm pole}^{\rm 1-loop} = 
m_{\overline{\rm MS}}(m_{\overline{\rm MS}})\,(1+4\alpha_s/(3\pi))$. 
$\bar{\Lambda}$ is then defined with respect to this mass definition, 
i.e.\ $\bar{\Lambda}=m_B-m_{\rm pole}^{\rm 1-loop}$. 
It is of order $m \alpha_s^2$, but we may not care about this, 
because working at one-loop order, we have left out other terms 
of order $m\alpha_s^2$. If this $\bar{\Lambda}$ is extracted from one 
process, it can be consistently used in another, also computed 
to one-loop order. In this procedure the value of $\bar{\Lambda}$ 
depends on the loop-order of the perturbative calculation and 
is meaningless without this specification. This is indeed a 
viable solution, provided the series for the pole mass does
not yet diverge and provided the pole mass is really 
relevant for the observable under consideration. 
This, of course, is just the usual 
problem of how to combine 
a divergent series with a power correction consistently, in particular 
in a purely perturbative context. What makes the problem more 
severe here is the fact that the divergent behaviour is particularly 
violent and, hence, relevant already at rather low orders, 
perhaps two-loop order, in perturbation theory. The procedure 
described here is not viable for short-distance quantities,  
which are less sensitive to long-distances than the pole mass and 
therefore have better behaved series. In this case, introducing 
$\bar{\Lambda}$ as an input parameter instead of a 
short-distance quark mass leads to large perturbative coefficients, 
the origin of which is obscured by using $\bar{\Lambda}$ as an 
input parameter. We return to this point in a less abstract context in 
Section~\ref{incb}.

Of course, we can avoid $\bar{\Lambda}$ altogether by eliminating 
it from the relations of physical observables, but in practice 
this is often not an option that is easy to implement. (It is for the 
same reason that one usually works with renormalized rather than bare 
parameters, although all divergences would drop out in the relation 
of physical quantities.)

{\em The kinetic energy.} The matrix element of the 
chromomagnetic operator, $\lambda_2$ in (\ref{energy}), is related 
to the mass difference of the vector and pseudoscalar meson 
in the heavy quark limit and therefore physical and 
unambiguous. For the matrix element $\lambda_1$ of the kinetic 
energy operator $\bar{h}_v\,(i D_\perp)^2 h_v$, the situation is 
not obvious. Curiously enough, there is no IR renormalon 
at $u=1$ in the Borel transform of the pole mass (\ref{poleagain}) 
and therefore it follows from 
(\ref{energy}) that there is no 
ambiguity in $\lambda_1$ at this order in the flavour 
expansion. \cite{BBZ94} speculated 
that the kinetic energy may be protected by Lorentz invariance 
from quadratic divergences, just as Lorentz invariance protects this 
operator from logarithmic divergences to all orders in 
perturbation theory \cite{LM92}. The problem was discussed 
further in \cite{MNS96} and it seems to have been settled 
finally by \cite{Neu97}, who showed that even for a Lorentz-invariant 
cut-off a quadratic divergence exists at the two-loop order. 
The kinetic energy operator mixes with the operator $\bar{h}_v h_v$ 
and its matrix element is not physical in the same sense as  
$\bar{\Lambda}$ is not physical.

{\em Exclusive semi-leptonic heavy hadron decays.} As already mentioned, 
the predictive power of HQET derives from the fact that the 
heavy quark symmetries relate different decays and reduce the number of 
independent form factors. According to our general understanding 
of IR renormalons in coefficient functions, they are related to 
the definition of non-perturbative matrix elements at subleading order in 
the $1/m$ expansion. Since there is a limited number of such 
matrix elements in semi-leptonic decays, the IR renormalon behaviour 
of the coefficient functions satisfies certain consistency relations, 
which simply express the fact that if the matrix elements are eliminated 
to a given order in $1/m$ and physical quantities are related directly, 
there should be no ambiguities left to that order in $1/m$. The 
decay $\Lambda_b\to \Lambda_c l\nu$ is a particularly simple example 
to illustrate this point, because the form factors, at subleading 
order in $1/m$, are proportional to the same Isgur-Wise function 
that appears at leading order. Hence, we can write (setting 
$m_b\to \infty$ and keeping $m_c$ large but finite)
\begin{equation}
\langle \Lambda_c(v')|\bar{c}\gamma^\mu b|\bar{\Lambda}_b(v)\rangle = 
\bar{u}(v') \left[F_1(w)\,\gamma^\mu+F_2(w)\,v^\mu+ F_3(w)\,v^{\prime
\mu}\right] u(v)
\end{equation}
with $w=v\cdot v'$ and 
\begin{equation}
F_i(w) = \xi(w)\,\left(C_i(m_{b,c},w) + D_i(m_{b,c},w)\,
\frac{\bar{\Lambda}}{m_c} + \,O(1/m_c^2)\right).
\end{equation}
The large-order behaviour of the series for the pole mass determines 
the renormalon ambiguity of $\bar{\Lambda}$. Because the unexpanded 
form factors $F_i$ are observables, the large-order behaviour of 
the series expansion of $C_i$ is determined by the requirement 
that its renormalon ambiguity matches with $D_i\bar{\Lambda}$. 
\cite{NS95z,LMS95} checked, in the large-$\beta_0$ limit, that this is 
indeed the case. The situation is more complicated for semileptonic 
$B\to D$ decays and has been considered in detail by \cite{NS95z}.

{\em Numerical results in the large-$\beta_0$ approximation.} 
The Borel transforms of some coefficient functions in HQET 
are known exactly in the large-$\beta_0$ approximation and we
give a brief overview of these results.

(i) The HQET Lagrangian reads
\begin{equation}
   {\cal L}_{\rm eff} = \bar h_v\,i v\!\cdot\!D\,h_v
   + \frac{1}{2 m}\,\bar h_v(i D)^2 h_v
   + \frac{C_{\rm mag}(\alpha_s)}{4 m}\,
   \bar h_v\sigma_{\mu\nu} g_sG^{\mu\nu} h_v + O(1/m^2). 
\label{Leff}
\end{equation}
It involves only one non-trivial coefficient function 
$C_{\rm mag}(\alpha_s)$ 
to this order in $1/m$. \cite{GN97} computed the coefficient 
function in the large-$\beta_0$ approximation. So far, this is the only 
exact result in the large-$\beta_0$ limit that involves 
diagrams with a three-gluon vertex. 
As expected, a rapid divergence of the series is found  
and an IR renormalon pole at $u=1/2$, which can be related to 
higher-dimension interaction terms in the effective Lagrangian. 
An interesting point is that the renormalization group (RG)
improved coefficient function requires both the anomalous dimension 
and matching relation to be computed. However, because 
anomalous dimensions are convergent series in the $\overline{\rm MS}$ 
scheme, while the expansion of matching coefficients is divergent, 
the contribution from the anomalous dimension is almost insignificant 
in higher orders. In sufficiently large orders the 
`leading logarithms' are in fact smaller than the factorially 
growing constant terms. This allows \cite{GN97} to conclude 
that the RG improved coefficient is already known accurately 
to next-to-next-to-leading order despite the fact that the 
three-loop anomalous dimension is not known. 

(ii) \cite{N95a,N95} considered the matching of $b\to c$ 
currents at zero recoil,
\begin{equation}
\bar{c}\Gamma_{V,A} b = \eta_{V,A}\,\bar{h}^c_v 
\Gamma_{V,A} h_v^b + \,O(1/m_{c,b}^2),
\end{equation}
for the vector and the axial-vector current. The decay rate for 
$B\to D^* l\bar{\nu}$ is proportional to 
$|V_{cb}|^2 {\cal F}(v\cdot v')$, where 
\begin{equation}
{\cal F}(1) = \eta_A (1+\delta_{1/m^2}).
\end{equation}
This leads to a precise determination of $|V_{cb}|$ from the 
measured rate near zero-recoil, provided the perturbative 
correction to $\eta_A$ and the leading power correction 
$\delta_{1/m^2}$ are under control. The large-$\beta_0$ 
approximation provided the first estimate of the second 
(and higher) order corrections to the matching coefficient 
and their magnitude was found to be moderate (with 
$\alpha_s$ normalized at the `natural scale' $\sqrt{m_b m_c}$). 
Meanwhile the two-loop correction is known exactly \cite{CM97a}. 
It turns out that the large-$\beta_0$ approximation 
to the two-loop coefficient is very accurate for the axial
vector case $\eta_A$, but not accurate at all for $\eta_V$. 
One reason seems to be that the $N_f\alpha_s^2$-term 
is anomalously small for $\eta_V$.\footnote{The exact one-loop and 
two-loop coefficients are both somewhat less than a factor $1/2$ smaller 
for $\eta_V$ than for $\eta_A$. However, the $N_f$-term 
is a factor of 15 smaller.} The main point, however, is that the 
two-loop correction to the matching of the currents is not large. 
But if there are no large corrections associated with the running 
coupling, no improvement due to a large-$\beta_0$ resummation 
should be expected. 

Further information on the 
form factor ${\cal F}(1)$ at zero recoil can be obtained from 
sum rules based on the spectral functions of heavy quark currents 
\cite{BSUV95}. In addition to the purely virtual form factors, 
real gluon emission has to be allowed for. The relevant calculation 
in the large-$\beta_0$ limit can be found in \cite{Ura97}.

\subsubsection{Inclusive $B$ decays}
\label{incb}
Inclusive semi-leptonic decays $B\to X_{u,c} l\bar{\nu}$, 
where $X$ is an inclusive final state without or with charm, can be 
treated in an expansion in $\Lambda/m_b$ \cite{CGG90,BUV92}. 
Contrary to exclusive decays the leading non-perturbative 
corrections are suppressed by two powers of $m_b$ and involve the 
parameters $\lambda_{1,2}$ introduced in (\ref{energy}). The 
leading term in the expansion is the hypothetical free quark decay. 
From a phenomenological point of view, the main result of the 
heavy quark expansion is to affirm that non-perturbative corrections 
are in fact 
small, less than $5\%$. Therefore the main uncertainty in the prediction 
of the decay rate, which for decays into charm can be used to 
measure the CKM element $|V_{cb}|$, comes from unknown perturbative 
corrections to the free quark decay beyond the one-loop order. 

Traditionally the free quark decay is expressed in terms of the 
quark pole mass. This seems indeed to be the natural choice for the 
decay of a free particle. The series expansion of the free quark 
decay is\footnote{For simplicity of notation we consider 
the decay into the massless $u$ quark in the general discussion.} 
\begin{equation}
\label{urate}
\Gamma(B\to X_u e \bar \nu)=\frac{G_F^2 |V_{ub}|^2 m_{b,\rm pole}^5}
{192\pi^3}\Bigg\{1
 - C_F\,g_0\,\frac{\alpha_s(m_b)}{\pi}\,\Big[1 + \Delta \Big]\Bigg\},
\end{equation}
where 
\begin{equation}
\Delta = \sum _{n=1} r_n \alpha_s(m_b)^n = \sum _{n=1} \alpha_s(m_b)^n 
\left[d_n (-\beta_0)^n+\delta_n\right]
\end{equation}
parametrizes perturbative corrections beyond one loop. The heavy quark 
expansion clarifies that it is not a good idea to use the pole 
quark mass parameter. When the pole mass is used, the series 
coefficients diverge rapidly due to an IR renormalon at $t=-1/(2\beta_0)$. 
It produces an ambiguity of order $\Lambda/m_b$ in summing the series, 
which does not correspond to any non-perturbative parameter in the 
heavy quark expansion. The resolution is that the rapid 
divergence appears only because the pole mass has been used. If 
we express the pole mass as a series times the $\overline{\rm MS}$ 
mass and eliminate it from (\ref{urate}), then the leading divergent 
behaviour of the $r_n$ cancels with the series that relates the 
pole mass to the $\overline{\rm MS}$ 
mass \cite{BSUV94,BBZ94}.\footnote{The explicit demonstration of this 
cancellation has now been extended, by purely algebraic methods, 
to two-loop order \cite{SAZ98}.}  
The left-over divergent behaviour corresponds to power corrections of 
order $\Lambda^2/m^2$, consistent with the heavy quark expansion. 
In the large-$\beta_0$ limit, it is found \cite{BBZ94,BBB95b} that 
the divergent behaviour left over corresponds in fact to a 
smaller power correction of order $\Lambda^3/m^3 \,\ln(\Lambda/m)$ 
consistent with the observation that in this approximation 
all matrix elements at order $1/m^2$ are unambiguous, see
section~\ref{hqet}. Numerically, the effect of the cancellation is 
dramatic beyond two-loop order and we have illustrated it in the 
large-$\beta_0$ limit (as explained in Section~\ref{lpc}) in 
Table~\ref{tab12}. Up to two-loop order we may note, however, that 
$g_0$ (see (\ref{urate})) in the 
$\overline{\rm MS}$ scheme is in fact larger than in the 
on-shell scheme and that $\overline{d}_1$ is not a small correction. 
\begin{table}[t]
\addtolength{\arraycolsep}{2pt}
\renewcommand{\arraystretch}{1.2}
$$
\begin{array}{l|cc|cc}
\hline\hline
n & d_n & 1+\Delta
& \overline{d}_n & 1+\bar{\Delta}\\ \hline\hline
0 & 1         & 1     & 1       & 1    \\ 
1 & 5.3381702 & 1.747 & 4.3163 & 1.604\\
2 & 34.409913 & 2.422 & 8.0992 & 1.763\\
3 & 256.48081 & 3.126 & 26.680 & 1.836\\
4 & 2269.4131 & 3.997 & 82.262 & 1.868\\
5 & 23679.005 & 5.271 & 421.33 & 1.890\\
6 & 289417.40 & 7.450 & 1656.1 & 1.903\\
7 & 4081180.2 & 11.75 & 12135 & 1.916\\
8 & 65496131. & 21.42 & 52862 & 1.924\\ \hline
\infty & - & 2.314\pm 0.615 &-& 1.925\pm 0.012\\ 
\hline\hline
\end{array}
$$
\addtolength{\arraycolsep}{-2pt}
\renewcommand{\arraystretch}{1}
\caption[]{\label{tab12}\small 
higher order coefficients to $b\to u$ decay in the 
large-$\beta_0$ approximation together with partial sums 
for $-\beta_0^{(N_f=4)}\alpha_s(m_b)=0.14$. 2nd and 3rd columns: 
Decay rate expressed in terms of the $b$ pole mass. 4th and 5th columns: 
Decay rate expressed in terms of the $b$ $\overline{\rm MS}$ mass. 
The last line gives the principal value Borel integral computed 
according to (\ref{sumrep}) together with an estimate of the 
uncertainty due to renormalon poles. In the $\overline{\rm MS}$ scheme 
this uncertainty is very small, because the leading term in the 
expansion of the one-loop correction with a massive gluon expanded 
in the gluon mass $\lambda$ is of order $\lambda^3/m_b^3 
\ln(\lambda^2/m_b^2)$. Table from \cite{BBB95b}.}
\end{table}

The example of inclusive $B$ decays illustrates the fact that pole masses 
are not useful bookkeeping parameters, say for the 
Particle Data Book. Either their value, extracted from some 
process (e.g. $B$ decays, if we knew the CKM matrix elements) 
would depend sensitively on the loop order of the theoretical input 
calculation or one would assign to it a large error due 
to higher order corrections. Another process predicted in terms of the 
pole mass would also seem to be poorly predicted, because of 
large higher order corrections. However, because the theoretical errors 
are correlated with those in the pole mass input parameter, the actual 
uncertainties are much smaller. It is preferable to use book-keeping 
parameters that do not introduce such correlations. The optimal choices 
are book-keeping parameters that themselves are less long-distance 
sensitive than any process in which one would use them. The 
$\overline{\rm MS}$ mass, which is basically a bare mass minus 
UV subtractions, is such a parameter, although only in 
a perturbative setting. 

We may also eliminate the quark mass 
in favour of the physical $B$ meson mass. In this case we get 
\begin{equation}
\Gamma(B\to X_u e \bar \nu)=\frac{G_F^2 |V_{ub}|^2 m_B^5}
{192\pi^3}\Bigg\{1
 - C_F\,g_0\,\frac{\alpha_s(m_b)}{\pi}\,\Big[1 + \Delta \Big] 
- 5\,\frac{\bar{\Lambda}}{m_B} + \,O(1/m_B^2)\Bigg\}.
\end{equation}
Now the large perturbative corrections in $\Delta$ are cancelled by the 
fact that $\bar{\Lambda}$ has to be specified as one-loop, two-loop etc.,  
and differs with loop order, such as to cancel the large corrections 
to $\Delta$. Apart from the fact that, beyond two loops, the magnitude 
of the so-defined $\bar{\Lambda}$ is far larger than that of $\Lambda$, the 
delicate cancellation of all $\Lambda/m_b$ effects that has to be 
arranged in this way seems a high price to pay, in comparison to 
using a quark mass definition without large long-distance sensitivity,  
together with a better-behaved series expansion.

\cite{BBB95b} performed a detailed numerical analysis of higher 
order corrections to inclusive semi-leptonic decays into charm 
in the large-$\beta_0$ limit in the 
$\overline{\rm MS}$ scheme and the on-shell scheme for the bottom and 
charm quark mass. They calculated the distribution function $T(\xi)$ 
that enters (\ref{sumrep}) analytically for $b\to u$ transitions 
and numerically for $b\to c$ transitions. 
The renormalon problem is less severe for 
$b\to c$ than for $b\to u$, because the leading IR renormalon 
cancels in the difference $m_b-m_c$ and, in the rate for $b\to c$, 
the quark masses appear numerically in this combination to a certain 
degree. \cite{BBB95b} found that, after higher order corrections 
are taken into account, one obtains values for $|V_{bc}|$, from 
the calculation in the $\overline{\rm MS}$ scheme and the on-shell scheme, 
which are consistent with each other, contrary to what is  
found in one-loop calculations \cite{BN94}. The corrections in 
the $\overline{\rm MS}$ scheme are not small at one- and two-loop order, 
which reflects the fact that the $\overline{\rm MS}$ mass at the 
scale of the mass, is relatively small and too far away from the 
natural range for a `physical' quark mass given by 
$m_{\rm pole}\pm \Lambda$. Instead of a full resummation of (some) 
higher order corrections we can also optimize the choice of scale 
and quark mass definition to avoid large corrections to  
some extent \cite{SU95,Ura95}.

Since the analysis of \cite{BBB95b} there has been some progress 
in the calculation of the exact 2-loop correction to inclusive 
$b\to c$ transitions and it is interesting to compare the 
large-$\beta_0$ limit with these results. Up to order 
$\alpha_s^3$ the series of radiative corrections in the on-shell scheme 
is \cite{BBB95b}\footnote{The second-order correction was first  
obtained by \cite{LSW95}. The difference to the 
value 15.1 quoted there comes from the fact that we use $N_f=4$ 
rather than $N_f=3$ (+1.14) and the remaining difference ($-$0.2) 
is probably due to numerical errors.}
\begin{eqnarray}
\label{crate}
\Gamma(B\to X_c e \bar \nu) &=&\frac{G_F^2 |V_{bc}|^2 m_{b,\rm pole}^5}
{192\pi^3}\,f_1(0.3)\,\Bigg[1-1.67\,\frac{\alpha_s(m_b)}{\pi} - 
14.2\left(\frac{\alpha_s(m_b)}{\pi}\right)^2
\nonumber\\
&&\hspace*{-1cm}
-173\left(\frac{\alpha_s(m_b)}{\pi}\right)^3+\ldots \Bigg],
\end{eqnarray}
where $f_1(m_c/m_b)$ is a tree level phase space factor,  
and the numerical values in square brackets assume $m_c/m_b=0.3$. 
($f_1(0.3)=0.52$.) The 
exact two-loop correction is still unknown. However, the differential 
decay rate $d\Gamma/dq^2$, where $q^2$ is the invariant mass of 
the lepton pair, has been computed analytically at three 
special kinematic 
points $q^2=(m_b-m_c)^2$ \cite{CM97a}, $q^2=0$ \cite{CM97b} and 
the intermediate point $q^2=m_c^2$ \cite{CM98}. 
The authors then interpolate the three points 
by a second-order polynomial in $q^2$ and obtain 
\begin{equation}
-(12.8\pm 0.4) \left(\frac{\alpha_s(m_b)}{\pi}\right)^2
\end{equation}
for the second-order correction, to be compared with $-14.2$ 
above. The large-$\beta_0$ limit has worked well in the 
on-shell scheme. Note that in this case the two-loop correction 
{\em is} large. This should be contrasted with the situation 
for exclusive semi-leptonic decays, see the end of section~\ref{hqet}.

The large-$\beta_0$ approximation has also been applied to the top decay 
$t\to W+b$ with the $W$ assumed to be on-shell in the approximation 
that $m_W/m_t=0$ \cite{BB95a} and for finite $m_W/m_t$ \cite{Meh97}. 
Not unexpectedly, the convergence of the series is again improved 
if one does not use the top pole mass in the decay rate, except 
for the hypothetical limit $m_W\to m_t$.

\subsubsection{Non-relativistic QCD}
\label{nrqcd}

Quarkonium systems, like heavy-light mesons, can be treated with 
effective field theory 
methods \cite{CL86,BBL95}. The effective theory is non-relativistic QCD 
(NRQCD). The expansion is done in $v^2$, where $v$ is 
the typical velocity 
of a heavy quark in an onium. This is somewhat different from 
the expansion we encountered before in HQET or DIS, which are 
expansions in $\Lambda/m$ and $\Lambda/Q$, respectively.

The renormalon structure in the matching 
of QCD currents on non-relativistic currents and in the velocity 
expansion of some quarkonium decays has been considered 
by \cite{BC97,BC98}. Since $v^2$ need not 
be connected with the QCD scale $\Lambda$ -- for example one could 
have the hierarchy $m\gg m v^2\gg \Lambda$ -- the situation is 
similar to the expansion (\ref{matchs}) considered 
in the limit $m\gg k\gg \Lambda$. \cite{BC97} showed that the 
IR renormalon structure of the short-distance coefficient is 
consistent with a unique relation for the ambiguities of 
NRQCD matrix elements. \cite{BC98} then considered the UV 
behaviour of these matrix elements and verified that the required 
relations are indeed satisfied. One then obtains a cancellation 
between coefficient functions and 
matrix elements in the matching relations similar to the cancellations 
that occur in HQET. 

There are two `peculiarities' in NRQCD compared to the examples 
discussed up to now, in particular the HQET examples. 
Consider the matching of the axial current 
matrix element (any other would do as well) in a spin-singlet state 
up to order $v^2$,
\begin{equation}
\langle 0|\bar{Q}\gamma^\mu\gamma_5 Q|\eta\rangle = 
\delta^{\mu 0} \,\left[C(\alpha_s)\,
\langle 0|\chi^\dagger\psi|\eta\rangle + 
\frac{1}{2 m^2}\,\langle 0|\chi^\dagger\,\vec{D}^{\,2}\psi|\eta
\rangle\right]
\end{equation}
where $\psi$ and $\chi$ are non-relativistic two-spinor fields.
A leading IR renormalon pole in the Borel transform 
of $C(\alpha_s)$ at $u=1/2$ is found, which corresponds to an ambiguity 
of relative order $\Lambda/m$. 

There is a UV renormalon pole at $u=1/2$ in the matrix element of the 
higher-dimension operator in square brackets. 
However, to obtain a complete cancellation 
of the singularity at $u=1/2$, one also has to take into account 
the fact that 
the first matrix element in square brackets has a renormalon ambiguity 
proportional to itself. This somewhat unfamiliar situation arises, 
because, due to insertions of higher-dimension operators in the NRQCD 
Lagrangian, the matrix element is expressed as a series in $v^2$, and 
there exist power-UV divergences from these insertions. 
In HQET, on the contrary, it is conventional to parametrize the 
contributions from insertions of higher-dimension operators in the 
Lagrangian as separate `non-local' operators. 

The second `peculiarity' is that the Borel transforms of the coefficient 
functions also have an {\em infrared} renormalon pole at negative $u=-1/2$. 
Recall that if a one-loop integral has the small-$k$ behaviour 
$\int d^4 k/k^{4+2 n}$, an IR renormalon pole at $u=n$ is obtained. 
The pole at $u=-1/2$ is therefore due to the fact that the integrals 
that contribute to the matching coefficient are linearly IR divergent. 
This divergence would be regulated by a small relative momentum of the 
heavy quark and anti-quark and then give rise to the Coulomb divergence 
$1/v$. To compute the coefficient function, the relative 
momentum is set 
exactly to zero. In dimensional regularization the power-like 
IR divergence is set to zero at every order in perturbation theory, 
but it leads to a Borel-summable 
IR renormalon at $u=-1/2$. (Recall that linear {\em ultraviolet} divergence 
gives rise to an unconventional non-Borel summable singularity at 
positive $u=1/2$.)


\section{Connections with lattice field theory}
\setcounter{equation}{0}
\label{lattice}

One may be surprised to find renormalons discussed in 
connection with lattice gauge theory, as we emphasized that renormalons 
are `artefacts' of performing a short-distance expansion. If the 
exact, non-perturbative result could be computed, one would never 
concern oneself with renormalons. The connection arises from the fact that  
it is difficult to simulate quantities on the lattice that involve 
two very different scales $Q\gg \Lambda$, because the lattice 
spacing and lattice size in physical units must satisfy 
$L^{-1}\ll \Lambda\ll Q\ll a^{-1}$, which requires larger lattices 
than computing resources may allow. In this situation one can 
use the short-distance expansion, compute the coefficient functions 
perturbatively, and use lattice simulations only to compute the 
non-perturbative matrix elements that involve only the scale $\Lambda$. 
Then the inverse lattice spacing acts as a `hard' factorization scale 
and the hierarchy of scales is $L^{-1}\ll \Lambda\ll a^{-1}\ll Q$. 
higher order terms in the short-distance expansion involve matrix elements 
of operators of high dimension, which have power divergences 
as $a\to 0$. For example, if ${\cal M}$ is the matrix element of 
an operator of dimension 1, whose `natural size' is $\Lambda$, then 
the unrenormalized matrix element computed on the lattice can 
be represented as
\begin{equation}
{\cal M} = \frac{1}{a}\,\sum_{n=0} c_n \left[\alpha_s(a^{-1})\right]^n + 
{\rm const}\cdot \Lambda + O(a).
\end{equation}
When the power divergence is subtracted perturbatively, as indicated 
in the equation, it is found that the series expansion is divergent 
and the ambiguity in summing it is of order $a \Lambda$. Hence the 
matrix element from which the linear divergence is subtracted 
is unambiguously defined only if a prescription is 
given on how to sum the series that multiplies the power divergence 
to all orders. The value of the subtracted matrix elements depends on 
this prescription. To our knowledge this point was discussed 
first by \cite{D84} in connection with lattice determinations of the 
gluon condensate \cite{dGR81}. 

It should be emphasized that in the context of effective theories 
there is no need to subtract the power divergence. The inverse 
lattice spacing acts as a factorization scale and the continuum 
limit should never be taken, because the factorization scale 
has to satisfy $a^{-1}\ll Q$. It is sufficient that $a^{-1}$ stays 
finite as $Q\to\infty$. It is important only that the matching 
conditions that specify the coefficient functions in the short-distance 
expansion be computed in a way that is consistent with the 
renormalization prescription for the matrix elements. 

In the following we will summarize two cases for which 
renormalons and the lattice calculation of power divergent 
quantities have been addressed recently: (i) $\bar{\Lambda}$ and 
quark masses in HQET (see Section~\ref{hqet}) 
and (ii) the gluon condensate.

\subsection{$\bar{\Lambda}$ and the quark mass from HQET}
\label{mhq}

That power divergences affect the non-perturbative parameters in 
subleading order of the $1/m$ expansion in HQET and require 
non-perturbative subtraction has been noted by \cite{MMS92} and 
related to renormalons in \cite{BB94a}. The problem is general, 
but in practice it has been discussed mainly for $\bar{\Lambda}$ 
and the kinetic energy parameter $\lambda_1$, see (\ref{energy}). 
For these two parameters \cite{MS95} proposed a prescription to 
subtract the power divergence non-perturbatively. Contrary to 
dimensional regularization, a linearly divergent residual mass 
term $\delta m$ is generated by quantum corrections in the lattice 
regularization of HQET. The residual mass counterterm can be defined 
non-perturbatively as\footnote{The relation of 
$\delta \bar{m}$ to $\delta m$ can be found in \cite{MS95}, but 
the distinction is not relevant to the present discussion.}
\begin{equation}
\delta\bar{m} = -\lim_{t\to\infty} \,\frac{1}{a}\,\ln\left(\frac{
\mbox{tr}\,S_h(\vec{x},t+a)}{\mbox{tr}\,S_h(\vec{x},t)}\right),
\end{equation}
where $S_h(\vec{x},t)$ is the static quark propagator in the Landau gauge 
at the point $(\vec{x},t)$ and the trace is over colour. In 
perturbation theory $\delta\bar{m}\sim \alpha_s/a$. The binding 
energy ${\cal E}$ of the ground state meson in a given channel 
is computed from the large-time behaviour of the two-point correlation 
function
\begin{equation}
\sum_{\vec{x}} \langle 0|J(\vec{x},t)\,J^\dagger(\vec{0},0)|0\rangle
\stackrel{t\to\infty}{=} Z^2\exp(-{\cal E} t),
\end{equation}
where $J$ is the heavy-light current in HQET with the appropriate 
quantum numbers. 
The binding energy is linearly divergent, but the linear divergence 
is the same to all orders in perturbation theory as that of 
$\delta\bar{m}$. Hence \cite{MS95} define 
\begin{equation}
\bar{\Lambda} \equiv {\cal E}-\delta\bar{m},
\end{equation}
which is finite as $a\to 0$ and of order $\Lambda$. The lattice 
calculation of \cite{CGMS95,GMS97} gives $\bar{\Lambda}=(180^{+30}_{-20})
\,$MeV. One can then define a `subtracted pole mass' 
$m_S=M_B-\bar{\Lambda}+\,O(\Lambda^2/m)$, which replaces the naive 
perturbative expression $m_{\rm pole}=M_B-\bar{\Lambda}_{\rm naive}$.
 
The subtracted pole mass is still a long-distance quantity, and 
useful only if it can be related to another mass definition such as the 
$\overline{\rm MS}$ mass 
$M_b=m_{\overline{\rm MS}}(m_{\overline{\rm MS}})$. But then 
$M_b$ can be computed directly from a lattice measurement of 
${\cal E}$. To see this, let 
$M_b=m_{\rm pole} (1+\sum_{n=0} c_n \alpha_s(M)^{n+1})$, then 
to a given order $N$ in perturbation theory, the relation is 
\begin{eqnarray}
\label{msrel}
M_b &=& \left(1+\sum_{n=0}^N c_n \alpha_s(M_b)^{n+1}\right)\left[
m_S-\delta\bar{m}(a) + \frac{1}{a}\sum_{n=0}^N r_n\alpha_s(a)^{n+1}
\right]
\nonumber\\
&=& \left(1+\sum_{n=0}^N c_n \alpha_s(M_b)^{n+1}\right)\left[
M_B-{\cal E}(a) + \frac{1}{a}\sum_{n=0}^N r_n\alpha_s(a)^{n+1}
\right],
\end{eqnarray}
where $\delta\bar{m}$ and ${\cal E}$ are evaluated non-perturbatively 
for a given $a$, and $\sum_{n=0} r_n\alpha_s(a)^{n+1}$ is the perturbative 
evaluation of the linear divergence of $\delta\bar{m}$ or ${\cal E}$. 
(They coincide.) The renormalon divergence cancels asymptotically 
between the two series in (\ref{msrel}) and the linear divergence 
also cancels up to order $\alpha_s^{N+1}$. However, because the series 
is truncated, one cannot take $a$ too small. Note that the subtraction 
is done perturbatively and it is not necessary to define 
$\bar{\Lambda}$ or $m_S$ to obtain $M$ as illustrated by the second 
line. But because the (leading) renormalons cancel, a non-perturbative 
subtraction is not necessary. In terms of Borel transforms the 
cancellation near the leading singularity at $u=1/2$ 
looks, schematically,
\begin{equation}
\frac{1}{1-2 u}\left[\left(\frac{m^2}{\mu^2}\right)^{-u}-
\frac{1}{m a}\left(\frac{1}{\mu^2 a^2}\right)^{-u}\right],
\end{equation}
if the Borel transform is taken with respect to $\alpha_s(\mu)$. 
In practice, the cancellation may be numerically delicate if 
$m$ and $a^{-1}$ are very different. Using the procedure explained 
here, \cite{GMS97} quote $M_b=(4.15\pm 0.05\pm 0.20)\,$GeV, where the 
second error has been assigned as a consequence of the unknown 
second order coefficient $r_1$. In physical units the inverse 
lattice spacings in these simulations are between $2\,$GeV and 
$4\,$GeV. There are corrections of order $\Lambda/M^2$ from 
higher dimension operators in HQET, see (\ref{energy}). These 
are smaller than the error due to the unknown perturbative subtraction 
terms. The important conclusion is that the $\overline{\rm MS}$ mass 
can be reliably determined from the $B$ meson mass and a lattice 
measurement of ${\cal E}(a)$, provided the $r_n$ are known to 
sufficiently high order in lattice perturbation theory.

An extended subtraction procedure for the kinetic energy 
\cite{MS95} has also been studied numerically \cite{CGMS95}, 
but the accuracy of the subtraction is not yet sufficient to reach 
physically interesting values. 

\subsection{The gluon condensate}

Power divergences are even more severe in the 
calculation of the gluon condensate, 
because the operator $\alpha_s G^{\mu\nu}G_{\mu\nu}$ is quartically 
divergent. On the lattice the gluon condensate is computed from the 
expectation value of the plaquette operator $U_P$. Classically, we have  
\begin{equation}
\frac{1}{a^4}\,\langle 1-\frac{1}{3}\,\mbox{tr}\,U_P\rangle 
\stackrel{a\to 0}{=} \frac{\pi^2}{36} \,\langle\frac{\alpha_s}{\pi} 
GG\rangle_{\rm latt}.
\end{equation}
Quantum fluctuations introduce corrections to the unit operator,  
and the above relation is modified to
\begin{equation} 
\label{plaquette}
\langle P\rangle\equiv \langle 1-\frac{1}{3}\,\mbox{tr}\,U_P\rangle
= \sum_{n=1}\frac{c_n^{\rm lat}}{\beta^n} + 
\frac{\pi^2}{36} \,C_{GG}(\beta)\,a^4\,\langle\frac{\alpha_s}{\pi} 
GG\rangle_{\rm latt} + \,O(a^6),
\end{equation}
where $\beta=6/(4\pi\alpha_s^0(1/a))$ denotes the lattice 
coupling at lattice spacing $a$ and $\alpha_s^0(1/a)$ the bare 
lattice coupling. Note that there is no term of 
order $a^2$, because there is no gauge-invariant operator of dimension 
2. For $a\Lambda\ll 1$, the first 
series is far larger than the gluon condensate, which one would like 
to determine and therefore has to be subtracted to high accuracy. 
Not only has it to be subtracted, it has to be defined in the 
first place. The series has an IR renormalon, and the coefficients 
$c_n^{\rm lat}$ are expected to diverge as 
\begin{equation}
\label{expect}
c_n^{\rm lat} \propto \left(-\frac{3\beta_0}{4\pi}\right)^n\,
\Gamma(n-2\beta_1/\beta_0),
\end{equation}
as follows from adapting (\ref{iras}) with $d=4$ to the present convention 
for the expansion parameter. The ambiguity or magnitude of the 
minimal term of the series is of order $(a\Lambda)^4$ as the 
gluon condensate term in (\ref{plaquette}) itself. Again we emphasize 
that in principle one need not subtract the power divergence and one 
can consider $a^{-1}$ as a hard factorization scale. 

Using the Langevin method \cite{PW81}, \cite{dRMO95} calculated 
the first eight coefficients $c_n^{\rm lat}$ in pure $SU(3)$ gauge 
theory to good accuracy:
\begin{equation}
\label{thecs}
c_n^{\rm lat} = \{1.998(2),\,1.218(4),\,2.940(16),\,9.284(64),\,
34.0(3),\,135(1),\,567(21),\,2505(103)\}.
\end{equation}
According to (\ref{expect}) the ratio of subsequent coefficients 
is expected to be $0.21n$ for large $n$. The coefficients 
(\ref{thecs}) of the series expressed 
in the lattice coupling grow much more rapidly than this. 

The behaviour of (\ref{expect}) is expected for series expressed 
in terms of an expansion parameter whose $\beta$-function is 
convergent, see Section~\ref{schemedep} on scheme-dependence 
of large-order estimates. We expect this to be true in the 
$\overline{\rm MS}$ scheme. We do not know the large-order behaviour of 
the $\beta$-function in the lattice scheme and we will assume 
that the  relation between the lattice and the $\overline{\rm MS}$ 
coupling does not diverge factorially. In this case (\ref{expect}) 
should hold in both schemes {\em asymptotically}. However, the 
lattice coupling is related to the $\overline{\rm MS}$ coupling 
by large finite renormalizations unrelated 
to renormalons. This causes series expansions in the lattice coupling 
to be badly behaved generally and to be irregular, basically 
because the scale parameter is unnaturally small in the lattice scheme:
$\Lambda_{\rm latt}=\Lambda_{\overline{\rm MS}}/28.8$. As a 
consequence it may be expected that the asymptotic behaviour 
(\ref{expect}) is obscured in low/intermediate orders 
of perturbation theory in the lattice scheme. \cite{dRMO95} 
suggest to assume that (\ref{expect}) holds in a well-behaved 
continuum scheme $R$ and then use a three-loop relation 
\begin{equation}
\label{lams}
\beta_R = \beta - r_1 - r_2/\beta 
\end{equation}
to express (\ref{expect}), assumed to hold for $\beta_R$, in 
terms of $\beta$. They find that the set of coefficients 
(\ref{thecs}) is well described if the continuum scheme 
is chosen such that $r_1=3.1$ and $r_2=2.0$ (values quoted 
from \cite{BdRMO97}). In the $\overline{\rm MS}$ scheme,  
with $\beta_{\overline{\rm MS}}$ normalized at $\pi/a$, we 
would have $r_1=1.85$ and $r_2=1.67$ \cite{LW95}. The preferred 
values of the fit can be understood as a change of scale: in 
terms of $\beta_{\overline{\rm MS}}(0.706/a)$ one obtains 
$r_1=3.1$ and $r_2=2.1$ in (\ref{lams}). 

Since IR renormalon divergence arises from large-size fluctuations, 
the asymptotic behaviour (\ref{expect}) does actually not 
appear on any finite lattice. According to the estimate 
(\ref{saddles}) the asymptotic behaviour is affected by 
finite volume effects at a critical order $n_{\rm cr}=4\ln N+
c$, where $N$ is the number of lattice points in each direction 
and $c$ is a constant in the limit of large $N$. For the values 
$N=8,12$ that pertain to the calculation of \cite{dRMO95} the 
precise value of $c$ is important to establish whether the 
IR renormalon contribution to the coefficients $c_n^{\rm lat}$ 
is already affected by the finite volume. An analysis of the 
situation in the $O(N)$ $\sigma$-model \cite{dRMO97} 
suggests that $c$ is large enough to leave the 8-loop 
coefficients unaffected.

The conclusion of \cite{dRMO95} is 
therefore that the factorial growth (\ref{expect}), with an 
ambiguity of order $(a\Lambda)^4$ corresponding to the 
gluon condensate, is confirmed by the pattern of the 
lattice coefficients $c_n^{\rm lat}$. 

Can the gluon condensate be obtained by subtracting the series 
to 8-loop order? \cite{Ji95b} suggested various procedures 
to extrapolate the 8-loop truncated series to a sum. Subtracting 
this sum from Monte Carlo data for the plaquette expectation value, 
he obtained the value 
$\langle (\alpha_s/\pi) GG\rangle \approx 0.2\,$GeV$^4$, 
which is at least a factor 10 larger than the `phenomenological 
value' quoted in (\ref{phengl}). \cite{BdRMO97} went further 
and examined the remainder as a function of $\beta$ (and hence 
$a$). The result is shown in Fig.~\ref{fig23}. The left plot 
shows Monte Carlo data of the plaquette expectation values from 
which the one-loop, two-loop etc., perturbative terms in (\ref{plaquette}) 
are consecutively 
subtracted. According to (\ref{plaquette}) one expects the 
remainder to scale as $(a \Lambda)^4$, if all terms in the 
perturbative series up to the minimal term are subtracted. 
In this case the series of curves in the left plot should 
approach the line marked $\Lambda^4/Q^4$ ($a\equiv 1/Q$) 
in the plot. Contrary to the expectation, the remainder 
approaches a clear $a^2=1/Q^2$ behaviour.\footnote{In fact, 
tentative evidence for an unexpected $a^2$ behaviour in the 
plaquette expectation value and a certain Creutz ratio derived from it 
was already reported by 
\cite{LM91} several years earlier. These authors had only second-order 
perturbation theory available.} The right plot checks 
that this is not due to the fact that not all terms up to the minimal 
have been subtracted. What is shown is a subtraction based on a 
Borel-type resummation of the higher order terms in the 
series, assuming that it follows 
the asymptotic behaviour (\ref{expect}). The resultant remainder 
has again a clear $a^2$ behaviour, despite the fact that 
such a term is not present in (\ref{plaquette}).

\begin{figure}[t]
   \vspace{0cm}
   \epsfysize=10cm
   \epsfxsize=12cm
   \centerline{\epsffile{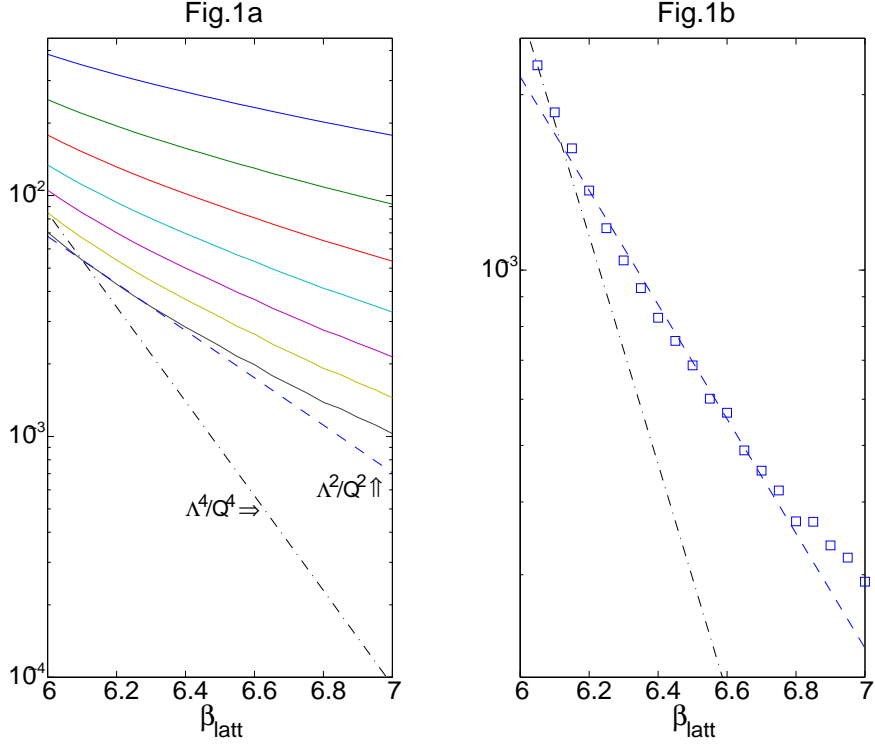}}
   \vspace*{0.2cm}
\caption[dummy]{\small \label{fig23} (a) The subtracted plaquette 
expectation value as a function of loop order compared to the 
scaling of a $1/Q^2$ and $1/Q^4$ term. (b) 
Comparison of the all-order subtracted plaquette MC data with 
the scaling of a $1/Q^2$ and $1/Q^4$ term. Figure taken from 
\cite{BdRMO97}.}
\end{figure}

The observation of $\Lambda^2/Q^2$ terms in the subtracted plaquette 
expectation value has led to speculations that there might be 
sources of power corrections of UV origin that give rise 
to $1/Q^2$ power corrections \cite{Gru97,AZ97}. Because they are 
of UV origin, they would not be in contradiction with the 
OPE according to which only a $a^4=1/Q^4$ 
term can appear in (\ref{plaquette}). These ideas can 
be motivated by considering the integral
\begin{equation}
\frac{1}{Q^4}\,\int\limits_0^Q d^4 k \,\alpha_{\rm eff}(k),
\end{equation}
where $\alpha_{\rm eff}(k)$ is supposed to be a physical definition 
of the coupling. The integral receives a contribution of order 
$\Lambda^4/Q^4$ from $k\sim \Lambda$ and of order $\alpha_{\rm eff}(Q)$ 
from $k\sim Q$. But if the effective coupling has a term of order 
$\Lambda^2/k^2$ in its own short-distance expansion, then this gives 
rise to a power correction of order $\Lambda^2/Q^2$ from {\em large} 
$k\sim Q$. The problem with the argument 
is that the definition of an effective coupling 
is to a large extent arbitrary and it is not clear how the argument could 
be applied to the lattice calculation above, where we assumed explicitly 
that the coupling definition does not contain power corrections. 
Furthermore, if one uses a coupling with larger power corrections 
than the observable under investigation, then one obtains  
additional power corrections not parametrized by matrix elements that 
appear in the short-distance expansion of that observable, but 
related only to the short-distance expansion of the coupling itself. These 
power corrections are, however, `standard'. One can always choose 
a coupling without power corrections by definition. Then the question 
is whether with such a definition of the coupling there exist 
power corrections that are 
not parametrized by matrix elements of operators in the OPE. An analysis 
of the $1/N$ expansion in the $\sigma$-model \cite{BBK98} finds a 
negative answer in that case. 

Before a definite conclusion can be drawn on the significance 
of lattice data above, one may consider the possibility that the 
observed $a^2$ scaling is a pure lattice artefact 
and does not indicate any unconventional power correction beyond the 
OPE. One point of concern is that (\ref{plaquette}),  
which is assumed in \cite{dRMO95,BdRMO97}, does not make the dependence 
of the plaquette expectation value on $a$ completely explicit. One can 
view lattice gauge theory at small values of $a\Lambda$ as an 
effective theory, i.e. an expansion around the continuum limit. 
The plaquette {\em operator} has the expansion
\begin{equation}
P\equiv 1-\frac{1}{3}\,\mbox{tr}\,U_P
= C_0(\ln a)\cdot 1 + C_{GG}(\ln a)\,a^4\,\frac{\alpha_s}{\pi} 
GG + \,O(a^6),
\end{equation}
in which there is no term of order $a^2$. This does not yet imply that 
the {\em matrix element} of the plaquette operator does not contain 
an $a^2$ term. The lattice Lagrangian in pure gauge theory 
can be expanded as 
\begin{equation}
\label{lag}
{\cal L}_{\rm latt}(a) = {\cal L}_{\rm cont} + a^2 \sum_i 
C_i(\ln a)\,{\cal O}^i_6 + \,O(a^4),
\end{equation}
with dimension-6 operators ${\cal O}_6^i$. Hence the vacuum expectation 
value of the plaquette has the small-$a$ expansion 
\begin{equation}
\label{newvev}
\langle P\rangle = C_0(\ln a)\,\langle 1\rangle + 
a^2\,\sum_i C_0(\ln a)\,C_i(\ln a)\int d^4 x\,\langle T(1,{\cal O}^i_6
(x))\rangle + \,O(a^4),
\end{equation}
where the vacuum expectation values are 
now taken in the $a$-independent 
vacuum of the continuum theory, contrary to the vacuum average in 
(\ref{plaquette}), which refers to the lattice vacuum. The $a^2$ 
correction in the form of 
a time-ordered product can be interpreted as a correction due to the 
fact that the vacua in the lattice and the continuum theory are 
different at order $a^2$. Such terms are not in contradiction 
with the {\em operator} product expansion of the plaquette operator. 
However, the connected part of the time-ordered product in (\ref{newvev}) 
is zero\footnote{I thank S.~Sharpe for this remark.}, and it remains 
unclear whether a higher-dimension operator in the effective lattice 
action is responsible for the remainder of order 
$a^2$, which \cite{BdRMO97} find after their subtraction procedure. 

In the continuum theory the dimension-6 operators in the Lagrangian 
are suppressed by the ultraviolet cut-off $\Lambda_{\rm UV}$ 
of QCD. Hence they are arbitrarily small in the operator product 
expansion in $\Lambda/Q$ of a physical process with 
$\Lambda\ll Q\ll \Lambda_{\rm UV}$. It is only  because in the lattice 
simulation one has identified $a^{-1}=\Lambda_{\rm UV}=Q$ that they 
become relevant. This conclusion is general and applies to the 
calculation of any power divergent quantity in lattice gauge 
theory.

Note that the dimension-6 operators on the right hand side of 
(\ref{lag}) can be eliminated by working with a (non-perturbatively) 
improved action. Thus a lattice simulation with an improved 
pure gauge theory action should find a reduced $a^2$ term, if it 
is due to higher-dimension operators in the effective lattice 
action.


\section{Conclusion}
\setcounter{equation}{0}

In this review we have described in detail the physics of 
renormalons from a predominantly phenomenological point of view. 
This has been a very active area of research over the past 
six years and the understanding of large-order behaviour and 
power corrections to particular processes in QCD has expanded 
enormously. In general, the renormalon phenomenon deals with 
the interface of perturbative and non-perturbative effects 
in observables that involve a large momentum scale compared 
to $\Lambda$. Such observables cannot be treated easily even in 
lattice gauge theory.

If we were forced to distill a single most important and 
general conclusion from the work reviewed here, it would be this: 
Since the conception of QCD the emphasis of perturbative QCD has 
been on constructing IR finite observables or to isolate 
the collinearly divergent contributions, for example in parton 
densities. This leads to perturbative expansions with finite 
coefficients. The study of IR renormalons and the 
power corrections associated with them calls on us to extend 
the notion of IR {\em finiteness} to the notion of IR 
{\em insensitivity}. For quantities that are perturbatively less 
sensitive to small loop momenta are not only expected to 
have smaller non-perturbative corrections, but also smaller 
higher order corrections in their perturbative expansions, and 
are therefore better predictable in a purely perturbative 
context. At the present times of precise experimental QCD 
studies, this is an issue of direct phenomenological relevance. 

The concept of IR insensitivity should be applied first of 
all to the fundamental parameters of the QCD Lagrangian, the 
coupling constant and the quark masses. In this respect we have 
concluded that the pole mass definition should be abandoned even 
for heavy quarks, because it is more sensitive to long distances 
than many processes involving heavy quarks. On the other hand, 
the $\overline{\rm MS}$ definition of the strong coupling, which 
has become the accepted standard for perturbative calculations, has 
very good properties from this point of view. 
The $\overline{\rm MS}$ scheme seems indeed to be a fortunate 
choice. In addition to fixed-sign 
IR renormalon divergence, which is related 
to physical and scheme-independent power corrections, there exist 
also UV renormalons related to irrelevant operators in the 
infinite UV cut-off limit. The corresponding divergent 
behaviour is universal, sign-alternating, and does not lead 
to physical power-suppressed effects. The minimal term of the 
series due to UV renormalons is scheme-dependent and it seems 
that in the $\overline{\rm MS}$ scheme the UV renormalon behaviour 
is generally suppressed and therefore of little relevance to 
accessible perturbative expansions in low or intermediate orders.  

Once infrared-insensitive input parameters are fixed, the infrared 
properties of any particular observable are manifest 
in its perturbative expansion. Perhaps one of the most interesting 
outcomes of IR renormalons is the prediction, based only on 
basic properties of QCD, that most observables that probe 
hadronic final states -- such as `event shape' observables 
in $e^+ e\to\,$hadrons -- have large $\Lambda/Q$ power corrections 
and large higher order perturbative corrections. The study 
of these power corrections has been pursued with vigour, theoretically 
and experimentally. Even though the theoretical interpretation of 
the results may turn out to be very difficult, the 
experimental studies are extremely important, not only to guide 
further theoretical developments. Since QCD has matured beyond the 
stage of qualitative `tests', the prediction of QCD (background?) 
processes with high precision has become crucial. Meeting this 
challenge requires the understanding of 
power corrections and higher order perturbative corrections. 

A review that leaves no open questions may be a cause of satisfaction 
for its author, but it would also reflect sad prospects for its 
subject. Because of this, we would like to conclude with 11 
problems, the solution of which we consider important (the numbers in 
square brackets refer to those sections relevant to the 
problem):

\begin{itemize}

\item[] \hspace*{-1.15cm} Formal and diagrammatic problems:

\item[I.] Is the expansion of the $\beta$-function in the $\overline{\rm 
MS}$ scheme convergent? [\ref{schemedep}]

\item[II.] Prove diagrammatically to all orders in $1/N_f$ that the 
large-order behaviour in QCD is determined by $\beta_0$ after a 
partial resummation of the flavour expansion. What is the explicit 
structure of singularities at next-to-leading order in the flavour 
expansion of QCD? [\ref{qcd}]

\item[III.] Can one classify the IR renormalon singularities of on-shell 
Green functions and min\-kows\-kian observables with the same generality as 
UV renormalon singularities? What are the universal elements in this 
classification? Determine the strength of IR renormalon singularities 
in on-shell Green functions. [\ref{irren}]

\item[IV.] Are there singularities in the Borel plane other than renormalon 
and instanton singularities? If not, why not? [\ref{borelplane}]

\item[] \hspace*{-1.15cm} Phenomenological questions:

\item[V.] Are there $1/Q$ corrections to Drell-Yan production 
beginning from two-loop order? [\ref{dy}]

\item[VI.] Which operators parametrize the $\Lambda/Q$ power 
correction to the longitudinal cross section in $e^+ e^-$ annihilation? 
[\ref{fragmod}]

\item[VII.] Can one construct `better' event shape variables, that is 
observables with reduced or no $\Lambda/Q$ power correction, which 
are sensitive to $\alpha_s$ at the same time? [\ref{eventshapes}]

\item[VIII.] Demonstrate that one can combine perturbative series 
at leading power and a lattice calculation of the first power 
correction with an accuracy better than the first power correction. 
[\ref{sumy}, \ref{lattice}]

\item[] \hspace*{-1.15cm}  Beyond renormalons:

\item[IX.] What is the large-order behaviour of the series of 
power corrections? There are compelling arguments \cite{Shi94} 
that this series also diverges factorially. But what is the precise 
behaviour in QCD?

\item[X.] Are there power corrections to time-like (minkowskian) 
processes related to the fact that parton-hadron duality is 
only approximate? Can one quantify `violations of parton-hadron 
duality'?

\item[XI.] If large-size ($\rho\sim 1/\Lambda$) instantons play an 
important role in the QCD vacuum, how do they affect properties of 
short-distance expansions \cite{CDSU97}?

\end{itemize}

\noindent 
We hope that the answers to these questions will some day necessitate 
another review.

\vspace*{0.4cm}

\noindent
{\em Acknowledgements.} I would like to thank first of all P.~Ball, 
N.~Kivel, L.~Magnea, V.A.~Smirnov, and in particular V.M.~Braun and 
V.I.~Zakharov for their collaboration on topics discussed in this 
report. In addition I have benefited from discussions with 
L.~Lellouch, G.~Marchesini and M.~Neubert, and many other colleagues. 
I thank V.M.~Braun,  G.~Buchalla, G.~Marchesini, M.~Neubert and 
C.T.~Sachrajda for their comments on the manuscript. 


\end{document}